\documentclass[a4paper, 11pt, oneside]{Thesis}  % Use the "Thesis" style, based on the ECS Thesis style by Steve Gunn
\graphicspath{Figures/}  % Location of the graphics files (set up for graphics to be in PDF format)

\usepackage{amsfonts}
\usepackage{amsmath}
\usepackage{mathtools}
\usepackage[separate-uncertainty=true]{siunitx}

\usepackage{bbold}
\usepackage{bm}

\usepackage[
    backend=biber, 
    natbib=true,
    style=nature,
    maxbibnames=12,
    sorting=none,doi=false,isbn=false,url=false,date=year
]{biblatex}
\AtEveryBibitem{\clearfield{month}\clearlist{language}\clearfield{note}}
\AtEveryCitekey{\clearfield{month}}
\AtNextBibliography{\small}

\newcommand{\THECOLOR}{RoyalBlue}

\newcommand{\COLORD}{\THECOLOR!100}
\DeclareBibliographyCategory{mypapers}
\colorlet{mypaperentry}{\THECOLOR}
\AtEveryBibitem{%
  \ifcategory{mypapers}%
    {\bfseries\color{\COLORD}}%
    {}%
  }

\addtocategory{mypapers}{%
  ronceray_variety_2011,
  ronceray_geometry_2012,
  ronceray_influence_2013,
  ronceray_multiple_2014,
  ronceray_connecting_2015,
  ronceray_favoured_2015,
  ronceray_liquid_2016,
  ronceray_free_2016,
  ronceray_fiber_2016,
  ronceray_suppression_2017,
  freeman_rosenzweig_eukaryotic_2017,
  han_cell_2018,
  carter_structural_2018,
  ronceray_range_2019,
  ronceray_stress-dependent_2019,
  ronceray_fiber_2019,
  shin_liquid_2018,
  gnesotto_learning_2020,
  xu_rigidity_2020,
  ronceray_stochastic_2020,
  bruckner_inferring_2020,
  frishman_learning_2020,
  bruckner_learning_2021,
  yang_physical_2021,
  schwarzendahl_self-organization_2021,
  shimobayashi_nucleation_2021,
  ronceray_stoichiometry_2022,
  broedersz_twenty-five_2022,
  ronceray_liquid_2022,
  berthier_nonlinear_2022,
  yang_local_2022,
  amiri_inferring_2023,
  ronceray_two_2023,
  barrat_soft_2023,
  koehler_how_2023,
}

\newcommand{\dd}{\mathrm{d}}
\newcommand{\av}[1]{\left\langle #1 \right \rangle}
\newcommand{\Fig}[1]{Figure~\ref{#1}}
\newcommand{\Figs}[1]{Figures~\ref{#1}}
\newcommand{\Eq}[1]{Equation~\ref{#1}}
\newcommand{\Eqs}[1]{Equations~\ref{#1}}

\newcommand{\Sec}[1]{Section~\ref{#1}}
\newcommand{\Chaps}[1]{Chapters~\ref{#1}}
\newcommand{\Chap}[1]{Chapter~\ref{#1}}

\DeclareMathOperator{\Tr}{\rm Tr}
\DeclareMathAlphabet\mathbfcal{OMS}{cmsy}{b}{n}

\newcommand{\Sdot}{ \Pi }

\usepackage{bm}

\usepackage{verbatim}  % Needed for the "comment" environment to make LaTeX comments
\usepackage{vector}  % Allows "\bvec{}" and "\buvec{}" for "blackboard" style bold vectors in maths
\begin{document}
\frontmatter      % Begin Roman style (i, ii, iii, iv...) page numbering

% Set up the Title Page
\title  {Learning dynamical models from stochastic trajectories}
\authors  {\texorpdfstring
            {{Pierre Ronceray}}
            {Pierre Ronceray}
            }
\addresses  {\groupname\\\deptname\\\univname}  % Do not change this here, instead these must be set in the "Thesis.cls" file, please look through it instead
\date       {
 \emph{Defense version - 03/06/2024}}
\subject    {}
\keywords   {}

\maketitle
%% ----------------------------------------------------------------

\setstretch{1.3}  % It is better to have smaller font and larger line spacing than the other way round

% Define the page headers using the FancyHdr package and set up for one-sided printing
\fancyhead{}  % Clears all page headers and footers
\rhead{\thepage}  % Sets the right side header to show the page number
\lhead{}  % Clears the left side page header

\pagestyle{fancy}  % Finally, use the "fancy" page style to implement the FancyHdr headers

\abstract{
  %\addtocontents{toc}{\vspace{1em}}  % Add a gap in the Contents, for aesthetics
  \begin{center}
    \setlength{\parskip}{0pt}
    {\large{\textbf{Jury members}} \par}
    %\bigskip
    %{\normalsize\bf \@title \par}
    \bigskip
  \end{center}

  \begin{center}
    \begin{tabular}{ r l l }
      Leticia Cugliandolo & Professor, LPTHE, Sorbonne Universit\'e & Reviewer \\ 
      J\"orn Dunkel & Professor, Massachusetts Institute of Technology & Reviewer \\  
      Roberto Cerbino & Professor, University of Vienna & Reviewer \\  
      Emmanuel Trizac & Professor, \'Ecole Normale Supérieure de Lyon  & President \\    
      Laurent Kodjabachian & Research Director, IBDM, Aix-Marseille Universit\'e   & Invited member \\    
      Kheya Sengupta & Research Director, CINaM, Aix-Marseille Universit\'e   & Tutor \\    
    \end{tabular}
  \end{center}
  
}

% The Abstract Page
\addtotoc{Abstract}  % Add the "Abstract" page entry to the Contents
\abstract{
  %\addtocontents{toc}{\vspace{1em}}  % Add a gap in the Contents, for aesthetics
  \begin{center}
    \setlength{\parskip}{0pt}
    {\large{\textbf{Abstract}} \par}
    %\bigskip
    %{\normalsize\bf \@title \par}
    \bigskip
  \end{center}

  The dynamics of biological systems, from proteins to cells to
  organisms, is complex and stochastic. To decipher their physical
  laws, we need to bridge between experimental observations and
  theoretical modeling.  Thanks to progress in microscopy and
  tracking, there is today an abundance of experimental trajectories
  reflecting these dynamical laws. Inferring physical models from
  noisy and imperfect experimental data, however, is challenging.
  Because there are no inference methods that are robust and
  efficient, model reconstruction from experimental trajectories is a
  bottleneck to data-driven biophysics. In this Thesis, I present a
  set of tools developed to bridge this gap and permit robust and
  universal inference of stochastic dynamical models from experimental
  trajectories. These methods are rooted in an information-theoretical
  framework that quantifies how much can be inferred from trajectories
  that are short, partial and noisy. They permit the efficient
  inference of dynamical models for overdamped and underdamped
  Langevin systems, as well as the inference of entropy production
  rates. I finally present early applications of these techniques, as
  well as future research directions.

}
  \vspace{2cm}
  \begin{center}
    \includegraphics[width=\textwidth]{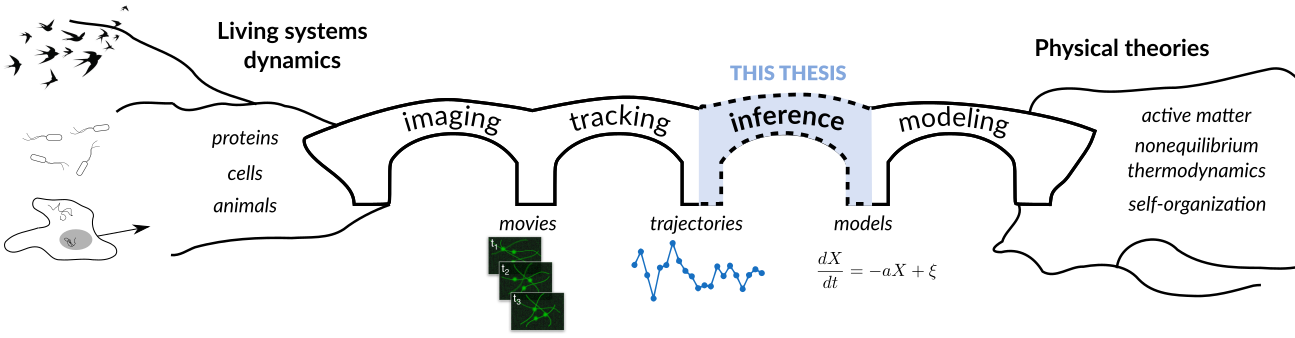}
  \end{center}

\clearpage  % Abstract ended, start a new page
%% ----------------------------------------------------------------

\dedicatory{Pour Arielle et H\'el\`ene.}

\setstretch{1.3}  % Reset the line-spacing to 1.3 for body text (if it has changed)

% The Acknowledgements page, for thanking everyone
\acknowledgements{
\addtocontents{toc}{\vspace{1em}}  % Add a gap in the Contents, for aesthetics

\vspace{1cm}

I would like to warmly thank the reviewers of this Thesis, Leticia
Cugliandolo, Roberto Cerbino and J\"orn Dunkel, as well as the other members of the jury, for their time and
comments.

The work presented here would not have been possible without its
funding sources, which have allowed me to carry independent research since my
PhD defense in 2016. In Princeton, as a postdoctoral fellow, I was
supported by a Princeton Center for Theoretical Science fellowship
(2016-2019), then by a Center for the Physics of Biological Function
fellowship (2019-2021). In Marseille, as a principal investigator, I
have been supported by the Turing Center for Living
Systems\footnote{\scriptsize{The project leading to this publication was supported
  by an institutional Institut Curie grant and received funding from
  France 2030, the French Government program managed by the French
  National Research Agency (ANR-16-CONV-0001) and from Excellence
  Initiative of Aix-Marseille University - A*MIDEX.}} (Centuri) and the
Centre National de la Recherche Scientifique. My group is currently
funded by Centuri and the European Research Council\footnote{Co-funded by the European Union (ERC SuperStoc 101117322). Views and opinions expressed are however those of the author(s) only and do not necessarily reflect those of the European Union or the European Research Council. Neither the European Union nor the granting authority can be held responsible for them.}.

I do not like working alone, and none of the projects presented here
would have been successful without my wonderful collaborators, friends
and inspirators. I want to thank in particular Chase Broedersz,
without whom I wouldn't even have considered working on stochastic
processes and who's been there all along since; Anna Frishman, for
going down the rabbit hole with me, blackboard after blackboard until
SFI was born; and David Br\"uckner, for the endless enthusiasm, crazy
ideas and collaborative prospects. This work has also benefitted from
discussions and informal exchanges with many friends and colleagues --
to name but a few: Andeas Mayer, Ricard Alert, Ben Machta,
Fr\'ed\'eric Van Wijland, David Lacoste, Ludwik Leibler, Kamesh
Krishnamurthy, Ricardo Martinez Garcia, Vincent D\'emery, among many others.
I also want to thank the many experimentalists that have exchanged
with me, helped me hone and tune the methods presented here so that
they have real-world applications, and collaborated with me (or
attempted to do so), in particular Yacine Amarouch\`ene, Christoph
Schmidt, Michael Murrell, Olivier Dauchot, C\'ecile Sykes, Sirine
Amiri, Pascal Martin, Laurent Cognet, Guga Gogia, Daniel Lee, Antoine
Aubret, J\'er\'emie Palacci, among many others again.

I thank Martin Lenz and Ned Wingreen for their mentorship and
continued support which was essential for my career and development. I
thank Centuri, and Thomas Lecuit in particular, for providing me with
the means to start my own research group. I thank the CINaM and PIV in
particular, as well as the Centuri community and admin team, for
providing a welcoming, friendly and stimulating work environment. I
thank the ICTP-SAIFR for hosting my numerous visits and providing me
with means to do research in S\~ao Paulo too.  And I thank all my
group members, past and current, for their trust, enthusiasm, ideas,
and all in all making this research an exciting and wholesome
collective adventure.

Finally, I would like to thank my friends, my family, and Elisa -- for
all the rest.
}
%\clearpage  % End of the Acknowledgements
%% ----------------------------------------------------------------

\pagestyle{fancy}  %The page style headers have been "empty" all this time, now use the "fancy" headers as defined before to bring them back

%% ----------------------------------------------------------------
\lhead{\emph{Contents}}  % Set the left side page header to "Contents"
\tableofcontents  % Write out the Table of Contents

\setstretch{1.3}  % Return the line spacing back to 1.3

%\pagestyle{empty}  % Page style needs to be empty for this page
%\dedicatory{For/Dedicated to/To my\ldots}

%\addtocontents{toc}{\vspace{2em}}  % Add a gap in the Contents, for aesthetics

% The Abstract Page
\addtotoc{Foreword}  % Add the "Abstract" page entry to the Contents
\lhead{\emph{Foreword}}  % Set the left side page header to "Contents"
\abstract{
%\addtocontents{toc}{\vspace{1em}}  % Add a gap in the Contents, for aesthetics
  \begin{center}
    \setlength{\parskip}{0pt}
    {\large{\textbf{Foreword}} \par}
    %\bigskip
    %{\normalsize\bf \@title \par}
    \bigskip
  \end{center}

\paragraph{About this Thesis.}
This Thesis presents a series of works that I initiated in 2018 during
my postdoctoral stay in Princeton, USA, in collaboration with Anna
Frishman, another postdoctoral fellow who is now Assistant Professor
at Technion, Israel. This collaboration led to a first major
publication in 2020 in \emph{Physical Review
  X}~\cite{frishman_learning_2020}, upon which
\Chaps{chap:capacity} and~\ref{chap:SFI} are based. While our
initial focus was on stochastic thermodynamics approaches and, in
particular, entropy production estimation, as the story unfolded I
became more and more interested in the data-driven aspects of it and,
in particular, model inference for stochastic processes.

This first publication was quickly followed by another on model
inference for underdamped systems~\cite{bruckner_inferring_2020}, in
collaboration with Prof. Chase Broedersz (now VU Amsterdam) and his
student David Br\"uckner (now postdoctoral fellow at IST Austria),
which is presented here in \Chap{chap:ULI}. This method was then
applied to the study of cell-cell
interactions~\cite{bruckner_learning_2021}, as discussed along with
other applications of the methods presented here in
\Chap{chap:applications}. Finally, \Chap{chap:BDB} describes results
towards our initial goal -- inferring entropy production -- which
includes results from Ref.~\cite{frishman_learning_2020} in
collaboration with Anna Frishman, as well as another article written
in collaboration with Chase Broedersz and his two students Federico
Gnesotto and Grzegorz Gradziuk dealing with track-free entropy
production inference~\cite{gnesotto_learning_2020}.

As this research program on stochastic model inference now forms the
core of the research of my group, it is still rapidly evolving. Rather
than a complete and final story, this Thesis is therefore a snapshot
of our state-of-the-art -- including the results in the
already-published corpus of
articles~\cite{gnesotto_learning_2020,frishman_learning_2020,bruckner_learning_2021,bruckner_inferring_2020,amiri_inferring_2023},
as well as some unpublished results and ideas for future research
directions. This work is complemented by Python implementations of the
algorithms developed and presented here.

\paragraph{What this Thesis is not about.}
I chose to present here a consistent series of works which resonates
the most with my current and projected research, rather than an
exhaustive description of my activities since I defended my PhD in
2016. In particular, in parallel with my research on stochastic
inference, I have worked in the past five years on the physical
properties of biomolecular condensates, a class of recently discovered
membrane-less cellular bodies. Through theoretical work and
collaborations with experimentalists, I have explored the role of
specific protein-protein interactions in their assembly and
dynamics~\cite{freeman_rosenzweig_eukaryotic_2017,xu_rigidity_2020,ronceray_stoichiometry_2022,
}, as well as their physical interactions with the surrounding
cellular
medium~\cite{shin_liquid_2018,shimobayashi_nucleation_2021,ronceray_liquid_2022}
and actively self-organizing
droplets~\cite{schwarzendahl_self-organization_2021}. In line with my
PhD work on the mechanics of biopolymer
networks~\cite{ronceray_connecting_2015,ronceray_fiber_2016}, I have
also continued working on their active contractile
properties~\cite{ronceray_stress-dependent_2019,ronceray_fiber_2019}
and their nonlinear mechanical
properties~\cite{han_cell_2018,berthier_nonlinear_2022,yang_local_2022}. My
Masters work on lattice models for local structures in supercooled
liquids~\cite{ronceray_variety_2011,ronceray_geometry_2012,ronceray_influence_2013,ronceray_multiple_2014,ronceray_favoured_2015,ronceray_liquid_2016,ronceray_free_2016}
was continued into a study of structural
competition~\cite{ronceray_suppression_2017,carter_structural_2018}
and geometrical frustration~\cite{ronceray_range_2019}.   Finally, I
have also developed an interest in
bioenergetics~\cite{yang_physical_2021}, active materials~\cite{barrat_soft_2023} and stochastic
thermodynamics~\cite{broedersz_twenty-five_2022,ronceray_two_2023}, and a long-standing project on frustrated self-assembly of particles is starting to bear fruits~\cite{koehler_how_2023}.

%% ----------------------------------------------------------------
\mainmatter	  % Begin normal, numeric (1,2,3...) page numbering
\pagestyle{fancy}  % Return the page headers back to the "fancy" style

\lhead{\leftmark}  % Set the left side page header to "Contents"

\chapter{Introduction}
\label{chap:intro}

\section{Stochastic models for the dynamics of complex systems}
\label{sec:stoc_intro}

\paragraph{Trajectories of biological systems contain precious
  information on their complex dynamics.}  From individual proteins to
motile cells to groups of animals, the way biological systems move,
interact and change shape in time is the subject of intense
biophysical research. The complexity of these systems makes it
challenging to understand their dynamics from first principles: such
bottom-up approaches must be complemented with data-driven approaches
to quantify their motion, identify underlying mechanisms, and discover
emergent laws. Thanks to the improvement of both imaging
instruments~\cite{manley_high-density_2008,lemon_live-cell_2020} and
tracking
algorithms~\cite{chenouard_objective_2014,amat_fast_2014,serge_dynamic_2008},
large amounts of data are now available at all scales. These data
consist in \emph{trajectories} of biological systems -- the time
traces of their position and configuration. To connect these
experimental datasets to physical theories of active biological
matter, one needs tools to infer models from these
trajectories. However, this task presents several challenges, due to
the stochasticity of the dynamics, the complexity of the models to
infer, and the limitations of available experimental data.

\paragraph{The dynamics of biological matter, at all scales, is
  stochastic.} Indeed, with living systems, reproducing an experiment
with the same initial conditions is unlikely to yield the exact same
trajectories. This stochasticity originates from multiple factors: at
the molecular scale, the dynamics of microscopic particles is
fundamentally random and Brownian, as they are subjected to thermal
noise. At the intermediate scale of organelles and cellular
structures, the relative importance of \emph{thermal agitation} is
lower, but further noise arises from the active processes of molecular
motors and enzymes, resulting in \emph{active fluctuations}. Finally,
at the larger scale of cells and organisms, up to groups of animals
such as flocks of birds, effective noise arises from the \emph{inner
  complexity} of these systems: for an observer that does not have
access to this full complexity, it thus appears stochastic. Models for
the dynamics of biological matter should thus account for
stochasticity.

\begin{figure}[bt]
  \centering
  \includegraphics[width=1.\columnwidth]{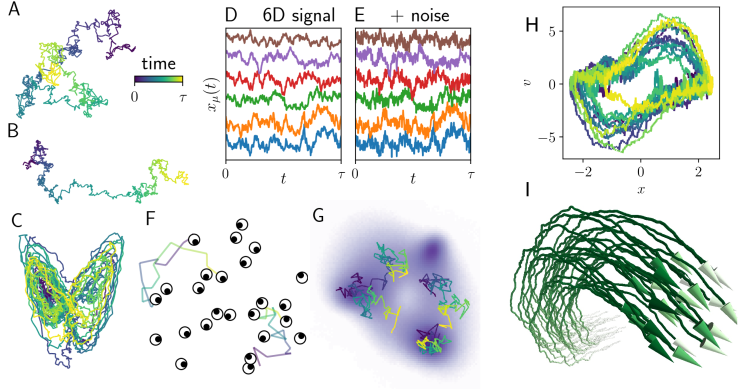}
  \caption{Simulated trajectories of example stochastic systems
    studied in this article. \textbf{A.} Pure Brownian motion in 2D,
    without forces. \textbf{B.} A drifted Brownian motion
    trajectory. \textbf{C.}  The stochastic Lorenz process, a complex
    three-dimensional system with a chaotic attractor. \textbf{D.}
    Time series of a 6D out-of-equilibrium Ornstein-Uhlenbeck
    process. \textbf{E.}  The same trajectories as in {D}, with
    additional time-uncorrelated measurement noise. \textbf{F.}
    Self-propelled active Brownian particles with soft repulsion and
    harmonic confinement.  \textbf{G.} Simulated single-molecule
    trajectories in a complex environment with space-dependent
    diffusion. \textbf{H.} A one-dimensional underdamped system in the
    position-velocity phase space: the van der Pol
    oscillator. \textbf{I.} A model of three-dimensional bird flock.}
  \label{fig:demo}
\end{figure}

\paragraph{Stochastic Differential Equations} \!\! (SDEs) provide a general
framework to model these dynamics in continuous space and time, by
splitting them into a deterministic term and a stochastic term. The
most important and widely used SDE describes \emph{Brownian
  dynamics}\footnote{Several other names are also used for this
  equation, such as It\^o process, diffusion process, and overdamped
  Langevin dynamics.}, and is the main focus of this proposal. It
consists in a first-order autonomous stochastic differential equations
of the form
\begin{equation}
  \label{eq:SDE1}
  \frac{\dd\mathbf{x}}{\dd t} = \mathbf{f}(\mathbf{x}_t) + \sqrt{2\mathbf{D}(\mathbf{x}_t)}\cdot\bm{\xi}(t)
\end{equation}
where $t$ is time, $\mathbf{x}_t$ is a real-valued $d$-dimensional
vector characterizing the system's state at time $t$ -- for instance
the position or orientations of one or many particles or organisms --
and $\bm{\xi}(t)$ is a $d$-dimensional Gaussian white noise,
\emph{i.e.} a random function obeying
$\av{\xi_\mu(t)\xi_\nu(t')} = \delta_{\mu\nu}\delta(t-t')$ with
$\av{\cdot}$ denoting ensemble averaging. The \emph{diffusion field}
$\mathbf{D}(\mathbf{x})$ is a positive definite tensor field
characterizing the noise strength.
Finally, the \emph{drift field}\footnote{The drift field is sometimes
  also termed ``force field'' by analogy with physical forces,
  although this can raise ambiguities when the noise is
  multiplicative, \emph{i.e.}  state-dependent. Here we use the It\^o
  convention of stochastic calculus for the definition of
  $\mathbf{f}$.}  $\mathbf{f}(\mathbf{x})$ is a vector field
characterizing the deterministic part of the dynamics. In practice, \Eq{eq:SDE1} can thus be
numerically implemented through the Euler-Maruyama algorithm, with
increments
\begin{equation}
  \label{eq:SDE1_discretized}
  \Delta\mathbf{x}_t\equiv \mathbf{x}_{t+\Delta t} - \mathbf{x}_t  = \mathbf{b}(\mathbf{x}_t)\Delta t + \sqrt{2\mathbf{D}(\mathbf{x}_t)} \sqrt{\Delta t} \bm{\zeta}_t
\end{equation}
where $\zeta_t$ are independent, identically distributed Gaussian
random variable of zero mean and unit variance. When the time
increments $\Delta t \to 0$, this algorithm approximates solutions of
\Eq{eq:SDE1}~\cite{kloeden_numerical_1992}. The scaling as
$\sqrt{\Delta t}$ of the stochastic term reflects the central
limit theorem: indeed, this term typically models the influence of
numerous fast and uncorrelated events. This implies that the
stochastic term dominates at short term, making the process highly
irregular and non-differentiable. In contrast, the drift field
$\mathbf{f}(\mathbf{x})$ has a negligible influence at very short
times, but shapes the long-time dynamics.

\paragraph{Brownian dynamics has a broad scope.}
The class of models defined by \Eq{eq:SDE1} is both simple and
powerful, and is used to statistically represent the dynamics of a
very broad variety of systems throughout the sciences. In its original
context of statistical physics, for instance for macromolecules or
colloidal particles in a solvent, it can be derived by taking the
overdamped limit $m/\gamma \to 0$ of the Langevin equation,
\begin{equation}
  \label{eq:Langevin}
  m\frac{\dd \mathbf{v}}{\dd t} = - \gamma\mathbf{v}(t)  + \mathbf{F}\left(\mathbf{x}(t),\mathbf{v}(t)\right) + \sqrt{2k_BT\gamma}\bm{\xi}(t)
\end{equation}
with $\mathbf{v}=\frac{\dd \mathbf{x}}{\dd t}$ the particle's
velocity, $m$ its mass, $\gamma$ the friction coefficient, $k_BT$ the thermal energy and $\mathbf{F}$ is
the external force acting on the particle. This equation corresponds to Newton's second
law, where the coupling with the solvent is modeled by the combination
of the damping force $- \gamma\mathbf{v}(t)$ and the stochastic forcing $ \sqrt{2k_BT\gamma}\bm{\xi}(t)$. In the high friction
limit, the inertial term becomes negligible, and \Eq{eq:Langevin}
simplifies into \Eq{eq:SDE1}.

The applications of Brownian dynamics, however, extend to a much
broader scope -- from climate~\cite{gottwald_ga_stochastic_2017},
population~\cite{lande_stochastic_2003} and evolutionary
dynamics~\cite{traulsen_stochastic_2012}, to
finance~\cite{el_karoui_backward_1997},
neuroscience~\cite{freeman_neurodynamics_2000} and theoretical
chemistry~\cite{van_gunsteren_computer_1990}, as well as living
systems at all scales: proteins~\cite{scheraga_protein-folding_2007},
cells~\cite{li_dicty_2011},
organisms~\cite{stephens_dimensionality_2008,bialek_statistical_2012}. In
all these cases, $\mathbf{x}_t$ is used to represent observable
variables, such as the position of one or many particles, the value of
a stock market index, or the fraction of an allele in a
population. The drift $\mathbf{f}$ models the slow, deterministic
trends in the dynamics of $\mathbf{x}$ that shapes its long-term
behavior, while the diffusion $\mathbf{D}$ models the coupling of the
observed system with fast, numerous, unobserved degrees of freedom:
for instance, shocks with solvent molecules, individual trading or
mating events. Here, however, we shall focus mostly on applications describing the
real-space dynamics of biological matter.

\paragraph{Model variants: other stochastic differential equations.}
While \Eq{eq:SDE1} is the simplest and most popularly used form of
SDE, many variants exist that extend the scope and applications of
this model. Both the drift and stochastic term can be explicitly
time-dependent, which then reads:
\begin{equation}
  \label{eq:SDE1_nonautonomous}
  \frac{\dd\mathbf{x}}{\dd t} = \mathbf{b}(t) + \sqrt{2\mathbf{D}(t)}\cdot\xi(t)
\end{equation}
where the time-dependency is now explicit. Such models are useful when
the system is non-stationary and its properties evolve with time --
for instance in externally driven systems, or in aging systems such as
a developing embryo.

Higher-order SDEs, while more rarely encountered, are sometimes
necessary to describe physical systems. In particular, second-order
SDEs are used to model the underdamped dynamics of systems of
particles where inertia is relevant (\Eq{eq:Langevin}), such as in a
gas or in a liquid at very short time scales. They are also sometimes
used as effective descriptions of overdamped systems for which an
effective inertia appears, for instance in migrating cells whose
polarization persists over
time~\cite{li_dicty_2011,bruckner_stochastic_2019,bruckner_learning_2023}. Such dynamics
generically reads:
\begin{equation}
  \label{eq:SDE2}
  \frac{\dd\mathbf{v}}{\dd t} = \mathbf{b}(\mathbf{x}_t,\mathbf{v}_t) + \mathbf{\sigma}(\mathbf{x}_t,\mathbf{v}_t)\cdot\xi(t) \qquad ; \qquad \frac{\dd\mathbf{x}}{\dd t} = \mathbf{v}_t
\end{equation}
where $\mathbf{v}_t$ is the velocity at time $t$. A further challenge
in the inference of such equations is that the velocity is generally
not directly observed, but must be estimated by differentiating the
data, which leads to additional noise and biases to which we recently
proposed a general solution~\cite{bruckner_inferring_2020}.

\Eqs{eq:SDE1}-\ref{eq:SDE2} are ordinary stochastic differential
equations where the system state is a finite-dimensional
vector. However, when modeling the time evolution of the shape of
deformable objects such as filaments or membranes, the dynamics of a
spatially heterogeneous reaction-diffusion system, or the slow change
in ecological patterns~\cite{martinez-garcia_spatial_2013}, the system
is better represented by a continuous field $\phi(\mathbf{r})$, with
$\mathbf{r}$ the spatial coordinate. To model the dynamics of such
systems, one can employ stochastic \emph{partial} differential equations
(SPDEs), of the form
\begin{equation}
  \label{eq:SPDE}
  \frac{\partial \phi}{\partial t}(\mathbf{r}) = \mathcal{B}[\phi_t](\mathbf{r}) + \Sigma[\phi_t] \mathbf{\xi}(t)
\end{equation}
where $\mathcal{B}$ and $\Sigma$ are now functionals of the field,
which can include spatially differential operators such as
$\Delta \phi$, as well as non-local functions of the field.

Finally, the noise can be colored (\emph{i.e.} involving memory
kernels), which results in non-Markovian
dynamics. %XXX, a subject that we discuss in \Sec{sec:colored}.
More complex memory kernels, such as those employed to model cells and
bacteria leaving chemical trails that modify their environment's
properties, can also be considered. The noise can also be
non-Gaussian, leading for instance to L\'evy flights and anomalous
diffusion, a subject of high current
interest~\cite{munoz-gil_objective_2021} that is not in the scope of
this Thesis.

\section{The inverse problem: Stochastic Inference}
\label{sec:stochinf_intro}

\paragraph{Complex systems require data-driven approaches.}
\Eq{eq:SDE1} provides a general framework for the dynamics of
biological matter. However, because of the complexity and
heterogeneity of these systems, it is generally not possible to fully
derive the drift and diffusion fields of a system from first-principle
modeling of its constituents. The appeal of data-driven techniques, in
contrast, is that they permit effective descriptions of complex
systems at a given scale, informed directly by experiments made at that scale without fully characterizing and
understanding what happens at the scales below. They are therefore
especially useful in biology, where full understanding of the
microscopic mechanisms remains far beyond reach. For instance, to
understand the collective dynamics of migrating cells in a tissue, it
can be more efficient to learn their individual dynamical laws and
interactions from data, rather than aiming at deriving them from their
molecular cytoskeletal dynamics and adhesion properties. To this aim,
one needs both dynamical data, and algorithms to infer models from
these data.

\paragraph{Dynamical data are increasingly abundant --- yet precious
  and imperfect.}
Tremendous progress in imaging and instrumentation, at all biological
scales, results in vast amounts of raw dynamical data -- \emph{i.e.}
movies of biological matter. Moreover, the advent of deep neural nets
has permitted fast advances in the treatment of these movies:
segmentation algorithms~\cite{ronneberger_u-net_2015}, as well as
tracking
algorithms~\cite{serge_dynamic_2008,chenouard_objective_2014}, are
progressing fast, resulting in an abundance of trajectories of systems
from single molecules to groups of animals.

Nevertheless, %with the exception of high-throughput experiments,
these trajectories do not generally constitute \emph{big data}: they
are the result of difficult and expensive experimental work.
Furthermore, in spite of technical advances in resolution and frame
rate, these trajectories are imperfect reflections of the physical
reality of the biological system: time discretization leads to missing
information between data points, while the measurement device and
tracking algorithm induce random experimental noise on the trajectory.
\textbf{Reliable model inference from trajectories of biological
  systems must therefore be done in a data-efficient manner that is
  robust to data imperfections.} There is currently a lack of
algorithms that can infer stochastic models from trajectories in such
a robust and data-efficient manner. Addressing this need is the main goal of this Thesis.

\paragraph{Definition and goals of stochastic inference.}
More precisely, my goal is to solve the \emph{inverse problem} of
Brownian dynamics: that is, inferring a model of the type described by
\Eq{eq:SDE1} from a time series of states of the
system. Schematically:  \vspace{-3mm}
\begin{equation}
  \label{eq:inference_schematic}
  \mathrm{input\ trajectory:} \quad \{x_0, x_{\Delta t}, x_{2\Delta t}, \dots, x_{N\Delta t} \} \quad \rightarrow\quad   \frac{\dd\mathbf{x}}{\dd t} = \hat{\mathbf{f}}(\mathbf{x}_t) + \sqrt{2\hat{\mathbf{D}}(\mathbf{x}_t)}\cdot\bm{\xi}(t)
  \quad\mathrm{inferred\ model  }
  \vspace{-2mm}\end{equation}
where $\hat{\mathbf{f}}$ and $\hat{\mathbf{D}}$ are the inferred drift
and diffusion fields. Importantly, SDEs are a mathematical
idealization that can only ever capture an \emph{approximation} of the
dynamics of the system. This is particularly true in complex and
biological systems, for which they are generally a heuristic that
effectively describes the system's dynamics over a given range of time
scales, rather than the reflect of a fundamental physical law. Quite
generally, the inference will be considered as successful if the
resulting model, once simulated, yields trajectories that are
statistically indistinguishable from the original data -- both on
short- and long-term features.

More specifically, the quality of an inference method can be assessed
by the following criteria:
\begin{itemize}
\item \textbf{consistency:} on simulated data where the ground truth is
  known, in the limit of long trajectories $N\Delta t \to\infty$ and high sampling rate
  $\Delta t \to 0$, the inferred fields $\hat{\mathbf{f}}$ and
  $\hat{\mathbf{D}}$ statistically converge to the fields used for the simulation.
\item \textbf{control:} the inference error (both fluctuating error
  and systematic bias) can be assessed self-consistently with the data only.
\item \textbf{efficiency:} this convergence requires as little data as possible,
  \emph{i.e.} the fluctuating error is as small as possible.
\item \textbf{robustness:} the error remains small even with imperfect
  data, including large $\Delta t$ and measurement noise.
\end{itemize}
The former four points ensures that the inferred model represents the
data faithfully and in a controlled manner. For applications involving
quantitative or predictive modeling, this can be
satisfactory. However, to obtain a physical understanding of the
system's dynamics, a fifth criterion is important:
\begin{itemize}
\item \textbf{interpretability:} the resulting model is physically
  intelligible and its parameters can be interpreted.
\end{itemize}
My long-term research goal is to \textbf{design, implement and
  distribute consistent, controlled, efficient, robust, and
  intepretable stochastic inference methods for the dynamics of soft
  biological matter.}

\section{State of the art of stochastic inference}

Reflecting the vast diversity of fields where Brownian dynamics is
used and of important questions that can be addressed using Brownian
inference, a flurry of methods has been proposed in the past decades
to perform this inference. No clear winner emerges among these
methods, with most approaches being either purely formal, or
specialized to a specific type of task or system. A complete review
shall not be attempted here; instead, I will focus on methods
developed and used for soft and biological matter. This includes my previous work on developing
Stochastic Force Inference, an important first step towards the
robustness goal. I will then discuss aspects of methods
from other fields that could be imported to strengthen and improve
stochastic inference for biological matter.

\paragraph{Kramers-Moyal approaches and maximum likelihood.}
The simplest and most intuitive inference method consists in
attempting to directly extract the drift and diffusion from increments
of the dynamics. To this aim, one can define the so-called
Kramers-Moyal (KM) coefficients,
$\mathbf{M}_1(t) = \Delta \mathbf{x}_t/\Delta t$ and
$\mathbf{M}_2(t) = \Delta \mathbf{x}_t^2/2\Delta t$. In the
$\Delta t\to 0$ limit, these allow to separate the deterministic and
stochastic contributions:
$\av{\mathbf{M}_1(t) | \mathbf{x}_t = \mathbf{x} } =
\mathbf{f}(\mathbf{x}) + O(\Delta t)$, and
$\av{\mathbf{M}_2(t) | \mathbf{x}_t = \mathbf{x} } =
\mathbf{D}(\mathbf{x}) + O(\Delta t)$, where
$\av{\ \cdot\ |\mathbf{x}_t = \mathbf{x}}$ denotes ensemble averaging
conditioned on the initial position $\mathbf{x}$ at time $t$. Here,
separation of $\mathbf{f}$ and $\mathbf{D}$ exploits the fact that the
stochastic term dominates at short time scales but has zero mean.

A basic way of performing stochastic inference consists in grid
binning the process: defining a regular $d$-dimensional grid and
averaging the KM coefficients over each bin, thus approximating the
process with constant drift and diffusion terms in each bin. This
method, which is data-inefficient but has the advantage of great
simplicity, is still widely used to analyze experimental
data~\cite{bruckner_stochastic_2019,dalessandro_cell_2021,boudet_collections_2021}. A
more refined approach consists in fitting these coefficients with
state- and/or time-dependent functions. This can be done as a linear
regression or using nonlinear functions.  In the case of constant,
known stochastic term (\emph{i.e.} when noise is additive) and with
least squares fitting, this approach coincides with a naive
Maximum-Likelihood (ML) formalism, valid and efficient in the limit
$\Delta t \to 0$. Note however that a ``true'' ML approach would infer
the diffusion at the same time, which is often impractical -- here, as
in most references, we focus on the separate inference of the
diffusion and drift fields.

In practice, these ``pseudo-ML'' methods boil down to inferring fields
as linear combinations of pre-selected functions ${b}_i(\mathbf{x})$,
where the parameters are the linear coefficients
$\hat{\mathbf{f}}_i$. The exact solution of the ML inference for the
drift is
\begin{equation}
  \label{eq:ML}
  \hat{\mathbf{f}}(\mathbf{x})=\sum_i \hat{\mathbf{f}}_i {b}_i(\mathbf{x})\quad \mathrm{with} \quad \hat{\mathbf{f}}_i = \sum_j ({C}^{-1})_{ij} \av{ \frac{\mathbf{x}_{t+\Delta t} - \textcolor{red}{\mathbf{x}_t}}{\Delta t} {b}_j(\textcolor{red}{\mathbf{x}_t}) }_t \quad \mathrm{and}\quad {C}_{ij} = \av{ {b}_i(\mathbf{x}_t){b}_j(\mathbf{x}_t) }_t
\end{equation}
This approach, which encompasses grid binning (for constant-by-parts
$c(\mathbf{x})$), forms the basis of numerous works on
inference~\cite{kloeden_numerical_1992,kutoyants_statistical_2004,comte_penalized_2007,boninsegna_sparse_2018,perez_garcia_high-performance_2018}. It
is consistent, efficient and can be
controlled~\cite{frishman_learning_2020}. However, it is not robust:
indeed, in the presence of measurement error~\footnote{One should be
  careful to distinguish \emph{dynamical noise} -- which acts on
  $d\mathbf{x}/dt$ and thus affects the dynamics -- from
  \emph{measurement error}, which is induced by the measurement device
  and data processing, and acts on the measurements $\mathbf{x}$. The
  former is intrinsic to the system and of physical interest, while
  the latter is extrinsic and should be filtered out. Separating these
  two types of stochasticity is a key challenge of stochastic
  inference methods. } with mean-square $\Lambda$, the correlations
between these errors in the red $\textcolor{red}{\mathbf{x}_t}$
results in a diverging bias of order $\Lambda / \Delta t$. With small
$\Delta t$, this precludes inference, while a large $\Delta t$ leads
to large variations of the drift over a single time step. For this
reason, this approach has limited use in biological physics. Other
approaches based on adaptive-grid Bayesian
estimation~\cite{beheiry_inferencemap:_2015,laurent_tramway_2022}
explicitly account for measurement noise, but still result in large
biases.

\paragraph{Recent developments and neural networks.} Since the start
of this project, the field of stochastic inference has been rapidly
developing. In particular, a number of recent articles propose to use
deep neural networks for the inference. These can be split in two
categories: one the one hand, some papers use neural networks to
efficiently learn coefficients of the drift and diffusion fields with
a given functional form~\cite{yu_extracting_2022} (in particular
linear combinations of functions, as in \Eq{eq:ML}): here the neural
net takes as input a series of trajectory descriptors, which includes
the estimators we derived, and is trained on large amounts of
simulated data to relate the descriptors to the true coefficients. In
this case, the deep learning consists essentially in learning how to
correct biases from other methods. On the other hand, numerous recent
articles~\cite{ruiz-garcia_discovering_2022,ren_statistical_2024,bae_inferring_2024}
propose to use a deep neural net as fitting function -- training it to
associate the observed state of the system $\mathbf{x}$ to the
time-derivative of the state $\Delta \mathbf{x}/\Delta t$. While
promising, this approach is still at its early stages, and has yet to
prove its robustness to data imperfection -- as, in particular, it
inherits the imperfections from \Eq{eq:ML}.

\paragraph{Other approaches.} Finally, other, more heuristic
approaches have also been proposed to the inference problem. These
include variational inference
approaches~\cite{archambeau_variational_2007,batz_approximate_2018,opper_variational_2019}
and kernel regularization
methods~\cite{lamouroux_kernel-based_2009,batz_variational_2016,yildiz_learning_2018}. However,
these approaches do not provide any proof or rate of convergence; as
such, they are neither principled not controlled, and, to my
knowledge, they have not been widely used in practice.

\section{Plan of this Thesis}
\label{sec:plan}

Here, I present my past and current work towards designing stochastic
inference methods. This thesis is organized as follows:

\begin{itemize}
\item In Chapter~\ref{chap:capacity}, I present an
  information-theoretical interpretation of Brownian dynamics showing
  that such processes have a finite information rate. This provides a
  baseline for the efficiency of the inference methods. I also discuss
  the connection between this information rate and stochastic
  thermodynamics quantities.
\item In Chapter~\ref{chap:SFI}, I present Stochastic Force Inference
  (SFI), a robust method to infer Brownian dynamics equations from
  data, along with proofs of convergence and estimation of the
  error. I also discuss the ways to use this method to learn entropy
  production rates by estimating out-of-equilibrium currents in the
  system.
\item In Chapter~\ref{chap:ULI}, I present Underdamped Langevin
  Inference (ULI), an extension of SFI to underdamped Langevin
  dynamics.
\item In Chapter~\ref{chap:BDB}, I present an application of SFI to
  estimate entropy production rates directly from raw movies, without
  tracking.
\item In Chapter~\ref{chap:applications}, I present two recent
  applications of these methods to experimental data on cell
  migration.
\item Finally, in Chapter~\ref{chap:perspectives}, I present some
  perspectives for this work, first coming back on the scientific
  process that underlies this Thesis, then discussing future
  development of these ideas.
\end{itemize}

\chapter{The information content of stochastic trajectories}
\label{chap:capacity}

\emph{In this Chapter, we study the information-theoretical constraints that
exist on the stochastic inference problem. Quite generally, a time
series with finite amounts of data contains finite information about
the underlying dynamics. While obvious, this assertion is crucial when
attempting to perform model inference: indeed, the information
available limits the complexity of the model that can be inferred
without overfitting. It is therefore important to assess the
information content of stochastic trajectories. Here we employ
communication-theoretical arguments to provide an upper bound to this
information content. Importantly, we show that contrarily to deterministic systems where only the measurement setup limits information acquisition, Brownian dynamics contains a fundamentally finite information rate about its drift field, even with perfect data. This rate is further reduced with imperfect data.}

\vspace{5mm}

Adapted from: \\
\textsc{Learning Force Fields from Stochastic Trajectories}\\
Anna Frishman and Pierre Ronceray   \\
Physical Review X 10, 021009 (2020).

\section{The capacity of Brownian dynamics}
\label{sec:capacity_intro}

\paragraph{Capacity of drifted Brownian motion.}

We propose to interpret Brownian dynamics (\Eq{eq:Langevin}) as a
noisy transmission channel. where the force is the encoded signal and
$\sqrt{2\mathbf{D}} \xi$ is the noise (\Fig{fig:schematic}). Information can be read out
from such a channel at a maximal rate $C$, called the channel
capacity, which relates to the signal-to-noise ratio of the
input~\cite{cover_elements_2006}. This fundamentally limits the
ability to infer forces by monitoring the dynamics. To build up
intuition, consider the simplest case of a spatially constant force
with isotropic diffusion, corresponding to drifted Brownian
motion with isotropic diffusion (\Fig{fig:driftedBM}):
\begin{figure*}[t]
  \centering
  \includegraphics[width=0.5\textwidth]{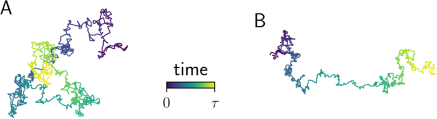}
  \caption{ \label{fig:driftedBM} A drifted Brownian motion with (A)
    zero drift and zero information, (B) a constant drift and 12.1
    bits of information.}
\end{figure*}

\begin{align}
  \label{eq:drifted_BM}
  \dot{\mathbf{x}}  = \mathbf{F} + \sqrt{2 D } \xi(t) 
\end{align} 
We can interpret this equation using the Shannon-Hartley theorem for
continuous-time Gaussian channels, with $\mathbf{F}$ the signal
transmitted and $\sqrt{2 D } \xi(t) $ the noise. In this framework,
the capacity is then given by $C=F^2/4D$ (expressed in natural
information units, or nats, per time unit --- 1 nat = $1/\log 2$
bits).  The force to infer is here equal to the persistent velocity,
which can be estimated as $\hat{F}_\mu = \Delta x_\mu / \tau$, where
$\Delta \mathbf{x}$ is the end-to-end vector along the trajectory of
duration $\tau$. The relative error on this estimator due to random
diffusion is
$\av{||\hat{\mathbf{F}} - \mathbf{F}||^2/F^2} = 2dD/\tau F^{2} =
d/2I$, where $d$ is the space dimension. We have identified here
$I = C\tau$, defining it as the information in the trajectory.
Persistent motion thus starts to emerge from the noise if the
trajectory duration $\tau$ is longer than $d/C$, corresponding to the
diffusive-to-persistent transition for the mean-squared
displacement. Equivalently, the force starts to be resolved if
$I > d$, \emph{i.e.} if more than one bit of information is available
for each degree of freedom $\hat{F}_\mu$ to infer.

\paragraph{Capacity of general Brownian dynamics.}
We now address more generally the question of quantifying the rate at which
information about the force can be read out, or is encoded in a trajectory. We
assume that the system follows the overdamped Langevin equation,
\begin{align}
  \label{eq:langevin}
  \dot{x}_{\mu}  = F_{\mu}(\mathbf{x}) + \sqrt{2 D }_{\mu\nu} \xi_{\nu} && \av{\xi_{\mu}(t)\xi_{\nu}(t')}= \delta(t-t').
\end{align}
Throughout we use Einstein's convention over repeated indices.

\begin{figure*}[t]
  \centering
  \includegraphics[width=\textwidth]{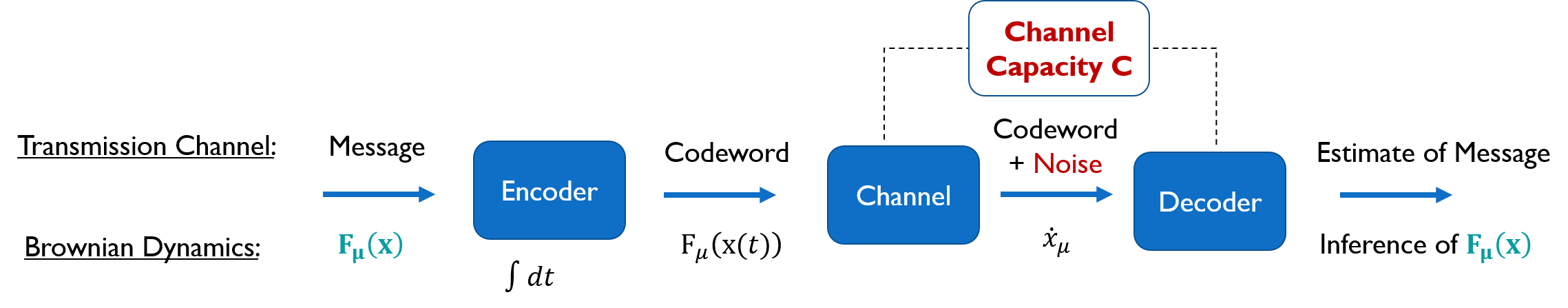}
  \caption{ \label{fig:schematic} The dynamics of an overdamped system
    can be seen as a noisy data transmission channel, encoding
    information about the force field, with a rate bounded by the
    channel capacity $C$ as defined in \Eq{eq:capacity}. Note that
    this definition does not include the information loss stemming
    from the measurement device.}
\end{figure*}

The complete force field is characterized by an infinite number of
degrees of freedom, and thus in principle contains an infinite amount
of information (the value of the force components at each location in
phase space). It is therefore pertinent to ask if there is a bound to
the rate at which this information can be read off from the trajectory. We consider an infinite length trajectory, from which, in principle, all information about the force field can be recovered.  We
argue that indeed there is such a maximal rate, given by the capacity
(in natural information units, or nats)
\begin{equation}
C = \frac{1}{4} D^{-1}_{\mu\nu}  \int  F_{\mu}(\mathbf{x}) F_{\nu}(\mathbf{x})  P(\mathbf{x})\dd \mathbf{x}
 \label{eq:capacity}
\end{equation}

To explain this formula, let us first focus on a one dimensional
system. A trajectory which satisfies the dynamics given by \Eq{eq:langevin}
encodes the information about the force field in the form of a
continuous time signal $F(x(t))$ corresponding to the values of the
force field at the points $x(t)$ that the trajectory visits. However,
what can actually be read out from the trajectory is $\dot{x}$, \emph{i.e.}
the signal $F(x(t))$ with noise $\xi$ added to it (\Fig{fig:schematic}). 
Thus, we can think
of the dynamics \Eq{eq:langevin} as a noisy communication channel, with Gaussian
white correlated noise, where the information about the force is
transmitted in the form of a codeword $F(x(t))$ which satisfies
$\lim_{\tau\to \infty}1/\tau\int_0^\tau F^2 dt =\int F^2(x)  P(x)\  \dd x$.
In communication theory, such a channel is called an infinite
bandwidth Gaussian channel~\cite{cover_elements_2006}. 
It has a well
defined capacity, \emph{i.e.} a maximal rate of information transmission: for
codewords of duration $\tau$ that satisfy the so-called ``power constraint''
$1/\tau\int_0^\tau dt F^2(t) \leq {\cal P}$, and a white noise with
amplitude $2D$ the capacity is given by ${\cal P}/(4D)$ nats per
second.  Information cannot be transmitted through the channel at a
faster rate. Stated differently, the capacity quantifies the (exponential) rate with which the maximal number of distinguishable signals grows with the amount of time the channel is used for, in particular as $\tau \to \infty$. In our case, the capacity is related to the distinguishability of different force fields with the same power constraint. The maximal rate is obtained for a signal which saturates the power constraint, so that the relevant constraint to consider is ${\cal P}=\lim_{\tau\to \infty}1/\tau\int_0^\tau F^2 dt$. 
Thus, our trajectory which has
$\lim_{\tau\to \infty}1/\tau\int_0^\tau F^2 dt =\int F^2 P(x)\ \dd x$ cannot
produce information about the force field at a rate faster than the
capacity as defined in \Eq{eq:capacity}. Note that in contrast to the
usual communication theory setting, we do not control the codeword
through which the force field is encoded, only the decoding
scheme---the code word is determined by the dynamics, the force field
being sampled according to the probability density function (pdf) $P(x)$.  To go from the capacity
for a one dimensional process to that of a $d$ dimensional process, \Eq{eq:capacity}, we
have decomposed the channel into $d$ parallel channels and added up
their capacities. Indeed, let us first go into the basis where the noise is diagonal and normalize its amplitude to two, such that all components of the new force $D^{-1/2}_{\mu\nu}F_\nu$ have the same units ($t^{-1/2}$). 
The components of the noise become independent, and the $d$
components in that basis become parallel channels, with signals measured in the same units, whose capacities sum
up to \Eq{eq:capacity}.

\paragraph*{The Shannon-Hartley formula and infinite bandwidth channels.}
The infinite-bandwidth capacity of Brownian dynamics, as presented in
\Eq{eq:capacity}, corresponds to that of the continuous dynamics. It
can also be seen as the $\Delta t \to 0$ limit of a discrete signal
(\emph{i.e.} a finite bandwidth signal) such as can be acquired in
practice. The capacity of such a discrete Gaussian channel is given by
the Shannon-Hartley formula~\cite{cover_elements_2006}
\begin{equation}
C= \frac{1}{2 \Delta t} \log\left(1+\frac{{\cal P}\Delta t}{{\cal N}}\right)
\end{equation}
where we consider as before power-limited signals, where
${\cal P}\Delta t/{\cal N}$ is the signal-to-noise ratio: ${\cal P}$
is the signal power (note that it is not the power of the system in
the energetic sense, only in the signal theory sense), and
${\cal N}/\Delta t$ the noise power.  When the bandwidth is taken to
infinity, \emph{i.e.} $\Delta t \to 0$, we get
\begin{equation}
C_0= \frac{{\cal P}}{2{\cal N}} \log_2e \text{        bits per second}
\end{equation}
which corresponds to \Eq{eq:capacity}.  For a finite but small
$\Delta t$ the expression for the capacity becomes
\begin{equation}
C= \frac{{\cal P}}{2{\cal N}}  - \frac{{\cal P}^2\Delta t}{4{\cal N}^2}+...\approx C_0\left(1-C_0 \Delta t\right)  
\end{equation} 
The first correction to the continuous-time capacity due to finite
rate of sampling is thus of relative order $C_0 \Delta t$, \emph{i.e.}
the information per sample: the loss of information when monitoring
Brownian dynamics at a finite rate is thus negligible provided that
the information per sample remains small. This has an important
practical consequence for experimental applications, where there is
often a trade-off between acquisition rate and duration of the
experiment (for instance due to photobleaching of fluorescent
proteins): when the information per sample becomes small, very little
can be learned about the force field by increasing the acquisition
frequency.

\section{The capacity as a stochastic thermodynamics quantity}
\label{sec:capacity_entropy}

\paragraph{Decomposition of the capacity.}

We now connect the capacity, introduced in an information-theoretical
setting in the previous Section, to known quantities in stochastic
thermodynamics. To this aim, we start with the steady-state
Fokker-Planck equation, which allows to decompose the force into a sum
of two terms,
\begin{equation}
  F_\mu = v_\mu + D_{\mu\nu} \partial_\nu \log P
  \label{eq:FP}
\end{equation}
where $v_\mu$ is the average phase space velocity, quantifying the
presence of irreversible currents, and
$D_{\mu\nu} \partial_\nu \log P$ quantifies reversible, diffusive
currents. Interestingly, this implies that the capacity defined in
\Eq{eq:capacity} decomposes into two non-negative parts, one related
to dissipation and the other to spatial structure, as
\begin{equation}
4C = \Sdot + G
\label{eq:C_decomposition}
\end{equation}
Here $\Sdot$ is the steady-state entropy production of the
process~\cite{seifert_stochastic_2012},
$\Sdot = \int v_\mu D^{-1}_{\mu\nu} v_\nu P(\mathbf{x})\dd \mathbf{x}$
(we set the Boltzmann constant $k_B=1$ throughout).  In the case of
thermal systems satisfying the Einstein relation, $\Sdot$ corresponds
to the rate at which the system dissipates heat into the bath, divided
by the temperature; in other cases, $\Sdot$ quantifies the
irreversibility of the dynamics. The second term, named \emph{inflow
  rate} $G = \int g_\mu D_{\mu\nu} g_\nu P(\mathbf{x})\dd \mathbf{x}$
with $g_\mu = \partial_\mu \log P$, was previously introduced and
studied in Ref.~\cite{baiesi_inflow_2015} in the case of Brownian
dynamics with homogeneous diffusion (and also for discrete Markov
processes, not discussed here). It reflects the amount of information
that the force field injects into the system in order to maintain
probability gradients against diffusion, and is positive even at
equilibrium. The inflow rate quantifies
the fact that in steady state, the system dwells in convergent regions
of the force field: an equivalent expression for it is
indeed~\cite{baiesi_inflow_2015}
$G = - \int \partial_\mu F_\mu(\mathbf{x}) P(\mathbf{x})\dd
\mathbf{x}$. In a deterministic system, it would thus correspond to
the average phase space contraction rate. As $G\geq 0$,
\Eq{eq:C_decomposition} provides a generic upper bound to the entropy
production in Brownian systems, $\Sdot \leq 4C$.

\paragraph*{ Physical interpretation of  the inflow rate. }
We now show that in a thought experiment where the force field would
be suddenly switched off, $G$ would correspond to the instantaneous
entropy production rate due to the relaxation of probability
gradients. Indeed, consider the entropy
$S(t) = -\int d\mathbf{x} P(x,t)\log P(x,t)$, after the force is set to
zero: $F_\mu=0$, denoting that instant by $t=0$. At that instant one
has $\partial_t P = \partial_{\mu} [ D_{\mu\nu}\partial_\nu P ]$. Then
\begin{equation}
\begin{split}
\partial_t S |_{t=0} = -\int d\mathbf{x}  \log P(\mathbf{x})\partial_{\mu} (D_{\mu\nu}\partial_\nu P(\mathbf{x}) )+\int d\mathbf{x} \partial_{\mu} (D_{\mu\nu}\partial_\nu P )=   \int d\mathbf{x}  \frac{\partial_{\mu} P(\mathbf{x})}P D_{\mu\nu}\partial_\nu P(\mathbf{x}) \\
 =\int d\mathbf{x} P(\mathbf{x}) \partial_{\mu} \log P(\mathbf{x}) D_{\mu\nu}(\mathbf{x}) \partial_{\nu} \log P(\mathbf{x})= G
\end{split}
\end{equation}
where we have used integration by parts, assuming boundary terms vanish. 
We can define $ v_\mu^\mathrm{Fick}= -D_{\mu\nu} g_\nu$, a Fick velocity related to the current $j_{\mu}^{\mathrm{Fick}}=-D_{\mu\nu}\partial_{\nu} P$, that would result from diffusion of particles with an initial density profile $P(\mathbf{x})$ in the absence of forces. Indeed, in these notations $G$ has a similar form to the entropy production rate
\begin{equation}
  \label{eq:Sdot}
G = \int v^\mathrm{Fick}_\mu v^\mathrm{Fick}_\nu D^{-1}_{\mu\nu}  P(\mathbf{x}) d\mathbf{x}
\end{equation}

However, the inflow rate is nonzero even at equilibrium.
It measures the heterogeneity of the steady-state probability
distribution. Indeed, for an equilibrium process $F^{\mu} = D_{\mu\nu} \partial_\mu \log P$ (and $G= C$ trivially).
In a sense, it is the amount of information that the
force field needs to continuously inject into the system in order to
maintain its spatial structure; while the entropy production can be
seen as the amount of information the force field injects into the
system to maintain its currents. 

\paragraph*{ The inflow rate as a phase space contraction rate. }
The relation $D_{\mu\nu} g_\mu= F_\mu-v_\mu $ (which holds for a
space-dependent diffusion tensor) can be used to rewrite the inflow
rate as
\begin{equation}
\begin{split}
G = \int d\mathbf{x} P(\mathbf{x}) g_{\mu} D_{\mu\nu} g_{\nu} = \int d\mathbf{x} P(\mathbf{x})( \partial_\mu \log P)  (F_\mu - v_\mu )  
\\ =\int d\mathbf{x}  (\partial_\mu P(\mathbf{x}))  F_\mu  +  \int d\mathbf{x}  \partial_\mu(v_\mu P(\mathbf{x}) ) \log P = -\int d\mathbf{x} P(\mathbf{x}) \partial_\mu F_\mu    
\end{split}
\end{equation}
where in the second line the steady state relation
$\partial_\mu(v_\mu P(\mathbf{x}) ) =\partial_\mu j_\mu =0$ was employed.
We have thus obtained an expression for the inflow rate as (minus) the
average divergence of the force. In a deterministic dynamical system
this is equal to the average sum of the Lyapunov exponents and is
called the average phase space contraction rate.  It then corresponds
to the mean rate of entropy production in the
environment~\cite{chetrite_fluctuation_2008}. For non-deterministic
systems it was mentioned in~\cite{chetrite_fluctuation_2008} as a
"natural entropy production". It is worth stressing the difference
between the deterministic case and overdamped Brownian dynamics in
this context. While for a deterministic system at equilibrium,
\emph{i.e.} a Hamiltonian system, the divergence of the force is
identically zero due to the symplectic structure (there is no entropy
production), for an equilibrium overdamped system that divergence is
nonzero. Indeed the inflow rate (which does not correspond to an
actual entropy production in this case) is positive, as discussed
above.

\paragraph{Relation  between capacity and traffic.}
We finally relate the capacity introduced here to other stochastic
thermodynamics quantities. The trajectory-based expression for the
capacity, \Eq{eq:traj_capacity_D}, is related to the "dynamical
entropy" introduced in~\cite{maes_steady_2008}: it is equal to the
dynamical entropy per unit time in the limit $\tau\to \infty$, i.e to
a rate of dynamical entropy. In~\cite{maes_steady_2008} the dynamical
entropy was split into two contributions: a time anti-symmetric
contribution, equal to $\Sdot/2$ and a time symmetric contribution
$-{\cal T}$, where ${\cal T}$ is called the traffic (and is related to
the so-called \emph{frenesy} in Markov jump processes).  The relations
between the capacity, the inflow rate we have defined, the entropy
production and the steady state traffic ${\cal T}$ are
\begin{align}
C= -{\cal T}+\frac{1}2 \Sdot  && {\cal T} = ( \Sdot-G)/4
\end{align}

\section{Information at the trajectory level}
\label{sec:trajectories}

\paragraph{Introduction.}

Here relate the notion of capacity to trajectory-level quantities, and
relate it to other stochastic thermodynamics quantities: the entropy
production and the inflow rate. Indeed, the decomposition of the
information into dissipative and structural contributions introduced
in \Eq{eq:C_decomposition} can be expressed at the level of individual
trajectories in phase space. The entropy production rate corresponds
to the rate at which trajectories,
$\mathcal{C} = \{\mathbf{x}(t)\}_{t=0..\tau}$, become distinguishable
from their time-reversed version,
$-\mathcal{C} = \{\mathbf{x}(\tau-t)\}_{t=0..\tau}$, as quantified by
the Kullback-Leibler divergence rate~\cite{seifert_stochastic_2012}:
\begin{equation}
\Sdot = \lim_{\tau\to\infty} \frac{1}{\tau} \av{ \log
\frac{  \mathcal{P}(\mathcal{C}|F)}{ \mathcal{P}(\mathcal{-C}|F)}}_F.\label{eq:Sdot_traj}
\end{equation}
Here $\mathcal{P}(\mathcal{C}|F)$ is the probability that the system
follows a trajectory $\mathcal{C} $
under Brownian dynamics (\Eq{eq:Langevin}) in the force field $F$,
 and $\av{\ \cdot\ }_F$ corresponds to
averaging over all possible trajectories $\mathcal{C}$ with weight
$\mathcal{P}(\mathcal{C}|F)$. Time reversal
$(\mathcal{C},F)\mapsto (-\mathcal{C},F)$ changes the sign of the heat
produced along the trajectory, and thus connects dissipation and
irreversibility of the dynamics.  Interestingly, a similar expression
can be derived for the inflow rate~\cite{baiesi_inflow_2015}:
\begin{equation}
G = \lim_{\tau\to\infty} \frac{1}{\tau} \av{ \log
  \frac{\mathcal{P}(\mathcal{C}|F)}{\mathcal{P}(\mathcal{-C}|-F)}}_F,\label{eq:G-traj}
\end{equation}

where $-F$ corresponds to the reversed force field. Indeed, the
operation $(\mathcal{C},F)\mapsto (-\mathcal{C},-F)$ now leaves the
heat unchanged, but reverses the sign of the divergence of the
force. At equilibrium, this corresponds to inverting the energy
landscape: for a typical trajectory that dwells in potential wells,
the reverse trajectory is atypical in the force field $-F$, as it
spends time around unstable maxima of energy. Finally, the capacity
can be expressed as
\begin{equation}
4C = \lim_{\tau\to\infty} \frac{1}{\tau} \av{ \log \frac{
  \mathcal{P}(\mathcal{C}|F)}{ \mathcal{P}(\mathcal{C}|-F)}}_F\label{eq:C-traj}
\end{equation}
where the operation $(\mathcal{C},F)\mapsto (-\mathcal{C},-F)$ reverses both heat and force divergence. Intuitively,
there is information about the force in a trajectory if it allows to
distinguish the force field from its reverse. More naturally, the
capacity quantifies the rate at which a trajectory becomes
distinguishable from force-free Brownian motion: indeed, it can be
written as
$C = \lim_{\tau\to\infty} \frac{1}{\tau} \av{ I(\mathcal{C})}_F$, where we define 
\begin{equation}
   I(\mathcal{C}) = \log \frac{ \mathcal{P}(\mathcal{C}|F)}{ \mathcal{P}(\mathcal{C}|0)}
  \label{eq:information-traj}
\end{equation}
as the trajectory-wise information gain about the force field.

\paragraph{Path-integral formulation.}

We now proceed to prove formulas~\ref{eq:G-traj}, \ref{eq:C-traj} and \ref{eq:information-traj}.  We consider the general
case with not only a state-dependent force, but also a state-dependent
diffusion tensor. In that case, the noise is no longer additive: it
has a multiplicative component, and care must be taken to specify the
convention within which the Langevin equation is written. We use the
It\^o convention here, writing:
\begin{equation}
\dot{x}_\mu = \Phi_\mu(\mathbf{x}) + \sqrt{2  D(\mathbf{x})}_{\mu\nu}  \xi_\nu
\label{eq:Langevin_D_cont}
\end{equation}
where
$\Phi_\mu =F_\mu(\mathbf{x}(t_i)) +\partial_\nu D_{\mu\nu}
(\mathbf{x}(t_i))$ is the drift term~\cite{lau_state-dependent_2007}, and
$F_\mu(\mathbf{x}(t_i))$ equals the mobility matrix times the physical
force.

To relate the capacity to path-dependent quantities, we consider a
trajectory
${\cal C}^N =(\mathbf{x}(0),\mathbf{x}(\Delta t),..\mathbf{x}(N\Delta
t))$, with $t_i = i \Delta t$, and where we have defined the discrete
difference $\Delta x_{\mu}(t_i) = x_{\mu}(t_i+\Delta t)-x_{\mu}(t_i)$
and $\tau = N \Delta t$.  The path integral formula for the
probability density ${\cal P}({\cal C}^N|F)$ of a trajectory
${\cal C}^N$ in the force field $F$, written in the It\^o convention,
reads~\cite{risken_fokker-planck_1996}:
\begin{align}
&{\cal P}({\cal C}^N|F) =  \frac{P_0(\mathbf{x}(0))}{(4\pi)^{dN/2} }\prod_{i=0}^{N-1} \frac{1}{(\det D(\mathbf{x}(t_i))\Delta t)^{1/2}} \\
&\times\exp\left[-\frac{1}4 \Delta t\left(\frac{\Delta x_{\mu}(t_i)}{\Delta t} - F_{\mu}(\mathbf{x}(t_i))-\partial_\rho D_{\mu\rho}(\mathbf{x}(t_i))\right)D^{-1}_{\mu\nu} (\mathbf{x}(t_i))\left(\frac{\Delta x_{\mu}(t_i)}{\Delta t} - F_{\nu}(\mathbf{x}(t_i))-\partial_\sigma D_{\nu\sigma}(\mathbf{x}(t_i))\right)\right]
\label{eq:traj_probability_D_ito}
\end{align}
Note that in the limit of long trajectories, the initial point
probability becomes unimportant. We show here that the capacity of the system relates to
the Kullback-Leibler divergence rate between ${\cal P}({\cal C}^N|F)$ and
the probability density at zero force (but with the same diffusion
field), ${\cal P}({\cal C}^N|0) \equiv {\cal P}({\cal
  C}^N|F=0)$:
\begin{equation}
  C= \lim_{\tau\to \infty}\frac{1}{\tau}\int {\cal{D}}{{\cal C}^{\tau}} \  {\cal P}({\cal C}^{\tau}|F)\log\frac{ {\cal P}({\cal C}^{\tau}|F)}{{\cal P}({\cal C}^{\tau}|0)} =\av{\frac{1}4 F_\mu(\mathbf{x}(t)) D_{\mu\nu}^{-1}(\mathbf{x}(t))F_\nu(\mathbf{x}(t))}
\label{eq:traj_capacity_D}
\end{equation}
Indeed, for a constant diffusion coefficient the right hand side of the above equation reduces to the capacity discussed in \Sec{eq:capacity}, \Eq{eq:capacity}. Note that for systems with multiplicative noise,
to the best of our knowledge a formula for
the channel capacity, as defined in transmission
theory, has yet to be derived. Moreover, the interpretation from the standpoint of transmission theory is further complicated as, from
physical considerations, we wish to infer $F_\mu$ rather than
$ \Phi_\mu$. However, one may use the trajectory based formula in \Eq{eq:traj_capacity_D} as a general definition of the capacity for Brownian dynamics. Then, the generalization of
\Eq{eq:capacity} to systems with inhomogeneous diffusion is seen to be: 
\begin{equation}
C = \frac{1}{4}   \int  D^{-1}_{\mu\nu}(\mathbf{x}) F_{\mu}(\mathbf{x}) F_{\nu}(\mathbf{x})  P(\mathbf{x})\dd \mathbf{x} 
 \label{eq:capacity_nonH_SI}
\end{equation}

Let us proceed to show \Eq{eq:traj_capacity_D},
\begin{align}
C &= \lim_{\tau\to \infty}\frac{1}{\tau}\int {\cal{D}}{{\cal C}^{\tau}} \  {\cal P}({\cal C}^{\tau}|F)\log\frac{ {\cal P}({\cal C}^{\tau}|F)}{{\cal P}({\cal C}^{\tau}|0)} \\
&= \lim_{\tau\to \infty}\frac{1}\tau   \av{\frac{1}2\int^{\text{It\^o}} dt \ \dot{x}_\mu D_{\mu\nu}^{-1}F_\nu(\mathbf{x}(t)) -\frac{1}2 \int_0^{\tau} dt (\partial_\rho D_{\rho\mu}) D_{\mu\nu}^{-1}F_\nu(\mathbf{x}(t)) -\frac{1}4 \int_0^{\tau} dtF_\mu D_{\mu\nu}^{-1}F_\nu(\mathbf{x}(t))} \\
& =\av{\frac{1}4 F_\mu(\mathbf{x}(t)) D_{\mu\nu}^{-1}(\mathbf{x}(t))F_\nu(\mathbf{x}(t))}
\end{align}
where we have used that $\av{\int^{\text{It\^o}} dt\dot{x}_\mu D_{\mu\nu}^{-1}F_\nu(\mathbf{x}(t))}=\av{\int_0^{\tau}dt(F_\mu+ \partial_\rho D_{\rho\mu})D_{\mu\nu}^{-1}F_\nu(\mathbf{x}(t))}$.
Note that passing between the first and second line in the above equation is equivalent to deriving the Girsanov formula for diffusions.

\paragraph*{ Trajectory based interpretation of the inflow rate. }
Here we prove that an equivalent expression for the inflow rate is
\begin{equation}
G = \lim_{\tau\to \infty}\frac{1}{\tau}\int {\cal{D}}{{\cal C}^{\tau}} \  {\cal P}({\cal C}^{\tau}|F)\log\frac{ {\cal P}({\cal C}^{\tau}|F)}{{\cal P}({-\cal C}^{\tau}|-F)} =  \lim_{\tau\to \infty}\frac{1}{\tau}\av{ \log\frac{ {\cal P}({\cal C}^{\tau}|F)}{{\cal P}(-{\cal C}^{\tau}|-F)}}_F 
\label{eq:inflow_trajectory}
\end{equation}

The simplest way to do that is to express the probability density of a trajectory (\Eq{eq:traj_probability_D_ito}) in an alternative form, as we now show. We begin with the expression for the probability of a transition to the point $\mathbf{x}$ from the point $\mathbf{x'}$ in an infinitesimal time $\Delta t$~\cite{risken_fokker-planck_1996}
\begin{equation}
\begin{split}
P(\mathbf{x},t+\Delta t|\mathbf{x'},t)=\frac{1}{\sqrt{(4\pi)^d \det D(\mathbf{x})\Delta t}}  \exp\left[\Delta t \left\{  -\partial_\mu \Phi_\mu(\mathbf{x}) +\partial_\mu\partial_\nu D_{\mu\nu}(\mathbf{x}) \right.\right.\\
\left.\left.-\frac{1}4 \left(\frac{x_\mu-x'_\mu}{\Delta t}- \Phi_\mu(\mathbf{x}) +2\partial_\rho D_{\mu\rho}(\mathbf{x})\right) D^{-1}_{\mu\nu}(\mathbf{x})\left(\frac{x_\nu-x'_\nu}{\Delta t}- \Phi_\nu(\mathbf{x}) +2\partial_\sigma D_{\nu\sigma}(\mathbf{x})\right)\right\}\right]
\end{split}
\end{equation}
Note that here the diffusion coefficient and $\Phi_\mu$ are both evaluated at the point $\mathbf{x}$ to which the system transitions. The probability of a trajectory is then simply given by a product of such transition probabilities, and the distribution of the initial point. Using that  $\Phi_\mu = F_\mu+\partial_\nu D_{\mu\nu}$ we then get
\begin{equation}
\begin{split}
{\cal P}({\cal C}^N|F) = \frac{P_0(\mathbf{x}(0))}{(4\pi)^{dN/2} }\prod_{i=0}^{N-1} \frac{1}{(\det D(\mathbf{x}(t_{i+1}))\Delta t)^{1/2}} \exp\left[  -\partial_\mu F_\mu(\mathbf{x}(t_{i+1}))\Delta t \right.\\
\left.-\frac{1}4 \Delta t\left(\frac{\Delta x_{\mu}(t_{i})}{\Delta t} - F_{\mu}(\mathbf{x}(t_{i+1}))+\partial_\rho D_{\mu\rho}(\mathbf{x}(t_{i+1}))\right)D^{-1}_{\mu\nu} (\mathbf{x}(t_{i+1}))\left(\frac{\Delta x_{\nu}(t_{i})}{\Delta t} - F_{\nu}(\mathbf{x}(t_{i+1}))+\partial_\sigma D_{\nu\sigma}(\mathbf{x}(t_{i+1}))\right)\right]
\end{split}
\label{eq:traj_probability_D_aito}
\end{equation}

It follows that the probability of the time reversed trajectory $-{\cal C}^N=\{\mathbf{x}(t_N),\mathbf{x}(t_{N-1})...,\mathbf{x}(t_0)\}$ can be written in the form
\begin{equation}
\begin{split}
{\cal P}(-{\cal C}^N|F) = \frac{P_0(\mathbf{x}(N\Delta t))}{(4\pi)^{dN/2} }\prod_{i=0}^{N-1} \frac{1}{(\det D(\mathbf{x}(t_{i}))\Delta t)^{1/2}} \exp\left[  -\partial_\mu F_\mu(\mathbf{x}(t_{i}))\Delta t \right.\\
\left.-\frac{1}4 \Delta t\left(\frac{-\Delta x_{\mu}(t_{i})}{\Delta t} - F_{\mu}(\mathbf{x}(t_{i}))+\partial_\rho D_{\mu\rho}(\mathbf{x}(t_{i}))\right)D^{-1}_{\mu\nu} (\mathbf{x}(t_{i}))\left(\frac{-\Delta x_{\nu}(t_{i})}{\Delta t} - F_{\nu}(\mathbf{x}(t_{i}))+\partial_\sigma D_{\nu\sigma}(\mathbf{x}(t_{i}))\right)\right]
\end{split}
\label{eq:-traj_probability_D_aito}
\end{equation}

Now, it becomes straightforward to evaluate \Eq{eq:inflow_trajectory}, dividing term by term in the product in \Eq{eq:traj_probability_D_ito} by the product in ${\cal P}(-{\cal C}^N|-F)$, using \Eq{eq:-traj_probability_D_aito} with the reversed sign for the force. Indeed, we notice that all terms cancel out except for the divergence of $F_\mu$, which yields (we ignore the terms related to the initial and final distributions whose contribution vanishes in the limit of $\tau\to \infty$)
\begin{equation}
\begin{split}
G= \lim_{\tau\to \infty}\frac{1}{\tau}\int {\cal{D}}{{\cal C}^{\tau}} \  {\cal P}({\cal C}^{\tau}|F)\log\frac{ {\cal P}({\cal C}^{\tau}|F)}{{\cal P}(-{\cal C}^{\tau}|-F)} =   -\lim_{\tau\to \infty} \int_0^\tau \frac{dt}\tau \av{\partial_\mu F_\mu(\mathbf{x}(t)) }
\end{split}
\end{equation}

\paragraph{Decomposition of the capacity at the trajectory level.}

The decomposition of the capacity that we have presented in the main text can also be presented as the sum of time symmetric and anti-symmetric parts, but  corresponding to a different trajectory-based expression for the capacity:
\begin{align}
4 C &=  \lim_{\tau\to \infty}\frac{1}{\tau}\int {\cal{D}}{{\cal C}^{\tau}} \  {\cal P}({\cal C}^{\tau}|F)\log\frac{ {\cal P}({\cal C}^{\tau}|F)}{{\cal P}({\cal C}^{\tau}|-F)} =  \lim_{\tau\to \infty}\frac{1}{\tau}\av{ \log\frac{ {\cal P}({\cal C}^{\tau}|F)}{{\cal P}({\cal C}^{\tau}|-F)}}_F\\
& =\lim_{\tau\to \infty}\frac{1}\tau   \av{\int^{\text{It\^o}} dt\ \dot{x}_\mu D_{\mu\nu}^{-1}F_\nu(\mathbf{x}(t)) - \int_0^{\tau} dt (\partial_\rho D_{\rho\mu}) D_{\mu\nu}^{-1}F_\nu(\mathbf{x}(t))}\\
& =  \lim_{\tau\to \infty}\frac{1}\tau\av{\int^{\text{Strat}}  \!\!\!  \!\!\! dt\ \dot{x}_\mu D_{\mu\nu}^{-1}F_\nu(\mathbf{x}(t)) } -  \lim_{\tau\to \infty}\frac{1}\tau\av{\int_0^\tau dt D_{\mu\rho} \partial_\rho (D_{\mu\nu}^{-1}F_\nu)(\mathbf{x}(t)) - \int_0^{\tau} dt \partial_\rho (D_{\rho\mu}) D_{\mu\nu}^{-1}F_\nu(\mathbf{x}(t))} \\
& = \lim_{\tau\to \infty}\frac{1}\tau\av{\underbrace{\int^{\text{Strat}} \!\!\! dt\ \dot{x}_\mu D_{\mu\nu}^{-1}F_\nu(\mathbf{x}(t)) }_{\text{time anti-symmetric}}} + \lim_{\tau\to \infty}\frac{1}\tau\av{\underbrace{-\int_0^{\tau} dt \partial_\mu F_\mu(\mathbf{x}(t))}_\text{ time symmetric}} -\lim_{\tau\to \infty}\frac{1}\tau \av{ \int_0^{\tau} dt F_\nu\underbrace{\partial_\rho (D_{\rho\mu} D_{\mu\nu}^{-1})}_0}
\label{eq:capacity_trajectory2}
\end{align}
Indeed, the first term in the last line is time anti-symmetric, and is equal to the entropy production rate, and the second term is time symmetric and is equal to the inflow rate.  

As a conclusion, one can think of the decomposition of the capacity into $\Sdot$ and $G$ as decomposing the influence of the force field into two types of ``orders'': ``\emph{go
  there!}'' -- corresponding to a dissipative, irreversible motion
quantified by $\Sdot$ -- and ``\emph{stay there!}'' -- corresponding
to a nondissipative, reversible motion fighting thermal diffusion, and
quantified by $G$.

\section{Towards a comprehensive quantification of the capacity}
\label{sec:capacity_real}

In the previous Sections, we have derived a bound for the information
rate -- the \emph{capacity} -- of mathematically ideal trajectories,
by drawing a quantitative analogy between Brownian dynamics and
communication channels. However, experimental trajectories are
non-ideal, as they have a finite frame rate and are corrupted by a
level of measurement noise. A future direction for this research will
be to expand these results to quantify the capacity of real
trajectories, with finite time step and measurement noise.
Schematically:
\begin{figure}[h!]
  \makebox[\textwidth][c]{\includegraphics[width=1.2\linewidth]{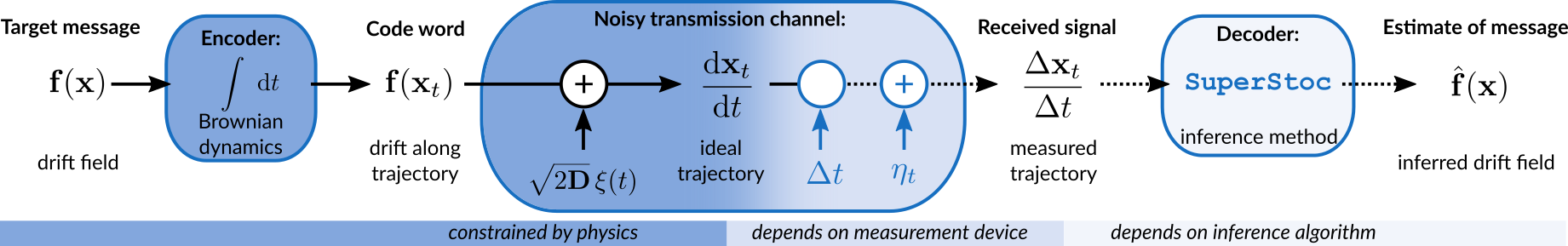}}    
\end{figure}

There is information loss in the noisy channel that converts the drift
into the trajectory increments, due both to intrinsic Brownian noise
and extrinsic measurement devices. The previous sections have established that the intrinsic capacity with ideal signals is 
\begin{equation}
  \label{eq:Cideal}
  C_{\mathbf{f},\mathrm{ideal}} = \lim_{\Delta t \to 0} \av{\frac{1}{2\Delta t} \log\left( 1 + \frac{\mathbf{f}^2(\mathbf{x}_t)\Delta t^2}{2\mathbf{D}(\mathbf{x}_t)\Delta t}\right)}_t =\frac{1}{4} \av{ \mathbf{f}(\mathbf{x}_t)\cdot\mathbf{D}^{-1}(\mathbf{x}_t)\cdot\mathbf{f}(\mathbf{x}_t)}_t
\end{equation}
which forms a strict upper bound to the amount of information about
the drift field that can be extracted from the
trajectories. Importantly however, in practice, this capacity is not
achieved for long trajectories, as it would require an optimal
encoding. The role of the encoder here is played by the dynamics
integrator, on which there is no control, and which is
sub-optimal. Therefore, it is possible that tighter upper bounds may
be established in general for the actual information rate of Brownian
systems.

Furthermore, a future goal will expand and adapt \Eq{eq:Cideal} to
estimate the capacity of real signals, including finite $\Delta t$ and
measurement noise, for both the drift and diffusive terms. This could
be done by employing stochastic It\^o-Taylor expansions to derive
finite-$\Delta t$ corrections, as well as by including measurement
error as a second noise source.  Importantly, while single increments
contain all the information for ideal signals, it may not be the case
for real trajectories, and multi-point estimates will have to be
considered. Finally, for very large time increments, It\^o-Taylor
expansions might fail, and alternative tools inspired from other
fields such as data
augmentation~\cite{papaspiliopoulos_nonparametric_2012,papaspiliopoulos_data_2013}
or operator eigenpair inference~\cite{crommelin_diffusion_2011} will
have to be used to strengthen information bounds.  All in all, these
future projects will help derive quantitative estimates of the amount
of information available for inference in real data.

\chapter{Stochastic Force Inference}
\label{chap:SFI}

\emph{This Chapter presents the core result of this Thesis: a general
  method, Stochastic Force Inference (SFI), to reconstruct overdamped
  Langevin equations from data, including in high dimensions and in
  the presence of measurement and multiplicative noise.  It is adapted
  from the second and third parts of
  Ref.~\cite{frishman_learning_2020}, as well as its appendices. It
  first re-introduces the maximum-likelihood approach (\Eq{eq:ML})
  from the perspective of projecting the dynamics onto a vector space,
  then shows that a modification makes it robust to measurement noise,
  yielding the SFI estimator. \Sec{sec:SFI} presents the method
  focusing on the simpler case of additive noise, where the dynamics reads:
\begin{equation}
  \dot{x}_\mu = F_\mu(\mathbf{x}) + \sqrt{2D}_{\mu\nu} \xi_\nu,
  \label{eq:LangevinSFI}
\end{equation}
while \Sec{sec:inhomogeneous} generalizes to multiplicative noise and
\Sec{sec:SFI_error} provides elements of proof of convergence and
error estimation. I finally discuss perspectives in
\Sec{sec:SFI-discussion}, adapted from the conclusion of the article.}

\vspace{5mm}

Adapted from: \\
\textsc{Learning force fields from stochastic trajectories} \\
Anna Frishman$^*$ and Pierre Ronceray$^*$ \\
Physical Review X 10, 021009 (2020). 

\vspace{5mm}

\section{Principle of the method}
\label{sec:SFI}

A trajectory of finite duration contains finite information,
quantified as we have seen in the previous Chapter by:
\begin{equation}
   I(\mathcal{C}) = \log \frac{ \mathcal{P}(\mathcal{C}|F)}{ \mathcal{P}(\mathcal{C}|0)}.
  \label{eq:information}
\end{equation}
We now show how to use this
information in practice and reconstruct the force field through
Stochastic Force Inference (SFI).  In contrast with the drifted
Brownian motion, a spatially variable force field is in principle
characterized by an infinite number of degrees of freedom: the force
value at each point in space. With a finite trajectory, only a finite
number of combinations of degrees of freedom can be estimated.

\paragraph{Projecting the force field.}
It is therefore natural to approximate the force field as a linear
combination of a finite basis of $n_b$ known functions
$b = \{b_\alpha(\mathbf{x})\}_{\alpha = 1..n_b}$. The force can, in
principle, be approximated arbitrarily well by using a large enough
set of functions from a complete basis, such as polynomials or Fourier
modes. Alternatively, a limited number of functions might suffice if
an educated guess for the functional form of the force field can be
made. We propose to perform this approximation by projecting the force
field onto the space spanned by $b_\alpha(\mathbf{x})$ using the
steady-state probability distribution function $P$ as a measure. This
corresponds to a least-squares fit of the force field by linear
combinations of the $b_\alpha$'s. To this aim, we define the projector
$c_\alpha(\mathbf{x}) = B^{-1/2}_{\alpha\beta} b_\beta(\mathbf{x})$,
where $B_{\alpha\beta}$ is an orthonormalization matrix such that
$\int c_\alpha c_\beta P(\mathbf{x})\dd \mathbf{x} =
\delta_{\alpha\beta}$.  Our approximation of the force field is then
$F_\mu(\mathbf{x}) \approx F_{\mu\alpha} c_\alpha(\mathbf{x})$ with
the projection coefficient
\begin{equation}
  \label{eq:F_moments}
  F_{\mu\alpha} = \int F_\mu(\mathbf{x}) c_\alpha(\mathbf{x}) P(\mathbf{x})\dd \mathbf{x}.
\end{equation}
This is akin to projecting the dynamics onto a finite-dimensional
sub-channel of capacity
$C_b = \frac{1}{4}D^{-1}_{\mu\nu} F_{\mu\alpha} F_{\nu\alpha} < C $.
Similarly, we can define the projection $v_{\mu\alpha}$ of the phase
space velocity. The corresponding entropy production
$\Sdot_b = D^{-1}_{\mu\nu} v_{\mu\alpha} v_{\nu\alpha}$ is then a
lower bound to the total entropy production. Interestingly, for a
system obeying Brownian dynamics (\Eq{eq:LangevinSFI}) but where only a
subset of degrees of freedom can be observed, our framework gives the
force averaged over hidden variables, and provides a lower bound on
the entropy production limited to the observable
currents.

\paragraph{Inferring the projection coefficients.}
The projected force field has a finite number of degrees of freedom
$N_b = d n_b$, one per element of the $d \times n_b$ tensor
$F_{\mu\alpha}$, and corresponds to a finite capacity $C_b$. Inferring
the approximate force with a finite trajectory is thus in principle
possible if the information $I_b = \tau C_b > N_b$. However, the force
coefficients introduced in \Eq{eq:F_moments} are not directly
accessible from experimental data. Indeed, neither the force nor the
probability distribution function $P$ are known, the latter being also
required in the definition of the orthonormal projectors
$c_\alpha$. Instead, the available data is typically a discrete time
series $\mathbf{x}(t_i)$ of phase space positions, at sampling times
$t_i = i \Delta t$. We thus propose to estimate phase space averages
by discrete time integrals along the trajectory. The empirical
projectors are defined as
\begin{equation}
\hat{c}_\alpha = \hat{B}^{-1/2}_{\alpha\beta} b_\beta \qquad \text{with} \qquad
\hat{B}_{\alpha\beta} = \sum_i b_\alpha(\mathbf{x}(t_i))
b_\beta(\mathbf{x}(t_i)) \frac{\Delta t}{\tau}.\label{eq:B_empirical}
\end{equation}
Furthermore, the
force can be expressed in terms of a local It\^o average of
$\dot{\mathbf{x}}$~\cite{risken_fokker-planck_1996}: a local estimator
for the force at $\mathbf{x}(t_i)$ is thus
$\Delta \mathbf{x}(t_i)/\Delta t$, with
$\Delta \mathbf{x}(t_i) = \mathbf{x}(t_{i+1}) - \mathbf{x}(t_i)$.
Combining these two insights yields an operational definition for the
estimator of \Eq{eq:F_moments} in terms of a discrete It\^o
integral,%~\ref{XXXsec:trajectory}
\begin{equation}
  \label{eq:F_moments_empirical}
  \hat{F}_{\mu\alpha} = \frac{1}{\tau} \sum_i \Delta x_\mu(t_i) \hat{c}_\alpha(\mathbf{x}(t_i)) 
\end{equation}
which is the discretized version of the It\^o integral
$\frac{1}{\tau}\int_0^\tau \hat{c}_\alpha(\mathbf{x}(t)) \dd
x_\mu(t)$.  Indeed, discretizing \Eq{eq:LangevinSFI} yields
$\Delta \mathbf{x}(t_i) = \mathbf{F}(\mathbf{x}(t_i)) \Delta t +
\sqrt{2\mathbf{D}}\Delta \xi_i$, where $\Delta \xi_i$ is independent
of $\mathbf{x}(t_i)$: in the long trajectory limit, the main
contribution comes from the force, while the noise averages to
zero. Equation~\ref{eq:F_moments_empirical} corresponds to a linear
regression of the local force estimator, previously suggested for
one-dimensional systems~\cite{comte_penalized_2007}, and coincides
with the maximum-likelihood estimator of the force projection
coefficients.

\paragraph{Estimating the inference error.}
The typical squared relative error on the inferred
coefficients due to the diffusive noise can be estimated in practice
as
$\delta \hat{F}^2 / \hat{F}^2 \sim
N_b/2\hat{I}_b$, where
$\hat{I_b} = \frac{\tau}{4}D^{-1}_{\mu\nu} \hat{F}_{\mu\alpha}
\hat{F}_{\nu\alpha}$ is the empirical estimate of information
contained in the trajectory. This formula, which we derive more formally in \Sec{sec:SFI_error}, indicates that again, in
order to resolve the force coefficients, the information in the
trajectory should exceed the number of inferred parameters.  Another
source of error stems from the fact that the force varies over a
finite time step $\Delta t$.

\begin{figure}[b]
  \centering
  { \includegraphics[width=1.\columnwidth]{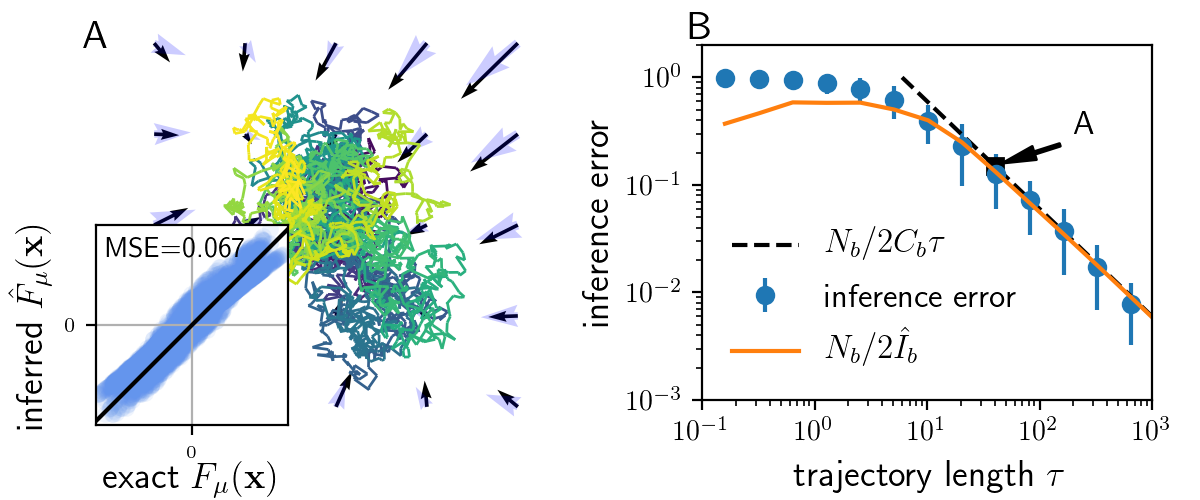}}
  \caption{ Stochastic force inference for a 2D Ornstein-Uhlenbeck
    process, with force field
    $F_\mu(\mathbf{x}) = -\Omega_{\mu\nu} x_\nu$ and isotropic
    diffusion. \textbf{A.} An example trajectory. The inferred force
    field for this trajectory, using SFI with functions
    $b=\{1,x_\mu\}$ (blue arrows), is compared to the exact force
    field (black arrows). \emph{Inset:} the inferred force components
    along the trajectory \emph{versus} the exact force components,
    with normalized mean-squared error (MSE).  \textbf{B.} The average of the
    relative error
    $[(\hat{F}_{\mu\alpha}-F^{\tau}_{\mu\alpha})D^{-1}_{\mu\nu}(\hat{F}_{\nu\alpha}-F^{\tau}_{\nu\alpha})]/[\hat{F}_{\mu\alpha}D^{-1}_{\mu\nu}\hat{F}_{\nu\alpha}]$
    on the inferred projection coefficients $\hat{F}_{\mu\alpha}$ and
    its self-consistent estimate $N_b/2\hat{I}_b$ both converge to
    $N_b/2I_b$, as expected from theory. Here
    $F^{\tau}_{\mu\alpha} = \int F_\mu(\mathbf{x}(t))
    \hat{c}_\alpha(\mathbf{x}(t)) \frac{\dd t}{\tau}$ is the
    projection of the exact force on the empirical projectors. }
  \label{fig:2dOU}
\end{figure}

\paragraph{Proof of concept on simulated data.}

We now demonstrate the utility of our method using simulated data of
simple models. The simplest spatially varying force field is a
harmonic trap, \emph{i.e.} an Ornstein-Uhlenbeck process (\Fig{fig:2dOU}). We
benchmark our method by using a first-order polynomial basis,
$b=\{1,x_\mu\}$, which can capture the exact force field. The 2D
trajectory displayed in \Fig{fig:2dOU}A has an information content of $I=27.6$
bits, while this linear channel has $N_b=6$ degrees of freedom,
allowing precise inference of the projected force field (\Fig{fig:2dOU}A). Indeed, the squared relative error on the force coefficients is
$0.15$; this is consistent with the operational estimate of this
error, $N_b/2\hat{I_b} = 0.16$. The force along the trajectory is thus
inferred to a good approximation (\Fig{fig:2dOU}A, \emph{inset}). Furthermore,
the projected force field
$\hat{F}_{\mu\alpha} \hat{c}_\alpha(\mathbf{x})$ provides an ansatz
that can be extrapolated beyond the trajectory (\Fig{fig:2dOU}A), which works
equally well here as the functional form of the force field is fully
captured by our choice of basis. More quantitatively, we confirm the
predicted behavior for the squared relative error by studying an
ensemble of trajectories (\Fig{fig:2dOU}B).

\paragraph{Inferring currents and entropy production.}

In the case of out-of-equilibrium Brownian systems, our method also
permits the approximation of phase space currents and entropy
production. Indeed, the phase space velocity $\mathbf{v}$ can be
expressed in terms of a local Stratonovich average of
$\dot{\mathbf{x}}$, reflecting the fact that it is odd under time
reversal~\cite{chetrite_eulerian_2009}. Our estimator for the
projection coefficients of the phase space velocity is thus
\begin{equation}
  \hat{v}_{\mu\alpha} = \frac{1}{\tau} \sum_i \Delta x_\mu(t_i)\hat{c}_\alpha\left(\frac{\mathbf{x}(t_{i+1}) +
    \mathbf{x}(t_i)}{2}\right) \label{eq:v}
\end{equation}
which is the discretized version of the Stratonovich integral
$\frac{1}{\tau}\int_0^\tau \hat{c}_\alpha(\mathbf{x}(t)) \circ \dd
x_\mu(t) $. This allows the inference of the entropy production rate:
\begin{equation}
\hat{\Sdot}_b = D^{-1}_{\mu\nu} \hat{v}_{\mu\alpha}
\hat{v}_{\nu\alpha}\label{eq:Sdot_inferred}
\end{equation}
associated to the observed currents.  This is a biased estimator of
the entropy production, with an error that can be self-consistently
controlled as
$\hat{\Sdot}_b = \Sdot_b + 2N_b/\tau +
O((2\hat{\Sdot}_b/\tau+(2N_b/\tau)^2)^{1/2})$: the entropy
production rate in the channel can thus be inferred using a single
trajectory provided that several $k_B$'s per degree of freedom have
been dissipated.

\begin{figure}[t]
  \centering
{ \includegraphics[width=1.\columnwidth]{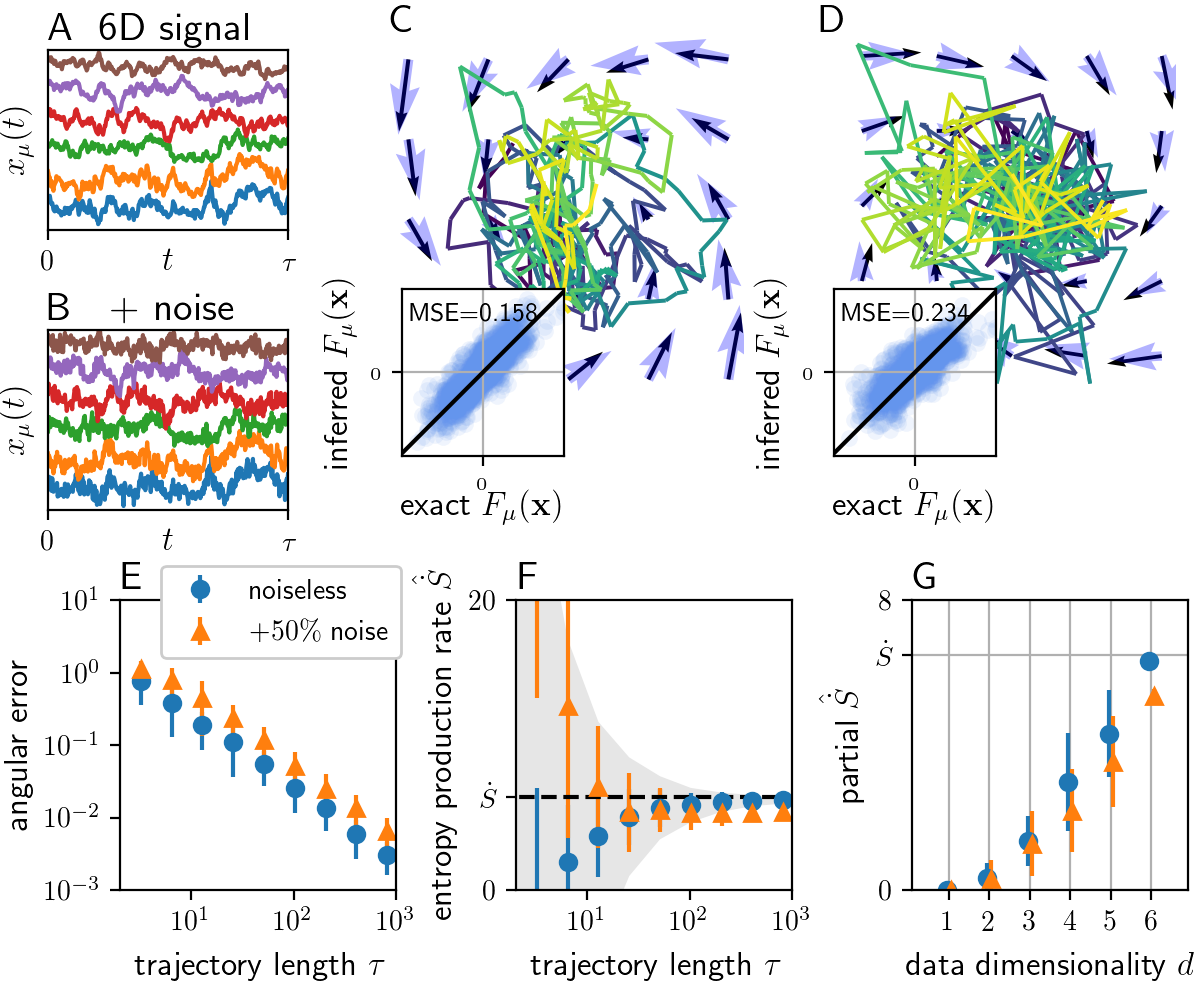}}
\caption{\textbf{A.}  Time series of a 6D out-of-equilibrium
  Ornstein-Uhlenbeck process, with anisotropic harmonic confinement
  and diffusion tensor, and circulation. The force field is
  $F_\mu(\mathbf{x}) = -\Omega_{\mu\nu} x_\nu$. The matrix $\Omega$
  and the diffusion matrix are chosen from a random ensemble. The
  antisymmetric part of $D^{-1}\Omega$ has rank 2, thus inducing
  circulation in a randomly chosen plane.  \textbf{B.}  The same
  trajectories as in {D}, with additional time-uncorrelated
  measurement noise.  \textbf{C.} SFI for the trajectory in {A} allows
  precise identification of the plane of circulation and
  reconstruction of the force along the trajectory. \textbf{D.} SFI
  applied to the trajectory in {B}, with measurement noise. It can
  still detect forces accurately. \textbf{E.}  Convergence of the
  angular error for cycle detection with increasing trajectory length,
  for the process shown in {D-E}. \textbf{F.}  Inferred entropy
  production rate for this process, with and without measurement noise
  (we subtracted here the systematic bias $2N_b/\tau$). The shadowed
  area indicates the self-consistent confidence interval for the
  inferred entropy production. The dotted line shows the exact value
  of the entropy produced; for the noisy process SFI underestimates
  this value due to blurring of the currents. \textbf{G.} Entropy
  production captured when observing a $d$-dimensional projection of
  the trajectory, averaged over direction of observation, for long
  trajectories.  In plots {E,F,G}, error bars indicate standard
  deviation over an ensemble of $32$ trajectories.  }
  \label{fig:6D}
\end{figure}

\paragraph{Entropy production, cycling frequencies, and area enclosing rate.}
The simplest structure for phase space currents corresponds to cyclic
circulation around a point. The detection of such features in active biological
systems has been the focus of a number of recent studies, which employ
phase space
coarse-graining~\cite{battle_broken_2016,gnesotto_broken_2018,seara_entropy_2018}. This
method is however limited to low-dimensional systems, and even then
requires large amounts of data: indeed, the capacity per degree of
freedom is low, as each grid cell is visited infrequently. In
contrast, our method provides a way to detect circulation in any
dimension with minimal data. Using the centered linear basis
$b_\alpha(\mathbf{x})=\bar{x}_\alpha=x_\alpha-\int x_\alpha\frac{\dd
  t}{\tau}$, we can infer the velocity coefficients
$\hat{v}_{\mu\alpha}$, which have a matrix structure. This matrix
reads $\hat{v}_{\mu\alpha}=C^{-1/2}_{\alpha\beta} A_{\beta\mu}$, where
$C_{\mu\nu} = \int \bar{x}_\mu \bar{x}_\nu\frac{\dd t}{\tau}$ is the
covariance matrix, and the antisymmetric part of $A_{\mu\nu} $ is
$A_{\{\mu\nu\}} = \frac{1}{2\tau}\int \bar{x}_\mu\dd x_\nu -
\bar{x}_\nu\dd x_\mu$, which is the rate at which the process
encircles area in the $(\mu,\nu)$
plane~\cite{ghanta_fluctuation_2017,gonzalez_experimental_2019}. This
rate, sometimes called probability angular
momentum~\cite{shkarayev_exact_2014,zia_manifest_2016}, intuitively
quantifies circulation and closely connects to cycling
frequencies~\cite{gladrow_broken_2016,mura_nonequilibrium_2018}. Indeed,
the eigenvectors of $A_{\{\mu\nu\}}$ can be used to define cycling
planes. The entropy production rate due to cycling reads
$\hat{\Sdot}_b = D^{-1}_{\mu\nu} A_{\nu\rho} C^{-1}_{\rho\sigma}
A_{\sigma\mu}$.

We demonstrate the potency of our cycle-detection method on a
challenging dataset: a short trajectory of an out-of-equilibrium
Ornstein-Uhlenbeck process in dimension $d=6$ (\Fig{fig:6D}A), which
is equivalent to popularly used bead-spring
models~\cite{battle_broken_2016,mura_nonequilibrium_2018,li_quantifying_2019}. Our
method identifies the principal circulation plane accurately, together
with the force field (\Fig{fig:6D}C). Quantitatively, we demonstrate
that the angular error in the identification of this plane vanishes
with increasing trajectory length (\Fig{fig:6D}E), concomitant with
the convergence of $\hat{\Sdot}_b$ to the exact value
(\Fig{fig:6D}F). The entropy production inferred is associated to the
observable currents: if only a fraction of the degrees of freedom can
be observed, $\hat{\Sdot}_b$ is a lower bound to the total entropy
production of the system (\Fig{fig:6D}G), as some currents are not
observable. In particular, if only one degree of freedom can be
measured, this technique will yield $\hat{\Sdot}_b = 0$; alternative
techniques based on the non-Markovianity of the dynamics are better
suited to inferring entropy production in this
case~\cite{roldan_arrow_2018}.

\paragraph{Dealing with measurement noise.}
A major challenge in the inference of dynamical properties of
stochastic systems from real data is time-uncorrelated measurement
noise, which dominates time derivatives of the signal. Indeed, in our
inference scheme, \Eq{eq:F_moments_empirical} is highly sensitive to
such noise. In contrast, the time-reversal antisymmetry of the
velocity coefficients $\hat{v}_{\mu\alpha}$ makes them robust against
measurement noise. Exploiting this symmetry, we obtain
an unbiased estimator for the force by using the relation between
It\^o and Stratonovich integration,
\begin{equation}
  \hat{F}_{\mu\alpha} = \hat{v}_{\mu\alpha} + D_{\mu\nu}\hat{g}_{\nu\alpha}
\label{eq:F_corrected_estimator}
\end{equation}
where
$\hat{g}_{\mu\alpha} = - \sum_i \frac{\Delta t}{\tau}\partial_\mu
\hat{c}_\alpha(\mathbf{x}(t_i))$ is an estimator for the projection of
$g_\mu=\partial_\mu \log P$ onto the basis (note that while
$\hat{g}_\mu(\mathbf{x}) \equiv
\hat{g}_{\mu\alpha}\hat{c}_\alpha(\mathbf{x})$ is an estimate of
$\partial_\mu\log P(\mathbf{x})$, it is not a gradient, and thus
cannot be integrated to estimate $P(\mathbf{x})$). The modified
estimator proposed in \Eq{eq:F_corrected_estimator} can only be computed if the
projection basis is smooth, and would not apply to grid
coarse-graining, for instance. It requires knowledge
of the diffusion tensor $D_{\mu\nu}$, as discussed in
\Sec{sec:inhomogeneous}. Using this modified force estimator
allows precise reconstruction of the force field, circulation and
entropy production even in the presence of large measurement noise
(\Fig{fig:6D}B,D-G). The limiting factor on force inference due to
measurement noise then becomes the blurring of the spatial structure
of the process. For observations with a finite time step $\Delta t$,
the currents are also blurred by time discretization, introducing an
additional bias in the force estimator, and resulting
in an underestimate of the entropy production. Note however that this
finite $\Delta t$ effect only induces a bias on $\hat{v}_{\mu\alpha}$:
for an equilibrium, time-reversible process,
$\hat{v}_{\mu\alpha}\to 0$ and the force estimator reduces to
$D_{\mu\nu}\hat{g}_{\nu\alpha}$, which is independent of the
time-ordering of the data.

\begin{figure}[tbp]
  \centering
  \includegraphics[width=0.8\columnwidth]{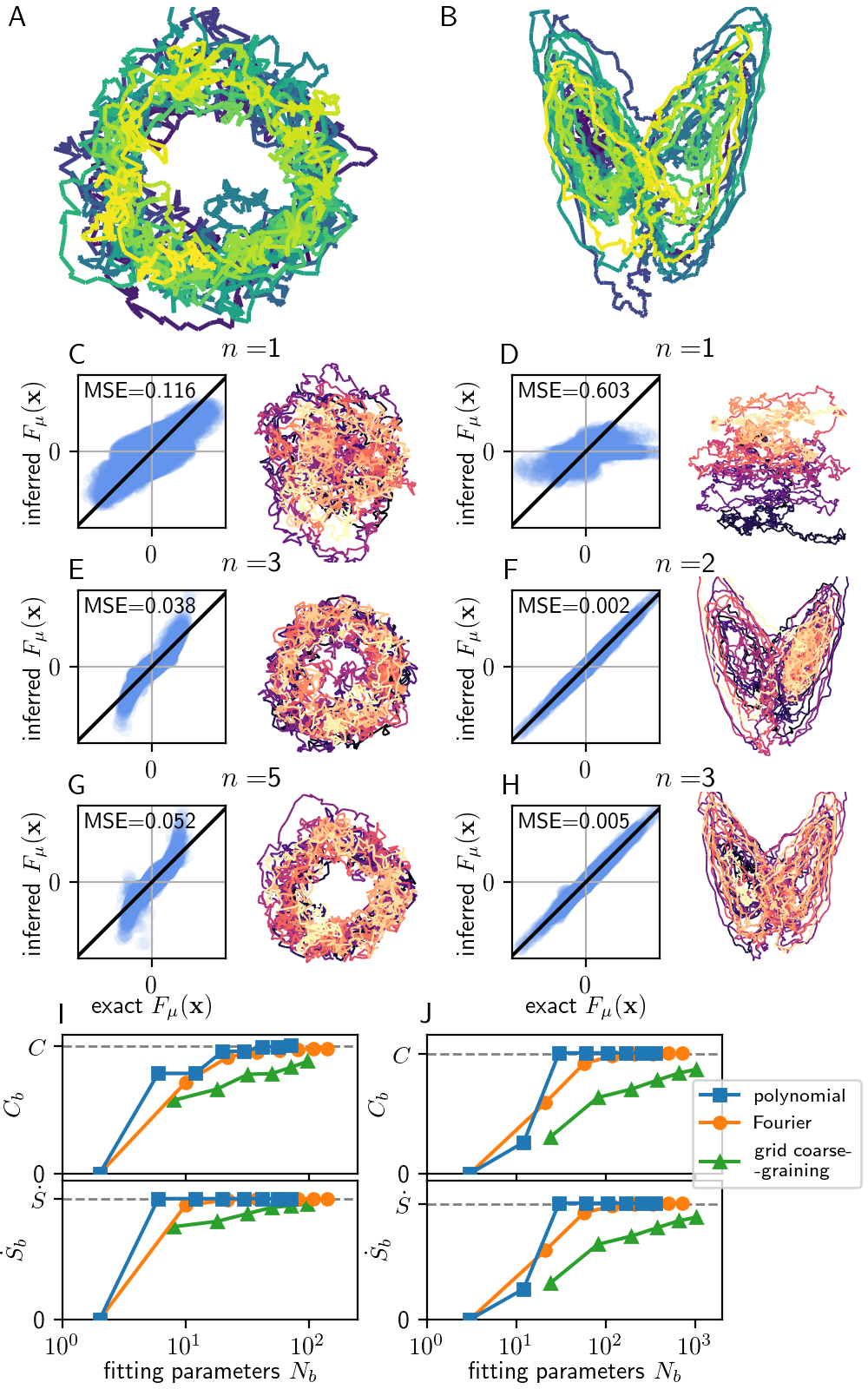}
  \caption{Stochastic force inference with non-linear force
    fields. \textbf{A.} Trajectory of an out-of-equilibrium process
    with harmonic trapping and circulation, and a Gaussian repulsive
    obstacle in the center. The force field is given by
    $F_\mu(\mathbf{x}) = -\Omega_{\mu\nu} x_\nu + \alpha
    e^{-x^2/2\sigma^2} x_\mu$ where $\Omega$ has both a symmetric and
    antisymmetric part. \textbf{B.} Trajectory of the stochastic
    Lorenz process, a 3D process with a chaotic attractor. The force
    field is $F_x = s(y-x),\ F_y = rx-y-zx,\ F_z = xy - bz$, where we
    choose $r=10$, $s=3$, and $b=1$. \textbf{C-H.} SFI for these two
    trajectories, respectively with polynomials of order $n=1,3,5$ and
    $n=1,2,3$: inferred force versus exact force (\emph{left}) and
    bootstrapped trajectory using the inferred force field
    (\emph{right}). \textbf{I-J.} Capacity (top) and entropy
    production (bottom) of each process projected on different bases
    for an asymptotically long trajectory, as a function of the number
    of degrees of freedom $N_b$ in the basis. These bases are
    polynomial and Fourier functions with order $n=0\dots 7$, and a
    coarse-grained approximation with a variable number of grid cells
    $n=2 \dots 7$ in each dimension.  }
  \label{fig:nonlinear}
\end{figure}

\paragraph{Inferring nonlinear force fields.}

We have so far considered only the case of linear systems projected
onto linear functions. In general, force fields are nonlinear, which
can result in a complex spatial structure. We illustrate this in \Figs{fig:nonlinear}A-B for processes with, respectively, non-polynomial forces and a
complex attractor~\cite{allawala_statistics_2016}.  For such
processes, SFI with a linear basis captures the covariance of the data and
the circulation of its current. However, it fails to reproduce finer
features, as evident by inspecting bootstrapped trajectories generated
using the inferred force field (\Fig{fig:nonlinear}C-D).  A better approximation of
the force can be obtained by expanding the projection basis, for
instance by including higher-order polynomials
$\{x_\mu x_\nu\},\{x_\mu x_\nu x_\rho\}\dots$ (\Fig{fig:nonlinear}E-H) or Fourier
modes. The captured fraction of the capacity and entropy production
increases monotonically when expanding the basis (\Fig{fig:nonlinear}I-J),
corresponding to finer geometrical details: the force field is well
resolved if the measured capacity does not increase upon further
expansion of the basis.  However, expanding the basis also results in
an increase in the number of parameters to infer, which eventually
leads to overfitting. 

\paragraph{Adjusting the basis size to the data.}

For a finite trajectory, there is therefore a trade-off between the
precision of the inferred force and the completeness of the force
field representation. This is demonstrated in \Fig{fig:overfitting}A-B
by plotting the force inference error along the trajectory as a
function of the number $N_b$ of degrees of freedom in the basis. At
small $N_b$, this error decreases, as it mostly originates from
underfitting. At large $N_b$, the error increases, as all
statistically significant information is already captured and adding
new functions primarily fits the noise.  This is reflected in the
inferred information $\hat{I}_b$ which steadily increases with the
number of fitting parameters $N_b$: the increase is initially mainly
due to the increase in the captured information $I_b$, but as $N_b$
grows, so does the typical error on $\hat{I}_b$,
$\delta \hat{I}_b \approx \sqrt{2\hat{I}_b + N_b^2/4 }$, and this error eventually overwhelms the
gain in $I_b$.  As a practical criterion to optimize between under-
and overfitting and best estimate the force along the trajectory, we
thus propose to use the basis $b$ which maximizes the information
$I_b$ that can be statistically resolved.  In practice, we find that
choosing the basis size that maximizes $\hat{I}_b - \delta \hat{I}_b$
(\emph{i.e.} the inferred information minus one standard deviation)
robustly selects the optimal basis size for a given trajectory (star
symbols on \Fig{fig:overfitting}A-B).  An alternative optimization
procedure, based on a similar balance, was suggested
in~\cite{comte_penalized_2007} for one-dimensional processes. We
empirically observe that when using this criterion to adapt the basis
to the trajectory, the typical squared error on force inference scales
as $\tau^{-1/2}$ with the trajectory duration $\tau$
(\Fig{fig:overfitting}C-D). There is an exception to this scaling:
when the force field can be exactly represented by a finite number of
functions of the basis, such as the Lorenz process with order 2
polynomials, this same criterion selects the smallest adapted basis:
further adding functions does not resolve more information. This
results in a faster convergence of the force field as $\tau^{-1}$
(\Fig{fig:overfitting}D), which is the rate of convergence of the
force projections for a given basis size.

\begin{figure}[t!]
  \centering 
  \includegraphics[width=\columnwidth]{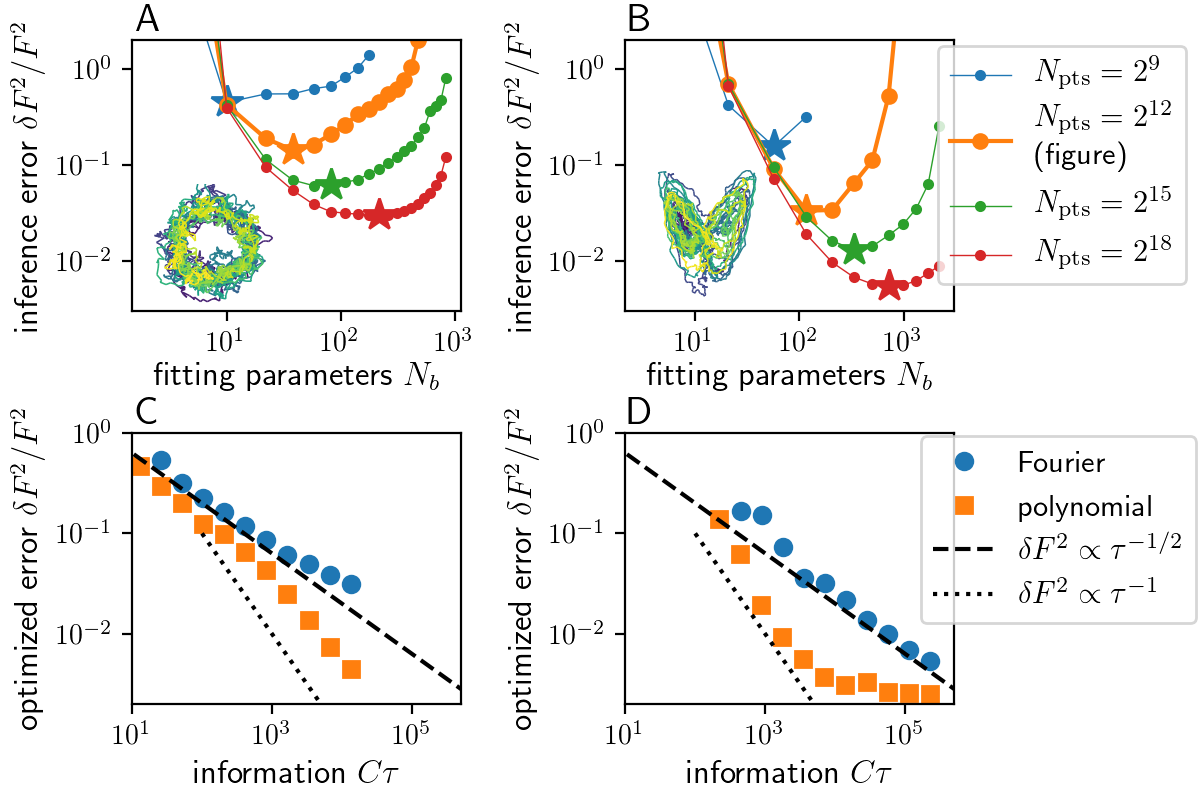}
  \caption{ Influence of the size of the basis on the precision of
    SFI. \textbf{A-B.} SFI error as a function of the number of fit
    parameters, respectively for the models presented in
    \Fig{fig:nonlinear}A-B, with a Fourier basis, and for different
    numbers of time steps in the trajectory.  Specifically, the
    $y$-axis is the mean squared relative error on the inferred force
    along the trajectory,
    $\av{(\hat{F}_\mu-F_\mu) D^{-1}_{\mu\nu} (\hat{F}_\nu-F_\nu)}
    / \av{\hat{F}_\mu D^{-1}_{\mu\nu} \hat{F}_\nu}$. The
    crossover from under- to overfitting is apparent, and takes place
    at larger $N_b$ and lower error with longer trajectories. The star
    symbols indicate the optimal basis size predicted by our
    self-consistent criterion of maximizing
    $\hat{I}_b - \delta \hat{I}_b$. \textbf{C-D}. The squared error as
    a function of the amount of information $C\tau$ in a trajectory of
    duration $\tau$, for the optimal basis, averaged over $n=3$
    trajectories. For the Lorenz process with a polynomial basis (D,
    orange squares), the convergence is fast as the basis is adapted
    to the exact force field, and the saturation of the error to a
    lower plateau is due to the finite time step.  }
  \label{fig:overfitting}
\end{figure}

\begin{figure}[t]
  \centering
  \includegraphics[width=\columnwidth]{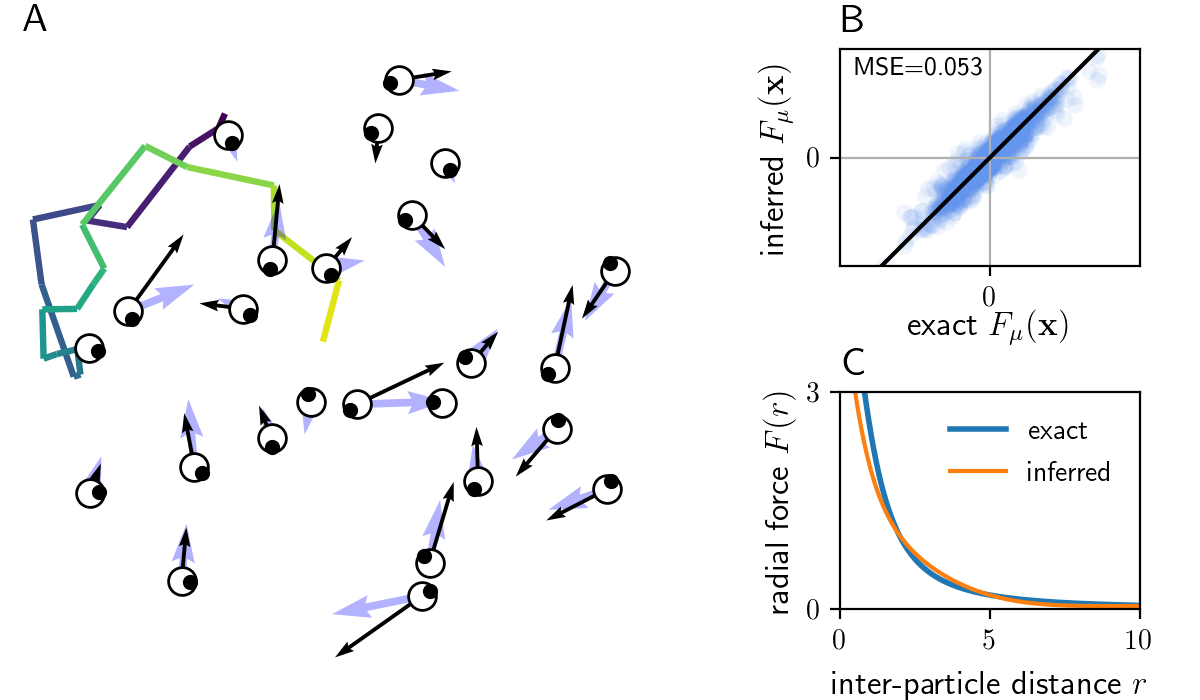}
  \caption{Stochastic force inference for harmonically trapped
    active Brownian particles with soft repulsive interactions
    $F(r) = 1/(1+r^2)$ between particles at distance $r$. \textbf{A.}
    Snapshot of a configuration for $25$ active particles. The black
    dots indicate the direction of self-propulsion. We perform SFI on
    a trajectory of only $25$ frames, blurred to mimic measurement
    noise. Background shows the trajectory of one particle, and force
    on each particle, inferred (blue arrows) and exact (black
    arrows). The fitting basis for SFI consists in a combination of
    harmonic trapping, constant velocity self-propulsion and radial
    interactions between particles with the form $r^k e^{-r/r_0}$ with
    $k = 0...5$ and $r_0$ a typical nearest-neighbour distance between
    particles. \textbf{B.} Inferred \emph{versus} exact components of
    the force on all particles along the trajectory. \textbf{C.}
    Inferred radial force between interacting particles, compared to
    the exact force.}
  \label{fig:particles}
\end{figure}

\paragraph{Systems of particles.}

Systems with many degrees of freedom, such as active interacting
particles (\Fig{fig:particles}A), are challenging to treat. Indeed,
with limited data, the criterion $\hat{I}_b \gg N_b$ precludes even
the inference of gross features of the force field. In such cases
however, the use of symmetries can make the problem tractable. For
instance, treating particles as identical implies that forces are
invariant under particle exchange, which greatly reduces the number of
parameters to infer. Forces can then be expanded as one-particle
terms, pair interactions, and higher orders, by choosing an
appropriate basis. With this scheme, a large
number of particles actually results in enhanced statistics, allowing
accurate inference of the force components (\Fig{fig:particles}A-B)
and reconstruction of the pair interactions (\Fig{fig:particles}C)
with a limited amount of data. This method could be straightforwardly
extended to include, \emph{e.g.}, alignment interactions between
particles. In contrast to standard methods to infer pair interaction
potentials, we do not rely here on an equilibrium assumption.

\section{Inference with multiplicative noise}
\label{sec:inhomogeneous}

\paragraph{It\^o drift and physical force.}

We have so far assumed that the diffusion tensor does not depend on
the state of the system. While this is a natural first approximation,
it is rarely strictly the case: for instance, the mobility of colloids
depends on their distance to walls and other colloids due to
hydrodynamic interactions~\cite{lau_state-dependent_2007}. In order to
mathematically describe Brownian dynamics in the presence of an
inhomogeneous diffusion tensor $D_{\mu\nu}(\mathbf{x})$,
\Eq{eq:LangevinSFI} should be modified into
\begin{equation}
  \dot{x}_\mu = \Phi_\mu(\mathbf{x})  +  \sqrt{2D(\mathbf{x})}_{\mu\nu} \xi_\nu,
  \label{eq:LangevinSFI_inhomogenous}
\end{equation}
written in the It\^o convention, \emph{i.e.} evaluating
$\mathbf{D}(\mathbf{x})$ at the start of the step. Here $\Phi_\mu$ is
the drift, which relates to the physical force through
\begin{equation}
  \label{eq:drift}
  \Phi_\mu(\mathbf{x}) = F_\mu(\mathbf{x}) + \partial_\nu
  D_{\mu\nu}(\mathbf{x}).
\end{equation}
The additional term $\partial_\nu D_{\mu\nu}$, sometimes called
``spurious force'', combines with the noise term to ensure that the
dynamics does not induce currents and probability gradients in the
absence of forces~\cite{lau_state-dependent_2007}. To our knowledge,
the only way to infer the physical force is to infer both terms in
\Eq{eq:drift} independently, and involves taking gradients of the
inferred diffusion. Here we show how to infer both the diffusion field
and the drift field, following the same idea as in \Sec{sec:SFI}.

\begin{figure}[ht]
  \centering
  \includegraphics[width=\columnwidth]{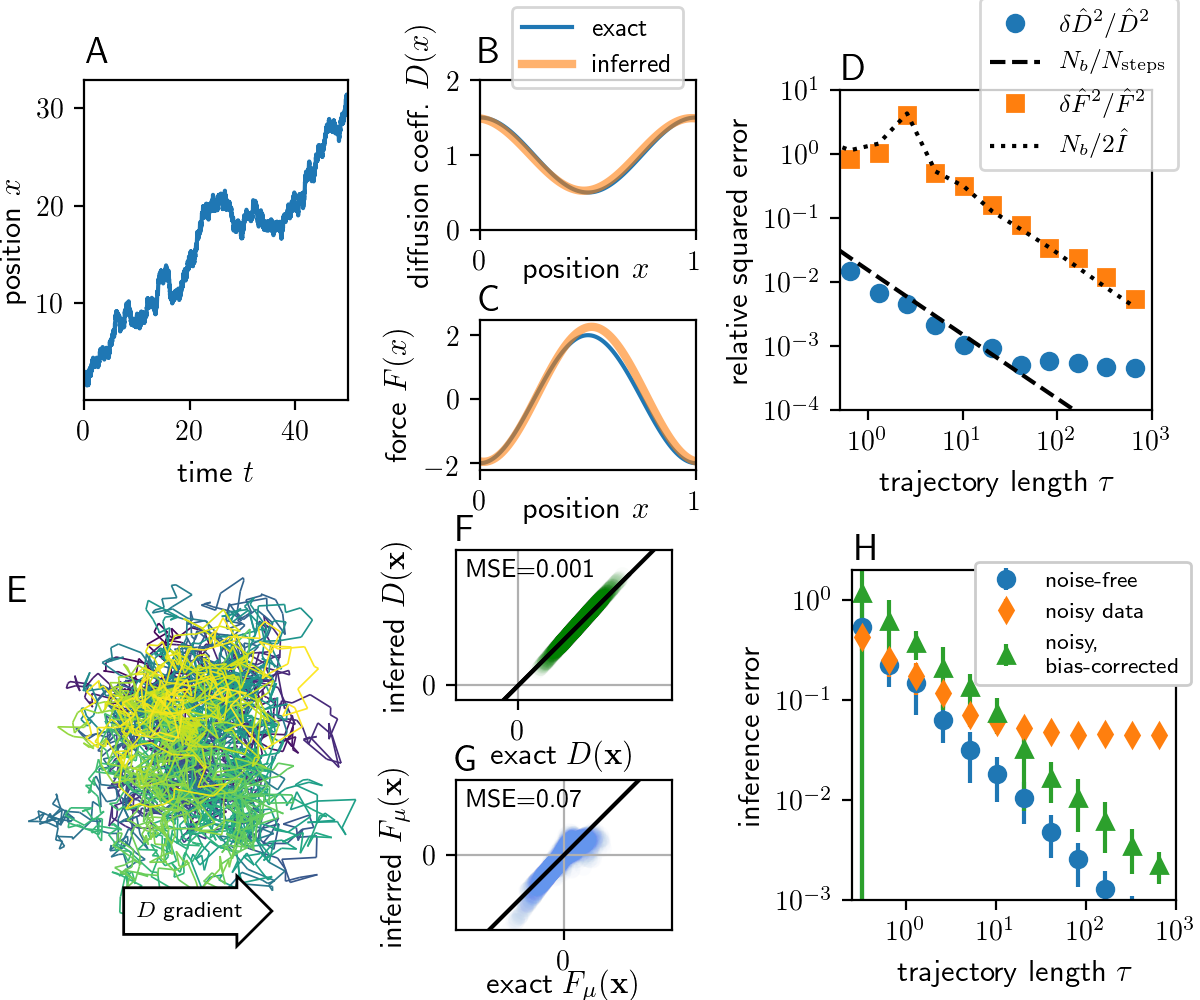}
  \caption{Stochastic inference of inhomogeneous diffusion and
    forces. \textbf{A.} A trajectory of a 1D ratchet model with
    $F(x)= F_0\cos(2\pi x)$ and $D(x) = 1 + a \cos(2\pi x)$, with
    periodic boundary conditions. \textbf{B-C} For the trajectory
    presented in A, inferred and exact diffusion coefficient (using
    \Eq{eq:D}) and force field (using \Eq{eq:F_D}) as a function of
    position. We use a $1$st order Fourier basis to infer both force
    and diffusion.  \textbf{D.} Analysis of the convergence of the
    diffusion (blue) and force (orange) estimators, as a function of
    trajectory duration, for the process presented in A. The dotted
    and dashed black lines are the self-consistent estimates for the
    squared error, respectively for the diffusion and the force. The
    plateau for the diffusion inference is due to the finite time
    step. \textbf{E.} A trajectory of a minimal 2D model, an isotropic
    harmonic trap at equilibrium,
    $F_\mu (\mathbf{x}) = - D_{\mu\nu}(\mathbf{x}) x_\nu$, in a
    constant gradient of isotropic diffusion,
    $D_{\mu\nu}(\mathbf{x}) = (1 + a_\rho x_\rho) \delta_{\mu\nu}$.
    \textbf{F-G.} Inferred \emph{versus} exact diffusion coefficient
    (using \Eq{eq:D}) and force components (using \Eq{eq:F_D}) along
    trajectory A. A linear polynomial basis was used to fit the
    diffusion coefficient, and a quadratic basis to fit $F_{\mu}$.
    \textbf{D.} Convergence of the diffusion projection estimator
    (normalized by the average diffusion tensor) to its exact value
    for the process shown in A. Circles: using \Eq{eq:D}, diamonds:
    using \Eq{eq:D} in the presence of time-uncorrelated measurement
    noise; triangles: using the bias-corrected local estimator. Error
    bars represent the standard deviation over $64$
    samples.   }
  \label{fig:diffusion}
\end{figure}

\paragraph{Inference of a state-dependent diffusion tensor.}

We propose to approximate $D_{\mu\nu}(\mathbf{x})$ by its projection
as a linear combination of known functions,
$D_{\mu\nu}(\mathbf{x}) \approx D_{\mu\nu\alpha} c_\alpha(\mathbf{x})$
with
$ D_{\mu\nu\alpha} = \int D_{\mu\nu}(\mathbf{x}) c_\alpha(\mathbf{x})
P(\mathbf{x})\dd \mathbf{x}$.  As before, we can estimate the projectors $\hat{c}_\alpha$
using trajectory averages; the only missing ingredient is a local
estimate $\hat{d}_{\mu\nu}(t_i)$ for the diffusion tensor
$D_{\mu\nu}(\mathbf{x}(t_i))$. Such an estimator can be constructed as
$\hat{d}_{\mu\nu}(t_i) = \Delta x_\mu(t_i) \Delta x_\nu(t_i) / 2\Delta t$, so that our
estimator for $D_{\mu\nu\alpha}$ reads
\begin{equation}
  \hat{D}_{\mu\nu\alpha} = \frac{1}{\tau} \sum_i \hat{d}_{\mu\nu}(t_i) \hat{c}_\alpha (\mathbf{x}(t_{i})) \Delta t  
  \label{eq:D}
\end{equation}
The relative error on these projection coefficients is of order
$\sqrt{N_b \Delta t/\tau}$. Similarly to \Eq{eq:F_moments_empirical} for the force field, \Eq{eq:D} corresponds to a linear regression of $\hat{d}_{\mu\nu}(t_i)$, and was previously suggested for one-dimensional systems in~\cite{comte_penalized_2007}. We test this estimator using
two minimal models: a one dimensional ratchet process with sinusoidal
force and diffusion coefficient, inspired by the B\"uttiker-Landauer
model~\cite{buttiker_transport_1987,landauer_motion_1988}
(\Fig{fig:diffusion}A-D); and a two-dimensional process in a harmonic
trap with a constant diffusion gradient (\Fig{fig:diffusion}E-H). We
quantitatively recover the diffusion coefficient as a function of
position (\Fig{fig:diffusion}B,F) and confirm that the error vanishes
in the limit of long trajectories
(\Fig{fig:diffusion}D,H).

\paragraph{Correcting measurement noise biases.}

Importantly, the estimator introduced in
\Eq{eq:D} is biased in the presence of noise on the measured
$\mathbf{x}$, and becomes effectively useless if this noise is larger
than the typical $\Delta \mathbf{x}$. Inspired by the estimator
proposed by Vestergaard \emph{et al.}~\cite{vestergaard_optimal_2014}
for homogeneous, isotropic diffusion, we define a bias-corrected local
estimator
\begin{equation}
 \hat{\mathbf{d}}(t_i) = \frac{(\Delta \mathbf{x}(t_{i-1}) +\Delta
\mathbf{x}(t_i))^2}{ 4\Delta t} +  \frac{\Delta \mathbf{x}(t_i) \Delta \mathbf{x}(t_{i-1})}{ 2\Delta t}
\label{eq:Vestergaard}
\end{equation}
where tensor products are implied.  Modifying \Eq{eq:D} accordingly
thus corrects measurement noise bias (\Fig{fig:diffusion}H), at the
price of an increased relative error for short
trajectories.

\paragraph{Inferring the It\^o drift.}
We also approximate the drift as a linear combination of functions,
$\Phi_\mu(\mathbf{x}) = \Phi_{\mu\alpha}
c_\alpha(\mathbf{x})$. Equation~\ref{eq:F_moments_empirical} provides
an estimator for the projection coefficients $\Phi_{\mu\alpha}$ in
terms of an It\^o integral. This estimator is however impractical for
experimental data, as even moderate measurement noise induces large
errors in these coefficients. As in \Eq{eq:F_corrected_estimator}, we
exploit the It\^o-to-Stratonovich conversion to obtain an estimator
that is not biased by measurement noise:
\begin{equation}
  \label{eq:phi_strato}
    \hat{\Phi}_{\mu\alpha} = \hat{v}_{\mu\alpha} - \frac{1}{\tau} \sum_i  \hat{d}_{\mu\nu}(t_i)  \partial_\nu 
\hat{c}_\alpha(\mathbf{x}(t_i)) \Delta t
\end{equation}
where $\hat{v}_{\mu\alpha}$ is the velocity projection coefficient
(\Eq{eq:v}), and $\hat{d}_{\mu\nu}(t_i)$ can either be the
local biased-corrected estimator (\Eq{eq:Vestergaard}) or another
estimator of $D_{\mu\nu}(\mathbf{x}_i)$. The convergence properties of
$\hat{\Phi}_{\mu\alpha}$ to its asymptotic value are similar to those
of \Eq{eq:F_moments_empirical}.

\paragraph{Reconstructing the physical force.}

In many cases of interest, inference of the It\^o drift
(\Eq{eq:phi_strato}) and diffusion tensor (\Eq{eq:D}) provides a
satisfactory characterization of the system: it is, for instance,
sufficient to simulate new trajectories. In strongly
out-of-equilibrium cases with non-thermal noise, such as in cell
dynamics, or large-scale stochastic systems such as climate
fluctuations and financial markets, this is amply sufficient. However,
in soft matter systems with thermal noise obeying a
fluctuation-dissipation relation, knowledge of the physical force --
and of the underlying potential -- is important. To estimate it, we
can combine our diffusion (\Eq{eq:D}) and drift (\Eq{eq:phi_strato})
projection estimators to reconstruct the force field,
\begin{equation}
  \label{eq:F_D}
    \hat{F}_{\mu}(\mathbf{x}) = \hat{\Phi}_{\mu\alpha}c_\alpha(\mathbf{x}) - \hat{D}_{\mu\nu\alpha} \partial_\nu \hat{c}_\alpha(\mathbf{x})
\end{equation}
using \Eq{eq:drift}. This estimator allows for quantitative inference
of the force provided that the divergence of the diffusion coefficient
is well approximated. We demonstrate this (\Fig{fig:diffusion}C,D,G)
for the simple processes presented in \Fig{fig:diffusion}A,E using an
adapted basis to fit the diffusion coefficient.  Note however that
since \Eq{eq:F_D} requires taking the divergence of the fitted
diffusion field, we have no control over the inference error in this
estimator, which somewhat hinders the practicality of this estimator.

\section{Estimating the error of SFI}
\label{sec:SFI_error}

In this Section, we derive the core results of our article: how to
perform SFI in practice, and self-consistently estimate the error in
the inference. 

\paragraph{The force as a trajectory average}

To be able to deduce the force from the trajectory one
first needs an expression for the force in terms of measurable
quantities along the trajectory. We have 
\begin{equation}
\mathbf{F}(\mathbf{x})= \lim_{\epsilon\to0}\av{\left.\frac{(\mathbf{x}(t+\epsilon)-\mathbf{x}(t))}{\epsilon}\right|\mathbf{x}(t) = \mathbf{x}} =\av{\dot{\mathbf{x}}^{+}|\mathbf{x}(t)} = \av{\delta(\mathbf{x}(t)-\mathbf{x})\dot{\mathbf{x}}^{+}}/P(x)
\label{eq:full_force_from_traj}
\end{equation}
where $\av{\left. \cdot \right|\mathbf{x}(t) = \mathbf{x}}$ means
averaging over realizations of the noise, conditioned on being at
position $\mathbf{x}$ at time $t$. We have defined here
$\dot{\mathbf{x}}^{+}$ as the right hand derivative, corresponding to
It\^o calculus (see Appendix A of \cite{chetrite_eulerian_2009}).  The
coefficients of the force field in its decomposition with respect to
the phase space projector $c_\alpha(\mathbf{x})$ are:
\begin{equation}
\begin{split}
F_{\mu\alpha} =  \int d\mathbf{x} P(\mathbf{x}) F_{\mu}(\mathbf{x}) c_\alpha(\mathbf{x})=\int d\mathbf{x}  \av{\delta(\mathbf{x}(t)-\mathbf{x})\dot{x}_{\mu}^{+}}c_\alpha(\mathbf{x}) \\
=  \av{\int d\mathbf{x} \delta(\mathbf{x}(t)-\mathbf{x})\dot{x}_{\mu}^{+}c_\alpha(\mathbf{x})} = \av{\dot{x}_{\mu}^{+}c_\alpha(\mathbf{x})}
\end{split} 
\label{eq:Fn}
\end{equation}
Because of this last expression, the force projection coefficient
$F_{\mu\alpha}$ can be expressed as an average quantity along an
infinitely long trajectory, which can thus be estimated by computing
it on a finite trajectory.

Note that, similarly to the force, the phase space velocity can also be defined through an average of $\mathbf{\dot{x}}$, where the time derivative is taken in the Stratonovich sense:
\begin{align}
\mathbf{v}(\mathbf{x}) = & \lim_{\epsilon\to0}\av{\left.\frac{(\mathbf{x}(t+\epsilon)-\mathbf{x}(t-\epsilon))}{2\epsilon}\right|\mathbf{x}(t)=\mathbf{x}} =\left.\av{\frac{1}2(\dot{\mathbf{x}}^{+}+\dot{\mathbf{x}}^{-})\right|\mathbf{x}(t)=\mathbf{x}} \\  = & \av{\delta(\mathbf{x}(t)-\mathbf{x})\frac{1}2(\dot{\mathbf{x}}^{+}+\dot{\mathbf{x}}^{-})}/P(\mathbf{x})
\end{align}
(see Appendix A of \cite{chetrite_eulerian_2009}).
The phase space velocity in its decomposition with respect to the phase space basis $c_\alpha(\mathbf{x})$ is, analogously to the force,:
\begin{equation}
\begin{split}
v_{\mu\alpha}  = \av{\frac{1}2(\dot{x}_{\mu}^{+}+\dot{x}_{\mu}^{-})c_\alpha(\mathbf{x})}
\end{split}
\label{eq:vn}
\end{equation}

\paragraph{Projection on the empirical basis}
The second difficulty in evaluating Eq.2 of the main text in practice
is that the phase space measure $P(\mathbf{x})$ is unknown in
practice. As a consequence, the phase space basis,
$c_\alpha(\mathbf{x})$ is not known either, as it is the
orthonormalized basis derived from $b$ using $P$ as the measure.  Our
approach consists in approximating $P(\mathbf{x})$ by the empirical
measure
\begin{equation}
\hat{P}_{\tau}(\mathbf{x}) = \frac{1}{\tau}\int_0^{\tau} \delta(\mathbf{x} - \mathbf{x}(t))dt
\end{equation}
corresponding to a time average along the trajectory.

We then define the empirical projector $\hat{c}_\alpha$ with respect
to this measure, as in the main text:
\begin{equation} 
\hat{c}_\alpha(\mathbf{x}) = \hat{B}^{-1/2}_{\alpha\beta} b_\beta(\mathbf{x}) \qquad \mathrm{with} \qquad
\hat{B}_{\alpha\beta} = \int b_\alpha(\mathbf{x}) b_\beta(\mathbf{x})
\frac{\dd t}{\tau}.
\label{eq:c_empirical}
\end{equation}
In the long-trajectory limit, these ``empirical projectors''
$\hat{c}_\alpha(\mathbf{x})$ converge to the phase-space projectors
${c}_\alpha(\mathbf{x})$; more precisely, we expect that for typical
trajectories
$\hat{c}_\alpha(\mathbf{x}) = {c}_\alpha(\mathbf{x}) +
O(\sqrt{\tau_0/\tau})$, where $\tau$ is the duration of the trajectory
and $\tau_0$ is a relaxation time of the system.  In the case of the
polynomial basis for instance, the convergence of the basis at order
$n$ is related to the convergence of the $n$-th cumulant of the
probability distribution function. We do not seek to make this
statement more mathematically precise here.

As an intermediate variable for this calculation, we define the
projection coefficients $F^{\tau}_{\mu\alpha}$ of the (exact) force
onto these empirical projectors. These coefficients are trajectory
dependent; however, $\hat{c}_\alpha$ are directly accessible from the trajectory, as is the empirical measure with respect to which they are projectors, so that obtaining the coefficients $F^{\tau}_{\mu\alpha}$ precisely, would result in an accurate approximation of the force field $F_\mu \approx F^{\tau}_{\mu\alpha}\hat{c}_\alpha$ along the trajectory. For this reason, we focus here on how the estimator
$\hat{F}_{\mu\alpha}$ as defined in \Eq{eq:F_moments} of the main text converges
to $F^{\tau}_{\mu\alpha}$.  The relative errors presented in the main text also refer to this convergence (rather than the convergence to the phase-space projection $F_{\mu\alpha}$). Recall that our estimator is given by
\begin{align}
\hat{F}_{\mu\alpha} & = \frac{1}{\tau}\int^{\text{It\^o}} \hat{c}_\alpha(\mathbf{x}) d\mathbf{x}_t^{\mu} \\
& =  \underbrace{\frac{1}{\tau}\int_0^{\tau} \hat{c}_\alpha(\mathbf{x}) F_{\mu}(\mathbf{x})dt}_{F^\tau_{\mu\alpha}} +\underbrace{\frac{1}{\tau}\int^{\text{It\^o}}\hat{c}_\alpha(\mathbf{x}) \sqrt{2} D^{1/2}_{\mu\nu}d\xi_t^{\nu}}_{Z_{\mu\alpha}} 
\label{eq:Fn_estimator}
\end{align}
using the Langevin equation (\ref{eq:langevin}).  Since
$F^{\tau}_{\mu\alpha}$ is what we wish to infer, we propose to study
now the statistics of
$Z_{\mu\alpha} = \hat{F}_{\mu\alpha} - F^{\tau}_{\mu\alpha}$,
\emph{i.e.} its mean and variance.

\paragraph{Statistics of the error in the inference of the projection coefficients}
\label{sec:stats_Z}

We thus study the first and second moment of the random tensor
$Z_{\mu\alpha}$, \emph{i.e.} respectively the systematic bias and the
typical error of $\hat{F}_{\mu\alpha}$ as an estimator of
$F^{\tau}_{\mu\alpha}$. To make the norm of these moments meaningful,
it is necessary here to go to dimensionless coordinates: indeed,
different phase space coordinates can have different dimensions (such
as, for instance, a phase space comprising both distances and angles,
as in \Fig{fig:particles} of the main text), and thus different
coordinates of $Z_{\mu\alpha}$ cannot be compared or summed. To this
end, we define $W_{\mu\alpha} = D^{-1/2}_{\mu\nu} Z_{\nu\alpha}$, all
the coordinates of which have the dimension of $t^{-1/2}$.

First recall that we defined both phase-space and empirical
projectors as a linear combination of the basis functions $b$,
$c_\alpha=B_{\alpha\beta}^{-1/2} b_{\beta} $ and
$\hat{c}_\alpha =\hat{B}_{\alpha\beta}^{-1/2} b_{\beta} $, where
\begin{align} \label{eq:B}
B_{\alpha\beta} = \int d\mathbf{x} P(\mathbf{x}) b_{\beta}(\mathbf{x}) b_{\alpha}(\mathbf{x}) && \hat{B}_{\alpha\beta} = \int_0^{\tau}\frac{dt}\tau b_{\beta}(\mathbf{x}(t)) b_{\alpha}(\mathbf{x}(t))
\end{align}
Thus we have
$\lim_{\tau\to \infty} \hat{B}_{\alpha\beta}^{-1/2} =
B_{\alpha\beta}^{-1/2}$ and $\av{\hat{B}_{\alpha\beta}}=B_{\alpha\beta}$.
Let us denote
$\Delta_{\alpha\beta} = B_{\alpha\gamma}^{1/2}\hat{B}_{\gamma\beta}^{-1/2}-\delta_{\alpha\beta} 
$
the dimensionless error on the orthonormalization matrix (indeed, the
basis functions $b_\alpha$ can in principle have a dimension). We have
$\lim_{\tau\to \infty}\Delta_{\alpha\beta} =0$; typically, we'll have
more precisely $\Delta_{\alpha\beta} = O(1/\sqrt{\tau})$,
corresponding to the convergence of trajectory integrals to
phase-space integrals in \Eq{eq:B}. We then have
\begin{equation} \label{eq:Z_decomp}
Z_{\mu\alpha} \equiv \frac{1}{\tau}\int^{\text{It\^o}}\hat{c}_\alpha(\mathbf{x}) \sqrt{2} D^{1/2}_{\mu\nu}d\xi_t^{\nu} = B^{-1/2}_{\alpha\beta} \sqrt{2} D^{1/2}_{\mu\nu} \frac{1}{\tau}\int^{\text{It\^o}} b_\beta(\mathbf{x})d\xi_t^{\nu}+B^{-1/2}_{\alpha\beta}\Delta_{\beta\gamma} \sqrt{2} D^{1/2}_{\mu\nu} \frac{1}{\tau}\int^{\text{It\^o}} b_\gamma(\mathbf{x})d\xi_t^{\nu}.
\end{equation}
For the remainder of this Section we will denote the It\^o integral by a regular integration: $\int^{\text{It\^o}} d\xi_t^{\nu}=\int_0^{\tau} d\xi_t^{\nu}$.
We now put an upper bound on the first moment of $Z_{\mu\alpha}$,
\emph{i.e.} on the systematic bias.  Note that the first term in
\Eq{eq:Z_decomp} has zero average, as it is linear in the noise. In
contrast, due to possible correlations between the noise and the
random variable $\Delta_{\alpha\beta}$, the second term may not
average to zero.  Going to dimensionless coordinates, we use the
Cauchy-Schwarz inequality to bound the norm of this bias:
\begin{equation}
\begin{split}
\|\av{W_{\mu\alpha} }\|^2 =& \left\|\av{ B^{-1/2}_{\alpha\beta}\Delta_{\beta\gamma} \frac{1}{\tau}\int_0^{\tau}  b_\gamma(\mathbf{x}) D^{-1/2}_{\mu\nu}\sqrt{2}D^{1/2}_{\nu\rho}  d\xi_t^{\rho} }\right\|^2  \leq 2 B^{-1}_{\beta\delta}\av{\Delta_{\beta\rho}\Delta_{\rho\delta}}\av{\frac{1}{\tau^2}  \int_0^{\tau}  b_\gamma(\mathbf{x}) d\xi_t^{\mu} \int_0^{\tau}  b_\gamma(\mathbf{x})d\xi_{t'}^{\mu}} 
\end{split}
\end{equation}
We can then use the It\^o isometry
relation~\cite{gardiner_stochastic_2009} to prove that
\begin{equation}
  \av{\int_0^{\tau}  b_\alpha(\mathbf{x})d\xi_t^{\mu} \int_0^{\tau}  b_\beta(\mathbf{x})d\xi_{t'}^{\mu}} =  \av{ \int_0^{\tau}  b_\alpha(\mathbf{x}(t))   b_\beta(\mathbf{x}(t))dt} = \av{\hat{B}_{\alpha\beta} }
\label{eq:isometry}
\end{equation}
which implies that 
\begin{equation}
  \|\av{W_{\mu\alpha} }\|^2  \leq \frac{2}{\tau} B^{-1}_{\beta\delta}\av{\Delta_{\beta\rho}\Delta_{\rho\delta}}\av{\hat{B}_{\gamma\gamma} }
\end{equation}
Since $\Delta_{\alpha\beta} = O(\tau^{-1/2})$, we thus have
$\av{W_{\mu\alpha} } = O(1/\tau)$, which corresponds to a fast
convergence of the bias towards zero: the bias is negligible compared
to the fluctuating part of inference error, which goes as
$O(\tau^{-1/2})$.

Indeed, let us now compute the second moment of $W_{\mu\alpha}$.  We have
\begin{equation}
\av{W_{\mu\alpha}W_{\nu\beta} } = \frac{2}{\tau^2}   \av{ \hat{B}^{-1/2}_{\alpha\gamma} \hat{B}^{-1/2}_{\beta\delta} \int_0^{\tau} \int_0^{\tau} d\xi_t^{\mu} d\xi_{t'}^{\nu} b_\gamma(\mathbf{x}(t))  b_\delta(\mathbf{x}(t')) } 
\end{equation}
As $\hat{B}^{-1/2}_{\alpha\gamma}$ depends on all values of $t$, it is
not \emph{adapted} to the Wiener process $d\xi_t^{\mu}$, and thus we
cannot apply the It\^o isometry. However, we have
$\hat{B}^{-1/2}_{\alpha\gamma} =
{B}^{-1/2}_{\alpha\beta}(\delta_{\beta\gamma}+\Delta_{\beta\gamma})$.
Applying the It\^o isometry
(\Eq{eq:isometry}) yields:
\begin{align} \label{eq:varW}
\av{W_{\mu\alpha}W_{\nu\beta} } & = \frac{1}{\tau^2} \delta_{\mu\nu}  {B}^{-1/2}_{\alpha\gamma} {B}^{-1/2}_{\beta\delta}  2 \tau \av{\hat{B}_{\gamma\delta} } + R_{\mu\alpha\nu\beta}    \\
& = \frac{2}{\tau} \delta_{\mu\nu}\delta_{\alpha\beta} + R_{\mu\alpha\nu\beta} 
\end{align}
where we have defined the remainder 
\begin{equation}
R_{\mu\alpha\nu\beta} = \frac{2}{\tau^2}   \av{\left({B}^{-1/2}_{\alpha\gamma}{B}^{-1/2}_{\beta\lambda}\Delta_{\lambda\delta}+ {B}^{-1/2}_{\alpha\lambda}\Delta_{\lambda\gamma} \hat{B}^{-1/2}_{\beta\delta} \right)\int_0^{\tau} \int_0^{\tau} d\xi_t^{\mu} d\xi_{t'}^{\nu} b_\gamma(\mathbf{x}(t))  b_\delta(\mathbf{x}(t')) } 
\end{equation}
which is, as we show now, subleading in \Eq{eq:varW}. We now wish to bound the amplitude of the remainder $|\av{W_{\mu\alpha}W_{\mu\alpha} }-\frac{2}{\tau} N_b| =|R_{\mu\alpha\mu\alpha}|$. Since for typical trajectories $\Delta_{\alpha\beta} = O(\tau^{-1/2})$, we can bound every element of the matrix $|{B}^{-1/2}_{\alpha\gamma}{B}^{-1/2}_{\alpha\lambda}\Delta_{\lambda\delta}+{B}^{-1/2}_{\alpha\lambda}\Delta_{\lambda\gamma} \hat{B}^{-1/2}_{\alpha\delta}| \leq R\cdot O_{\gamma\delta}$ for such trajectories, where $R = O(1/\sqrt{\tau})$ is a (non-fluctuating) number and $O_{\gamma\delta}$ is the matrix with ones at all places. We get
\begin{equation}
\begin{split}
|\av{W_{\mu\alpha}W_{\mu\alpha} }-\frac{2}{\tau} N_b| = \frac{2}{\tau^2}  \left|\av{ \left({B}^{-1/2}_{\alpha\gamma}{B}^{-1/2}_{\alpha\lambda}\Delta_{\lambda\delta}+{B}^{-1/2}_{\alpha\lambda}\Delta_{\lambda\gamma} \hat{B}^{-1/2}_{\alpha\delta}\right) \int_0^{\tau} \int_0^{\tau}   d\xi_t^{\mu} d\xi_{t'}^{\mu} b_\gamma(\mathbf{x}(t))  b_\delta(\mathbf{x}(t')) } \right|  \\
\leq  \frac{2}{\tau^2}  \av{ \left|\left({B}^{-1/2}_{\alpha\gamma}{B}^{-1/2}_{\alpha\lambda}\Delta_{\lambda\delta}+{B}^{-1/2}_{\alpha\lambda}\Delta_{\lambda\gamma} \hat{B}^{-1/2}_{\alpha\delta}\right)\right|\left| \int_0^{\tau} \int_0^{\tau}d\xi_t^{\mu} d\xi_{t'}^{\mu}  b_\gamma(\mathbf{x}(t))  b_\delta(\mathbf{x}(t'))\right| }
 \\
\leq  \frac{2}{\tau^2} R\cdot O_{\gamma\delta}\av{\left| \int_0^{\tau} \int_0^{\tau} d\xi_t^{\mu} d\xi_{t'}^{\mu}  b_\gamma(\mathbf{x}(t))  b_\delta(\mathbf{x}(t'))\right|} \\
\leq \frac{2}{\tau^2} R\cdot O_{\gamma\gamma}\av{ \left|\int_0^{\tau} \int_0^{\tau} d\xi_t^{\mu} d\xi_{t'}^{\mu}  b_\delta(\mathbf{x}(t))  b_\delta(\mathbf{x}(t'))\right|}
\\
= \frac{2}{\tau^2} R\cdot O_{\gamma\gamma}\av{  \int_0^{\tau}   d\xi_t^{\mu}  b_\delta(\mathbf{x}(t))  \int_0^{\tau}  d\xi_t^{\mu}  b_\delta(\mathbf{x}(t)) } = \frac{1}{\tau^2} R\cdot O_{\gamma\gamma} 2\tau \av{\hat{B}_{\delta\delta}} = O(1/\tau^{3/2}).
\end{split} 
\end{equation}
In the fourth line we have used that for two semi-definite matrices $M_{\alpha\beta}$ and $N_{\alpha\beta}$, $M_{\alpha\beta}N_{\beta\alpha} \leq\sqrt{ M^2_{\alpha\alpha}N^2_{\beta\beta}}\leq M_{\alpha\alpha}N_{\beta\beta}$, an identity based on the Cauchy-Schwarz inequality. In the fifth line we employed the the It\^o isometry
(\Eq{eq:isometry}).
Again, this subleading term originates from the convergence of the
empirical projected basis to its long-trajectory limit.

\paragraph{Self-consistent estimate of the error on the projected
  force}

The previous error estimates are rigorous, but require knowledge of
the exact force field to assess their amplitude. The goal of this
section is to provide approximate estimates of the typical error that
can be obtained using only the inferred force field, and are thus
useful in practical situations. Now that we know the statistical
properties of the dimensionless error term $W_{\mu\alpha}$, we can
write the covariance of the inferred force projection coefficients
explicitly:
\begin{equation}
\av{\left(\hat{F}_{\mu\alpha}-F^{\tau}_{\mu\alpha}\right)\left(\hat{F}_{\nu\alpha}-F^{\tau}_{\nu\alpha}\right)} = \frac{2 D_{\mu\nu}}{\tau}\delta_{\alpha\beta}(1+O(1/\sqrt{\tau}))
\label{eq:force_cov}
\end{equation}

Now, let us define the information along the trajectory by
\begin{equation}
I^{\tau}_b = \frac{1}{4}\tau F^{\tau}_{\mu\alpha}D^{-1}_{\mu\nu}F^{\tau}_{\nu\alpha}.
\end{equation}
In the long time limit, the rate of information $I^{\tau}_b /\tau$ converges to the capacity we had discussed previously. Similarly, we define the empirical estimate of the information along the trajectory,
\begin{equation}
\hat{I}_b=  \frac{\tau}{4}\hat{F}_{\mu\alpha}D^{-1}_{\mu\nu}\hat{F}_{\nu\alpha} =I^\tau_b + \frac{1}2\tau F^{\tau}_{\mu\alpha}D^{-1}_{\mu\nu}Z_{\nu\alpha}+\frac{1}4\tau Z_{\mu\alpha}D^{-1}_{\mu\nu}Z_{\nu\alpha}=I^\tau_b+\frac{1}{2}\tau\hat{F}_{\mu\alpha}D^{-1}_{\mu\nu}Z_{\nu\alpha} -\frac{1}4\tau Z_{\mu\alpha}D^{-1}_{\mu\nu}Z_{\nu\alpha}.
\end{equation}
so that
\begin{equation}
I^{\tau} _b= \hat{I}_b -\frac{1}2\tau\hat{F}_{\mu\alpha}D^{-1}_{\mu\nu}Z_{\nu\alpha}+\frac{1}{4} \tau Z_{\mu\alpha}D^{-1}_{\mu\nu}Z_{\nu\alpha}
\end{equation}

We can also relate the average of the empirical information to the
trajectory information:
\begin{equation} 
\av{\hat{I}_b}-I_b^{\tau} = \frac{1}{2} N_b
\end{equation}
at leading order. The estimator $\hat{I}_b$ is thus biased, with bias
$\frac{1}2 N_b$.  The variance of this estimator is well approximated
by $\av{(I^{\tau}_b - \hat{I}_b)^2} \approx 2 \av{\hat{I}_b} + N_b^2/4$.

In practice, the ``true'' force field is not known -- inferring it is
the goal here. It is therefore important to provide an estimate of the
inference error using only the inferred quantities.  \Eq{eq:force_cov}
allows us to propose such a self-consistent estimate of the
error. Indeed, it can be interpreted as the (squared) typical error on
the force projection coefficients, its right-hand-side can be
estimated using only trajectory-dependent quantities (again, we assume
that the diffusion matrix is known). We can also combine these
quantities in a single number quantifying the relative inference
error, as
\begin{equation}
  \label{eq:self_consistant_error}
\frac{(F^{\tau}_{\mu\alpha}-\hat{F}_{\mu\alpha}) D^{-1}_{\mu\nu} (F^{\tau}_{\nu\alpha}-\hat{F}_{\nu\alpha})}{\hat{F}_{\mu\alpha} D^{-1}_{\mu\nu}\hat{F}_{\nu\alpha}} \sim  N_b/2\hat{I}.
\end{equation}
Thus $N_b/2\hat{I}$ provides a self-consistent estimate of the
relative error. Note that in the absence of forces,
$\av{\hat{I}} = N_b/2$, corresponding to an inferred error of $1$,
which is consistent. Similarly, based on our estimate of the variance
of $\hat{I}_b$, we define a self-consistent confidence interval around
this inferred information as $\delta\hat{I}_b^2 = 2 \hat{I}_b + N_b^2/4$.

\paragraph{The force estimator and maximum likelihood}
Here we show that the estimator we propose in \Eq{eq:Fn_estimator} is also the maximum log-likelihood estimator for $F_{\mu\alpha}$. Indeed, given a measured trajectory $C^{\tau}$, we use the expression for the probability of a trajectory, \Eq{eq:traj_probability_D_ito}, to calculate
\begin{equation}
0=\frac{\partial\log {\cal P}({\cal C}^{\tau}|F)}{\partial F^{\tau}_{\mu\alpha} }= \int d\vec{x}\frac{\partial\log {\cal P}({\cal C}^{\tau}|F)}{\partial F_\nu(\vec{x})}\frac{\partial F_\nu(\vec{x})}{\partial F^{\tau}_{\mu\alpha}}.
\end{equation}
We have
\begin{equation}
\frac{\partial\log {\cal P}({\cal C}^{\tau}|F)}{\partial F_\nu(\vec{x})} = \frac{1}2\int_0^\tau dt D^{-1}_{\nu\mu} (\dot{{x}}_\mu(t) -F_\mu(\vec{x}(t))) \delta(\vec{x} -\vec{x}(t))
\end{equation}
Next, the empirical projectors $\hat{c}_\alpha$, corresponding to the trajectory, give the decomposition of the force as
\begin{equation}
F_\nu(\vec{x}) = F^{\tau}_{\nu\alpha} \hat{c}_\alpha(\vec{x})+ F^{\perp}_\nu
\end{equation}
so that
\begin{equation}
\frac{\partial F_\nu(\vec{x})}{\partial F^{\tau}_{\mu\alpha}} = \hat{c}_\alpha(\vec{x})\delta_{\mu\nu}
\end{equation}
and 
\begin{equation}
0=\int d\vec{x}\frac{\partial\log {\cal P}({\cal C}^{\tau}|F)}{\partial F_\nu(\vec{x})}\frac{\partial F_\nu(\vec{x})}{\partial F^{\tau}_{\mu\alpha}}=\int d\mathbf{x}  \hat{c}_\alpha(\vec{x})\int_0^\tau dt ({\vec{x}}_\nu (t)-F_\nu(\vec{x}(t))) \delta(\vec{x} -\vec{x}(t))
\end{equation}
resulting in
\begin{equation}
 \int_0^\tau dt \dot{x}_\nu (t) \hat{c}_\alpha(\vec{x}(t))\underbrace{\int d\vec{x} \delta(\vec{x} -\vec{x}(t))}_1= \int d\vec{x}  \hat{c}_\alpha(\vec{x})F_\nu(\vec{x}) \underbrace{\int_0^\tau dt\delta(\vec{x} -\vec{x}(t)) }_{\tau\hat{P}(\vec{x})} = \tau F^\tau_{\nu\alpha}
\end{equation}
which is solved by our estimator in \Eq{eq:Fn_estimator}. 
This estimator indeed maximizes the log-likelihood, since $  \hat{c}_\alpha(\vec{x})$ is independent of $F_{\mu\alpha}^\tau$ so that 
\begin{equation}
\begin{split}
\frac{\partial\log {\cal P}({\cal C}^{\tau}|F)}{\partial F^{\tau}_{\mu\alpha}\partial F^{\tau}_{\nu\beta} } = \frac{\partial}{\partial F^{\tau}_{\nu\beta} }  \int d\vec{x}\frac{1}2\int_0^\tau dt D^{-1}_{\mu\rho} ({\vec{x}}_\rho(t) -F_\rho(\vec{x}(t))) \delta(\vec{x} -\vec{x}(t))\hat{c}_\alpha(\vec{x})\\
 =- \int d\vec{x}\frac{1}2\int_0^\tau dt D^{-1}_{\mu\nu} \delta(\vec{x} -\vec{x}(t))\hat{c}_\alpha(\vec{x}(t)) \hat{c}_\beta(\vec{x}(t)) = -\frac{\tau}2 \delta_{\alpha\beta}D^{-1}_{\mu\nu}
\end{split}
\end{equation}
which is a negative definite matrix.

\section{Discussion}
\label{sec:SFI-discussion}

In this Chapter, we have introduced Stochastic Force Inference, a
method to reconstruct force and diffusion fields and measure entropy
production from Brownian trajectories. Based on the communication
theory notion of capacity, we have shown that such trajectories
contain a limited amount of information. With finite data, force
inference is thus limited by the information available per degree of
freedom to infer. SFI uses this information to fit the force field
with a linear combination of known functions. We have demonstrated its
utility on a variety of model systems and benchmarked its accuracy
using data comparable to current experiments.

\begin{figure}[ht]
  \centering
  \includegraphics[width=\columnwidth]{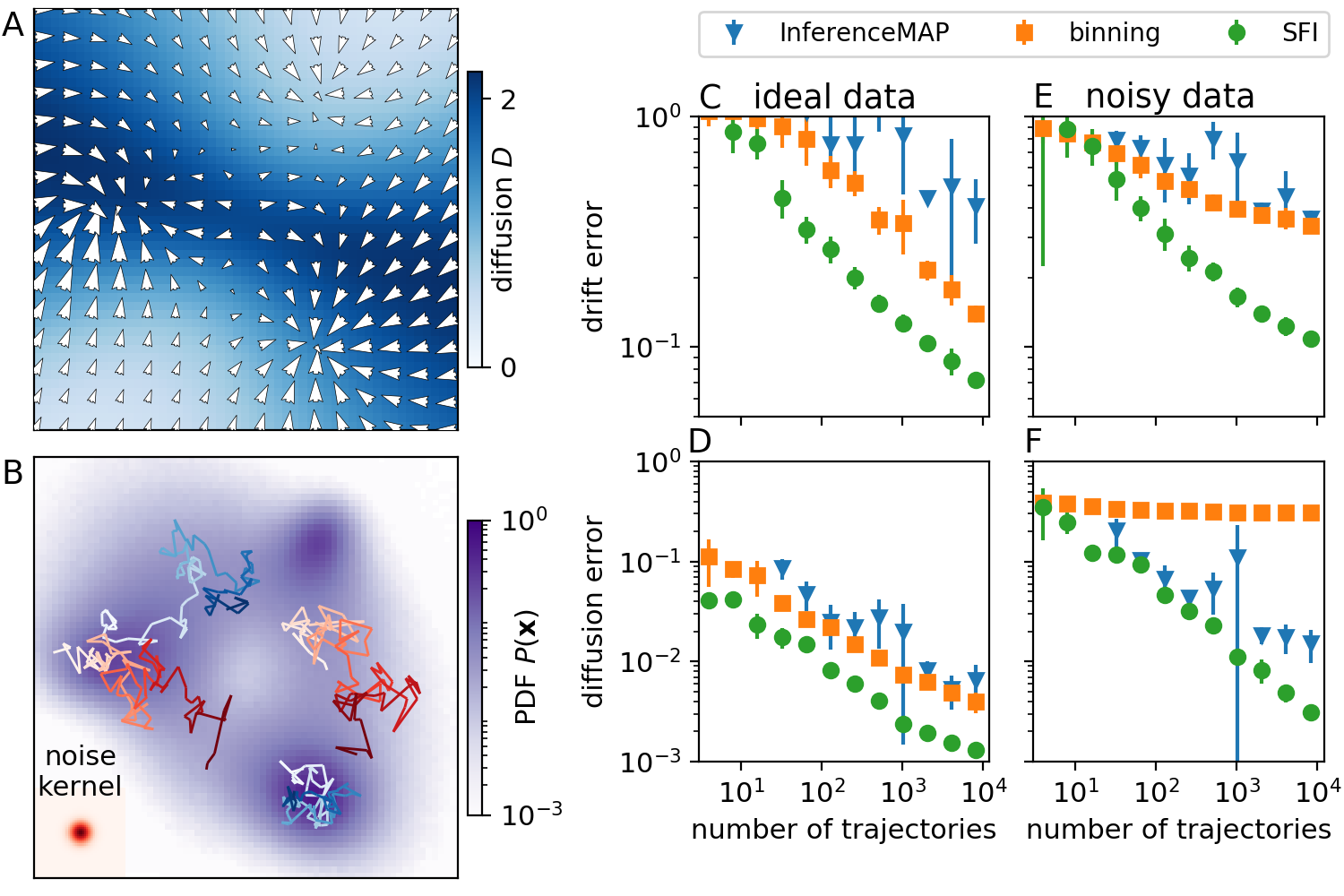}
  \caption{Quantitative comparison of SFI with other methods, on a
    simulated system mimicking 2D single molecule trajectories in a
    complex cellular environment with multiple potential wells,
    out-of-equilibrium circulation, and space-dependent isotropic
    diffusion. \textbf{A.} The diffusion field (blue gradient) and
    drift field (white arrows, scaled as $|\Phi|^{1/2}$ for better
    legibility). \textbf{B.} The steady-state probability distribution
    function of the process. The blue traces show two representative
    trajectories with $n=100$ time steps. The red traces show
    trajectories blurred by moderate Gaussian measurement error (with
    amplitude shown as a red kernel). \textbf{C-F.} Comparison of the
    performance of SFI with adaptive Fourier basis (green circles) and
    two widely used inference methods:
    InferenceMAP~\cite{beheiry_inferencemap:_2015}, a Bayesian method
    for single molecule inference (blue triangles), and grid-based
    binning with maximum-likelihood
    estimation~\cite{hoze_heterogeneity_2012,friedrich_approaching_2011}
    (\Eq{eq:F_moments_empirical}) and an adaptive mesh size (orange
    squares). We evaluate the performance of these methods on the
    approximation of the drift field (C,E) and diffusion field (D,F),
    as a function of the number $N$ of single-molecule trajectories
    (similar to those in B) used, with ideal data (C,D) and in the
    presence of measurement noise (E,F). The performance is evaluated
    as the average mean-squared error on the reconstructed field along
    trajectories. SFI outperforms both other methods in all cases; for
    noisy data, SFI is the only one that provides an unbiased
    estimation of the drift. Details and parameters in Ref.~\cite{frishman_learning_2020}. }
  \label{fig:comparison}
\end{figure}

\paragraph{Comparison of SFI to other inference methods.}
SFI combines the ability to infer arbitrary force fields, for
non-equilibrium processes, in high dimensions and in the presence of
measurement noise. In contrast, many previous methods essentially rely
on a specific linear~\cite{penland_prediction_1993} or
parametric~\cite{bishwal_parameter_2008} form for the force, or are
specific to one-dimensional
systems~\cite{papaspiliopoulos_nonparametric_2012,kutoyants_statistical_2004,comte_penalized_2007}. Other
approaches include spectral
methods~\cite{crommelin_diffusion_2011,gobet_nonparametric_2004},
Bayesian
methods~\cite{turkcan_bayesian_2012,ruttor_approximate_2013,yildiz_learning_2018,beheiry_inferencemap:_2015},
maximum likelihood techniques~\cite{sarfati_maximum_2017} or methods
that rely on coarse graining through
constant-by-parts~\cite{hoze_heterogeneity_2012,friedrich_approaching_2011,hoffmann_adaptive_1999}
or linear-by-parts~\cite{perez_garcia_high-performance_2018}
approximations. However, these techniques become inefficient as the
system's dimensionality increases. Furthermore, none offers a generic
unbiased estimator in the presence of measurement noise. Few of these
general methods are being used on experimental data in soft matter and
biological systems. We quantitatively compare SFI to two of the most
popular such
methods~\cite{hoze_heterogeneity_2012,friedrich_approaching_2011,beheiry_inferencemap:_2015}
that rely on spatial binning (\Fig{fig:comparison}). Our method
significantly outperforms them for a two-dimensional process
simulating single molecule dynamics in a complex cellular environment,
in particular in the presence of realistic measurement noise.

\paragraph{Comparison to other entropy production inference techniques.}
An important by-product of SFI is the ability to quantify the
irreversibility of a system by measuring the entropy production
associated to its currents. Alternative methods to estimate entropy
production also exist, either by coarse-graining trajectories to
estimate currents~\cite{lander_noninvasive_2012,battle_broken_2016,seara_entropy_2018}, by
measuring cycling 
frequencies~\cite{gladrow_broken_2016,mura_nonequilibrium_2018}, by
using non-Markovian signatures of irreversibility in hidden
variables~\cite{roldan_arrow_2018}, or by using thermodynamic bounds
on the fluctuations of dissipative
currents~\cite{barato_thermodynamic_2015,li_quantifying_2019}. These
methods are however inherently limited to relatively low-dimensional
systems with homogeneous diffusion, and even then require large
amounts of well-resolved data; SFI, in contrast, performs well in high
dimensions -- even with trajectories too short to resolve the
steady-state density -- and in the presence of measurement noise and
inhomogeneous diffusion.

\paragraph{Limits of SFI.}
We have limited our scope here to systems whose dynamics is described
by \Eq{eq:LangevinSFI} or \ref{eq:LangevinSFI_inhomogenous}, with a
time-independent force field and white-in-time noise.  When the force
field varies in time, for instance due to the dynamics of unobserved
variables, SFI captures the average projection of the force onto the
observed variables. Furthermore, SFI could be extended
to capture an explicit time-dependence of the force by using a
time-dependent basis. Finally, force inference is notably complicated
by non-Markovian terms in the dynamics~\cite{daldrop_butane_2018},
such as colored noise; however, in such cases, our projection approach
to estimate phase-space velocities (\Eq{eq:v}) remains useful and
valid.

\paragraph{Perspectives.}
Our approach, all in all, proposes a solution to the inverse problem
of Brownian dynamics: inferring the force and diffusion fields from
trajectories. This method consists in a few intelligible equations,
and provides a powerful data analysis framework that could be used on
a broad class of stochastic systems where inferring effective forces
and currents from limited noisy data is of interest.  Our work thus
applies to microscopic systems where thermal noise is relevant, such
as single molecules~\cite{hoze_heterogeneity_2012}, active
colloids~\cite{palacci_living_2013,bricard_emergence_2013} and
cytoskeletal filaments~\cite{gladrow_broken_2016,seara_entropy_2018}.
Beyond thermal systems, for stochastic dynamical systems that can be
effectively modeled by Brownian dynamics, applications of our
framework range from the behavior of
cells~\cite{celani_bacterial_2010,li_dicty_2011,bruckner_stochastic_2019}
and animals~\cite{stephens_dimensionality_2008}, to modeling of
climate
dynamics~\cite{hasselmann_stochastic_1976,wheeler_all-season_2004,penland_prediction_1993}
and trend finding in financial data~\cite{oksendal_stochastic_2003}.
Our method could be combined with sparsity-promoting techniques, as
used to infer dynamical equations in deterministic
systems~\cite{brunton_discovering_2016}, to go from force fitting to
identifying the simple rules governing the dynamics.

\chapter{Underdamped Langevin Inference}
\label{chap:ULI}

\emph{In this Chapter, we extend the stochastic inference method
  introduced in \Chap{chap:SFI} for overdamped systems into a method
  for underdamped systems. This extension is useful both for truly
  inertial systems, such as schools of fish and flocks of birds, as
  well as for \emph{effectively} inertial systems such as migrating
  cells. In the latter case, when the polarity of the cell is not
  observed, it induces a persistence in the motion that can be
  interpreted as effective inertia -- more formally, this is a
  consequence of \emph{embedding theorems}: when observing only part
  of a system that obeys a first-order differential equation as
  dynamical law, the hidden degrees of freedom effectively increase
  the order of the dynamics of the observed ones.}

\emph{Inferring such second-order stochastic differential equations
  presents one major challenge compared to first-order ones: the
  velocity, upon which the force and diffusion may depend, is not
  directly observed, but has to be deduced from data. The error in
  this estimate correlates with the acceleration, which incurs severe
  biases in the force estimates. Here we identify this bias and
  propose a practical method that addresses it.  This work was
  published as an article in Ref~\cite{bruckner_inferring_2020},
  titled ``Inferring the Dynamics of Underdamped Stochastic Systems'',
  that we reproduce here. Note that I have chosen not to include the
  Supplementary Information of the article here, which includes
  lengthy proofs and extensive calculations.}

\vspace{5mm}

Adapted from: \\
\textsc{Inferring the dynamics of underdamped stochastic systems} \\
David B Br\"uckner$^*$, Pierre Ronceray$^*$, Chase P Broedersz \\
Physical Review Letters 125, 058103 (2020).

Across the scientific disciplines, data-driven methods are used to
unravel the dynamics of complex systems. These approaches often take
the form of inverse problems, aiming to infer the underlying
governing equation of motion from observed trajectories. This problem
is well understood for deterministic
systems~\cite{crutchfield_equations_1987,daniels_automated_2015,brunton_discovering_2016}. For a broad
variety of physical systems, however, a deterministic description is
insufficient: fast, unobserved degrees of freedom act as an effective
dynamical noise on the observable quantities.
Such systems are described by Langevin dynamics, and inferring their equation of motion is notoriously harder: one must then disentangle the stochastic from the deterministic contributions, both of which contribute to shape the trajectory. In molecular-scale systems described by the overdamped Langevin equation, a first-order stochastic differential equation, recently developed
techniques make it possible to efficiently reconstruct the dynamics
from observed trajectories~\cite{siegert_analysis_1998,ragwitz_indispensable_2001,beheiry_inferencemap:_2015,perez_garcia_high-performance_2018,frishman_learning_2020}.
Many complex systems at larger scales, however, exhibit stochastic dynamics governed by the \emph{underdamped} Langevin equation, a second-order stochastic differential equation. 
Examples include cell
motility~\cite{selmeczi_cell_2005,li_dicty_2011,sepulveda_collective_2013,dalessandro_contact_2017,bruckner_stochastic_2019}, 
postural dynamics in animals~\cite{stephens_dimensionality_2008,stephens_emergence_2011},
movement in interacting swarms of fish~\cite{gautrais_deciphering_2012,katz_inferring_2011,jhawar_noise-induced_2020},
birds~\cite{cavagna_scale-free_2010,attanasi_information_2014}, and
insects~\cite{buhl_disorder_2006,attanasi_collective_2014}, as well as dust particles in a
plasma~\cite{gogia_emergent_2017}.
Due to recent advances in tracking
technology, the diversity, accuracy, dimensionality, and size of these
behavioral data-sets is rapidly
increasing~\cite{brown_ethology_2018}, resulting in a growing
need for accurate inference approaches for high-dimensional underdamped stochastic systems.
However, there is currently no rigorous method to infer the dynamics of such underdamped stochastic systems.

Inference from underdamped stochastic systems suffers from a major
challenge absent in the overdamped case. In any realistic application, the accelerations
of the degrees of freedom must be obtained as discrete second
derivatives from the observed position trajectories, which are sampled
at discrete intervals $\Delta t$. Consequently, a straightforward generalization of the estimators for the force and noise fields of overdamped systems fails: these estimators do not converge to the correct values, even in the limit $\Delta t \to 0$~\cite{pedersen_how_2016,ferretti_building_2020}. To make matters worse, real
data is always subject to measurement errors, leading to divergent
biases in the discrete estimators~\cite{lehle_analyzing_2015}. These problems have so far precluded
reliable inference in underdamped stochastic systems.

Here, we introduce a general framework, Underdamped Langevin Inference (ULI), that conceptually explains the origin of these
 biases, and provides an operational scheme to reliably infer the equation of motion of underdamped stochastic systems governed by non-linear force fields and multiplicative noise amplitudes. To
provide a method that can be robustly applied to realistic
experimental data, we rigorously derive estimators that converge to
the correct values for discrete data subject to measurement errors. We demonstrate the power of our
method by applying it to experimental trajectories of single
migrating cells, as well as simulated complex high-dimensional data
sets, including flocks of active particles with Viscek-style alignment
interactions.

%%%%%%%%%%%%%%%%%%%%%%%%%%%%%%%%%%%%
%%RESULTS
%\section{II. Results}
We consider a general $d$-dimensional stationary stochastic process $\mathbf{x}(t)$ with components $\{ x_\mu(t) \}_{1 \leqslant \mu \leqslant d}$ governed by the underdamped Langevin equation
\begin{align}
\label{eqn:process}
\begin{split}
\dot{x}_\mu &= v_\mu \\
\dot{v}_\mu &= F_\mu(\mathbf{x},\mathbf{v}) + \sigma_{\mu \nu}(\mathbf{x},\mathbf{v}) \xi_\nu(t)
\end{split}
\end{align}
which we interpret in the It\^o-sense. Throughout, we employ the Einstein summation convention, and $\xi_\mu(t)$ represents a Gaussian white noise with the properties $\langle \xi_\mu(t) \xi_\nu(t') \rangle = \delta_{\mu\nu}\delta(t-t')$ and $\langle \xi_\mu(t)\rangle =0$. Our aim is to infer the force field $F_\mu(\mathbf{x},\mathbf{v})$ and the noise amplitude $\sigma_{\mu \nu}(\mathbf{x},\mathbf{v})$ from an observed finite trajectory of the process~\footnote{Since we interpret eqn.~\eqref{eqn:process} in the It\^o-sense, the inferred force field $F_\mu(\mathbf{x},\mathbf{v})$ corresponds to this convention.}. 

%%%%%%%%%%%%%%%%%%%%%%%%%%%%%%%%%%%%
%FIGURE 1

\begin{figure*}[ht]
\centering
	\includegraphics[width=\textwidth]{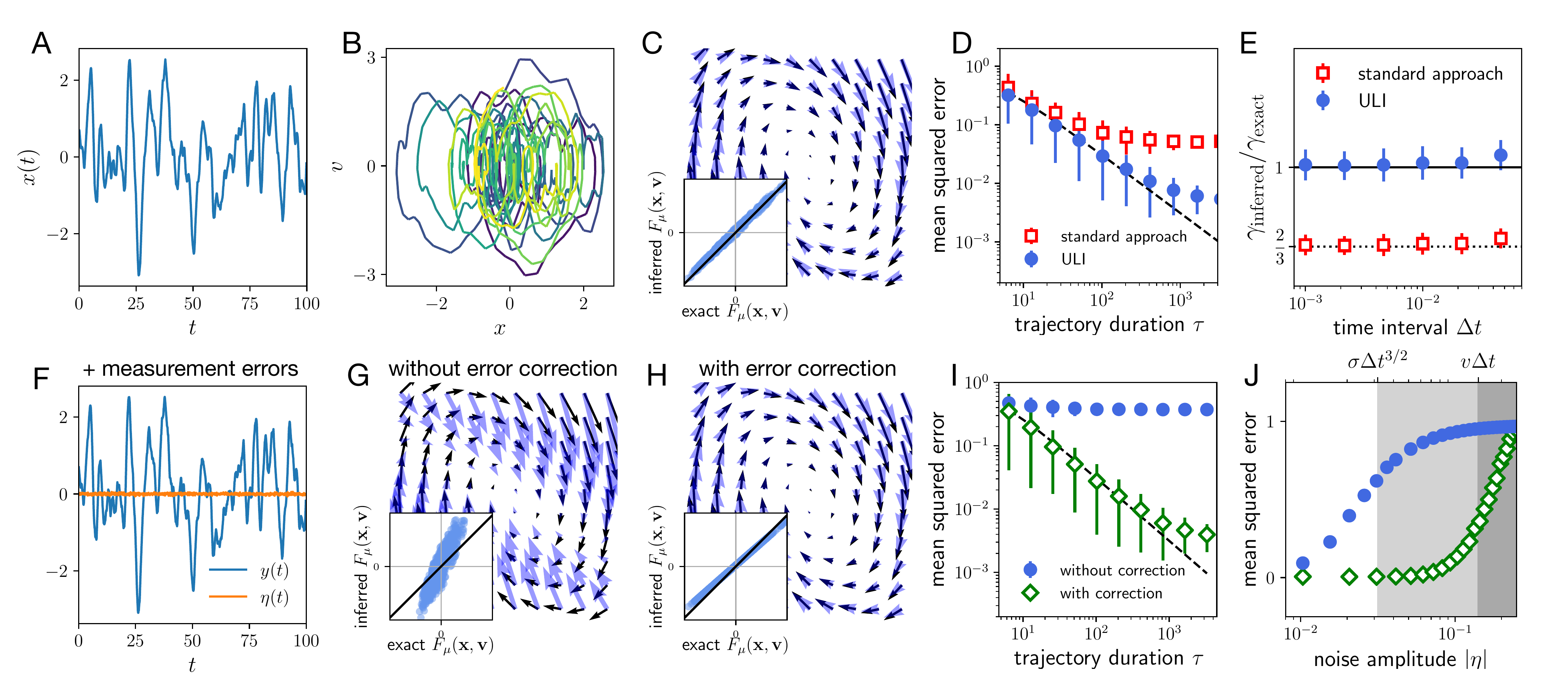}
	\caption{
		\textbf{Inference from discrete time series subject to measurement error.} 
		\textbf{A.} Trajectory $x(t)$ of a stochastic damped harmonic oscillator, $F(x,v)=-\gamma v - kx$.
		\textbf{B.} The same trajectory represented in $xv$-phase space. Color coding indicates time.
		\textbf{C.} Force field in $xv$-space inferred from the trajectory in A using ULI with basis functions $b = \{ 1, x, v \}$ (blue arrows), compared to the exact force  field (black arrows). \textit{Inset:} inferred components of the force along the trajectory \textit{versus} the exact values.
		\textbf{D.} Convergence of the mean squared error of the inferred force field, obtained using ULI (circles) and with the previous standard approach~\cite{bruckner_stochastic_2019,stephens_dimensionality_2008,pedersen_how_2016,lehle_analyzing_2015} (squares). Dashed lines indicate the predicted error $\delta \hat{F}^2/\hat{F}^2 \sim N_b/2\hat{I}_b$.
		\textbf{E.} Inferred friction coefficient $\gamma$ divided by the exact one as a function of the sampling time interval $\Delta t$, comparing the previous standard approach to ULI.
		\textbf{F.} Trajectory $y(t) = x(t) + \eta(t)$ (blue) corresponding to the same realization $x(t)$ in A, with additional time-uncorrelated measurement error $\eta(t)$ (orange) with small amplitude $|\eta|=0.02$. 		
		\textbf{G,H.} Force field inferred from $y(t)$ using estimators without and with measurement error corrections, respectively.
		\textbf{I.} Inference convergence for data subject to measurement error using estimators without (circles) and with (diamonds) measurement error corrections.
		\textbf{J.} Dependence of the inference error on the noise amplitude $|\eta|$ (same symbols as in I).
		}
	\label{fig1ULI}
\end{figure*}

%%%%%%%%%%%%%%%%%%%%%%%%%%%%%%%%%%%%

We start by approximating the force field as a linear combination of $n_b$ basis functions $b = \{  b_\alpha(\mathbf{x},\mathbf{v})  \}_{1 \leqslant \alpha \leqslant n_b}$, such as polynomials, Fourier modes, wavelet functions, or Gaussian kernels~\cite{stephens_dimensionality_2008}. From these basis functions, we construct an empirical orthonormal basis $\hat{c}_\alpha(\mathbf{x},\mathbf{v}) = \hat{B}_{\alpha \beta}^{-1/2}b_\beta(\mathbf{x},\mathbf{v})$ such that $\langle \hat{c}_\alpha \hat{c}_\beta \rangle = \delta_{\alpha \beta}$, an approach that was recently proposed for overdamped systems~\cite{frishman_learning_2020}. Here and throughout, averages correspond to time-averages along the trajectory. We can then approximate the force field as $F_\mu(\mathbf{x},\mathbf{v}) \approx F_{\mu \alpha} \hat{c}_\alpha(\mathbf{x},\mathbf{v})$. Similarly, we perform a basis expansion of the noise amplitude $\sigma^2_{\mu \nu}(\mathbf{x},\mathbf{v})$. Thus, the inference problem reduces to estimating the projection coefficients $F_{\mu \alpha}$ and $\sigma^2_{\mu \nu \alpha}$.

\section{Dealing with discreteness}

In practice, only the configurational coordinate $\mathbf{x}(t)$ is
accessible in experimental data, sampled at a discrete time-interval
$\Delta t$. We therefore only have access to the discrete estimators
of the velocity
$\mathbf{\hat{v}}(t)= [\mathbf{x}(t) - \mathbf{x}(t-\Delta t)]/\Delta
t$ and acceleration
$\mathbf{\hat{a}}(t) = [\mathbf{x}(t+\Delta t) - 2\mathbf{x}(t) +
\mathbf{x}(t-\Delta t)]/{\Delta t^2}$. Our goal is to derive an
estimator $\hat{F}_{\mu \alpha}$, constructed from the discrete
velocities and accelerations, which converges to the exact projections
$F_{\mu \alpha}$ in the limit $\Delta t \rightarrow 0$.

An intuitive approach would be to simply generalize the estimators for
overdamped systems~\cite{frishman_learning_2020} and calculate the projections
of the accelerations
$\langle \hat{a}_\mu \hat{c}_\alpha(\mathbf{x},\mathbf{\hat{v}})
\rangle$. This expression has indeed previously been used for
underdamped
systems~\cite{bruckner_stochastic_2019,stephens_dimensionality_2008,pedersen_how_2016,lehle_analyzing_2015}. We derive the correction term to
this estimator by expanding the basis functions
$\hat{c}_\alpha(\mathbf{x},\mathbf{\hat{v}}) =
\hat{c}_\alpha(\mathbf{x},\mathbf{v}) + (\partial_{v_\mu}
\hat{c}_\alpha) (\hat{v}_\mu - v_\mu) + ...$, where the leading order
contribution to the second term is a fluctuating (zero average) term
of order $\Delta t^{1/2}$. Similarly, we perform a stochastic
It\^o-Taylor expansion of the discrete acceleration
$\mathbf{\hat{a}}(t)$, which has a leading order fluctuating term of
order $\Delta t^{-1/2}$. Thus, while each of these terms individually
averages to zero, their product results in a bias term with non-zero
average of order $\Delta t^{0}$:
$\langle \hat{a}_\mu \hat{c}_\alpha(\mathbf{x},\mathbf{\hat{v}})
\rangle = F_{\mu\alpha} + \frac{1}{6} \left\langle \sigma^2_{\mu \nu}
  \partial_{v_\nu}c_\alpha(\mathbf{x},\mathbf{v}) \right\rangle +
\mathcal{O}(\Delta t)$~\footnote{See Supplemental Material at [URL
  will be inserted by publisher] for detailed derivations of the
  correction terms and estimators.}. As expected, this bias vanishes
in the limit $\sigma \to 0$, and therefore does not appear in
deterministic systems. However, it poses a problem wherever a second derivative of a stochastic
signal is averaged conditioned on its first derivative. The occurrence of such a bias was observed in linear
systems~\cite{pedersen_how_2016,ferretti_building_2020}. Specifically, for a linear
viscous force $F(v)=-\gamma v$, it was found that
$\langle \hat{a} c(\hat{v}) \rangle = -\frac{2}{3}\gamma +
\mathcal{O}(\Delta t)$, which is recovered by our general expression for the systematic bias~\cite{noauthor_see_nodate-1}. 

Previous approaches to correct for this bias rely on \textit{a priori} knowledge of the observed stochastic
process~\cite{pedersen_how_2016}, are limited to simple parametric forms~\cite{ferretti_building_2020}, or perform an \textit{a
  posteriori} empirical iterative scheme~\cite{bruckner_stochastic_2019}. In
contrast, by simply deducting the general form of the bias, we obtain
our Underdamped Langevin Inference (ULI) estimator:
\begin{align}
\label{eqn:guesstimator}
\hat{F}_{\mu \alpha} 
	= \langle \hat{a}_\mu \hat{c}_\alpha(\mathbf{x},\mathbf{\hat{v}}) \rangle -
		 \frac{1}{6} \left\langle \widehat{\sigma^2}_{\mu \nu}(\mathbf{x},\mathbf{\hat{v}}) \partial_{v_\nu} \hat{c}_\alpha(\mathbf{x},\mathbf{\hat{v}}) \right\rangle
\end{align}
The presence of the derivative of a basis function in the estimator highlights the importance of projecting the dynamics of underdamped systems onto a set of \textit{smooth} basis functions, in contrast to the traditional approach of taking conditional averages in a discrete set of bins~\cite{siegert_analysis_1998,ragwitz_indispensable_2001}, equivalent to a basis of non-differentiable top-hat functions. 

Similarly to the force field, we expand the noise amplitude as a sum of basis functions, and derive an unbiased estimator for the projection coefficients~\cite{noauthor_see_nodate-1}
\begin{align}
\label{eqn:difftimator}
\widehat{\sigma^2}_{\mu \nu \alpha} = \frac{3 \Delta t}{2} \langle \hat{a}_\mu \hat{a}_\nu \hat{c}_\alpha(\mathbf{x},\mathbf{\hat{v}}) \rangle
\end{align}
To test our method, we start with a simulated minimal example, the stochastic damped harmonic oscillator $\dot{v}=-\gamma v - kx+\sigma \xi$ (Fig.~\ref{fig1ULI}A-E). Indeed, we find that even for such a simple system, the intuitive acceleration projections $\langle \hat{a}_\mu \hat{c}_\alpha(\mathbf{x},\mathbf{\hat{v}}) \rangle$ yield a biased result (Fig.~\ref{fig1ULI}E). In contrast, ULI, defined by Eqs.~\eqref{eqn:difftimator} and~\eqref{eqn:guesstimator}, provides an accurate reconstruction of the force field (Fig.~\ref{fig1ULI}C,E). To test the convergence of these estimators in a quantitative way, we calculate the expected random error due to the finite length $\tau$ of the input trajectory, 
$\delta \hat{F}^2/\hat{F}^2 \sim N_b/2\hat{I}_b$, 
where we define $\hat{I}_b = \frac{\tau}{2}\hat{\sigma}_{\mu\nu}^{-2}\hat{F}_{\mu\alpha}\hat{F}_{\nu\alpha}$ as the empirical estimate of the information contained in the trajectory, and $N_b=d n_b$ is the number of degrees of freedom in the force field~\cite{frishman_learning_2020}. We confirm that the convergence of our estimators follows this expected trend, in contrast to the biased acceleration projections (Fig.~\ref{fig1ULI}D). Therefore, ULI provides an operational method to accurately infer the dynamical terms of underdamped stochastic trajectories.

\section{Treatment of measurement errors}  A key challenge in stochastic inference from real data is the unavoidable presence of time-uncorrelated random measurement errors $\boldsymbol{\eta}(t)$, which can be non-Gaussian: the observed signal in this case is $\mathbf{y}(t) = \mathbf{x}(t) + \boldsymbol{\eta}(t)$. This problem is particularly dominant in underdamped inference, where the signal is differentiated twice, leading to a divergent bias of order $\Delta t^{-3}$~\cite{noauthor_see_nodate-1}. Thus, for small $\Delta t$, even small measurement errors can lead to prohibitively large systematic inference errors, which cannot be rectified by simply recording more data.

To overcome this challenge, we derive estimators which are robust against measurement error. These estimators are constructed such that the leading-order bias terms cancel. For the force estimator, we find that this is achieved by using the local average position $\mathbf{\overline{x}}(t)=\frac{1}{3}(\mathbf{x}(t-\Delta t) + \mathbf{x}(t) +  \mathbf{x}(t+\Delta t))$ and the symmetric velocity $\mathbf{\hat{v}}(t)= [\mathbf{x}(t+\Delta t) - \mathbf{x}(t-\Delta t)]/(2\Delta t)$ in Eq.~\eqref{eqn:guesstimator}~\footnote{Note that due to the change of definition of $\mathbf{\hat{v}}$, the prefactor of the correction term in Eq.~\eqref{eqn:guesstimator} changes from $1/6$ to $1/2$.}. Similarly, we derive an unbiased estimator for the noise term, which is constructed using a linear combination of four-point increments~\cite{noauthor_see_nodate-1}.

Remarkably, these modifications result in a vastly improved inference performance in the presence of measurement error (Fig.~\ref{fig1ULI}F-J). Specifically, while the bias becomes dominant at an error magnitude $|\eta| \sim \sigma \Delta t^{3/2}$ in the standard estimators, the bias-corrected estimators only fail when the measurement error becomes comparable to the displacement in a single time-step, $|\eta| \sim v \Delta t$ (Fig.~\ref{fig1ULI}J)~\cite{noauthor_see_nodate-1}. Thus, our method has a significantly larger range of validity extending up to the typical displacement in a single time-frame. 

%%%%%%%%%%%%%%%%%%%%%%%%%%%%%%%%%%%%
%FIGURE 2
\begin{figure}[]
	\includegraphics[width=\textwidth]{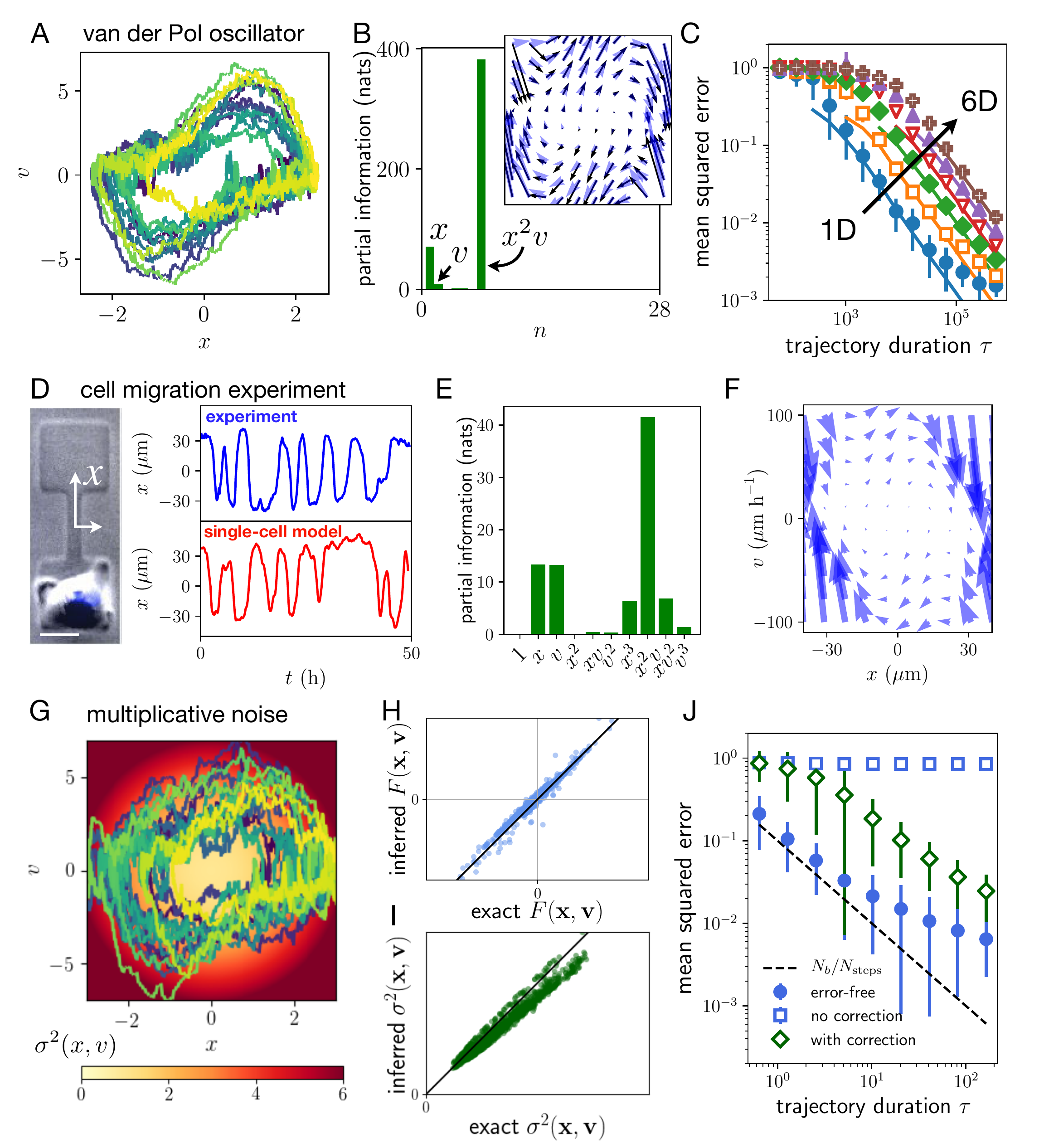}
	\caption{
		\textbf{Inferring non-linear dynamics and multiplicative noise.} 
		\textbf{A.} $xv$-trajectory of the stochastic Van der Pol oscillator, $F(x,v)=\kappa(1-x^2)v-x$ with measurement error.
		\textbf{B.} Partial information of the 28 basis functions of a 6th order polynomial basis in natural information units (1 nat $=1/\log2$ bits), inferred from the trajectory in A. \textit{Inset:} Corresponding force field reconstruction. 		
		\textbf{C.} Convergence of the inference error for the $d$-dimensional Van der Pol oscillator $F_\mu(\mathbf{x},\mathbf{{v}})=\kappa_\mu(1-x_\mu^2)v_\mu-x_\mu$ (no summation, $1 \leqslant \mu \leqslant d$) with $d=1...6$, using a third-order polynomial basis.
		%cell shown is j_cell = 31
		\textbf{D.} Microscopy image of a migrating human breast cancer cell (MDA-MB-231) confined in a two-state micropattern (scale bar: $20 \mu$m). Experimental trajectory of the cell nucleus position, recorded at a time-interval $\Delta t = 10$ min (blue), and simulated trajectory using the inferred model (red). 
		\textbf{E.} Partial information for the experimental trajectory in D, projected onto a third-order polynomial basis.
		\textbf{F.} Deterministic flow field inferred from the experimental trajectory in D.		
		\textbf{G.} Trajectory of a Van der Pol oscillator with multiplicative noise $\sigma^2(x,v) = \sigma_0 + \sigma_x x^2+\sigma_v v^2$ (colormap).
		\textbf{H,I.} Inferred \textit{versus} exact components of the force and noise term, respectively, for the trajectory in G.
		\textbf{J.} Inference convergence of the multiplicative noise amplitude, using Eq.~\eqref{eqn:difftimator} without measurement error (circles), with measurement error (squares), and using the error-corrected estimator (diamonds). The error saturation at large $\tau$ is due to the finite time-step. Dashed line: predicted error $\delta \widehat{\sigma^2}/\widehat{\sigma^2} \sim \sqrt{N_b \Delta t/\tau}$~\cite{frishman_learning_2020}.
		}
	\label{fig2ULI}
\end{figure}
%%%%%%%%%%%%%%%%%%%%%%%%%%%%%%%%%%%%

\section{Non-linear dynamics} Since our method does not assume linearity, we can expand the projection basis to include higher order functions to capture the behavior of systems with non-linear dynamics. As a canonical example, we study the stochastic Van der Pol oscillator $\dot{v}=\kappa(1-x^2)v-x + \sigma \xi$, a common model for a broad range of biological dynamical systems~\cite{kruse_oscillations_2005}. We simulate a short trajectory of this process, with added artificial measurement error (Fig.~\ref{fig2ULI}A). Indeed, we find that ULI reliably infers the underlying phase-space flow (Fig.~\ref{fig2ULI}B). This is not limited to one-dimensional systems, as shown by studying convergence of higher-dimensional oscillators (Fig.~\ref{fig2ULI}C). Importantly, this good performance does not rely on using a polynomial basis to fit a polynomial field: employing a non-adapted basis, such as Fourier components, yields similarly good results~\cite{noauthor_see_nodate-1}. 

To capture the Van der Pol dynamics, only the three basis functions
$\{ x,v, x^2v\}$ are required. But can these functions be identified
directly from the data without prior knowledge of the underlying
force field? To address this question, we introduce the concept of partial
information. We can estimate the information contained in a finite
trajectory as
$\hat{I}_b(n_b) =
\frac{\tau}{2}\hat{\sigma}_{\mu\nu}^{-2}\hat{F}_{\mu\alpha}\hat{F}_{\nu\alpha}$,
where $\hat{F}_{\nu\alpha}$ are the projection coefficients onto the
basis $b$ with $n_b$ basis functions~\cite{frishman_learning_2020}. To assess
the importance of the $n^\mathrm{th}$ basis function in the expansion,
we calculate the amount of information it contributes:
\begin{equation}
\label{eqn:part_info}
\hat{I}_b^{\mathrm{(partial)}}(n) = \hat{I}_b(n) - \hat{I}_b(n-1)
\end{equation}
which we term the partial information contributed by the basis function $b_n$. This approach successfully recovers the relevant terms in large basis sets (Inset Fig.~\ref{fig2ULI}B). Thus, the partial information provides a useful heuristic for detecting the relevant terms of the force field.

To illustrate that ULI is practical and data-efficient, we apply it to experimental trajectories of cells migrating in two-state confinements (Fig.~\ref{fig2ULI}D). Within their lifetime, these cells perform several transitions between the two states, resulting in relatively short trajectories. Previously, we inferred dynamical properties by averaging over a large ensemble of trajectories~\cite{bruckner_stochastic_2019,fink_area_2020,bruckner_disentangling_2020}. In contrast, with ULI, we can reliably infer the governing equation of motion from single cell trajectories. Here, $F(x,v)$ corresponds to the deterministic dynamics of the system, and not to a physical force. We employ the partial information to guide our basis selection: indeed, it recovers the intrinsic symmetry of the system, suggesting a symmetrized third order polynomial expansion is a suitable choice (Fig.~\ref{fig2ULI}E). Using this expansion, we infer the deterministic flow field of the system (Fig.~\ref{fig2ULI}F), which predicts trajectories similar to the experimental ones (Fig.~\ref{fig2ULI}D). Importantly, the inferred model is self-consistent: re-inferring from short simulated trajectories yields a similar model~\cite{noauthor_see_nodate-1}. Using ULI, we can thus perform inference on small data sets, enabling "single-cell profiling'', which could provide a useful tool to characterize cell-to-cell variability~\cite{bruckner_disentangling_2020}.

To demonstrate the broad applicability of our approach, we evaluate its performance in the presence of multiplicative noise amplitudes $\sigma_{\mu \nu}(\mathbf{x},\mathbf{v})$, which occur in a range of complex systems~\cite{friedrich_how_2000,bruckner_stochastic_2019,stephens_dimensionality_2008}. ULI accurately recovers the space- and velocity-dependence of both the force and noise field, and the estimators converge to the exact values, even in the presence of measurement errors (Fig.~\ref{fig2ULI}G-J). To summarize, we have shown that ULI performs well on short trajectories of non-linear data sets subject to measurement errors, and can accurately infer the spatial structure of multiplicative noise terms. 

%%%%%%%%%%%%%%%%%%%%%%%%%%%%%%%%%%%%
%FIGURE 3
\begin{figure}
	\includegraphics[width=\textwidth]{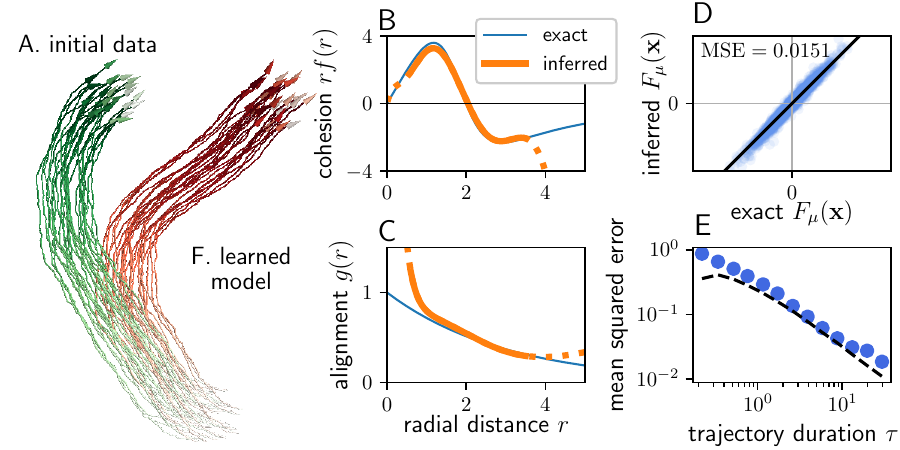}
	\caption{
		\textbf{Interacting flocks.} 
		\textbf{A.} Trajectory (green) of $N=27$ Viscek-like particles (Eq.~\ref{eqn:aligning}) in the flocking regime (1000 frames). We perform ULI on this trajectory using a translation-invariant basis of pair interaction and alignment terms, both fitted with $n=8$ exponential kernels.
                \textbf{B.}~Exact (blue) and inferred (orange) cohesion $rf(r)$. Exact form includes short-range repulsion and long-range attraction, $f(r) = \epsilon_0 (1 - (r/r_0)^3)/((r/r_0)^6+1)$. Dotted inference dependence indicates distances not sampled by the initial data.
                \textbf{C.} Exact and inferred alignment kernel $g(r)$. Exact form: $g(r) = \epsilon_1 \exp(-r/r_1)$.
                \textbf{D.} Inferred \emph{versus} exact components of the force field.
                \textbf{E.} Convergence of the inferred force as a function of trajectory length. Dashed line is the predicted error $\delta \hat{F}^2/\hat{F}^2 \sim N_b/2\hat{I}_b$.
                \textbf{F.} Simulated trajectory (red) employing the inferred force and noise, showing qualitatively similar flocking behavior.
		}
	\label{fig3ULI}
\end{figure}
%%%%%%%%%%%%%%%%%%%%%%%%%%%%%%%%%%%%

\section{Collective systems} A major challenge in stochastic inference is the treatment of interacting many-body systems. In recent years, trajectory data on active collective systems, such as collective cell migration~\cite{sepulveda_collective_2013,dalessandro_contact_2017} and animal groups~\cite{cavagna_scale-free_2010,lukeman_inferring_2010,buhl_disorder_2006,attanasi_collective_2014,attanasi_information_2014}, have become readily available. Previous approaches to such systems frequently focus on the study of correlations~\cite{cavagna_scale-free_2010,cavagna_physics_2018,bialek_statistical_2012} or collision statistics~\cite{katz_inferring_2011,lukeman_inferring_2010,dalessandro_contact_2017}, but no general method for inferring their underlying dynamics has been proposed. The collective behavior of these systems, ranging from disordered swarms \cite{attanasi_collective_2014} to ordered flocking \cite{cavagna_scale-free_2010}, is determined by the interplay of active self-propulsion, cohesive and alignment interactions, and noise. Thus, disentangling these contributions could provide key insights into the physical laws governing active collective systems.

We  consider a simple model for the dynamics of a 3D flock with Viscek-style alignment interactions~\cite{vicsek_novel_1995,gregoire_moving_2003,chate_modeling_2008,sepulveda_collective_2013}, \begin{align}
\label{eqn:aligning}
\mathbf{\dot{v}}_{i} = \mathbf{p}_i + \sum_{j\neq i} \left[ f(r_{ij})\mathbf{r}_{ij} + g(r_{ij})\mathbf{v}_{ij} \right]  + \sigma \boldsymbol{\xi}_i
\end{align}
where $\mathbf{v}_{i}=\dot{\mathbf{r}}_{i}$, $\mathbf{r}_{ij} = \mathbf{r}_{j}-\mathbf{r}_{i}$, $\mathbf{v}_{ij} = \mathbf{v}_{j}-\mathbf{v}_{i}$, and $\mathbf{p}_i = \gamma(v_0^2-|\mathbf{v}_{i}|^2)\mathbf{v}_{i}$ is a self-propulsion force acting along the direction of motion of each particle $i$. Here, $f$ and $g$ denote the strength of the cohesive and alignment interactions, respectively, as a function of inter-particle distance $r_{ij}$. This model exhibits a diversity of behaviors, including flocking (Fig.~\ref{fig3ULI}A). Intuitively, one might expect that ULI should fail dramatically in such a system: a 3D swarm of $N$ particles has $6N$ degrees of freedom, and ``curse of dimensionality'' arguments make this problem seem intractable. However, by exploiting the particle exchange symmetry and radial symmetry of the interactions~\cite{noauthor_see_nodate-1}, we find that ULI accurately recovers the cohesion and alignment terms (Fig.~\ref{fig3ULI}B-C), and captures the full force field (Fig.~\ref{fig3ULI}D,E). Furthermore, simulating the inferred model yields trajectories with high similarity to the input data (Fig.~\ref{fig3ULI}F). This example illustrates the potential of ULI for inferring complex interactions from trajectories of stochastic many-body systems.

In summary, we demonstrate how to reliably infer the force and noise fields in complex underdamped stochastic systems. We show that the inevitable presence of discreteness and measurement errors result in systematic biases that have so far prohibited accurate inference. To circumvent these problems, we have rigorously derived unbiased estimators, providing an operational framework, Underdamped Langevin Inference, to infer underdamped stochastic dynamics~\footnote{A readily usable \textsc{Python} package to perform Underdamped Langevin Inference is available at https://github.com/ronceray/UnderdampedLangevinInference.}. Our method provides a new avenue to analyzing the dynamics of complex high-dimensional systems, such as assemblies of motile cells~\cite{sepulveda_collective_2013,dalessandro_contact_2017}, active swarms~\cite{cavagna_scale-free_2010,lukeman_inferring_2010,buhl_disorder_2006,attanasi_collective_2014}, as well as non-equilibrium condensed matter systems~\cite{baldovin_langevin_2019,gogia_emergent_2017,kruse_oscillations_2005}.

\chapter{Tracking-free inference of entropy production}
\label{chap:BDB}

\emph{In this Chapter, we leverage the method introduced in
  \Chap{chap:SFI} to infer currents, forces and dissipation from
  stochastic trajectories -- \emph{i.e.} from \emph{tracked} degrees
  of freedom of the system -- into a methodology to infer entropy
  production directly from movies -- thus bypassing the tracking step
  by generically performing a component analysis of the movie.  }

\vspace{5mm}

Adapted from:\\
\textsc{Learning the non-equilibrium dynamics of Brownian movies}\\
Federico S Gnesotto, Grzegorz Gradziuk, Pierre Ronceray$^\dagger$, Chase P Broedersz$^\dagger$ \\
Nature Communications 11, 5378 (2020).

\vspace{5mm}

\newenvironment{greytext}{\transparent{0.5}}{\ignorespacesafterend}
\newcommand{\intr}{\int\limits_{\rm [r]}^{}}
\newcommand{\intl}{\int\limits_{\rm [l]}^{}}

\newcommand{\Av}[1]{\overline{ #1 }}
\newcommand{\AER}{\mathbf{A}}
\newcommand{\smallav}[1]{\langle #1 \rangle}
\newcommand{\avc}[1]{\left.\left\langle #1 \right|\bar{c}^n_{\alpha}\right\rangle}
\newcommand{\Var}{\mathrm{Var}}
\newcommand{\Sdothat}{\widehat{\Sdot}}
\newcommand{\dP}{P(\mathbf{x}) \dd \mathbf{x}}
\renewcommand{\vec}[1]{\mathbf{#1}}
\newcommand{\heat}{\Pi} 

\newcommand{\czus}[2]{\genfrac(){0pt}{0}{#1}{#2}}
\newcommand{\la}{\langle}
\newcommand{\ra}{\rangle}
\newcommand{\ec}{\overline{c}}
\newcommand{\eC}{\overline{\mathbf{C}}}
\newcommand{\eCr}{\overline{\mathbf{C}}_{[{\rm r},{\rm r}]}}
\newcommand{\mean}[1]{\left\langle #1 \right\rangle}
\newcommand{\cic}{\rm cic}
\newcommand{\pca}{\rm pca}
\newcommand{\scic}{\rm scic}
\newcommand{\omn}{\omega_{\text{\tiny{NN}}}}
\newcommand{\omb}{\omega_{\text{\tiny{2B}}}}
\newcommand{\omemt}{\omega_{\text{\tiny{EMT}}}}
\newcommand{\percR}{\mathcal{P_{\omega_{\text{r}}}}}
\newcommand{\perc}{\mathcal{P_{\omega_{\text{\tiny{NN}}}}}}
\newcommand{\percb}{\mathcal{P_{\omega_{\text{\tiny{2B}}}}}}
\newcommand{\percemt}{\mathcal{P_{\omega_{\text{\tiny{EMT}}}}}}
\newcommand{\x}{\mathbf{x}}
\newcommand{\y}{\mathbf{y}}
\newcommand{\xr}{\mathbf{x}_{\rm r}}
\newcommand{\xl}{\mathbf{x}_{\rm l}}

\newcommand{\Ar}{\mathbf{A}_{[{\rm r},{\rm r}]}}
\newcommand{\Aeff}{\mathbf{A}_\textup{eff}}
\newcommand{\D}{\boldsymbol{D}}
\newcommand{\Di}{\boldsymbol{D}}

\newcommand{\Fr}{\mathbf{f}_{\rm r}}
\newcommand{\I}{\mathbfcal{I}}
\newcommand{\Ds}{\mathbfcal{D}}
\newcommand{\Vs}{\mathbfcal{V}}
\newcommand{\Fs}{\mathbfcal{F}}
\newcommand{\Fhat}{\widehat{\mathbfcal{F}}}

\newcommand{\Om}{\mathbf{\Omega}}
\newcommand{\Omr}{\mathbf{\Omega}_{\rm r}}
\newcommand{\pdx}{\widetilde{\partial}_{1}^2}
\newcommand{\pdy}{\widetilde{\partial}_{2}^2}
\newcommand{\rstar}{r^*}
\newcommand{\ch}[1]{{\color{red}{#1}}}
\newcommand{\pierre}[1]{{\color{green}{#1}}}
\newcommand{\fede}[1]{{\color{blue}{\bf #1}}}
\newcommand{\grz}[1]{{\color{magenta}{\bf #1}}}
\newcommand{\faded}[1]{{\color{grey}{\bf #1}}}
\newcommand{\mcal}{\mathcal}
\renewcommand{\cvec}{\mathbf{c}}
\newcommand{\sigmaF}{\sigma^2_{\widehat{\mathcal{F}}}}
\newcommand\chout{\bgroup\markoverwith{\textcolor{red}{\rule[0.5ex]{2pt}{1.0pt}}}\ULon}
\newcommand\fedout{\bgroup\markoverwith{\textcolor{blue}{\rule[0.5ex]{2pt}{1.0pt}}}\ULon}
\newcommand\grzout{\bgroup\markoverwith{\textcolor{magenta}{\rule[0.5ex]{2pt}{1.0pt}}}\ULon}

\newcommand{\omegastar}{\omega_{r^*}}
\newcommand{\omegaR}{\omega_r}
%\DeclareMathOperator{\Tr}{\rm Tr}
%\DeclareMathAlphabet\mathbfcal{OMS}{cmsy}{b}{n}
\newcommand{\vecxi}{\boldsymbol{\xi}}
\newcommand{\p}{\mathbf{pc}}
\newcommand{\Fb}{\mathbf{F}}
\newcommand{\Db}{\mathbf{D}}

\newcommand{\Fex}{\mathbf{F}_{\rm ex}}
\newcommand{\Div}{\nabla \cdot}
\newcommand{\area}{\dot{\bm{A}}}
\newcommand{\boldxi}{\bm{\xi} }

\begin{figure*}
\centering
\includegraphics[width=\textwidth]{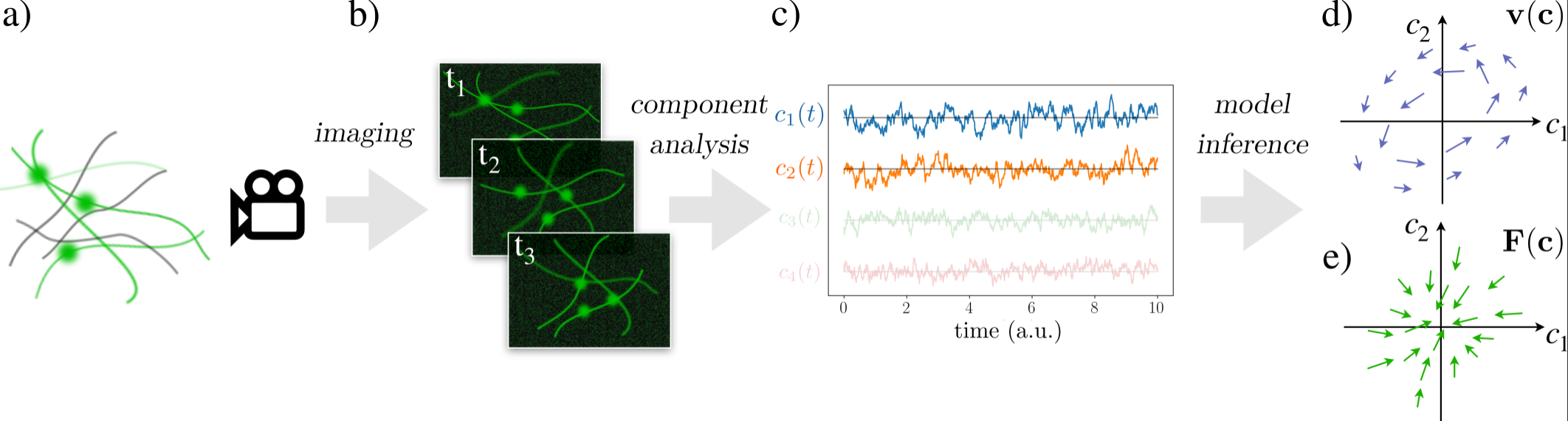}
\caption{\textbf{Schematic illustration of our approach to learn non-equilibrium dynamics from a Brownian movie}. a) Sketch of a network of biopolymers (black) with embedded fluorescent filaments and beads (green). b) Image-frames of the fluorescent components in panel a) at three successive time points. c) The time trajectories of the projection coefficients $c_1(t),c_2(t), \cdots$: the coefficients and respective trajectories discarded by the dimensional reduction are faded. Sketch of the the inferred velocity $\mathbf{v}(\cvec)$ (d) and of the force field $\Fb(\cvec)$ (e) in the space $\{c_1,c_2\}$.}
\label{Fig1BDB}
\end{figure*}

\noindent Over the last two centuries, fundamental insights have been gleaned about the physical properties of biological and soft
matter systems by using microscopes to image their dynamics~\cite{stephens_light_2003,sahl_fluorescence_2017}. 
At the micrometer scale and below, however, this dynamics is inherently stochastic, as ever-present thermally driven Brownian fluctuations give rise to 
short-time displacements~\cite{brown_brief_1828, einstein_uber_1905,von_smoluchowski_zur_1906, frey_brownian_2005}. This random motion makes such ``Brownian
movies'' appear jiggly and erratic; this randomness is further exacerbated by measurement noise and
limited resolution intrinsic to, e.g., fluorescence microscopy~\cite{waters_accuracy_2009}. In light of all these
sources of uncertainty, how can one best make use of measured Brownian movies of a systems dynamics, to learn the underlying physics of the fluctuating and persistent forces? 

In addition to thermal effects, active processes can strongly impact the stochastic dynamics of a system~\cite{mackintosh_active_2010,gnesotto_broken_2018,aranson_active_2013,cates_motility-induced_2015,fodor_how_2016}. Recently, there has been a  growing interest in quantifying and characterizing the
non-equilibrium nature of the stochastic dynamics in active soft and living systems~\cite{martinez_inferring_2019, guo_probing_2014,fakhri_high-resolution_2014,turlier_equilibrium_2016,battle_broken_2016,gladrow_broken_2016,mura_nonequilibrium_2018, seara_entropy_2018,ma_active_2014,li_quantifying_2019,sanchez_spontaneous_2012, frishman_learning_2020, roldan_estimating_2010}.
In cells,  molecular-scale activity, powered for instance by ATP hydrolysis,
controls mesoscale non-equilibrium processes in assemblies such as cilia~\cite{sanchez_cilia-like_2011,battle_intracellular_2015}, flagella~\cite{riedel-kruse_how_2007}, chromosomes~ \cite{weber_nonthermal_2012}, protein droplets~\cite{brangwynne_active_2011} or cytoskeletal networks~\cite{mizuno_nonequilibrium_2007,brangwynne_cytoplasmic_2008,koenderink_active_2009,brangwynne_nonequilibrium_2008}. 
The irreversible nature of such non-equilibrium processes can lead to measurable dissipative currents in a phase space of mesoscopic degrees of freedom~\cite{battle_broken_2016,gladrow_broken_2016,gnesotto_broken_2018, paijmans_discrete_2016,kimmel_inferring_2018,wan_time_2018,zia_manifest_2016}.
Such dissipative currents can be quantified by the entropy production rate~\cite{seifert_stochastic_2012}, which is a measure of the irreversibility of the dynamics~\cite{crooks_entropy_1999}. New approaches have been developed to measure this rate in real systems~\cite{li_quantifying_2019, frishman_learning_2020}, shedding light onto the structure of dissipative processes~\cite{ mura_nonequilibrium_2018} and their impact on the dynamics of living matter~\cite{seara_entropy_2018}. However, it remains an outstanding challenge to accurately infer the entropy production rate by analyzing Brownian movies of such systems. 

Traditional approaches to measure microscopic forces and analyze
time-lapse microscopy data typically rely on tracking the position or
shape of well-defined probes such as tracer beads, fluorescent
proteins and filaments, or simply on exploiting the natural contrast
of the intracellular medium to obtain such
tracks~\cite{mizuno_nonequilibrium_2007,turlier_equilibrium_2016,
  crocker_methods_1996,battle_broken_2016,
  weber_nonthermal_2012,guo_probing_2014,
  fakhri_high-resolution_2014,brangwynne_nonequilibrium_2008,levine_one-_2000,sawada_force_2006,
  grashoff_measuring_2010}. The tracer trajectories can be studied
through stochastic analysis techniques to extract an effective model
for their dynamics and infer quantities like the entropy
production rate~\cite{mura_nonequilibrium_2018,
  mura_mesoscopic_2019,frishman_learning_2020,seara_entropy_2018,bruckner_stochastic_2019,selmeczi_cell_2005,stephens_dimensionality_2008,li_quantifying_2019}.
There are, however, many cases in which tracking is
impractical~\cite{seara_dissipative_2019,edera_differential_2017}, due
to limited resolution or simply because there are no recognizable
objects to use as tracers. Another, more fundamental limitation of
tracking is that one then mostly learns about the dynamics of the
tracked object---not of the system as a whole. Indeed, the dissipative
power in a system might not couple directly to the tracked variables,
and a priori, it might not be clear which coordinates will be
most informative about such dissipation. This raises the question how
one can identify which degrees of freedom best encode the forces and
non-equilibrium dissipation in a given system.

Here we propose an alternative to tracking: learning the dynamics and
inferring the entropy production rate directly from the unsupervised
analysis of Brownian movies. We first decompose the movie into generic
principal modes of motion, and predict which ones are the most likely
to encode useful information through a ``Dissipative Component
Analysis'' (DCA). This allows us to perform a dimensional reduction, leading
to a representation of the movie as a stochastic trajectory in this
component space. Finally, we employ a recently introduced method,
Stochastic Force Inference (SFI)~\cite{frishman_learning_2020}, to
analyze such trajectories. Our approach not only yields an estimate of
the entropy production rate of a Brownian movie, which is a controlled
lower bound to the system's total entropy production rate, but also
important dynamical information such as a time-resolved force map of
the imaged system. Thus, our approach may provide an alternative to methods that use microcopic force sensors~\cite{sawada_force_2006, grashoff_measuring_2010, lucio_chapter_2015, han_cell_2018}.
 In this article, we first present the method in its
generality, then benchmark it on a simple two-beads model. Finally, we demonstrate the potential of our approach on simulated semi-realistic
fluorescence microscopy movies of out-of-equilibrium biopolymer
networks.

\section{Principle of the method}

\noindent We begin by  describing a tracking-free method to infer
 the dynamical equations of a system from raw image
sequences. This approach allows us to determine a bound on the dissipation of a system, as well as the force-field in image space. 

Our starting point is the assumption that the physical system we
observe (Fig.~\ref{Fig1BDB}a)---such as a cytoskeletal network or a fluctuating
membrane---can be described by a configurational state vector $\mathbf{x}(t)$ at time $t$, undergoing
steady-state Brownian dynamics in an unspecified $d$-dimensional
phase space:
\begin{equation}
  \label{eq:Lang_initial} 
  \frac{\dd \mathbf{x}}{\dd t} = \bm{\Phi}(\mathbf{x})  + \sqrt{2 \D (\mathbf{x})} \boldxi(t),
\end{equation}
where $\bm{\Phi}(\mathbf{x})$ is the drift field,
$\D(\mathbf{x})$ is the diffusion tensor field, and throughout
this article $\boldxi(t)$ is a Gaussian white noise vector
($\la \boldxi(t) \ra=0$ and
$\la \xi_i(t) \xi_j (s) \ra =\delta_{ij}\delta(t-s) $). Note that when
diffusion is state-dependent, $\sqrt{2\D(\mathbf{x})} \boldxi(t)$
is a multiplicative noise term: we employ the It\^o convention for the
drift, \emph{i.e.} $\bm{\Phi}(\mathbf{x})= \mathbf{F(\mathbf{x})}+\nabla \cdot \D(\mathbf{x})$, where $\mathbf{F(\mathbf{x})}$ is the product of the mobility matrix and the physical force in the absence of Brownian noise~\cite{lau_state-dependent_2007,risken_fokker-planck_1996}.

Our goal is to learn as much as possible about the process described by Eq.~\eqref{eq:Lang_initial} from an experimental observation. In particular, we aim to measure if, and how far, the system is out-of-equilibrium by determining the irreversible nature of its dynamics. This irreversibility is quantified by the system's entropy production rate~\cite{seifert_stochastic_2012}
\begin{equation}
  \label{eq:SdotBDB}
  \Sdot_\mathrm{total} = \av{ \mathbf{v}(\mathbf{x}) \D^{-1}(\mathbf{x}) \mathbf{v}(\mathbf{x})  },
\end{equation}
where $\av{ \cdot }$ denotes a steady-state average, throughout this article we set Boltzmann's constant
$k_{\rm B} =1$, and 
$\mathbf{v}(\mathbf{x})$ is the mean phase space velocity field quantifying
the presence of irreversible currents. Specifically, using the steady-state Fokker-Planck
equation one can write
$\mathbf{v}(\mathbf{x}) = \mathbf{F}(\mathbf{x}) - \D(\mathbf{x}) \nabla \log P(\mathbf{x})$,
where $P(\mathbf{x})$ is the steady-state probability density function, and flux balance imposes that $\nabla \cdot (P \mathbf{v}) =0$.

The input of our method consists  of a discrete time-series of
microscopy images of the physical system $\{\I(t_0), \dots \I(t_N)\}$---a ``Brownian movie'' (Fig.~\ref{Fig1BDB}b). Each image $\I(t)$ is an imperfect
representation of the state $\mathbf{x}(t)$ of the physical system as
a bitmap, \emph{i.e.} a $L\times W$ array of real-valued pixel
intensities~\footnote{We neglect the discretization effect induced by
  the finite number of pixel intensities here.}. Specifically, we
model the imaging apparatus as a noisy nonlinear map $\I(t) = \bar{\I}(\mathbf{x}(t)) + \mathbfcal{N}(t)$, where $\mathbfcal{N}$
is a temporally uncorrelated random array representing measurement noise
(such as the fluctuations in registered fluorescence intensities), and
$\bar{\I}(\mathbf{x})$ is the ``ideal image'' returned on average by
the microscope when the system's state is $\mathbf{x}$. We assume that
the map $\mathbf{x}\mapsto \bar{\I}(\mathbf{x})$ is time-independent (i.e. that
the microscope settings are fixed and stable). 

Importantly, if no information is lost by the imaging process, the ideal image $\bar{\I}(t)$ undergoes a Brownian dynamics equation determined by the nonlinear transformation of Eq.~ \eqref{eq:Lang_initial} through the map $\mathbf{x}\mapsto\bar{\I}(\mathbf{x})$, as prescribed by 
It\^o's lemma~\cite{oksendal_stochastic_2003}. 
In general, however, there is information loss and this map is not invertible: due to finite optical resolution or because some elements are simply not visible, the imaging may not capture the full high-dimensional state of the system.  For
this reason, the dynamics in image space are not uniquely
specified by the ideal image value $\bar{\I}$; they also depend on
``hidden'' degrees of freedom $\x_{\rm h}$ not captured by the image. In this case, a
Markovian dynamical equation for $\bar{\I}$ alone does not exist, but by  including the dynamics of $\x_{\rm h}$, we can write 
\begin{equation}
\frac{\dd}{\dd t}(\bar{\I},\x{\rm _h}) =  \bm{\varphi}(\bar{\I},\x_{\rm h}) + \sqrt{2\Ds(\bar{\I},\x_{\rm h}) }\vecxi(t).
\label{eq:Langevin_images_new}
\end{equation}
Here $(\bar{\I},\x{\rm _h})$ is a column vector composed of pixel intensities $\bar{\I}$ and hidden degrees of freedom $\x{\rm _h}$, $\bm{\varphi}(\bar{\I},\x_{\rm h})$ and $\Ds(\bar{\I},\x_{\rm h})$ are  the drift field and diffusion tensor, respectively, in the combined space of pixel intensities and hidden variables. Our Brownian movie analysis allows us to infer the mean image drift $\bm{\varphi}(\bar{\I})\coloneqq\av{\bm{\varphi}_\I(\bar{\I},\x_{\rm h})|\bar{\I}}$ and mean image diffusion tensor $\Ds(\bar{\I})\coloneqq\av{\Ds_\I(\bar{\I},\x_{\rm h})|\bar{\I}}$, averaged over the degrees of freedom $\x_{\rm h}$ lost in the imaging process. From drift and diffusion fields we can directly obtain the mean image force field $\Fs(\bar{\I})=\bm{\varphi}(\bar{\I})-\nabla \cdot \Ds(\bar{\I})$. Similar to force and diffusion fields, the phase space velocity field
$\mathbf{v}(\mathbf{x})$ in the $d$-dimensional physical phase space,
transforms into a velocity field $\Vs(\bar{\I})$ in the
$L\times W$-dimensional image space---again, averaged over
unobserved degrees of freedom. The corresponding currents result in an apparent
entropy production rate associated to the image dynamics~\footnote{Note that
  we consider here only the entropy production rate associated to apparent
  currents. The irreversible dynamics of unobserved degrees of freedom
  has repercussion on non-Markovian effects in the dynamics, which
  result in other contributions to the entropy 
 production~\cite{roldan_estimating_2010}, which we
  neglect here.},
\begin{equation}
  \label{eq:Sdot_apparent}
  \Sdot_\mathrm{apparent} = \av{\Vs(\bar{\I}) \Ds^{-1}(\bar{\I}) \Vs(\bar{\I}) }.
  \end{equation}
Importantly, $\Sdot_\mathrm{apparent} \leq \Sdot_\mathrm{total}$:
  the apparent entropy production rate is a lower bound to the total
  one. Indeed, all transformations involved in the analysis process --
  imaging through the nonlinear map
  $\mathbf{x}\mapsto\bar{\I}(\mathbf{x})$, masking the hidden
  degrees of freedom, and averaging over their value -- have
  nonincreasing effects on the entropy production rate (see Supplementary Note~8). The measure of $\Sdot_\mathrm{apparent}$ thus provides
  direct insight into the dissipative processes in the physical
  system.

The goal of our method is to reconstruct the mean image-space dynamics
$(\Fs(\bar{\I}),\Ds(\bar{\I}))$, and in particular the
corresponding entropy production rate (Eq.~\eqref{eq:Sdot_apparent}). However, doing so in the high-dimensional image
space is unpractical and would require unrealistic amounts of data. We
therefore need to reduce the dimensionality of our system to a
tractable number of relevant degrees of freedom.

Because each image represents a physical state of the system, we expect that the ideal images $\bar{\I}(t)$ all share similar structural features. Consequently, the Brownian movie occupies only a smaller subspace in the space of all configurations of pixel intensities. To restrict ourselves to the manifold of images representing the physical states and to reduce the noise, we first perform a standard dimensionality reduction procedure: for simplicity, we employ Principal Component Analysis (PCA). As we shall see later, this standard procedure can be reinforced with an analysis that provides an additional basis transformation to select the most dissipative components. The
idea behind this approach is to find an appropriate basis, in which pairs of components can be hierarchically ordered according to how much they are expected to contribute to the total entropy production rate. It then becomes possible to truncate the basis and reduce the dimensionality of the problem, while retaining maximum information about the system's irreversible dynamics.

We truncate the basis of components according to two criteria: 1) Noise floor---due
  to the finite amount of data and the measurement noise present in
  the Brownian movie, some modes are indistinguishable from the measurement
  noise. We only keep modes that rise above this noise floor. 2) Time resolution of the dynamics---we only
  consider the components whose statistical properties are consistent
  with Brownian dynamics, i.e. such that the short-time
  diffusive behavior can be resolved through the noise. In low-dimensional systems, these criteria can be extended with an additional restriction based on estimating the dimensionality of the set of images in the Brownian movie.

Our task is now reduced to inferring the mean dynamics in  component space,
\begin{equation}
\bm{\Phi}(\cvec) \coloneqq \av{ \bm{\Phi}_{\cvec}(\cvec,\x_{\rm h})|\cvec} \ ,\  
\Di(\cvec) \coloneqq\av{\Di_{\cvec}(\cvec,\x_{\rm h})|\cvec}
\label{eq:Langevin_components}
\end{equation}
where $\cvec(t)=(c_1(t),c_2(t),\cdots,c_n(t))$ are the components obtained after a linear transformation of the images (see Fig.~\ref{Fig1BDB}c), $\Di_{\cvec}$ is the restriction of the diffusion tensor to the $\cvec$-space, and the hidden degrees of freedom $\x_{\rm h}$ now also include those present in the image, but left out after the components' truncation. This procedure has reduced the system's dynamics to that of a smaller
number of components, making it possible to learn $\bm{\Phi}(\cvec)$ and $\Di(\cvec)$. 

To this end, we employ a recently introduced method, Stochastic Force Inference~\cite{frishman_learning_2020}
(SFI), for the inverse Brownian dynamics problem. Briefly, this
procedure is based on a least-squares approximation of the diffusion
and drift fields using a basis of known functions (such as
polynomials). This method is data-efficient, not limited to
low-dimensional signals or equilibrium systems, robust against
measurement noise, and  provides estimates of the inference
error, making it well suited for our purpose. In practice, we use SFI in two ways: 1) we infer the velocity
field $\mathbf{v}(\mathbf{c})$ (Fig.~\ref{Fig1BDB}d) and the diffusion field
$\Di(\mathbf{c})$, which we use to measure
the entropy production rate. 2) We
 infer the drift field $\bm{\Phi}(\mathbf{c})$, compute the image force $\Fb(\mathbf{c})=\bm{\Phi}(\mathbf{c})-\nabla \cdot \Di(\cvec)$  (Fig.~\ref{Fig1BDB}e), and thus reconstruct the
dynamics of the components. To render this deterministic dynamics more intelligible,
we can transform $\Fb(\mathbf{c})$ back into image space by inverting the
$\I\mapsto \mathbf{c}$ linear transformation: this results in
a pixel force-map, which indicates at each time step
which pixel intensities tend to increase or decrease.  This provides,
we argue, a way to gain insight into the dynamics of Brownian
systems and disentangle deterministic forces from Brownian motion
without tracking.

Our analysis framework can thus be schematically summarized as:
imaging $\to$ component analysis $\to$ model inference (Fig.~\ref{Fig1BDB}). This procedure
allows the inference of entropy production rate and reconstruction of the
dynamical equations from image sequences of a Brownian system.

\section{A minimal example: two-beads Brownian movies}

\begin{figure*}
\centering
\includegraphics[width=\textwidth]{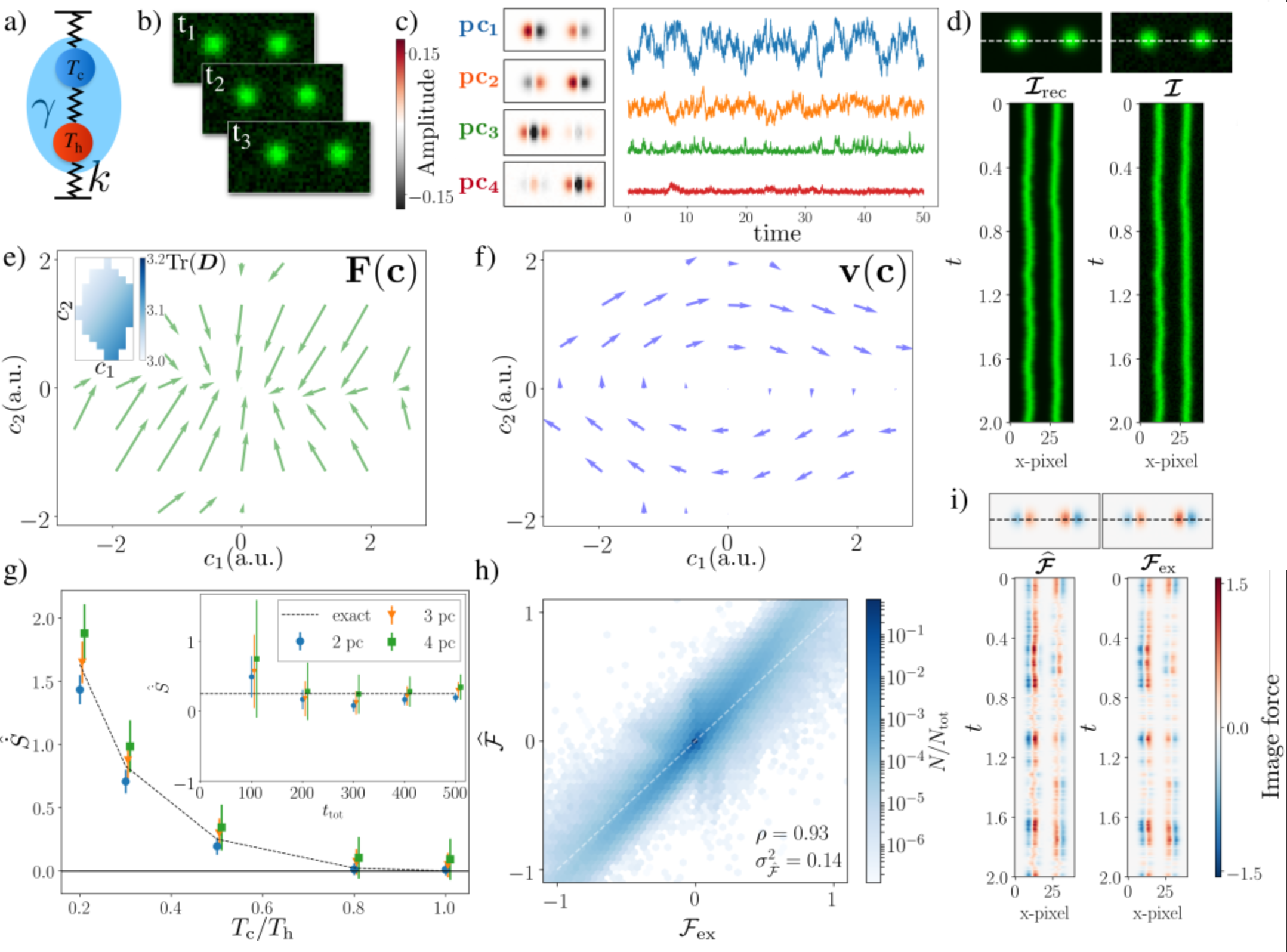}
\caption{\textbf{Benchmarking the Brownian movie learning approach with a simple toy model} a) Schematic of the two-bead model. The temperature of the hot bead $T_{\rm h}=1$ is fixed and the temperature of the cold bead $T_{\rm c} \leq 1$ is varied. b)  $40\times20$  Frames of the noisy ($10 \%$ noise) Brownian movie for the two bead-model at successive time-points c) The first 4 principal components (in arbitrary units) with  time-traces of respective projection coefficients. The color map displays negative values in black and positive values in red. d) Snapshot $\I_{\rm rec}$ of the reconstructed movie, reconstructed with the first four principal components, and snapshot $\I$ of the original movie (right), together with associated kymographs. Pixel intensity ranges from $0$ (black) to $1$ (bright green). We compare pixel intensities along the superimposed horizontal dashed line. Force field (e) and mean phase space velocity (f) in the space of the first two principal components  $\{c_1,c_2\}$. Arrows are scaled for visualization purposes. Inset e):  trace of diffusion tensor $\Tr(\Di)$ with the same axis scaling. g) Inferred entropy production rate $\widehat{\dot{S}}$ for varying temperature ratio $T_{\rm c}T_{\rm h}^{-1}$ and number of included principal components. Inset: $\widehat{\dot{S}}$ as a function of trajectory length for a fixed $T_{\rm c}T_{\rm h}^{-1}=0.5$. The error bars represent an estimate of the root-mean-square deviation between the true apparent entropy production rate and the inferred value (see Methods). h) Scatter plot of the elements of the exact image force field $\mathbfcal{F}_{\rm ex}$  vs. the inferred image force field $\Fhat$ for different pixels and time points (data has been binned for visualization purposes). Results are obtained using the first four principal components. i) Comparison of inferred $\Fhat$ and exact $\Fhat_{\rm ex}$ image-space force fields, together with associated kymographs.}
\label{Fig2BDB}
\end{figure*}

Next, we test the performance of our procedure on a simple
non-equilibrium model: two coupled beads moving in one
dimension. The beads are coupled by Hookean springs with stiffness $k$ and experience Stokes drag with friction coefficient $\gamma$, due to the surrounding fluid 
(Fig.~\ref{Fig2BDB}a).  In this two-bead model, the time-evolution of the
bead displacements $\x(t)=(x_1(t),x_2(t))$ obeys the overdamped
Langevin Eq.~\eqref{eq:Lang_initial}, with $\mathbf{F}(\x)=\boldsymbol{K} \x$ and
$K_{ij} = (1- 3\delta_{ij}) k \gamma^{-1}$. The system is driven out of
thermodynamic equilibrium by imposing different temperatures on the
two beads: $D_{ij}= \delta_{i j} k_{\rm B}T_i \gamma^{-1} $~\cite{crisanti_nonequilibrium_2012,berut_theoretical_2016, li_quantifying_2019, gnesotto_broken_2018,gnesotto_nonequilibrium_2019}.
 First, we obtain position trajectories for the two beads by
discretizing their stochastic dynamics using an Euler integration
scheme (see Supplementary Note~1). Then,  we use these position trajectories to 
construct a noisy Brownian movie (Fig.~\ref{Fig2BDB}b) (cf. Supplementary Note~2 and Supplementary Movie~1). Note that by construction, the
steady-state dynamics of the two-beads system in image space is
governed by a non-linear Langevin equation with multiplicative noise.

We seek to reduce the dimensionality of the data and to filter out measurement noise by finding
relevant components. To this end, we employ Principal Component
Analysis (PCA)~\cite{bishop_pattern_2006} and determine the basis of $n$ principal components
$\p_1, \p_2,\cdots, \p_n$ to expand each image around the
time-averaged image $\langle \I \rangle$:
$\mcal \I(t)= \langle \I \rangle+ \sum_{i=1}^n c_i(t) \p_i$. The dynamics of the projection
coefficients are on average governed by the drift field $\bm{\Phi}(\cvec)$
and diffusion tensor $\Di(\cvec)$ (see Eq.~\eqref{eq:Langevin_components}).

In the simulated data of the two-bead model,
the first four principal components satisfy criteria 1) and 2)
introduced above (Fig.~\ref{Fig2BDB}c). Interestingly, $\p_1$ and $\p_2$
resemble the in-phase and out-of-phase motion of the two beads,
respectively, and should suffice to reproduce the dynamics of $(x_1(t),x_2(t))$. The components $\p_3$ and $\p_4$ appear to mostly represent the isolated
fluctuations of the hot and cold beads and mainly account for the nonlinear details of the image representation. Together, the first four components allow for an adequate reconstruction of the
original images (Fig.~\ref{Fig2BDB}d, Supplementary Figure~1).

From the recorded trajectories in $\p_1 \times \p_2$ space we can already
infer key features of the system's dynamics using  SFI. Specifically, we infer the force and diffusion fields (Fig.~\ref{Fig2BDB}e). In the
phase space spanned by the first two principal components, we identify
a stable fixed point at $(0,0)$ (Fig.~\ref{Fig2BDB}e). As may be expected in this case, the $\p_1$-direction (in-phase motion) is less stiff than
the $\p_2$ direction (out-of-phase motion).

The temperature difference between the two beads results in
phase-space circulation, as revealed by the inferred mean velocity field
(Fig.~\ref{Fig2BDB}f). To quantitatively assess the irreversibility
associated with the presence of such phase space currents, we estimate
the entropy production rate of the system $\widehat{\dot{S}}$, which
converges for long enough measurement time
(Fig.~\ref{Fig2BDB}g-inset). Strikingly, already with two principal
components we find good agreement between the inferred and
the exact entropy production rate, capturing from 
$78 \pm 25\%$ at $T_{\rm c}T_{\rm h}^{-1}=0.5$) to $88 \pm 7 \%$
of the entropy production rate at $T_{\rm c} T_{\rm h}^{-1}=0.2$ (Fig.~\ref{Fig2BDB}g). Furthermore, the difference between
the exact and inferred entropy production rate is consistent with the
typical inference error predicted by SFI. As expected, the estimate of the entropy production rate increases with the number of included components. Note that including more modes than the dimension of the physical phase space (in this case 2) can lead to an overestimate of $\dot{S}$ (Fig.~\ref{Fig2BDB}g). In such low-dimensional systems, one can further restrict the number of included components based on estimating the dimensionality of the set of images in the Brownian movie.

We can also use the information contained in the first four
principal components to quantitatively infer forces in image-space via
the relation $\Fhat(\I(t))= \sum_{i=1}^4 \widehat{F}_i(\cvec(t)) \p_
i$. Note that while two modes were sufficient to infer $\widehat{\dot{S}}$, 
more modes are needed to reconstruct the full images and
image-force fields as a linear combination of modes.
When inferring forces we always subtract from the drift the
spurious force  $\nabla \cdot \Di(\cvec)$ arising in overdamped It\^o
stochastic differential equations with multiplicative noise~\cite{lau_state-dependent_2007, risken_fokker-planck_1996}. For
comparison purposes, the exact image force field is obtained directly
from the simulated data as:
$\Fhat_{\rm ex}(t)=\{\bar{\I}[\x(t)+\Fb(\x(t)) \Delta t]-\bar{\I}(\x(t))\}\Delta t^{-1}$. Remarkably, we find good qualitative
agreement between inferred and exact image force fields for specific
realizations of the system, as shown in the kymographs in
Fig.~\ref{Fig2BDB}i (see also Supplementary Movies~2 and 3). Moreover, we find a strong correlation (Pearson
correlation coefficient $\rho=0.93$) between inferred and exact
image-forces. To further quantify the performance of force inference, we compute the relative squared error on the inferred image force field $\sigmaF=\sum_t \lVert \Fhat(t)-\Fhat_{\rm ex}(t) \rVert^2 \left(
\sum_t \lVert \Fhat(t)\rVert ^2\right)^{-1}$, which in this case is modest, $\sigmaF=0.14$ (Fig.~\ref{Fig2BDB}h). 

Thus, with sufficient information, we can use our approach to accurately predict at any instant of time the physical force fields in image space from the Brownian movie, even if the system is out of equilibrium.
Moreover, the results for this simple two-bead system demonstrate the validity of our approach: we reliably infer the non-equilibrium dynamics of this system. Arguably, direct tracking of the two beads is, in this case, a more straightforward approach. However, this changes when considering more general soft assemblies comprised of many degrees of freedom. 

\section{Dissipative Component Analysis}
To expand the scope of our approach, we next consider  a more complex scenario inspired by cytoskeletal assemblies: a network of elastic filaments (Fig.~\ref{Fig3BDB}a). The filaments are modeled as Hookean springs represented as bonds connecting neighboring nodes of a triangular network. We randomly remove bonds to introduce spatial disorder in the system. The state of the network as a whole, represented by the set $\{\mathbf{x}_i\}$ of two-dimensional displacement of each node $i$,  undergoes  Langevin dynamics (Eq.~\eqref{eq:Lang_initial}). In this case, the force acting on node $i$ is $\Fb_i(\x)=-\sum_{j \sim i} \frac{k_{ij}}{ \gamma}(\lVert \x_{i,j}(t) \rVert-\ell_0)\hat{\x}_{i,j}$, where $k_{ij}=k$ if the bond is present, $k_{ij}=0$ if it is not, $\mathbf{x}_{i,j}=\mathbf{x}_{i}-\mathbf{x}_{j}$, $\hat{\mathbf{x}}_{i,j}$
is the corresponding unit vector, and the sum runs over the nearest-neighbor nodes $j$ of node $i$. Rigid boundary conditions are imposed to avoid rotations and diffusion of the system as a whole.
Finally, we drive the system out of equilibrium by randomly setting a fraction of the network nodes at an elevated temperature, as illustrated in Fig.~\ref{Fig3BDB}a. 

To study an experimentally relevant scenario, we generate a Brownian movie of a random filamentous network (Supplementary Note~2), which is only partially imaged  (black frame in Fig.~\ref{Fig3BDB}a) with measurement noise and at a limited optical resolution (Supplementary Note~6,~7). To simulate limited optical resolution, we blur the image-frames of the movie with a Gaussian filter (Fig.~\ref{Fig3BDB}b and Supplementary Movie~4). In this spatially extended system, generated from an underlying dynamics with $800$ degrees of freedom, it is not obvious based on the recorded Brownian movie ($80 \times 80$ pixels) how to select and analyze the relevant degrees of freedom. 

We start our movie-based analysis by employing PCA to reduce the dimensionality of the image data  (Fig.~\ref{Fig3BDB}c). For this set of simulation data, our truncation criteria indicate that the maximum number of retainable components is roughly $200$ (Supplementary Note~5 and Supplementary Figure 3). Although we greatly reduced dimensionality of the image data using this truncation, it is still intractable to infer dynamics in a $200$-dimensional space due to limited statistics. However, even a subset of these modes may suffice to glean useful information about the system's non-equilibrium dynamics. Therefore, as a first attempt, we infer the dynamics in increasingly larger PC-space via SFI. This allows us to infer the retained percentage of entropy production rate $\widehat{\Sdot}/ \Sdot_{\rm ex}$ in the observed region (See Supplementary Note~2) as a function of the number of principal components considered (Fig.~\ref{Fig3BDB}e). In contrast to the two-beads case, we observe that in this more realistic scenario we recover less than $4 \%$ of the  entropy production rate of the observed system with the first $30$ PCs. Indeed, PCA is designed to find modes that capture the most variance in the image data, and large variance does not necessarily imply large dissipation. Thus, in this case, PCA fails at selecting components that capture a substantial fraction of the entropy production rate. 

Our goal is to infer the system's non-equilibrium dynamics. We thus propose an alternative way of reducing data dimensionality that spotlights the time-irreversible contributions to the dynamics, which we term Dissipative Component Analysis (DCA). DCA represents a principled approach to determine the most dissipative pairs of modes for a linear system with state-independent noise (see Supplementary Note~3). For such a linear system, there exists a set of component pairs for which the entropy production rate can be expressed as a sum of independent positive-definite contributions, which can be ranked by magnitude. After a suitable truncation, this basis ensures that the components with the largest entropy production rate are selected. While the approach is only rigorous for a linear system with state-independent noise, we demonstrate below that this method  also performs well for more general scenarios.

\begin{figure*}[h!tb] 
\centering
\includegraphics[width=\textwidth]{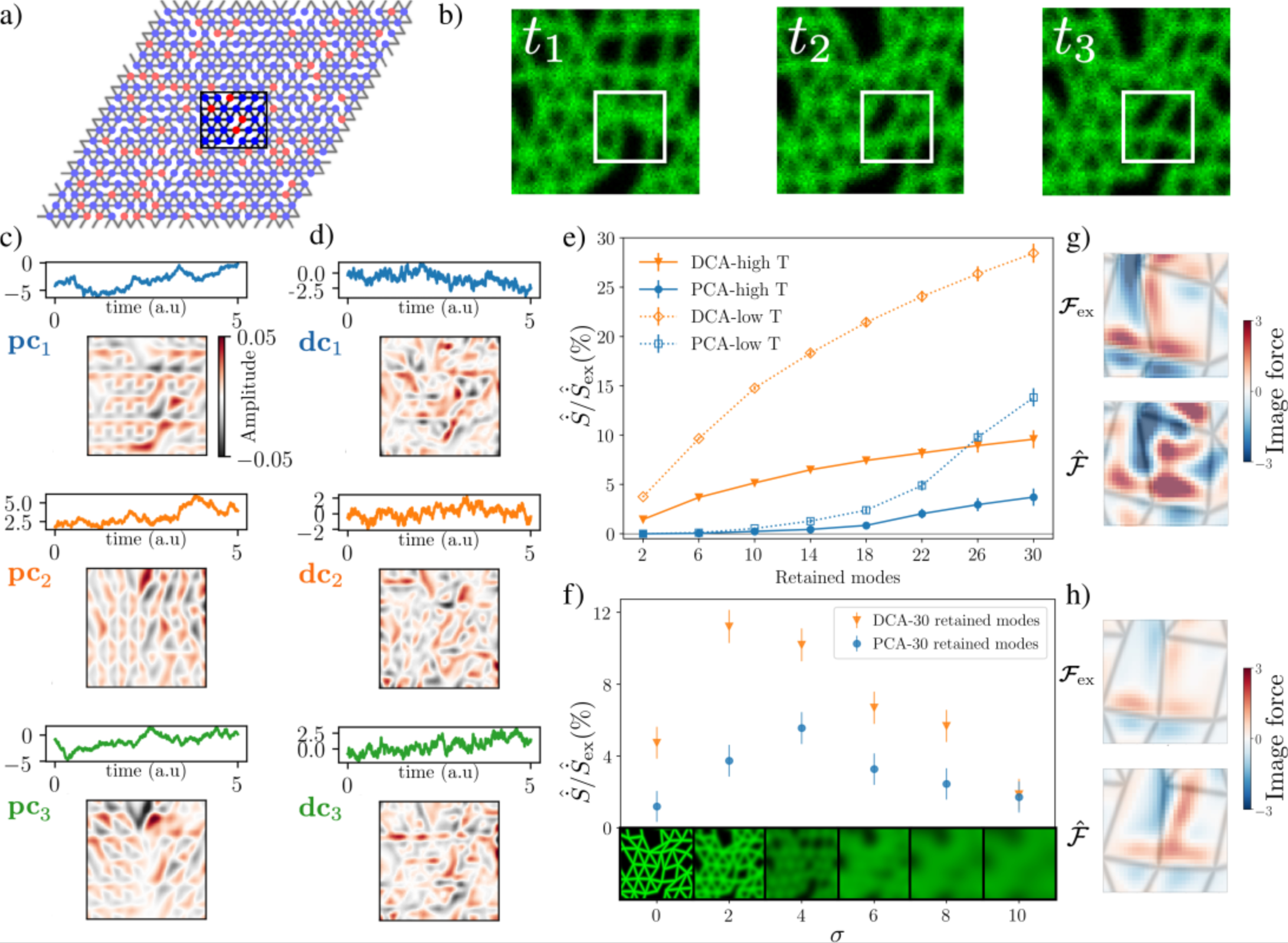}
\caption{\textbf{Learning the non-equilibrium dynamics of Brownian movies of simulated filamentous networks} a) The $20 \times 20$ filamentous network generated in the Brownian dynamics simulation with $20 \%$ random bond removal and heterogeneous temperatures: node temperatures are randomly set to $T_{\rm hot}$ with probability $0.2$, or else to $T_{\rm cold}=0.2 \, T_{\rm hot}$. The black frame indicates the observed region of the system which is analyzed with our movie-based method. b) Three time frames of the Brownian movie of the observed region of the system ($80 \times 80$ pixels, $T_{\rm hot}=0.25$). c-d) Trajectory of the projection coefficient $c_i$ in arbitrary units together with associated image-component for PCA (c) and DCA (d) for the observed region defined in panel a). Scale bar applies to all image-components. e) The recovered entropy production rate $\widehat{\dot{S}}/\dot{S}_{\rm ex}$ for the observed region as a function of the number of components included in the analysis. For the high and low temperature cases $T_{\rm hot}=0.25$ and $T_{\rm hot}=0.05$, respectively. See Supplementary Note~5 and  Supplementary Figure~2 for additional data at equilibrium and convergence of the estimates with total time. f): The recovered entropy production rate $\widehat{\dot{S}}/\dot{S}_{\rm ex}$ as a function of the blurring parameter $\sigma$ for $30$ retained PCs and DCs. We show a corresponding blurred  frame above every $x$-axis tick. The error bars in panels e,f) represent an estimate of the root-mean-square deviation between the true apparent entropy production rate and the inferred value (see Methods). g,h) Comparison of the exact image-force $\mathcal{F}_{\rm ex}$ to the inferred one $\hat{\mathcal{F}}$  at a selected instant of time for the region of interest in the white frame in panel b) for the high (g) and low (h) temperature cases. The underlying network structure is drawn in grey as a guide to the eye.}
\label{Fig3BDB}
\end{figure*}

DCA relies on the measurement of an intuitive trajectory-based non-equilibrium quantity: the area enclosing rate (AER) matrix $\area$ associated to a general set of coordinates $\y$. The elements of the AER matrix, in It\^o convention, are defined by~\cite{zia_manifest_2016,ghanta_fluctuation_2017,gonzalez_experimental_2019,gradziuk_scaling_2019,frishman_learning_2020}
\begin{equation}
\label{eq:AERBDB}
\dot{A}_{ij}=\frac{1}{2} \langle y_j \dot{y}_i- y_i\dot{y}_j \rangle,
\end{equation} 
where $y_i$ denotes the $i$-th coordinate centered around its mean value and $\langle \cdot \rangle$ a time average.
This non-equilibrium measure quantifies the average area enclosed by the  trajectory  in phase space per unit time. Importantly, the AER is tightly linked to the entropy production rate. Specifically, for a linear system $\dot{S}~=~\Tr (\area \bm{C}^{-1} \area^T \Di^{-1})$ where the covariance matrix ${C}_{ij} = \langle y_i y_j \rangle $. DCA identifies a basis of vector pairs
$\{(\mathbf{dc}_1,\mathbf{dc}_2); (\mathbf{dc}_3,\mathbf{dc}_4);\ldots\}$
 that simultaneously transforms $\bm{C}$ to the identity and diagonalizes $\area\area^{T}$ (see Supplementary Note~3). By doing so, DCA naturally separates the entropy production rate into independent contributions that can be readily ordered by magnitude, \emph{i.e.} $\dot{S}~=~\dot{S}_{\mathbf{dc}_1,\mathbf{dc}_2} + \dot{S}_{\mathbf{dc}_3,\mathbf{dc}_4} + \cdots$ with $\dot{S}_{\mathbf{dc}_1,\mathbf{dc}_2} > \dot{S}_{\mathbf{dc}_3,\mathbf{dc}_4} >~\cdots~$. Truncating the basis of dissipative components using the aforementioned criteria, allows us to identify a limited number of components that are assured to maximally contribute to the dissipation of the system. This is analogous to PCA, where the diagonalization of the covariance matrix $\bm C$ allows one to select the components which capture most of the variance.

To test the performance of DCA, we revisit the network simulations. We first perform PCA to reduce noise and dimensionality. Subsequently, we perform DCA with the first $200$ principal component coefficients as input.  The dissipative components   exhibit a different spatial structure than the principal components, as they aim to maximize different quantities (Fig.~\ref{Fig3BDB}d). Strikingly, DCA allows us to recover a larger portion of the entropy production rate of the observed region (almost $10 \%$ with $30$ components), performing consistently better than the PCA-based approach, as shown in Fig.~\ref{Fig3BDB}e. Finally, we note that the performance of our approach improves substantially in systems with smaller fluctuations in which the image-space dynamics is closer to linear (Fig.~\ref{Fig3BDB}e and Supplementary Movie~8).

In non-equilibrium systems our DCA-based method infers non-zero
  entropy production rates, even with poor optical resolution
  (Fig.~\ref{Fig3BDB}f, Supplementary Note~6, and Supplementary Figure~4) and with strong
  measurement noise (Supplementary Note~7 and Supplementary Figure~5). At the same time we
  measure no dissipation in equilibrium systems. Thus, this example
  illustrates the potential applicability of our approach to real
  experiments on biological assemblies.

Our inference approach reveals additional information about the dynamics in the system, such as force field estimates. These force fields provide insight into the spatial structure of the instantaneous deterministic forces in the system at a given configuration. In image space, these forces describe the dynamics of the pixel: positive and negative image forces represent a deterministic force acting to respectively raise and lower pixel values, which reflect the forces acting on the position and shape of the objects being imaged.
To investigate to what extent our movie-based learning approach reconstructs the elastic forces exerted by the network's filaments, we exploit the short range of the interactions in the system to facilitate extracting information about local forces from local  dynamics in image space. We consider a small region of interest (white frame in Fig.~\ref{Fig3BDB}b, Supplementary Movie~5)  and compare the inferred force field in image space to the exact one. For this purpose, we employ PCA in our dimensional reduction scheme, which can be used both in and out of equilibrium. Inferring image force fields with high accuracy for this complex example is challenging (Pearson correlation coefficient between exact and inferred images force  $\rho=0.37$ for the high temperature case and $\rho=0.56$  for the low temperature case). Nonetheless, despite the network disorder, large fluctuations, many hidden degrees of freedom, limited optical resolution, and measurement noise, we find that the inferred force field in image space can capture the basic features of the exact force field, as shown in Fig.~\ref{Fig3BDB}g,h (Supplementary Movies~6-11). Finally, we emphasize that our approach is scalable:  force inference on a small spatial region of interest can be applied to arbitrarily large systems, as long as the interactions are local.

\section{Discussion}
\noindent We considered the dynamics of movies of time-lapse microscopy data. Under the assumptions outlined in the first section of the Results, these movies undergo Brownian dynamics in image space: the image-field obeys an overdamped Langevin equation of the form of Eq.~\eqref{eq:Langevin_images_new}.  Rather than tracking  selected degrees of freedom, we propose to analyze the Brownian movie as a whole. 

Our approach is based on constructing a reduced set of relevant degrees of freedom to reduce dimensionality, by combining PCA with a new method that we term Dissipative Component Analysis (DCA). In the limit of a linear system with state-independent noise, DCA provides a principled way of constructing and ranking independent dissipative modes. 
The order at which we truncate is an important trade-off parameter of this method: on the one hand we wish to
  significantly reduce the dimensionality of the data, on the other
  hand we need to include enough components to retain the information
  necessary to infer the system's dynamics. After the dimensional reduction, we infer the stochastic dynamics of the system, revealing the force field, phase space currents, and the entropy production rate in this basis. This information can then be mapped back to image-space to provide estimators for the stochastic dynamics of the Brownian movie. We illustrated our approach on simulated data of a minimal two-beads model and on complex filamentous networks in both equilibrium and non-equilibrium settings, and showed that it is robust in the presence of  measurement noise and with limited optical resolution. Beyond providing controlled lower bounds of the entropy production rates directly from the Brownian movie, our approach yields estimates of the force-fields in image space for an instantaneous snapshot of the system and we demonstrated that this approach can be scaled up to large systems. Thus, we provide in principle an alternative to microscopic force and stress sensing \mbox{methods}~\cite{sawada_force_2006,grashoff_measuring_2010,lucio_chapter_2015,han_cell_2018}.
  
We  focused here on a class of soft matter systems termed ``active viscoelastic solids"~\cite{fletcher_active_2009,gnesotto_broken_2018}. Such systems include active biological materials such as cytoskeletal assemblies~\cite{mizuno_nonequilibrium_2007,jensen_mechanics_2015,koenderink_active_2009,brangwynne_nonequilibrium_2008}, membranes~\cite{turlier_equilibrium_2016,betz_atp-dependent_2009,ben-isaac_effective_2011}, chromosomes ~\cite{weber_nonthermal_2012}, protein droplets~\cite{brangwynne_active_2011}, as well as active turbulent solids ~\cite{hemingway_active_2015} and colloidal systems~\cite{aranson_active_2013}. Although these structures are constantly fluctuating both due to energy-consuming processes (\emph{e.g.} rapid contractions generated by molecular motors) and thermal motion, they do not exhibit macroscopic flow. 
Useful insights into the properties of such systems have been obtained via different non-invasive techniques. Typically, these techniques employ time traces of tracked objects to extract
information about the active processes governing the non-equilibrium behavior~\cite{mura_nonequilibrium_2018, gnesotto_nonequilibrium_2019, gladrow_broken_2016, battle_broken_2016, seara_entropy_2018,turlier_equilibrium_2016,betz_atp-dependent_2009}. Often, however, it is not a priori obvious which physical degrees of freedom should be tracked, how tracking can be performed in fragile environments, and to what extent the dynamical information about the system of interest is encoded in the measured trajectories~\cite{seara_dissipative_2019}. While tracking-free approaches have been proposed to obtain rheological information of a system under equilibrium conditions~\cite{edera_differential_2017}, our approach offers an alternative to tracking that can provide information on dissipative modes and the instantaneous force fields of a fluctuating non-equilibrium system. 

In summary, we presented a viable alternative to traditional analysis techniques of high-resolution video-microscopy of soft living assemblies. Indeed, we envision experimental scenarios where our approach may serve as a guide, providing insights by disentangling the deterministic and stochastic components of the dynamics, and by helping to identify the source of thermal and active forces as well as the dissipation in the system. Overall, our movie-based approach constitutes an adaptable tool that paves the road for a systematic, non-invasive and tracking-free analysis of time-lapse data of soft and living systems.\\

\chapter{Applications of stochastic inference to cell migration}
\label{chap:applications}

\emph{While the core of this Thesis is about mathematical method
  development, my main long-term interest lies in leveraging these
  methods to study biophysical systems and -- hopefully -- use them to
  discover new physics. While I am involved in many collaborations
  towards this goal, few of them have panned out yet. This Chapter
  presents two applications of the methods to confined cell migration
  problems, in which I was involved.}

Cell migration is a field of study where data-driven approaches are
particularly useful and relevant, as nicely summarized in a recent
review~\cite{bruckner_learning_2023}. There are multiple reasons for
this: first, it is important to obtain models for migration of
individual cells and their interactions, as these models are essential
to understand larger-scale phenomena of collective migration and
tissue organization. Second, such models cannot be derived from first
principles, as cells are tremendously complex objects: our theoretical
understanding of intracellular mechanisms is still very far from being
sufficient to derive quantitative models without empirical parameters
that have to be fitted from experiments. Third, while the
intracellular mechanisms are complex, the migration behavior of cells
is often sufficiently simple to be described as self-propelled objects
by stochastic differential equations -- with or without memory.

Here I present two applications of inference methods to cell
migration. In the first (\Sec{sec:2cells}), we employ underdamped
inference to study cell-cell interactions in a non-constricting
confined assay, where adhesion patches on a flat substrate confines
cells without exerting mechanical forces on them. This study was
published in Ref.~\cite{bruckner_learning_2021}; the experimental work
was performed by the group of Joachim R\"adler, while David Br\"uckner
and Chase Broedersz were leading the theoretical analysis, and my role
was mostly in the algorithmic development. Rather than my personal
work, it should thus be seen as a nice example of application of my
inference methods, to which I did a minor contribution.

The second application (\Sec{sec:translocation}) deals with the
mechanically confined migration of individual cells in tight
constrictions, and uses overdamped inference to quantify these
dynamics. It is a collaboration with the experimental group of
C\'ecile Sykes, and in particular her PhD student Sirine Amiri; the
theoretical analysis was performed by my postdoc Yirui Zhang. This
work is still under revision; I have reproduced it here in a slightly
adapted version of its current form~\cite{amiri_inferring_2023}.

\section{Overdamped or underdamped inference?}
\label{sec:overunder}

Cell trajectories vary from experiment to experiment, even in the same
conditions, reflecting the highly complex and actively fluctuating
nature of the internal structures of the cell. This effective
stochasticity is often modeled through a noise term by using
stochastic differential equations (SDEs). At the typical time scales
of cell migration experiments (seconds to hours), the motion of cells
is deeply overdamped: the friction with the environment largely
exceeds the inertia.  A SDE describing cell motility through the
position $X$ of its nucleus (which is the easiest structure to track
within the cell) is thus typically of the
form~\cite{bruckner_learning_2023} :
\begin{equation}
\dot{X}(t) = \underbrace{\Pi + f_{ext}(X)}_{\textrm{deterministic}} + \underbrace{\sqrt{2D_{X}}\cdot \eta_{X}(t)}_{\textrm{stochastic}}, 
\label{eqn:CM-x}
\end{equation}
which consists of two deterministic terms ($\Pi$ and $f_{ext}$)
reflecting the slow, predictable aspects of the dynamics and a
stochastic noise that models the coupling of the observed position
with fast, unobserved degrees of freedom. More specifically, in the
deterministic contribution of nucleus dynamics, $\Pi$ is a driving
term, also called \emph{polarity} of the cell, and captures the
asymmetry in the internal organization of the cell that generates the
motility \cite{drubin_origins_1996}. The other deterministic term,
$f_{ext}$, represents the direct effect of the environment on cell
nucleus dynamics. The noise term $\sqrt{2D_{X}} \cdot \eta_{X}$ has an
amplitude characterized by its diffusion coefficient $D_{X}$, which we
assume here to be state independent, and $\eta_{X}$ the noise, which
for simplicity we assume to be white and Gaussian, therefore,
$\langle \eta_{X}(t) \rangle = 0$ and
$\langle \eta_{X}(t) \eta_{X}(t')\rangle = \delta(t - t')$.

The polarity $\Pi$ itself is dynamical, and its dynamics describe the way cells sense their environment and actuate their self-propulsion accordingly. The dynamics of $\Pi$ follows an SDE of the form
\begin{equation}
\dot{\Pi}(X, t) = f_{\Pi}(X, \Pi) + \sqrt{2D_{\Pi}}\cdot \eta_{\Pi}(t), 
\label{eqn:overdamp-P}
\end{equation}
The drift term $f_{\Pi}(X, \Pi)$ encodes the internal dynamics of
$\Pi$ as well as the feedback of the nucleus polarity to the external
environment.  Note that there are thus two ways the environment
affects the dynamics: through direct forces on $X$ (term $f_{ext}(X)$
in Eq.\ref{eqn:CM-x}) and through indirect feedback ($f_{\Pi}(X, \Pi)$
in Eq.\ref{eqn:overdamp-P}) -- \emph{i.e.} mechanosensing. Here again,
fast internal processes of the cell are modeled as a Gaussian white
noise $\sqrt{2D_{\Pi}}\cdot \eta_{\Pi}(t)$ with diffusion coefficient
$D_{\Pi}$, which determines, for instance, the persistence length of
the free motion of the cell~\cite{li_dicty_2011}.

The class of cell motility models described by Eqs.~\ref{eqn:CM-x}
and~\ref{eqn:overdamp-P} is very general and widely used. However, a
key challenge to its applicability to experimental data is that the
polarity $\Pi$ is not directly measurable, as its molecular definition
remains unknown. To bypass this difficulty, previous studies have
relied on the use of underdamped dynamics: briefly, such approaches
consist of differentiating \Eq{eqn:CM-x} with respect to time, and
plugging into \Eq{eqn:overdamp-P} to eliminate $\Pi$, thus resulting
in an effectively second-order dynamics for
$X$~\cite{bruckner_learning_2023}. This type of \emph{embedding}
approach exploits Taken's theorem and is popularly used for
deterministic dynamical
systems~\cite{crutchfield_equations_1987}. While this approach has
been successful in quantifying, for instance, cell-cell interactions
from data~\cite{bruckner_learning_2021}, it has several
drawbacks. First, second-order inference is considerably more
difficult and demanding in terms of data quality and precision than
first-order inference -- this is actually what led me to develop ULI
(\Chap{chap:ULI}) in collaboration with David Br\"uckner and Chase
Broedersz following their first work on single cell
migration~\cite{bruckner_stochastic_2019} where they faced this
difficulty. Second, one has to neglect the noise on nucleus position
in order for this approach to work, which is not always
possible. Third, information about the nature of the polarity and its
feedback mechanisms is lost in the process. An alternative approach
was proposed recently, consisting of a model-driven treatment of data
where the polarity is explicitly included as a hidden
variable~\cite{bruckner_geometry_2022}, but this requires strong
assumptions on the motility mechanisms.

There are therefore two main ways to represent the dynamics of the
cell nucleus by SDEs: as an underdamped process where the inertia
emerges from the unobserved polarity (as we do in \Sec{sec:2cells}) or
as a higher-dimensional overdamped process where the polarity is
explicitly modeled and couples to the positional dynamics (which we
attempt in \Sec{sec:translocation}).

\section{Cell-cell interactions}
\label{sec:2cells}

Adapted from: \\
\textsc{Learning
the dynamics of cell–cell interactions in confined cell migration} \\
D. B. Brückner, N. Arlt,  A. Fink, P. Ronceray, J. O. Rädler and C.P. Broedersz \\
Proceedings of the National Academy of Sciences 118 (2021). \\

Collective cellular processes such as morphogenesis, wound healing, and cancer invasion, rely on cells moving and rearranging in a coordinated manner. For example, in epithelial wound healing, cells collectively migrate towards the injury and assemble to close the wound~\cite{poujade_collective_2007,stramer_live_2005,weavers_systems_2016}. In contrast, in metastasizing tumors, cancer cells migrate outwards in a directed fashion and invade surrounding tissue~\cite{friedl_tumour-cell_2003}. At the heart of these emergent collective behaviors lie contact-mediated cell-cell interactions ~\cite{weavers_systems_2016,carmona-fontaine_contact_2008,villar-cervino_contact_2013,theveneau_collective_2010,davis_emergence_2012,smeets_emergent_2016,stramer_mechanisms_2017}, which are apparent in two-body collisions of cells~\cite{stramer_mechanisms_2017,astin_competition_2010,teddy_vivo_2004,abercrombie_observations_1954}. These cellular interactions depend on complex molecular mechanisms, including cadherin-dependent pathways and receptor-mediated cell-cell recognition~\cite{moore_par3_2013,matthews_directional_2008,kadir_microtubule_2011}. At the cellular scale, this molecular machinery leads to coordinated, functional behaviors of interacting cells, which are highly variable and may take distinct forms in different biological contexts~\cite{abercrombie_contact_1979,milano_regulators_2016,li_coordination_2018,hayakawa_polar_2020}.

Achieving a quantitative understanding of the stochastic migratory dynamics of cells at the behavioral level could yield key insights into both the underlying molecular mechanisms~\cite{maiuri_actin_2015,lavi_deterministic_2016} and the biological functions~\cite{stramer_mechanisms_2017} associated to these behaviors. For non-interacting, single migrating cells, data-driven approaches have revealed quantitative frameworks to describe the behavior of free unconstrained migration~\cite{selmeczi_cell_2005,li_dicty_2011,pedersen_how_2016} and confined migration in structured environments~\cite{bruckner_stochastic_2019,bruckner_inferring_2020,fink_area_2020}. However, it is still poorly understood how the migratory dynamics of cells are affected by cell-cell interactions and a quantitative formalism for the emergent behavioral dynamics of interacting cells is still lacking~\cite{alert_physical_2020}. Indeed, it is unclear whether cellular collision behaviors follow a simple set of interaction rules, and if so, how these rules vary for different types of cells.

%%%%%%%%%%%%%%%%%%%%%%%%%%%%%%%%%%%%
%FIGURE 1

\begin{figure*}[ht]
\centering
	\includegraphics[width=0.8\textwidth]{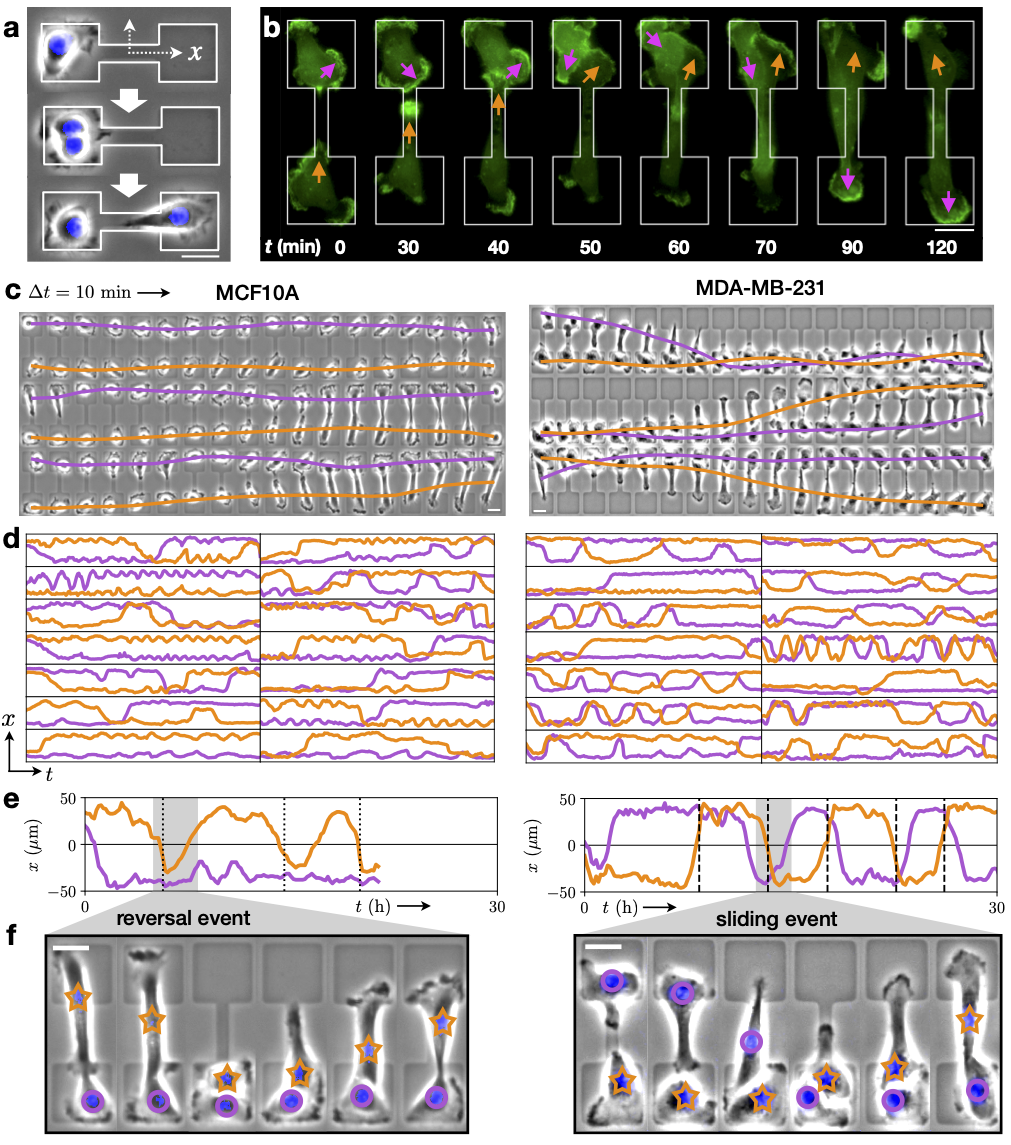}
	\caption{
		\textbf{Stochastic switching dynamics of confined cell pairs.} 
		\textbf{a.} Experimental design: single cells are confined to two-state micropatterns (white outline). We track cell pairs resulting from cell divisions. The stained nucleus is colored in blue.
		\textbf{b.} Time-series of two interacting MDA-MB-231 cells transfected with LifeAct-GFP. Arrows highlight regions of pronounced actin activity, and the arrow color indicates the cell identity.
		\textbf{c.} Brightfield image series with overlaid nuclear trajectories (orange, violet). Images are taken at a time interval $\Delta t$ = 10min.
		\textbf{d.} Sample set of nuclear trajectories $x_{1,2}(t)$ as a function of time, shown for 14 cell pairs. Axes limits are $0 < t < 30$ h and $-60 \ \si{\micro\meter} < x < 60 \ \si{\micro\meter}$, with $x = 0$ at the centre of the bridge. In total, we tracked 155 MCF10A cell pairs (corresponding to a total trajectory length of 3200 h) and 90 MDA-MB-231 cell pairs (2700 h).  
		\textbf{e.} Single cell-pair trajectory, with highlighted reversal (dotted lines) and sliding events (dashed lines).
		\textbf{f.} Key stages of the reversal and sliding events, corresponding to the sections highlighted in grey in \textbf{e}. Images are shown at 40 min time intervals for MCF10A, and 30 min intervals for MDA-MB-231. Orange stars and violet circles indicate the identities of the cells. 
		In panels \textbf{c}-\textbf{f}, the left column corresponds to MCF10A cells, and the right column to MDA-MB-231 cells. All scale bars correspond to $25 \si{\micro\meter}$.
		}
	\label{fig12C}
\end{figure*}

%%%%%%%%%%%%%%%%%%%%%%%%%%%%%%%%%%%%

The study of interacting cell dynamics is complicated by the complex settings in which they take place, confounding contributions of single-cell behavior, interaction with the local micro-environment, and cell-cell interactions. Thus, simplified assays have been developed where cells are confined by one-dimensional micro-patterned patches~\cite{huang_symmetry-breaking_2005,segerer_emergence_2015} or tracks~\cite{milano_regulators_2016,li_coordination_2018,desai_contact_2013,scarpa_novel_2013}, microfluidics~\cite{lin_interplay_2015}, and suspended fibers~\cite{singh_rules_2020}. In these systems, cells exhibit characteristic behaviors upon pair-wise collisions, including reversal, sliding and following events. Upon contact, many cell types exhibit a tendency to retract, repolarize and migrate apart - termed Contact Inhibition of Locomotion (CIL)~\cite{stramer_mechanisms_2017,abercrombie_observations_1954,mayor_keeping_2010}. Indeed, diverse cell types, including epithelial and neural crest cells, predominantly reverse upon collision~\cite{milano_regulators_2016,desai_contact_2013,scarpa_novel_2013}. In contrast, the breakdown of CIL is commonly associated with cancer progression~\cite{astin_competition_2010,abercrombie_observations_1954,milano_positive_2016}, and cancerous cells have been observed to move past each other more readily than non-cancerous cells~\cite{milano_positive_2016}. However, it is unclear how to describe these distinct collision behaviors in terms of physical interactions.

Models for collective cell migration often assume repulsive potentials or alignment terms~\cite{smeets_emergent_2016,alert_physical_2020,sepulveda_collective_2013,basan_alignment_2013,copenhagen_frustration-induced_2018,garcia_physics_2015}, but the form of these interactions is not derived directly from experimental data. Such data-driven approaches have been developed for single cell migration~\cite{selmeczi_cell_2005,li_dicty_2011,pedersen_how_2016,bruckner_stochastic_2019,bruckner_inferring_2020,fink_area_2020}, but have not yet been extended to interacting systems. The search for unifying quantitative descriptions of the dynamics of interacting cell trajectories is further complicated by their intrinsic stochasticity, resulting in highly variable migration and collision behavior~\cite{milano_regulators_2016,desai_contact_2013,scarpa_novel_2013,singh_rules_2020}. Thus, developing a system-level understanding of cell-cell interactions requires a quantitative data-driven approach to learn the full stochastic dynamics of interacting migrating cells.

Here, we develop a theoretical framework for the dynamics of interacting cells migrating in confining environments, inferred directly from experiments. Specifically, we confine pairs of migrating cells into a minimal 'cell collider': a two-state micropattern consisting of two square adhesive sites connected by a thin bridge. Both non-cancerous (MCF10A) and cancerous (MDA-MB-231) human breast tissue cells frequently migrate across the bridge, giving rise to repeated cellular collisions. In line with prior observations~\cite{milano_regulators_2016}, we find that while MCF10A cells predominantly reverse upon collision, MDA-MB-231 cells tend to interchange positions by sliding past each other. To provide a quantitative dynamical framework for these distinct interacting behaviors, we focus on a simplified, low-dimensional representation of these collision dynamics by measuring the trajectories of the cell nuclei. The cell collider experiments yield large data sets of such interacting trajectories, allowing us to infer the stochastic equation of motion governing the two-body dynamics of interacting cells. Our data-driven approach reveals the full structure of the cellular interactions in terms of the relative position and velocity of the cells. Specifically, the dynamics of MCF10A cells are captured by repulsion and friction interactions. In contrast, MDA-MB-231 cells exhibit novel and surprising dynamics, combining attractive and 'anti-friction' interactions, which have no equivalent in equilibrium systems. This inferred model quantitatively captures the key experimental observations, including the distinct collision phenotypes of both cell lines. Our framework can be generalized to provide a conceptual classification scheme for the system-level dynamics of cell-cell interactions, and is able to capture various previously observed types of cell-cell collision behaviors.

\begin{figure}[]
	\includegraphics[width=0.5\textwidth]{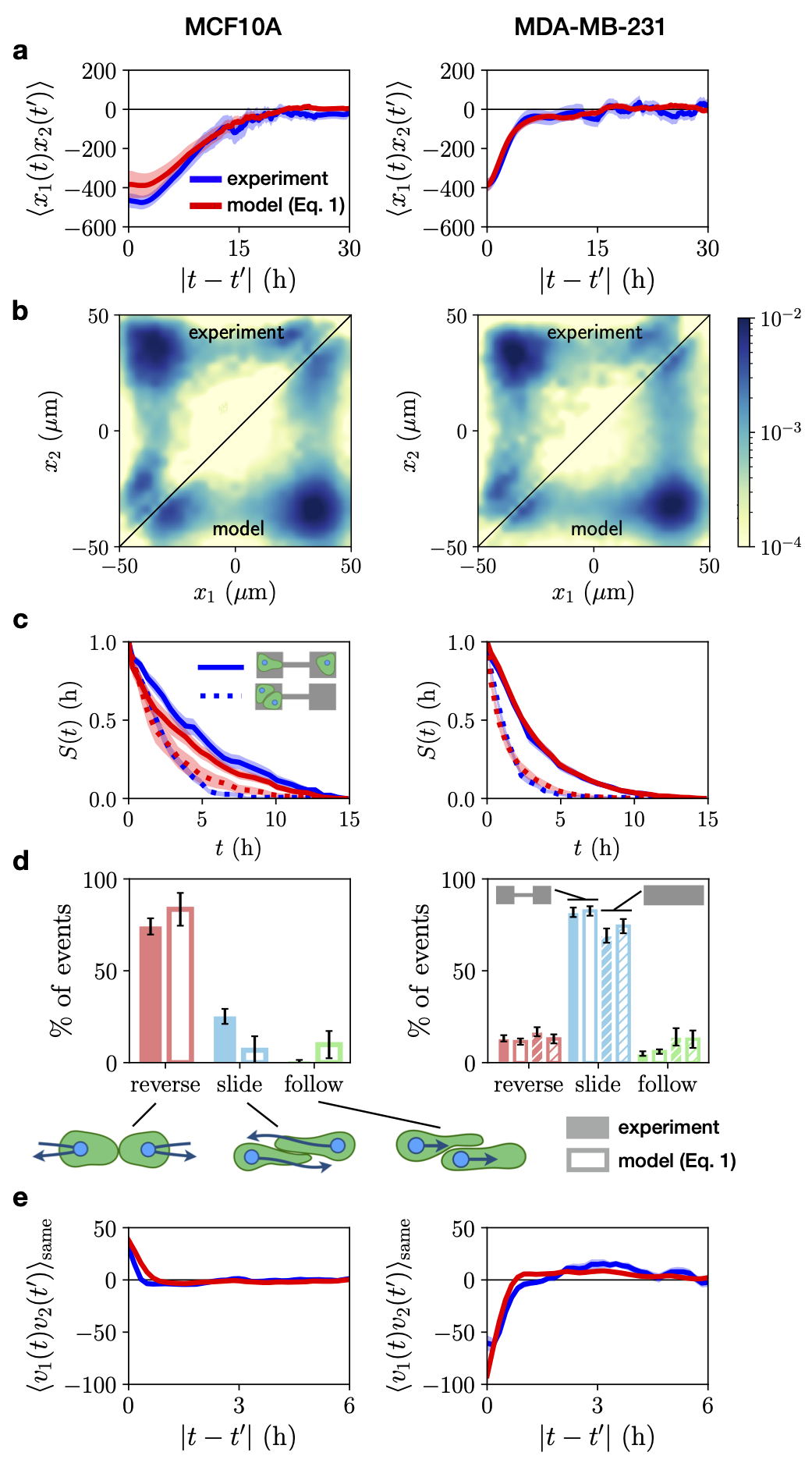}
	\caption{
	\textbf{Statistics of the stochastic interaction dynamics.} 
		\textbf{a.} Cross-correlation function of cell positions $\langle x_1 (t)x_2 (t')\rangle$.
		\textbf{b.} Joint probability distributions $p(x_1,x_2)$ of cell positions, plotted logarithmically. The top triangle of the symmetric distribution shows the experimental result, the bottom triangle shows the model prediction (for full distributions and linear plots, see Supplementary Fig. S12,13). 
		\textbf{c.} Probability distribution $S(t)$ giving the probability that a configuration switch has \textit{not} occurred after time $t$, for the opposite-side configuration (solid) and the same-side configuration (dotted). 
		\textbf{d.} Percentages of each of the three types of collision events observed, which are sketched below. For MDA-MB-231 cells, dashed bars correspond to data from cells on micropatterned tracks, with the corresponding model prediction obtained using a single-cell term inferred from single cells on a track, and interaction terms inferred from cell pairs on two-state patterns.
		\textbf{e.} Velocity cross-correlation function $\langle v_1 (t)v_2 (t')\rangle_\mathrm{same}$, calculated for times where the cells occupy the same island. 
		In panels \textbf{a} and \textbf{c}, experimental data are shown in blue, and model predictions (corresponding to Eqn.\eqref{eq1}) in red. Shaded regions and errorbars denote bootstrap errors (Supplementary Section S3).
		}
	\label{fig22C}
\end{figure}

%%%%%%%%%%%%%%%%%%%%%%%%%%%%%%%%%%%%

\paragraph{Two-state micropatterns provide minimal cell collider}
To investigate the two-body interaction dynamics of migrating cells, we designed a micropatterned system in which two cells repeatedly collide. The micropattern confines the cells to a fibronectin-coated adhesive region, consisting of a narrow bridge separating two square islands. Outside this dumbbell-shaped region the substrate is passivated with PLL-PEG, to which the cells do not adhere. We first confine single cells to these patterns, as described in previous work~\cite{bruckner_stochastic_2019}. Here, we identify cells which undergo division from which we obtain confined, isolated pairs of daughter cells (Fig.~\ref{fig12C}a). We employ phase-contrast time-lapse microscopy to study the homotypic interactions of pairs of non-cancerous (MCF10A) and cancerous (MDA-MB-231) human mammary epithelial cells. The confining bridge between the two islands leads to two well-defined configurations, with either both cells on the same island, or on opposite sides of the pattern, between which the system repeatedly switches (Fig.~\ref{fig12C}c,d and Supplementary Videos S1-4). During these switching events, the cells interact with each other. Therefore, our experimental setup offers a simple platform to study the interactions of confined migrating cells in a standardized manner: a minimal 'cell collider'.

Within this cell collider, cells are highly motile and exhibit actin-rich lamellipodia-like protrusions forming at the cell periphery (Fig.~\ref{fig12C}b, Supplementary Video S5). As a simplified, low-dimensional representation of the interaction dynamics, we use the trajectories of the cell nuclei, which reflect the long time-scale interacting behavior of the cells (Fig.~\ref{fig12C}c). These coupled cell trajectories are highly stochastic. Using this assay, we monitor the stochastic two-body dynamics of hundreds of cells over long periods of time (up to 40h per cell pair) in standardized micro-environments, yielding an unprecedented amount of statistics on cell-cell interactions (Fig.~\ref{fig12C}d). Importantly, we find that most of the interactive behavior is captured by the $x$ position along the long axis of the pattern (Supplementary Section S3). Thus, our cell-collider experiments provide a large data set of low-dimensional trajectories of interacting cells, allowing in-depth statistical analysis of the cellular dynamics.

\paragraph{Cell pairs exhibit mutual exclusion}
A key feature of the trajectories for both cell lines is the apparent preference for the configuration in which the cells are on opposite islands (Fig.~\ref{fig12C}d). Indeed, the positions of the two cells are strongly correlated: the cross-correlation function $\langle x_1 (t)x_2 (t')\rangle$ exhibits a pronounced negative long-time scale correlation for both cell lines (Fig.~\ref{fig22C}a). Correspondingly, the joint probability distribution of positions $p(x_1,x_2)$ exhibits prominent peaks where cells occupy opposite sides, and only faint peaks where they are on the same side (Fig.~\ref{fig22C}b), suggesting two distinct configurations. These configurations are connected by 'paths' in the probability density, along which transitions occur. We find that the cumulative probability $S(t)$ that a configuration switch has not occurred after time t decays more slowly for opposite-side than same-side configurations (Fig.~\ref{fig22C}c). Taken together, these results indicate that both MCF10A and MDA-MB-231 cells exhibit a mutual exclusion behavior.

\paragraph{MCF10A and MDA-MB-231 cells exhibit distinct collision behavior}
While the cells mutually exclude each other, they are also highly migratory and thus frequently transit the constricting bridge. This results in repeated stochastic collision events, providing statistics for how these cells interact during a collision. Following a collision, we observe three distinct types of behaviors: reversal events, where the cells turn around upon collision; sliding events, where the cells interchange positions by sliding past each other; and following events where the cells remain in contact and perform a joint transition (Fig.~\ref{fig12C}e,f, Supplementary Section S3). These three behaviors have been previously used as observables of cell-cell interactions in one-dimensional and fibrillar environments~\cite{milano_regulators_2016,desai_contact_2013,scarpa_novel_2013,singh_rules_2020,kulawiak_modeling_2016}.

To quantify the interaction behavior of MCF10A and MDA-MB-231 cells, we identify collision events and measure the percentage that result in reversal, sliding or following events (Fig.~\ref{fig22C}d). Both cell lines exhibit only a small fraction of following events. Remarkably however, we find that collisions of MCF10A cells predominantly result in reversals, while MDA-MB-231 cells typically slide past each other upon collision, in line with observations in other confining geometries~\cite{milano_regulators_2016}. To further explore the generality of this result, we perform additional experiments with  MDA-MB-231 cells on micropatterned tracks without constrictions, but the same overall dimensions of the two-state micropatterns. We find that sliding events similarly dominate for MDA-MB-231 cells on this pattern, with similar overall event ratios. The different responses to cell-cell contacts are also reflected by the velocity cross-correlation of the two cells when occupying the same side of the two-state micropatterns: $\langle v_1 (t)v_2 (t')\rangle_\mathrm{same}$: MCF10A cells exhibit a positive velocity correlation while MDA-MB-231 cells exhibit a negative velocity correlation (Fig.~\ref{fig22C}e). Taken together, these findings show that while both cell lines exhibit similar mutual exclusion behavior, there are clear differences in their collision dynamics. This raises a key question: is there an overarching dynamical description which captures both the similarities and differences of these interaction behaviors?

\emph{[A part on contact acceleration maps, in which I was not involved, was omitted here.]
}

\begin{figure*}[]
\centering
	\includegraphics[width=0.7\textwidth]{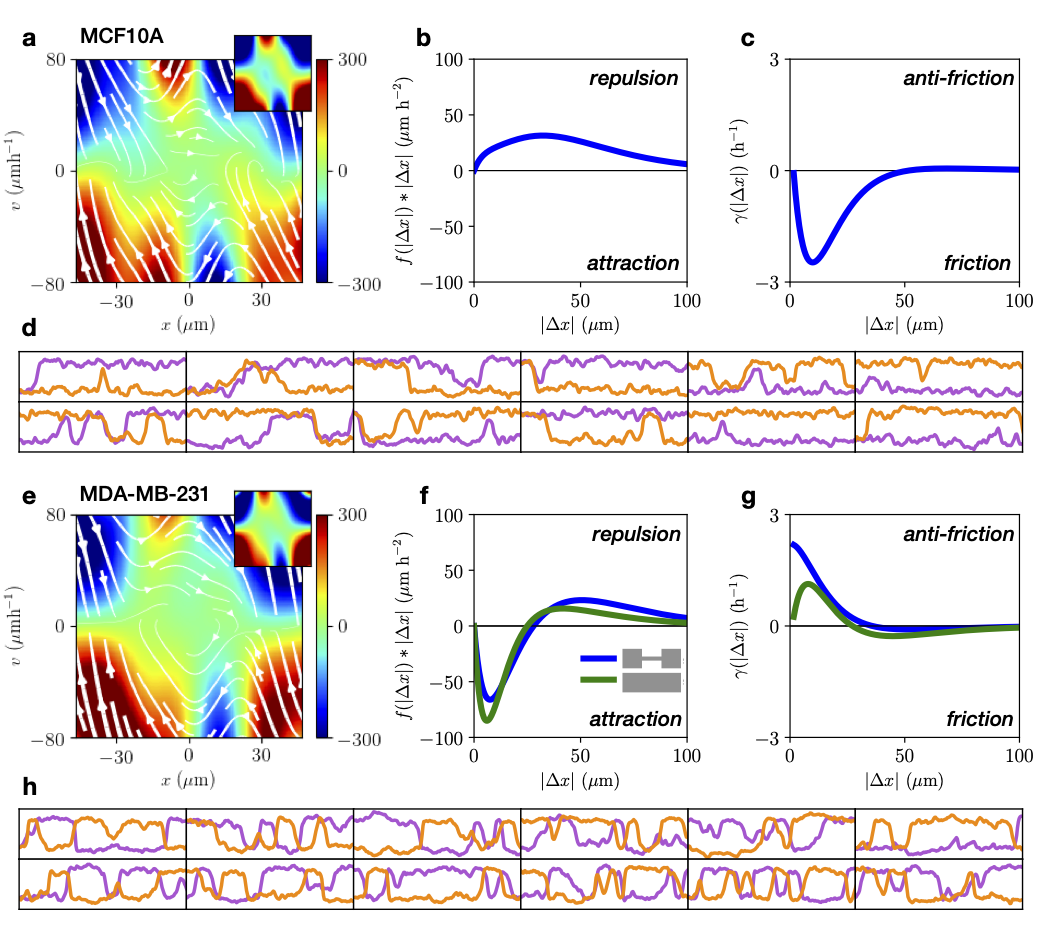}
	\caption{
		\textbf{Equation of motion for interacting cells.} 
		\textbf{a,e.} Single-cell contribution $F(x,v)$ to the interacting dynamics, measured in units of $\si{\micro\meter}/\mathrm{h}^{2}$. White lines indicate the flow field given by $(\dot{x},\dot{v})=(v,F(x,v))$. Inset: corresponding term inferred from experiments with single cells~\cite{bruckner_stochastic_2019}.
		\textbf{b,f.} Cohesive interaction term $f(|\Delta x|)\Delta x$. Positive values indicate repulsive interactions, while negative values correspond to attraction.
		\textbf{c,g.} Effective frictional interaction term $\gamma(|\Delta x|)$. Here, positive values indicate an effective anti-friction, and negative values an effective frictional interaction.
		\textbf{d,h.} Trajectories obtained from model simulations. Axes limits are 0 < $t$ < 30 h and -60 \si{\micro\meter} < $x$ < 60 \si{\micro\meter}.
		Panels \textbf{a}-\textbf{d} show data for MCF10A cells, and panels \textbf{e}-\textbf{h} for MDA-MB-231 cells. For MDA-MB-231 cells, green lines show the interactions inferred from cell pairs interacting on micropatterned tracks.
		}
	\label{fig42C}
\end{figure*}

%%%%%%%%%%%%%%%%%%%%%%%%%%%%%%%%%%%%

\paragraph{Interacting equation of motion captures experimental statistics}
To investigate whether the interacting dynamics of MDA-MB-231 and MCF10A cells can be described by the physical interactions implied by the contact acceleration maps, we consider a simple model for cell-cell interactions in confining environments. Motivated by the structure of the contact accelerations, we postulate that the dynamics of the cells can be described by a stochastic equation of motion of the form
\begin{equation}
\frac{\d v}{\d t}  = F(x,v)+ f(|\Delta x|)\Delta x + \gamma(|\Delta x|)\Delta v + \sigma \eta(t)
\label{eq1}
\end{equation}
Here, we assume that the interactions between each cell and the confinement can be described by a term $F(x,v)$, similar to single cell experiments~\cite{bruckner_stochastic_2019}. Furthermore, we assume that the interactions between the two cells can be separately described by two interaction terms: a cohesive term $f(|\Delta x|)\Delta x$, which captures repulsion and attraction; and an effective friction term $\gamma(|\Delta x|)\Delta v$ that may depend on the distance between the cells. The intrinsic stochasticity of the migration dynamics is accounted for by a Gaussian white noise $\eta(t)$, with $\langle \eta(t)\rangle=0$ and $\langle \eta(t)\eta(t') \rangle=\delta(t-t')$. Note that this equation of motion captures the effective dynamics that describe the cellular accelerations, rather than mechanical forces acting on the cell.

To investigate this model, we first require a systematic approach to infer the systems' stochastic dynamics and delineate single-cell (one-body) and interactive (two-body) contributions to the dynamics. Thus, we employ a rigorous inference method, Underdamped Langevin Inference (ULI)~\cite{bruckner_inferring_2020}, to infer the terms of this equation of motion from the experimentally measured trajectories. In this approach, the inferred terms are completely constrained by the short time-scale information in the measured trajectory, i.e. the velocities and accelerations of the cells (see Methods and Supplementary Section S4). 

Importantly, there is no a priori reason why \eqref{eq1} should provide a reasonable ansatz to correctly capture cell-cell interactions, which could require a more complex description. Thus, we investigate the predictive power of our model by testing whether it correctly captures experimental statistics that were not used to constrain the terms in \eqref{eq1}. Specifically, while the model is learnt on the experimental short time-scale dynamics, we aim to make predictions for long time-scale statistics such as correlation functions. To this end, we simulate stochastic trajectories of interacting cell pairs based on our model (Fig.~\ref{fig42C}d,h) to make a side-by-side comparison with the experiments. Remarkably, we find that the model performs well in predicting key experimental statistics for both cell lines, including the joint probability distributions (Fig.~\ref{fig22C}b), the distributions of switching times (Fig.~\ref{fig22C}c), the cross-correlations of positions and velocity (Fig.~\ref{fig22C}a,e), as well as the relative fractions of reversal, sliding and following events (Fig.~\ref{fig22C}d). In contrast, performing the same inference procedure with simpler models than \eqref{eq1}, e.g. with only cohesive or friction interactions, shows that simulated trajectories of these models do not capture the observed statistics (Supplementary Section S4). To further challenge our approach, we test whether we can use the interactions learnt from experiments on two-state micropatterns to predict the collision behavior in a different confinement geometry. Specifically, we use the single-cell term $F(x,v)$ inferred from single cell data of MDA-MB-231 cells migrating on micropatterned tracks, together with the interactions inferred from cell pair experiments on two-state micropatterns, to predict the collision ratios of cell pairs on tracks. We find that this model accurately predicts the observed event ratios (Fig.~\ref{fig22C}d), showing that the inferred interactions have predictive power also beyond the data set on which they are learnt.

Remarkably, our inference approach reveals that the inferred single-cell contributions $F(x,v)$ on two-state micropatterns are qualitatively and quantitatively similar to the equivalent term inferred from experiments with single cells for both cell lines~\cite{bruckner_stochastic_2019} (Fig.~\ref{fig42C}a,e, Supplementary Section S4). Also, the inferred noise amplitudes are similar to those inferred from single cell experiments for both cell lines, $\sigma \approx 50 \ \si{\micro\meter}/\mathrm{h}^{3/2}$. This suggests that the presence of another cell does not significantly alter the confinement dynamics experienced by one of the cells, and instead manifests in the interaction terms of the equation of motion. Our inference yields the spatial dependence of the cohesion term (Fig.~\ref{fig42C}b,f) and the effective friction term (Fig.~\ref{fig42C}c,g). Importantly, the functional dependence of the inferred terms is in accord with our interpretation of contact acceleration maps: MCF10A cells exhibit a repulsive cohesive interaction, and a regular effective friction, which reflects that cells slow down as they move past each other. In contrast, MDA-MB-231 cells interact through a predominantly attractive cohesion term, becoming weakly repulsive at long distances, and exhibit effective 'anti-friction'. We infer a similar 'anti-friction' interaction from MDA-MB-231 cell pairs migrating on micropatterned tracks, suggesting that this result is not sensitive to the presence of the constriction (Fig.~\ref{fig42C}f,g). This anti-friction generates sliding behavior, where cells on average accelerate as they move past each other with increasing relative speed. These results are robust with respect to the details of the inference procedure (Supplementary Section S4). Taken together, these findings demonstrate that the dynamics of interacting MCF10A and MDA-MB-231 cells on confining micropatterns are well described by our model (\eqref{eq1}) with distinct types of interactions for the two cell lines.

\paragraph{Interaction behavior space: a theoretical framework for cell-cell interactions}
To conceptualize the distinct interactions of MCF10A and MDA-MB-231 cells, we propose an \textit{interaction behavior space}, spanned by the amplitudes of the cohesive and frictional contributions (Fig.~\ref{fig52C}). Based on our inference, the two cell lines occupy diagonally opposed quadrants in this space. To investigate whether our model (\eqref{eq1}) is able to capture cellular interaction behaviors more broadly, we predict trajectories for various locations within this interaction map. For interactions consisting of repulsion and friction, we find that collisions predominantly result in reversal events, as we have observed for MCF10A cells. In contrast, for positive friction coefficients, corresponding to effective 'anti-friction', we find that sliding events dominate for all parameter values. This regime thus corresponds to the dynamics we have observed for MDA-MB-231 cells. Finally, attractive interactions with regular friction result in a dominance of following events. The interaction behavior space thus provides an insightful connection between the inferred interaction terms governing the instantaneous dynamics of the system, and the emergent macroscopic, long time-scale collision behavior.

%%%%%%%%%%%%%%%%%%%%%%%%%%%%%%%%%%%%
%FIGURE 5

\begin{figure*}[]
\centering
	\includegraphics[width=\textwidth]{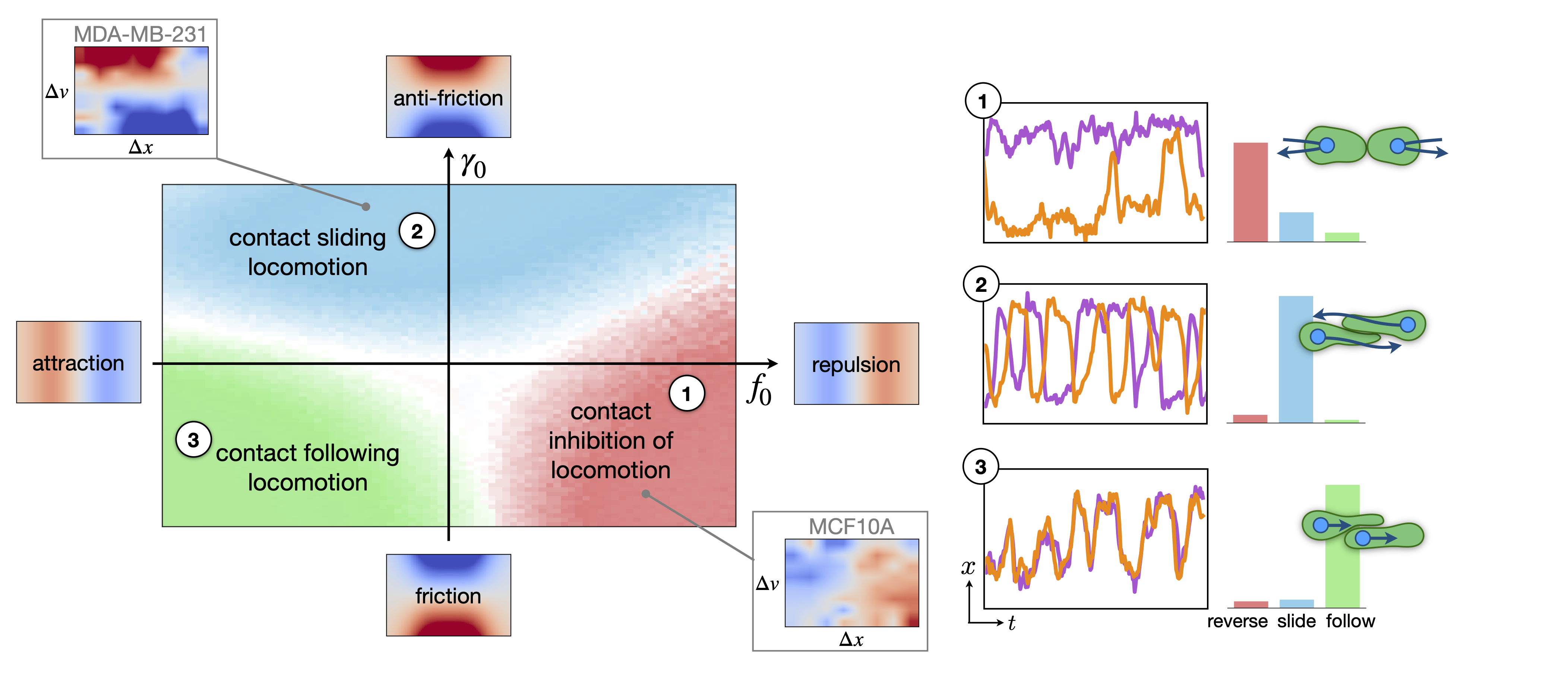}
	\caption{
		\textbf{Interaction behavior space.} 
We construct an interaction space by varying the amplitude of the cohesive and friction interactions, $f_0$ and $\gamma_0$, respectively. Contact acceleration maps for purely attractive, repulsive, frictional and anti-frictional interactions are indicated on the axes. Based on the inferred short-range interactions, we place MDA-MB-231 and MCF10A cells into diagonally opposed quadrants. Predicted behaviors in the interaction space are obtained by varying the cohesion and friction interactions in our model. Specifically, we simulate a model including the inferred MDA-MB-231 single-cell term $F(x,v)$ together with a cohesive term $f(|\Delta x|)=f_0 g_\mathrm{c}(|\Delta x|)$ and an effective friction term $\gamma(|\Delta x|)=\gamma_0 g_\mathrm{f}(|\Delta x|)$, for varying $f_0$ and $\gamma_0$. The distance-dependent functions $g_\mathrm{c,f}$ are positive and monotonically decreasing. These results do not sensitively depend on the specific choice of $F(x,v)$ or $g_\mathrm{c,f}$ ($g_\mathrm{c,f}=\exp[(-|\Delta x|/R_0)]$ is used here) (Supplementary Section S5). For each parameter combination, reversal, sliding and following events where identified. At each point, the dominant behavior is indicated by the color scheme, and white regions correspond to states where no single behavior contributes more than 50\% of events. Numbered insets show sample trajectories from different parts of the interaction map, and the corresponding percentages of reversal (red), sliding (blue), and following events (green).
		}
	\label{fig52C}
\end{figure*}

%%%%%%%%%%%%%%%%%%%%%%%%%%%%%%%%%%%%

\paragraph{Discussion}
In this study, we introduced a conceptual framework for the stochastic behavioral dynamics of interacting cells. To this end, we designed a micropatterned 'cell collider' in which pairs of cells repeatedly collide with each other, providing large amounts of statistics on the long time-scale interactions of migrating cell pairs. A key advantage of this setup is that it yields a large number of collisions under controllable conditions. Moreover, the dynamics of single cells migrating in this confinement is well understood~\cite{bruckner_stochastic_2019}, providing a benchmark for the dynamics inferred for interacting cells. We compare the homotypic interaction behavior of the non-malignant MCF10A and the metastatic MDA-MB-231 mammary epithelial cell lines. While phenomenological bottom-up models have been developed to describe cell-cell interactions~\cite{alert_physical_2020,segerer_emergence_2015,kulawiak_modeling_2016,camley_velocity_2014,lober_collisions_2015,vedel_migration_2013}, we propose an alternative, top-down approach to learn the interacting stochastic equations of motion governing cell migration from the experimentally observed trajectories. Such an effective model captures the emergent dynamics at the cellular scale which are driven by underlying mechanisms, including the intra-cellular polarity machinery. Our inferred models for interacting cells quantitatively capture the distinct behaviors of the two cell lines. This inference reveals that the dynamics can be decomposed into a one-body motility component, which qualitatively matches that observed in single cell experiments~\cite{bruckner_stochastic_2019}, and a two-body interaction term. 

The interaction terms we inferred from experiments take qualitatively different forms for the two cell lines: while MCF10A cells exhibit repulsion and effective friction, MDA-MB-231 cells exhibit attraction and a novel and surprising effective 'anti-friction' interaction. At the single-cell level, MDA-MB-231 cells are known to be more invasive than MCF10A cells~\cite{mak_microfabricated_2011,kraning-rush_microfabricated_2013}, and express lower levels of the cell-cell adhesion protein E-cadherin~\cite{milano_regulators_2016,sommers_cell_1991}, possibly underlying the different friction-like interactions we found for these cell lines. These two cell lines also display remarkably different collective behaviors~\cite{carey_leading_2013,lee_distinct_2020,kang_tumor_2020}: MCF10A cells in 2D epithelial sheets exhibit aligned, directed motion and form compact spheroids in 3D culture, with few invasive branches. In contrast, MDA-MB-231 cells in 2D epithelial sheets exhibit non-aligned, random motion and form invasive, non-contiguous clusters in 3D culture, with significant single-cell dispersion from the cluster. These differences in collective behavior may relate to the distinct types of interactions we have inferred from the two-body dynamics of these cell lines.

Based on the inferred equation of motion, we predict an interaction behavior space to link the interaction terms, which govern the instantaneous stochastic dynamics, to the emergent collision behaviors. The three distinct regimes emerging in our model correspond to specific behaviors observed in experiments for various cell types: predominant reversal behavior on 1D lines has been termed \textit{contact inhibition of locomotion}~\cite{desai_contact_2013,scarpa_novel_2013}, a common type of cell-cell interaction~\cite{carmona-fontaine_contact_2008,villar-cervino_contact_2013,davis_emergence_2012,smeets_emergent_2016,stramer_mechanisms_2017,abercrombie_observations_1954}. By inhibiting intracellular Rho-signalling in neural crest cells, this reversal-dominated behavior could be tuned to following-dominated behavior~\cite{scarpa_novel_2013}. Such following behavior has also been identified as an important mechanism in collective migration~\cite{teddy_vivo_2004,li_coordination_2018,hayakawa_polar_2020,fujimori_tissue_2019}, and was termed \textit{contact following locomotion}~\cite{li_coordination_2018}. Finally, previous work has shown that reducing the expression levels of E-cadherin enables otherwise reversing cells to mainly slide past each other~\cite{milano_regulators_2016}. For this regime of predominant sliding interactions, we propose the term \textit{contact sliding locomotion}. Based on our interaction behavior space, we find that the 'anti-friction' interactions we identified for MDA-MB-231 cells promote such sliding behavior. The interaction behavior space could thus provide a quantitative classification of distinct modes of interaction that may be achieved through molecular perturbations in experiments~\cite{milano_regulators_2016,scarpa_novel_2013}. On the other end of the scale, the 'anti-friction' interaction type we find here could play a role in collective systems such as the fluidization of tissue in tumor invasion~\cite{kang_tumor_2020,palamidessi_unjamming_2019,han_cell_2020}. The form of the interaction terms we inferred from experiments may thus inform models for collective cell migration~\cite{smeets_emergent_2016,alert_physical_2020,sepulveda_collective_2013,basan_alignment_2013,copenhagen_frustration-induced_2018,garcia_physics_2015}. Furthermore, the inference framework we have developed for the dynamics of interacting cell pairs can be extended to infer the dynamics of more complex collective systems, such as small clusters of cells~\cite{segerer_versatile_2016,copenhagen_frustration-induced_2018,dalessandro_contact_2017,dalessandro_contact_2017}, epithelial sheets~\cite{garcia_physics_2015,angelini_cell_2010}, or 3D organoids~\cite{palamidessi_unjamming_2019,han_cell_2020}. In summary, our model, which we rigorously derive directly from experimental data, could potentially describe the diversity of previously observed cell-cell interaction behaviors in a unifying quantitative framework. \\

\subsection{Inference method and model selection}
\label{sec_inference}

%------------------------------------------------------------
\paragraph{Application of Underdamped Langevin Inference}
\label{sec_uli}

To infer an interacting stochastic equation of motion for confined migrating cell pairs, we employ a rigorous inference method, Underdamped Langevin Inference (ULI)~\cite{bruckner_inferring_2020}. In this section, we lay out the details of applying ULI to our system. For further details on the method itself, see ref.~\cite{bruckner_inferring_2020}. Our inference ansatz is to postulate that the system can be described by the general equation of motion for cell $i$ with position $x_i(t)$, velocity $v_i(t) = \d x_i/\d t$, and acceleration $\dot{v}_i(t) = \d v_i/\d t$:
\begin{align}
\begin{split}
\label{eqn_eom}
\dot{x}_i &= v_i \\
\dot{v}_i &= F(x_i,v_i) + f(|\Delta x_{ij}|) \Delta x_{ij} + \gamma(|\Delta x_{ij}|) \Delta v_{ij} + \sigma \eta_i(t)
\end{split}
\end{align}
where $\Delta x_{ij}=x_i-x_j$, $\Delta v_{ij}=v_i-v_j$, $\langle \eta_i(t) \rangle =0$, and $\langle \eta_i(t) \eta_j(t')  \rangle = \delta_{ij}\delta(t-t')$. 

Using ULI, the stochastic equation of motion of such an interacting system can be reconstructed by projecting the dynamics onto a set of $n_b$ basis functions $\{ b_\alpha(x,v) \}_{\alpha=1...n_b}$, which are subjected to an orthonormalization scheme $\hat{c}_\alpha(x,v) = \sum_{\beta=1}^{n_b} \hat{B}_{\alpha \beta}^{-1/2}b_\beta(x,v)$ such that $\langle \hat{c}_\alpha \hat{c}_\beta \rangle = \delta_{\alpha \beta}$. The total deterministic contribution $F^\mathrm{(total)} = F(x_i,v_i) + f(|\Delta x_{ij}|) \Delta x_{ij} + \gamma(|\Delta x_{ij}|) \Delta v_{ij}$ of the system can then be approximated as a linear combination of these basis functions, $F^\mathrm{(total)} \approx \sum_{\alpha=1}^{n_b} F^\mathrm{(total)}_{\alpha} \hat{c}_\alpha(x,v)$. Using ULI, we estimate the coefficients of this expansion of the deterministic term $\hat{F}^\mathrm{(total)}_{\alpha}$ and the noise amplitude $\hat{\sigma}$ using the increments of the observed position trajectories $x_i(t)$.

For interacting systems, we separate single-particle and interaction contributions into separate sets of basis functions. We approximate the cohesion and friction terms $f(|\Delta x_{ij}|)$ and $\gamma(|\Delta x_{ij}|)$ using a set of interaction kernels $\{u_\alpha(|\Delta x_{ij}|)\}$ (see section~\ref{sec_basis}). We fit the single-cell term $F(x_i,v_i)$ with a basis consisting of Fourier components in $x_i$ and polynomials in $v_i$ including terms up to third order~\cite{bruckner_stochastic_2019}: 
\begin{equation}
\label{eqn_fourier}
F(x_i,v_i) \approx \sum_{n=0}^N \sum_{m=0}^M [ A_{nm} \cos(2\pi nx_i/w) + B_{nm} \sin(2\pi nx_i/w) ] v_i^m
\end{equation}  
where $N=M=3$ and $w=100  \ \si{\micro\meter}$. As we show in section~\ref{sec_basis}, our inference results are not sensitive to the precise choise of basis employed. 

A key assumption of our model (Eq.~\eqref{eqn_eom}) is that the noise $\eta_i(t)$ is uncorrelated in time. To self-consistently test this assumption, we calculate the trajectories of the noise increments $\Delta W_i(t) = \int_t^{t+\Delta t} \eta_i(s) \ \mathrm{d} s$. An empirical estimator for $\Delta W_i(t)$ is~\cite{selmeczi_cell_2005,stephens_emergence_2011,bruckner_stochastic_2019}:
\begin{equation}
\widehat{\Delta W}_i (t) \approx \frac{\Delta t}{\hat{\sigma}} \left[ \dot{v}_i(t) - \hat{F}^\mathrm{(total)}(x_i,v_i) \right]
\end{equation}  
Thus, we calculate the auto-correlation function of the noise as $\hat{\phi}_{\Delta W} = \langle \widehat{\Delta W}_i (t) \widehat{\Delta W}_i(t') \rangle$. We find that for both cell lines, the noise decays to zero within a single time-step, confirming our white noise assumption. The weak negative correlation at $|t-t'|=\Delta t$ is due to the presence of measurement errors in the positions, as discussed in refs.~\cite{pedersen_how_2016,bruckner_stochastic_2019}.

%%%%%%%%%%%%%%%%%%%%%%%%%%%%%%%%%%%%
%FIGURE 
\begin{figure}[h!]
	\includegraphics[width=0.5\textwidth]{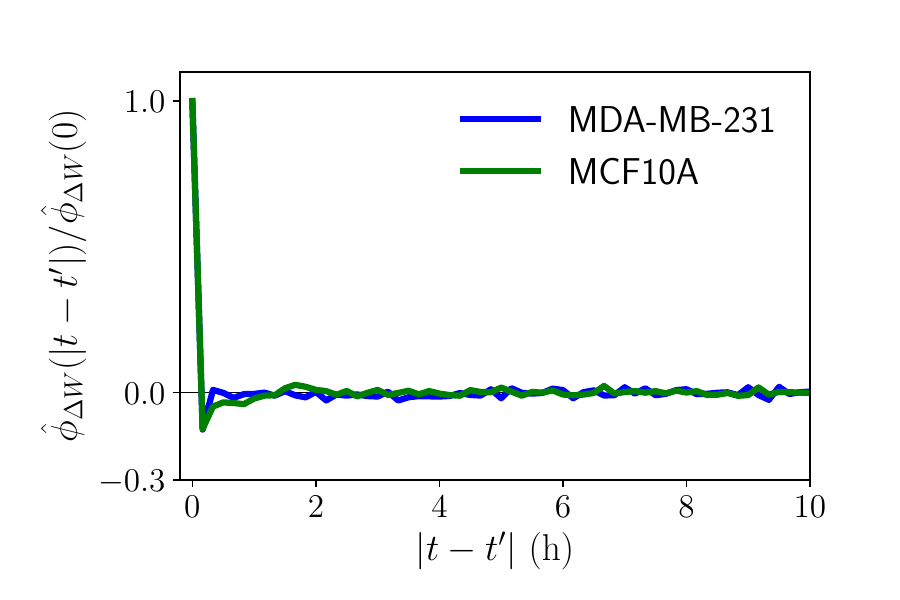}
	\centering
		\caption{
		\textbf{Inferred noise correlation functions.} The correlation functions are normalized by their value at $|t-t'|=0$. The blue curve corresponds to MDA-MB-231 cells, the green curve to MCF10A cells.
				}
	\label{noise_corr}
\end{figure}
%%%%%%%%%%%%%%%%%%%%%%%%%%%%%%%%%%%%

Three conditions for accurate inference from stochastic underdamped systems are (i) sufficiently long trajectories to constrain the $n_b$ parameters of the fitted model, (ii) a sufficiently small measurement time interval $\Delta t$ to resolve the dynamics and (iii) measurement errors on the positions that are smaller than the typical displacement in a single time-step:

\textbf{(i) Trajectory length} $-$ Inference from a finite data set relies on having sufficient information to accurately resolve the features of the underlying dynamical terms of the equation of motion. The information contained in a set of trajectories for a system of the type of Eq.~\eqref{eqn_eom} can be empirically estimated as $\hat{I}_b=\frac{\tau}{2 \hat{\sigma}^2} \sum_{\alpha=1}^{n_b} \left( \hat{F}^\mathrm{(total)}_{\alpha} \right)^2$, where $\tau$ is the total length of the trajectory. The parameters of the expansion can be accurately inferred if $\hat{I}_b \gg n_b$, where $\hat{I}_b$ is given in natural information units (1 nat $=1/\log2$ bits)~\cite{bruckner_inferring_2020}. Here, we employ a basis with $n_b=34$ parameters (28 parameters for the single cell term and 6 parameters for the interaction kernels). As shown in table~\ref{tab_info}, our data sets contain enough information to constrain these parameters.

\textbf{(ii) Discretization} $-$ To ensure a sufficiently accurate temporal sampling of the observed signal, we ensured that the measurement time interval $\Delta t$ should be small enough to resolve the time-scales of the collision dynamics, i.e. the switching time $\langle \tau_\mathrm{same} \rangle = (1.69 \pm 0.11) \ \mathrm{h}$ of MDA-MB-231 cells. Our measurement time interval is $\Delta t = 10 \ \mathrm{min}$, and thus sufficiently small to resolve this time-scale. Additionally, the time interval plays an important role as velocities and accelerations are obtained as discrete derivatives from the position trajectories $x_i(t)$. Indeed, even for small $\Delta t$, inference from underdamped systems exhibits systematic discretization biases~\cite{pedersen_how_2016,bruckner_inferring_2020,ferretti_building_2020}. The leading order term of the bias is removed through the construction of the ULI estimators~\cite{bruckner_inferring_2020}. We show empirically that higher order biases do not strongly affect our inference results by performing a self-consistency test (see section~\ref{sec_stability}).

\textbf{(iii) Measurement error} $-$ In any tracking experiment, the observed position trajectories are subject to time-uncorrelated measurement error $m(t)$, such that the observed signal is $y(t)=x(t)+m(t)$, where $x(t)$ is the true signal. ULI yields accurate inference results in the regime $|m| < v\Delta t$, where $v\Delta t$ is the typical displacement in a single time-step.  We can evaluate this condition from the data, using the average speed of the cells, and comparing it to the measurement error amplitude inferred from the trajectories~\cite{bruckner_inferring_2020}. As shown in table~\ref{tab_info}, this condition is fulfilled for both data sets.

\begin{table}
\begin{center}
 \begin{tabular}{|c || c | c | c | c | c | c |} 
 \hline
   & $N_\mathrm{pairs}$ & $N_\mathrm{pts}$ & $\hat{I}_b \ (\mathrm{nats})$ & $\hat{\sigma} \ (  \ \si{\micro\meter\hour}^{3/2})$ & $\langle |\hat{v}| \rangle \Delta t \ (  \ \si{\micro\meter})$ & $|\hat{m}| \ (  \ \si{\micro\meter})$ \\ [0.5ex] 
 \hline\hline
 MDA-MB-231  & 90 & 15,979 & $11,800 \pm 150$ & 51.4 & 2.6 & 1.3  \\ 
 \hline
 MCF10A  & 155 & 19,470 & $11,900 \pm 160$ & 47.9 & 2.4 & 1.4 \\ 
 \hline
\end{tabular}
\caption{\label{tab_info} \textbf{Statistics of the stochastic trajectory data sets for both cell lines}. From left to right, the columns denote: (i) The number of tracked cell pairs. (ii) The total number of recorded time-points. (iii) The empirical estimate of the information content of the data set, obtained by projecting the dynamics onto our standard basis choice (see section~\ref{sec_basis}). The error in the inferred information content is estimated as $\delta \hat{I}_b \approx [2\hat{I}_b+n_b^2/4]^{1/2}$ \cite{bruckner_inferring_2020}. (iv) The inferred noise amplitude. (v) The typical displacement in a single time-step. (vi) The inferred amplitude of the measurement error, which is in line with previous estimates for single cell migration in the same setup~\cite{bruckner_stochastic_2019}.}
\end{center}
\end{table}

%------------------------------------------------------------
\paragraph{Robustness with respect to the projection basis}
\label{sec_basis}

To infer the interaction terms of the dynamics, we approximate the cohesion and friction terms $f(|\Delta x_{ij}|)$ and $\gamma(|\Delta x_{ij}|)$ using a set of interaction kernels $\{u_\alpha(|\Delta x_{ij}|)\}$. Physically, we expect cell-cell interactions to be spatially local. Thus, to ensure accurate inference in the region of interest, we choose kernels which decay at large distances, $u_\alpha(|\Delta x_{ij}|\to\infty) \to 0$. A simple choice for such kernels is a set of exponentials $u_n(|\Delta x_{ij}|) = \exp(-|\Delta x_{ij}|/nr_0)$ with $n=1...N$. This basis therefore has two hyperparameters that have to be chosen, the number for kernels $N$ and the maximum decay length $r_\mathrm{max}=Nr_0$. Alternatively, we also test a basis consisting of Gaussian functions $u_n(|\Delta x_{ij}|) = \exp(-(|\Delta x_{ij}|-nr_0)^2/2W^2)$ with $n=1...N$. This basis therefore has three hyperparameters $N, r_\mathrm{max}=Nr_0,$ and $W$. While this inference scheme could be supplemented by an additional optimization of the hyperparameters, we find this not to be necessary in this case, as the inferred interaction terms do not sensitively depend on the choice of hyperparameters or basis functions (Figs. \ref{bases_expon},\ref{bases_gauss}). Furthermore, the predictive power of the inferred model is not sensitively affected by the choice of basis (Fig. \ref{bases_CF}). Throughout the main text, we choose an exponential basis with an intermediate value of $N=3$ functions and a maximum decay length $r_\mathrm{max}= 20  \ \si{\micro\meter}$ (black line in Figs. Figs. \ref{bases_expon},\ref{bases_gauss}).

%%%%%%%%%%%%%%%%%%%%%%%%%%%%%%%%%%%%
%FIGURE 
\begin{figure}[h!]
	\includegraphics[width=\textwidth]{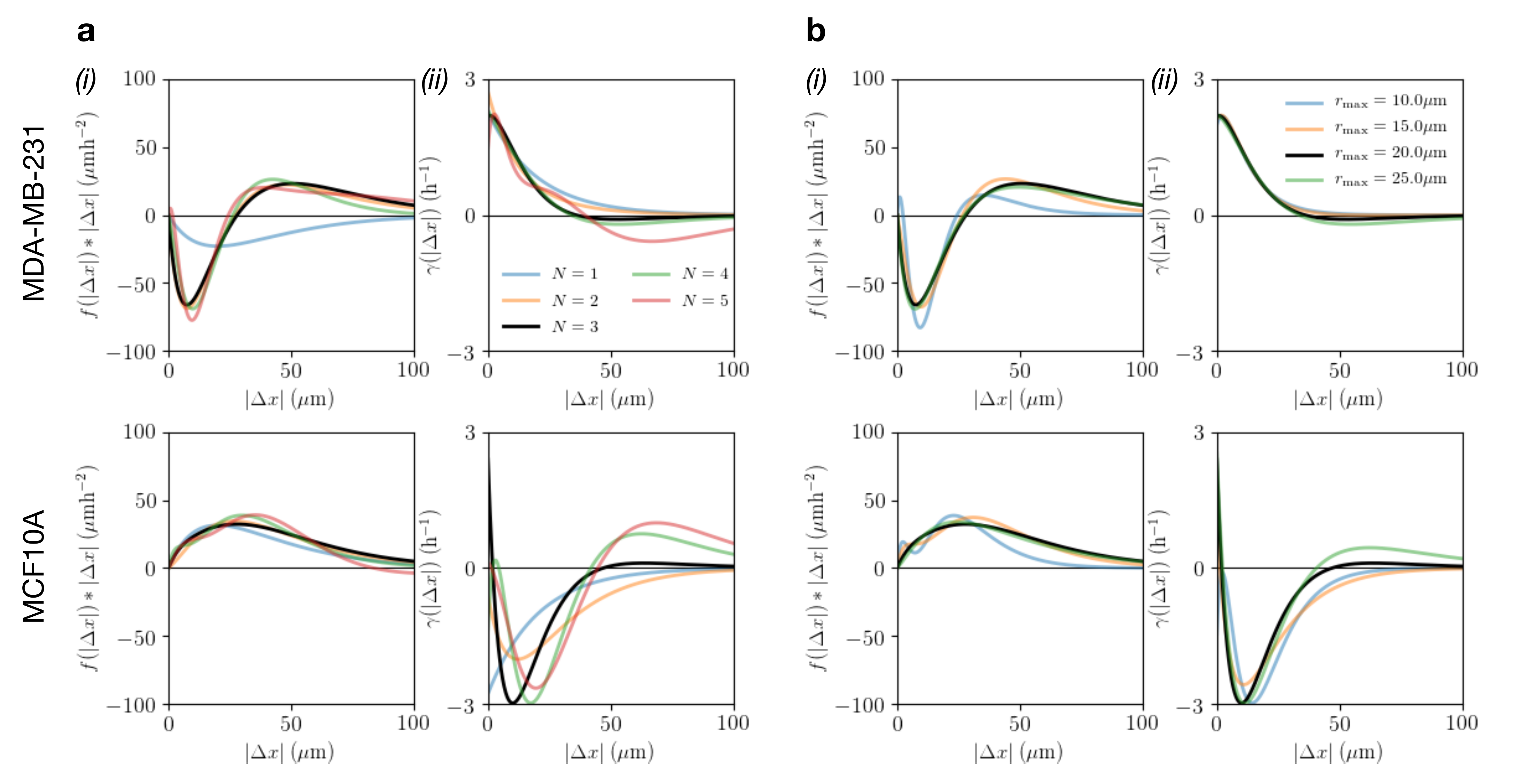}
	\centering
		\caption{
		\textbf{Inference results for exponential interaction kernels.}
		\textbf{a}, Varying the number of kernels $N$, using $r_\mathrm{max}= 20  \ \si{\micro\meter}$
		\textbf{b}, Varying the maximum decay length $r_\mathrm{max}$, using $N=3$.
		\textit{(i)}, Cohesive component $f(|\Delta x_{ij}|) |\Delta x_{ij}|$.
		\textit{(ii)}, Friction kernel $\gamma(|\Delta x_{ij}|)$.
		\textit{Top row:} MDA-MB-231 cells.
		\textit{Bottom row:} MCF10A cells. Black line corresponds to the curves shown in Fig. 4 of the main text, using $N=3$ and $r_\mathrm{max}= 20  \ \si{\micro\meter}$.
				}
	\label{bases_expon}
\end{figure}
%%%%%%%%%%%%%%%%%%%%%%%%%%%%%%%%%%%%

%%%%%%%%%%%%%%%%%%%%%%%%%%%%%%%%%%%%
%FIGURE 
\begin{figure}[h!]
	\includegraphics[width=\textwidth]{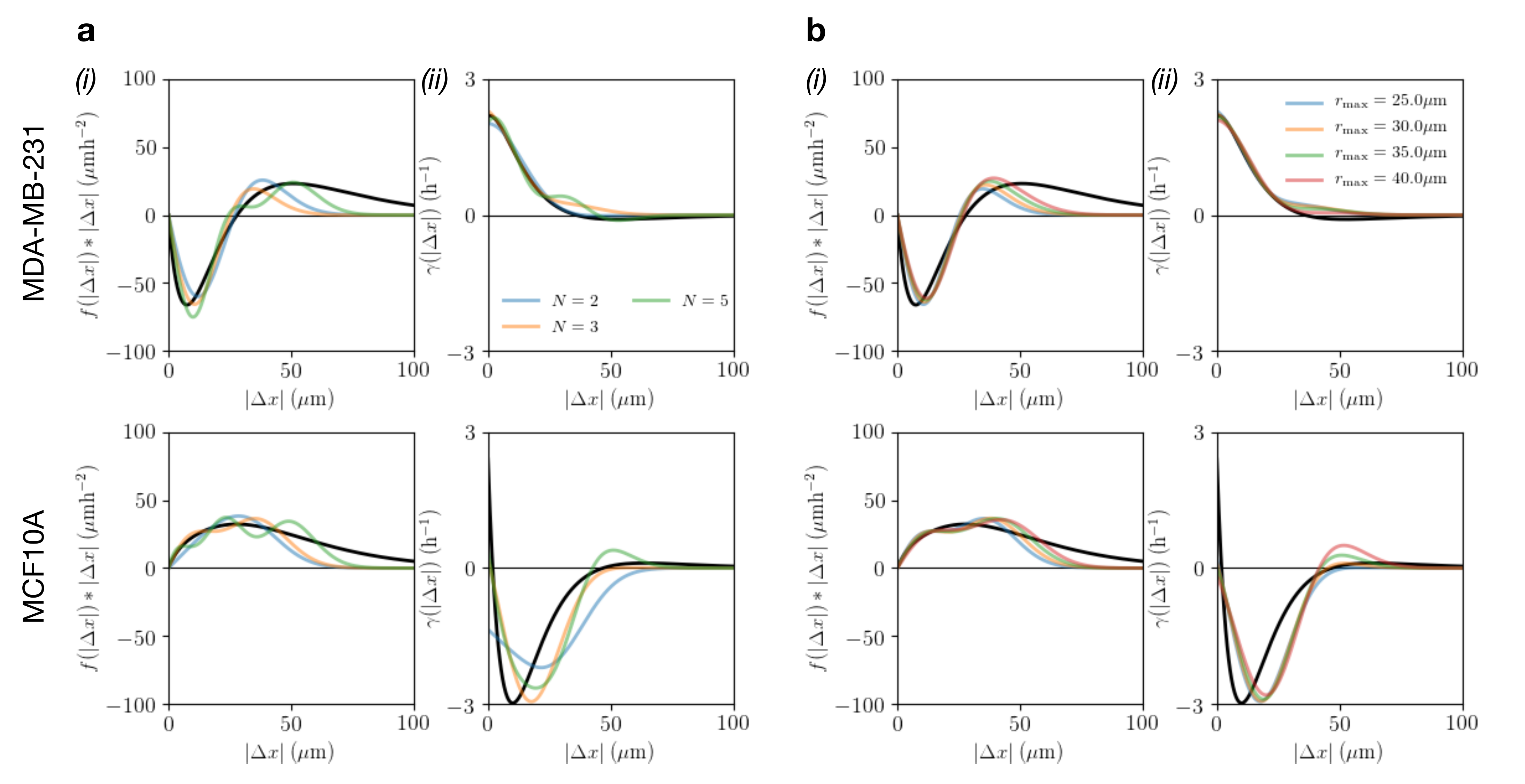}
	\centering
		\caption{
		\textbf{Inference results for Gaussian interaction kernels.} 
		\textbf{a}, Varying the number of kernels $N$, using $r_\mathrm{max}= 25  \ \si{\micro\meter}$ and $W = 4  \ \si{\micro\meter}$.
		\textbf{b}, Varying the maximum decay length $r_\mathrm{max}$, using $N=3$ and $W = 4  \ \si{\micro\meter}$.
		\textit{(i)}, Cohesive component $f(|\Delta x_{ij}|) |\Delta x_{ij}|$.
		\textit{(ii)}, Friction kernel $\gamma(|\Delta x_{ij}|)$.
		\textit{Top row:} MDA-MB-231 cells.
		\textit{Bottom row:} MCF10A cells. Black line corresponds to the curves shown in Fig. 4 of the main text, using an exponential basis with $N=3$ and $r_\mathrm{max}= 20  \ \si{\micro\meter}$.
				}
	\label{bases_gauss}
\end{figure}
%%%%%%%%%%%%%%%%%%%%%%%%%%%%%%%%%%%%

%%%%%%%%%%%%%%%%%%%%%%%%%%%%%%%%%%%%
%FIGURE 
\begin{figure}[h!]
	\includegraphics[width=0.9\textwidth]{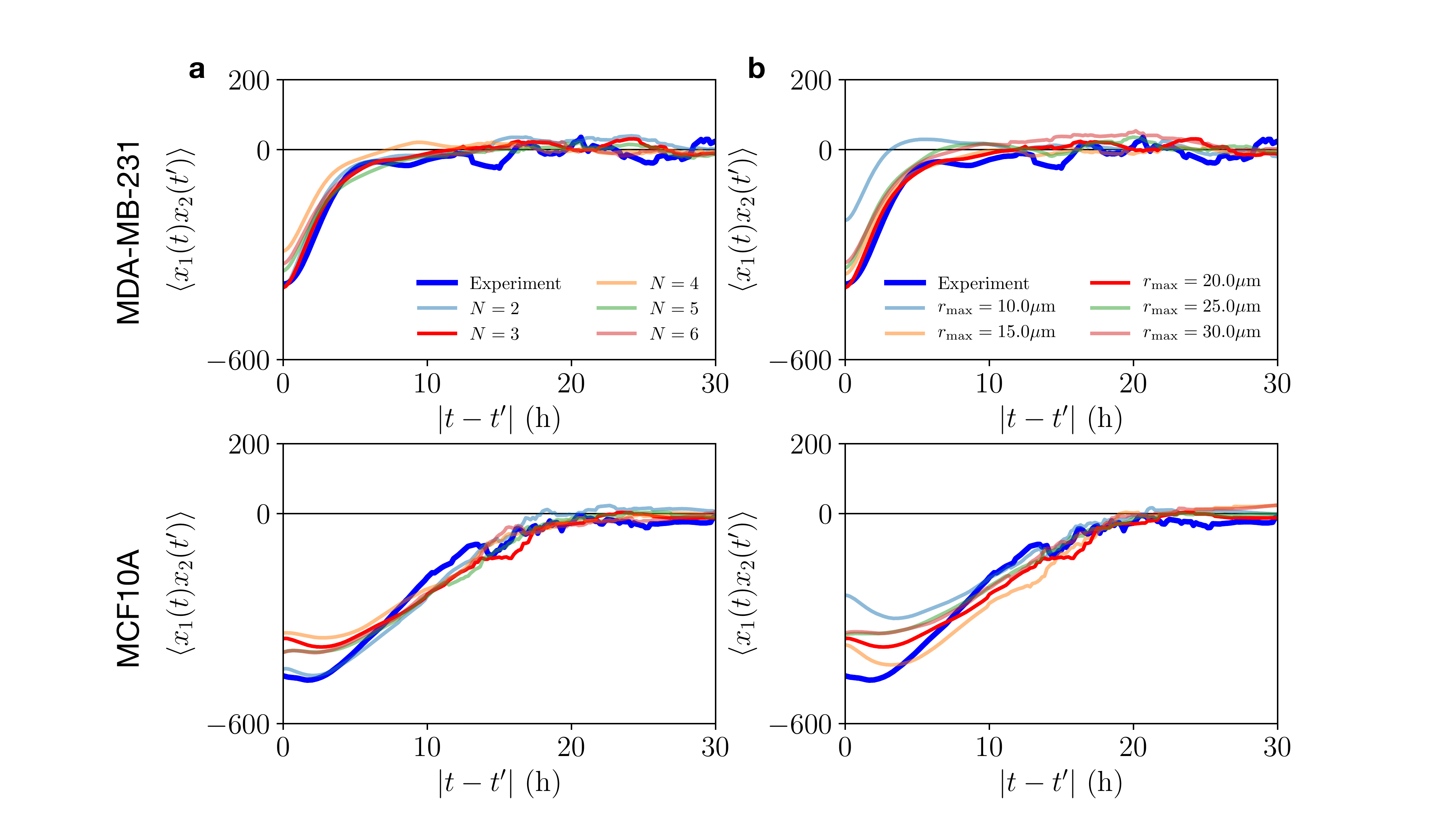}
	\centering
		\caption{
		\textbf{Predicted position cross-correlation functions for various exponential bases.} 
		\textbf{a}, Varying the number of kernels $N$.
		\textbf{b}, Varying the maximum decay length $r_\mathrm{max}$.
		\textit{Top row:} MDA-MB-231 cells.
		\textit{Bottom row:} MCF10A cells.
				}
	\label{bases_CF}
\end{figure}
%%%%%%%%%%%%%%%%%%%%%%%%%%%%%%%%%%%%

\newpage
%------------------------------------------------------------
\paragraph{Simulations of the inferred model}
\label{sec_sim}

An important step in performing inference from data is to test the inferred model by simulating stochastic trajectories based on the inferred model terms, and to compare their statistical properties to those observed experimentally. We simulate the dynamics using Verlet integration with a small time step $\mathrm{d}t$. To compare the statistics of these simulated trajectories to those observed experimentally, we sample the simulated trajectories with the same discrete time interval as in experiments, $\Delta t=10 \ \mathrm{min} \gg \mathrm{d}t$.

%------------------------------------------------------------
\subparagraph{Self-consistency test} 
\label{sec_stability}
First, we determine whether the inferred model is self-consistent: for a self-consistent inference, re-inferring a model from simulated trajectories should yield the same model. Here, we use the same number of simulated trajectories as experimentally observed trajectories, with a similar trajectory length, and the same sampling interval $\Delta t$ as in the experiment. We apply this test to the inferred models for MDA-MB-231 and MCF10A cells, and find that the re-inferred model accurately matches the original inferred model (Fig.~\ref{stability}), showing that our inference approach is numerically stable.

%%%%%%%%%%%%%%%%%%%%%%%%%%%%%%%%%%%%
%FIGURE 
\begin{figure}[h!]
	\includegraphics[width=0.85\textwidth]{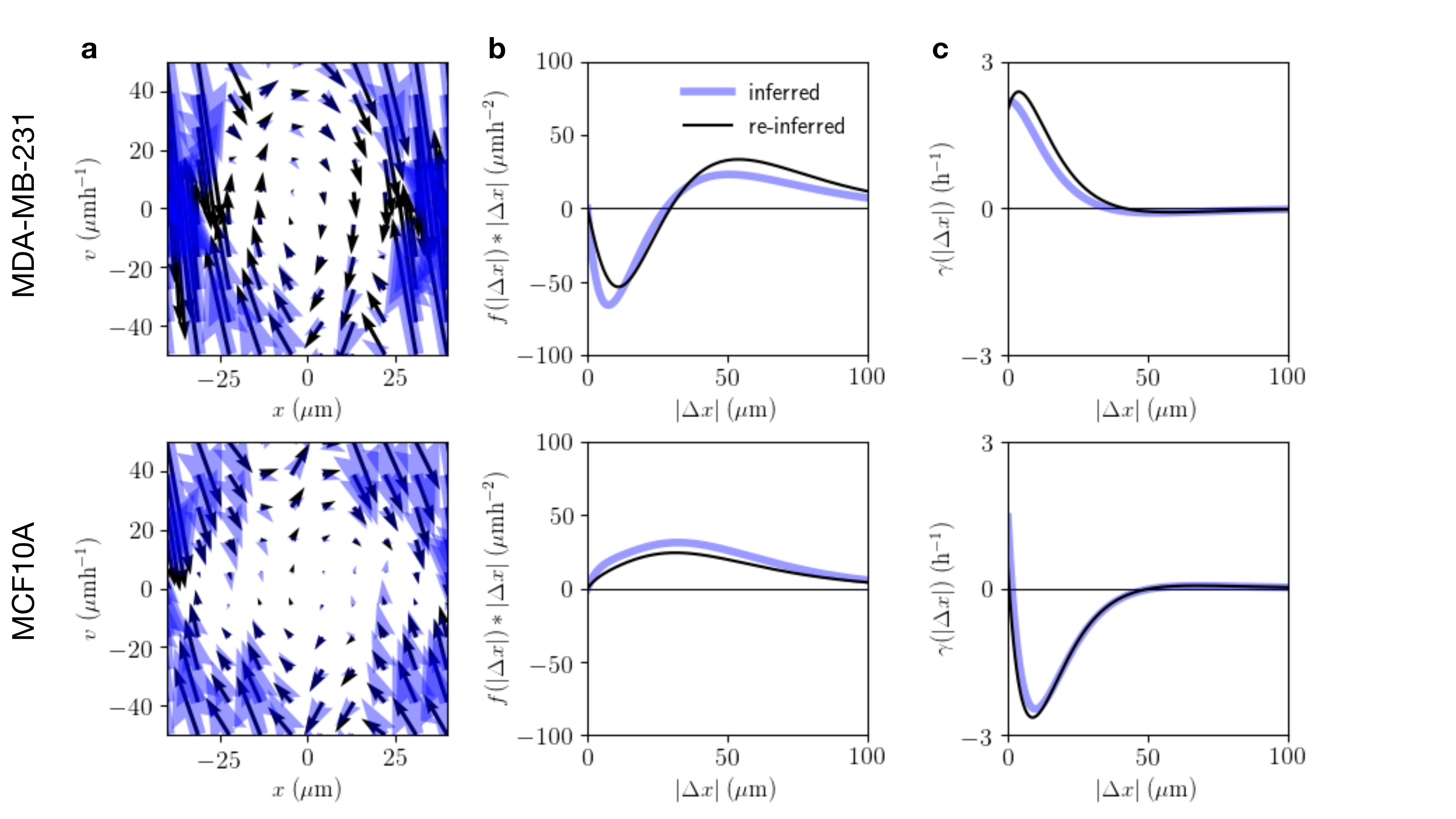}
	\centering
		\caption{
		\textbf{Stability test of the inferred model.} 		
		\textbf{a}, Flow field of the confinement term $F(x_i,v_i)$. Blue arrows: inferred from experimental data, black arrows: re-inferred from simulated trajectories.
		 \textbf{b}, Cohesive component $f(|\Delta x_{ij}|) |\Delta x_{ij}|$.
		 \textbf{c}, Friction kernel $\gamma(|\Delta x_{ij}|)$.
		\textit{Top row:} MDA-MB-231 cells.
		\textit{Bottom row:} MCF10A cells.
			}
	\label{stability}
\end{figure}
%%%%%%%%%%%%%%%%%%%%%%%%%%%%%%%%%%%%

%------------------------------------------------------------
\subparagraph{Testing the predictive power of the model}
\label{sec_predictions}

%%%%%%%%%%%%%%%%%%%%%%%%%%%%%%%%%%%%
%FIGURE 
\begin{figure}[h!]
	\includegraphics[width=0.8\textwidth]{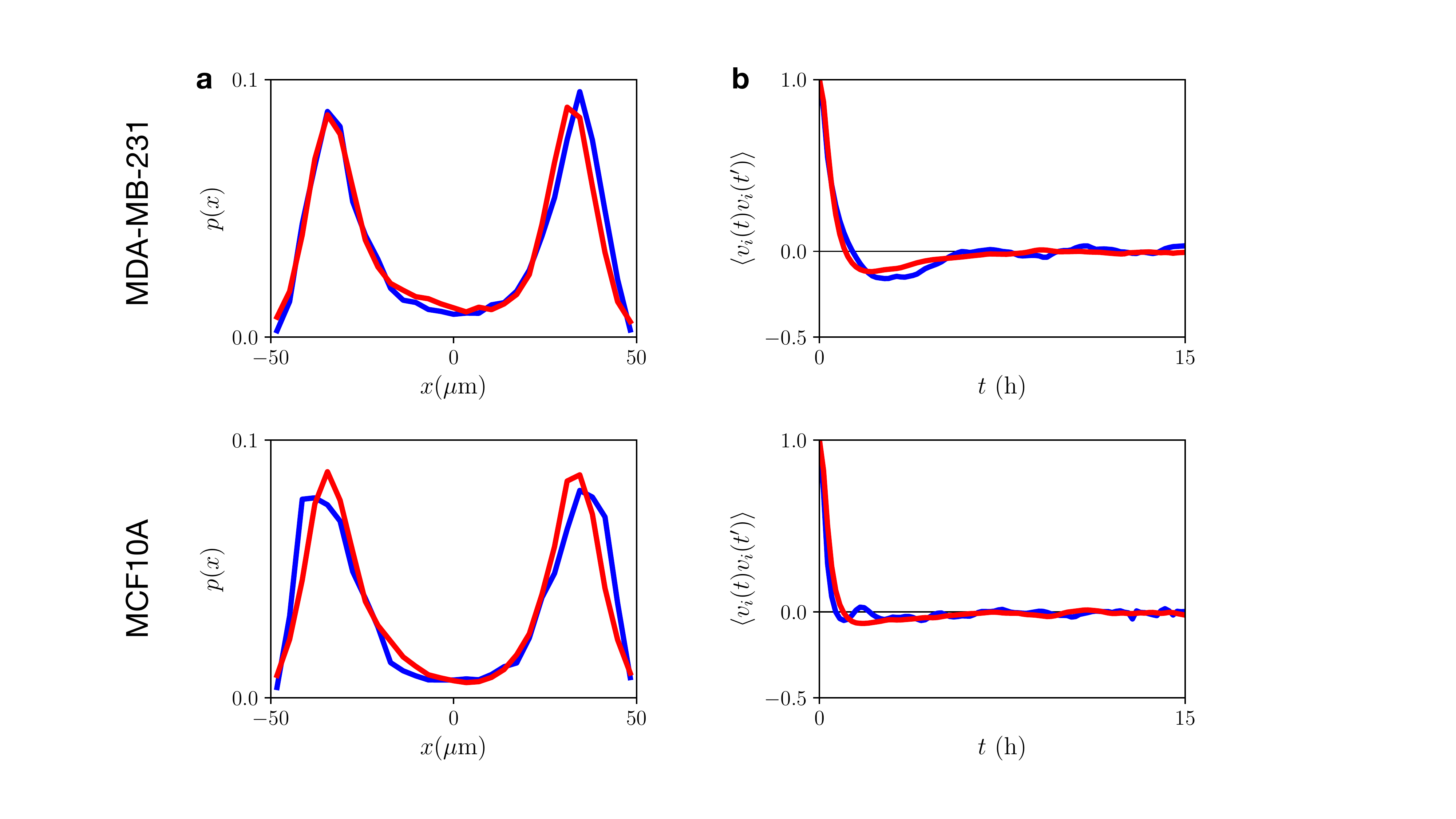}
	\centering
		\caption{
		\textbf{Experimental and predicted dynamics of the inferred model (Eq. 1 in the main text).} 
		\textbf{a}, Probability distribution of all cell positions $p(x)$ (experiment shown in blue, model predictions in red).
		 \textbf{b}, Normalized velocity auto-correlation function $\langle v_i(t) v_i(t') \rangle$.
		\textit{Top row:} MDA-MB-231 cells. \textit{Bottom row:}  MCF10A cells.
				 }
	\label{fitting_otherstats}
\end{figure}
%%%%%%%%%%%%%%%%%%%%%%%%%%%%%%%%%%%%

To test the predictive power of the model, we apply the same analysis routines that were applied to the experimental data to our simulated data (results shown in Fig. 2 of the main text). The inferred model is fully constrained by the short time-scale accelerations of the dynamics. Thus, comparing the predicted long time-scale features such as correlation functions to the experimental data provides an independent test of the model. In addition to the statistics shown in the main text, here we show several additional statistics to test experiment-model agreement. To test how accurately the model captures the dynamics at the single-cell level, we plot the distribution of all cell positions $p(x)$, and the velocity auto-correlation function $\langle v_i(t) v_i(t') \rangle$. As shown in Fig.~\ref{fitting_otherstats}a,b, these statistics are well captured by the model.

In our model, we assume that the cell-cell interactions separate into a cohesive contribution $f(|\Delta x|) \Delta x$ and an effective linear friction $\gamma(|\Delta x|) \Delta v$. This choice is motivated by the observation that the $\Delta v$-dependent component of the contact acceleration maps is linear (Fig. 3c,f in the main text). We find that the contact acceleration maps predicted by the model are qualitatively very similar to those inferred from experiments (Fig.~\ref{assumptions_CAM}), indicating that this assumption is valid.

%%%%%%%%%%%%%%%%%%%%%%%%%%%%%%%%%%%%
%FIGURE 
\begin{figure}[h!]
	\includegraphics[width=0.8\textwidth]{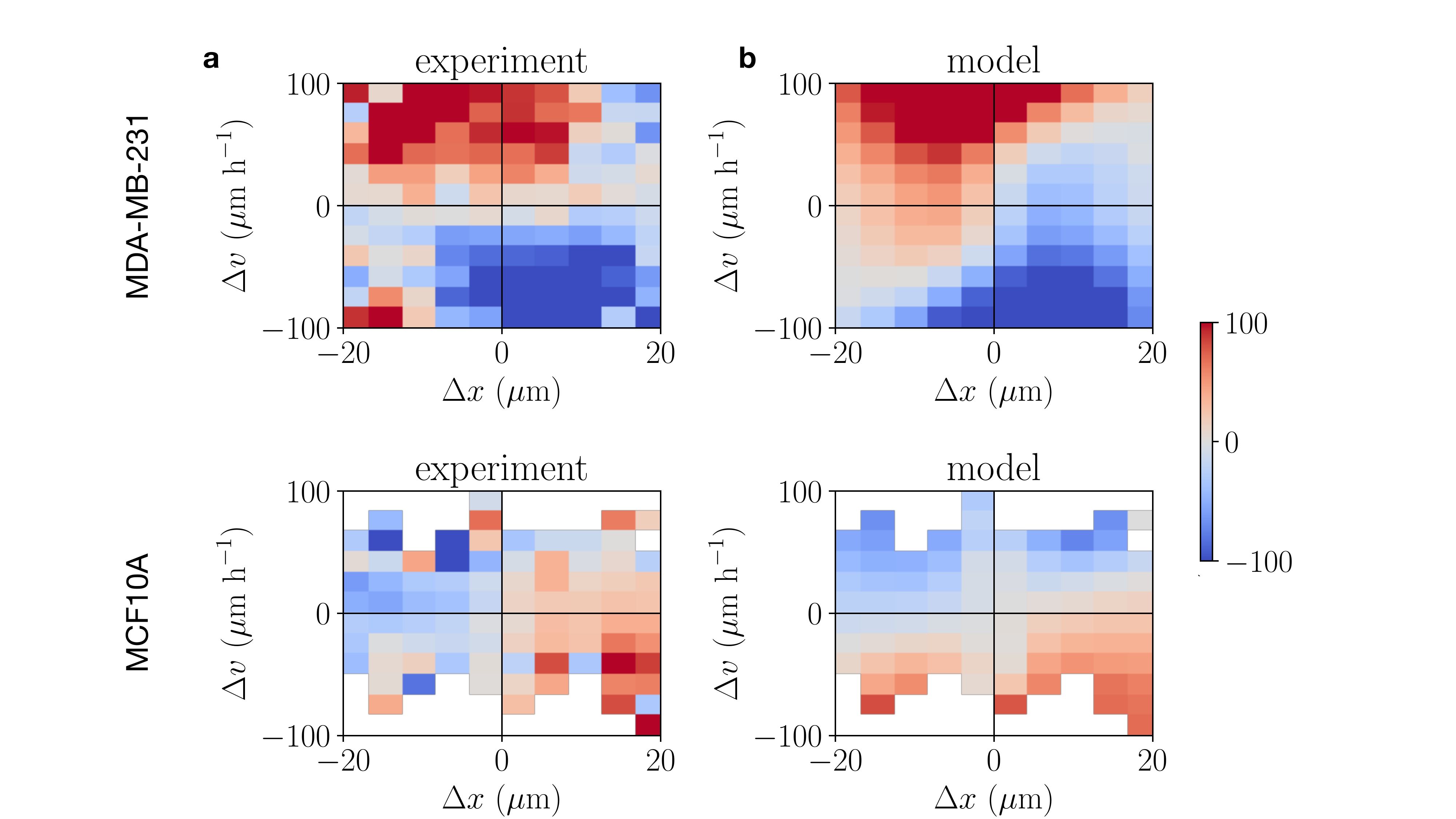}
	\centering
		\caption{
		\textbf{Experimental and predicted contact acceleration maps.} 
		\textbf{a}, Experimentally measured contact acceleration map.
		 \textbf{b}, Contact acceleration map measured from simulation data, plotted for the same region of phase space sampled in the experiment.
		\textit{Top row:} MDA-MB-231 cells. \textit{Bottom row:}  MCF10A cells.
				 }
	\label{assumptions_CAM}
\end{figure}
%%%%%%%%%%%%%%%%%%%%%%%%%%%%%%%%%%%%

%%%%%%%%%%%%%%%%%%%%%%%%%%%%%%%%%%%%
%FIGURE 
\begin{figure}[h!]
	\includegraphics[width=0.9\textwidth]{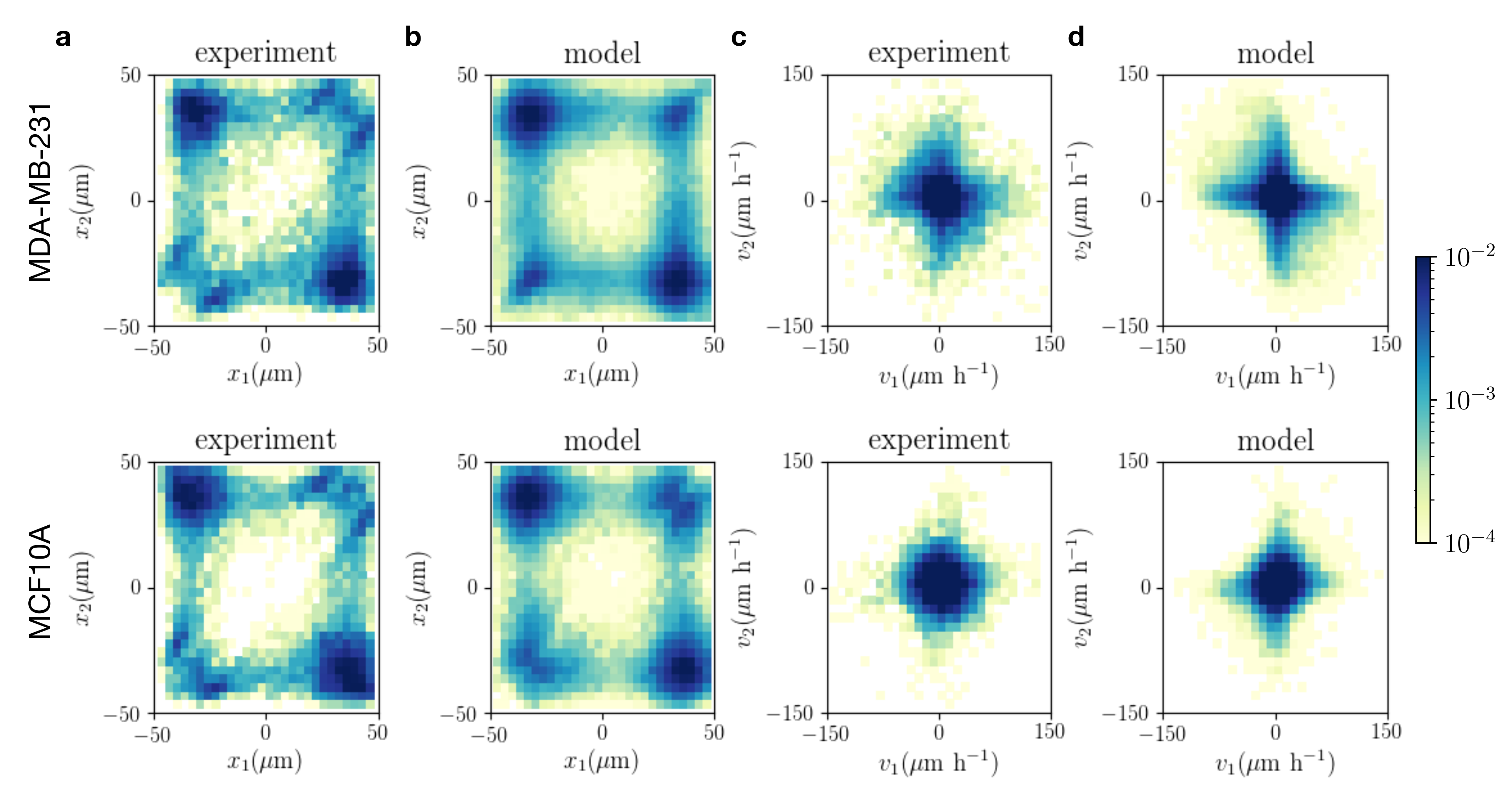}
	\centering
		\caption{
		\textbf{Experimental and predicted joint probability distributions.} 
		\textbf{a,b.} Experimental and predicted joint probability distribution of positions $p(x_1,x_2)$.
		\textbf{c,d.} Experimental and predicted joint probability distribution of velocities $p(v_1,v_2)$.
		\textit{Top row:} MDA-MB-231 cells. \textit{Bottom row:}  MCF10A cells.
				 }
	\label{fitting_fulldists}
\end{figure}
%%%%%%%%%%%%%%%%%%%%%%%%%%%%%%%%%%%%

%%%%%%%%%%%%%%%%%%%%%%%%%%%%%%%%%%%%
%FIGURE 
\begin{figure}[h!]
	\includegraphics[width=0.9\textwidth]{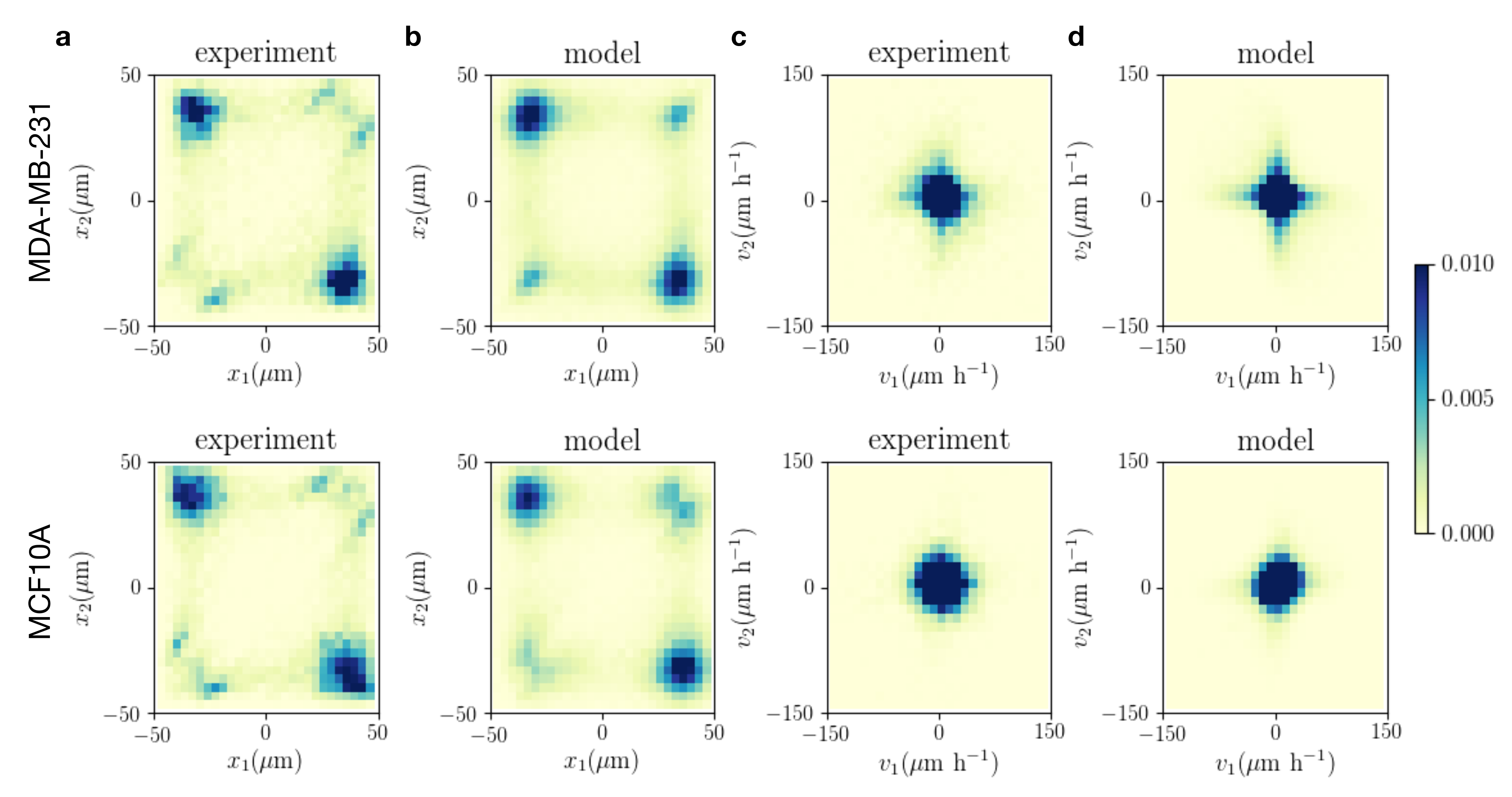}
	\centering
		\caption{
		\textbf{Experimental and predicted joint probability distributions, plotted on a linear scale.} Same panels as shown in Fig.~\ref{fitting_fulldists}, but with a linear colour scale.
				 }
	\label{fitting_fulldists_lin}
\end{figure}
%%%%%%%%%%%%%%%%%%%%%%%%%%%%%%%%%%%%

Next, we show side-by-side comparisons of the full joint probability distribution of positions $p(x_1,x_2)$ and velocities $p(v_1,v_2)$ (Fig.~\ref{fitting_fulldists}). The experimental distributions $p(x_1,x_2)$ exhibit several features (Fig.~\ref{fitting_fulldists}a). First, there is a clear minimum around $(0,0)$, corresponding to both cells occupying the connecting bridge. Second, we find peaks where each cell occupies one island, and fainter peaks where both cells occupy the same island. This reflects the mutual exclusion behavior exhibited by these cells. These peaks are connected by horizontal and vertical 'paths', indicating that during transitions, typically, only one cell performs a transition at a time. Finally, we find that the peaks corresponding to both cells occupying the same island are 'split', and exhibit two distinct close-by maxima. Our model captures almost all of these features, including the relative occupation of the same- and opposite-side configuration, and the path-structure of the map (Fig.~\ref{fitting_fulldists}b). However, the model does not exhibit the same splitting of the same-side probabilities, which may be due to movement in the second dimension (the short axis of the micropattern, $y$), which is not captured by the model. Our model further captures the structure of the velocity distributions $p(v_1,v_2)$ (Fig.~\ref{fitting_fulldists}c,d).

\subparagraph{Ruling out simpler models}
\label{sec_simpler}
 
We arrived at our model (Eq.~\eqref{eqn_eom}) by first excluding simpler alternatives. First, we consider the non-interacting case, consisting only of the single-cell term:
\begin{equation}
\dot{v}_i = F(x_i,v_i) + \sigma \eta_i(t)
\end{equation}
As expected, such a model is unable to capture the correlations in the system, and can therefore be ruled out (Fig.~\ref{fitting_only_noint}). This model is still able to capture the distinct minimum in the joint probability density around $(x_1,x_2)=(0,0)$, suggesting that this feature is due to the single-cell term: due to the confinement very little occupancy is expected near the center of the connecting bridge.

%%%%%%%%%%%%%%%%%%%%%%%%%%%%%%%%%%%%
%FIGURE 
\begin{figure}[h!]
	\includegraphics[width=0.9\textwidth]{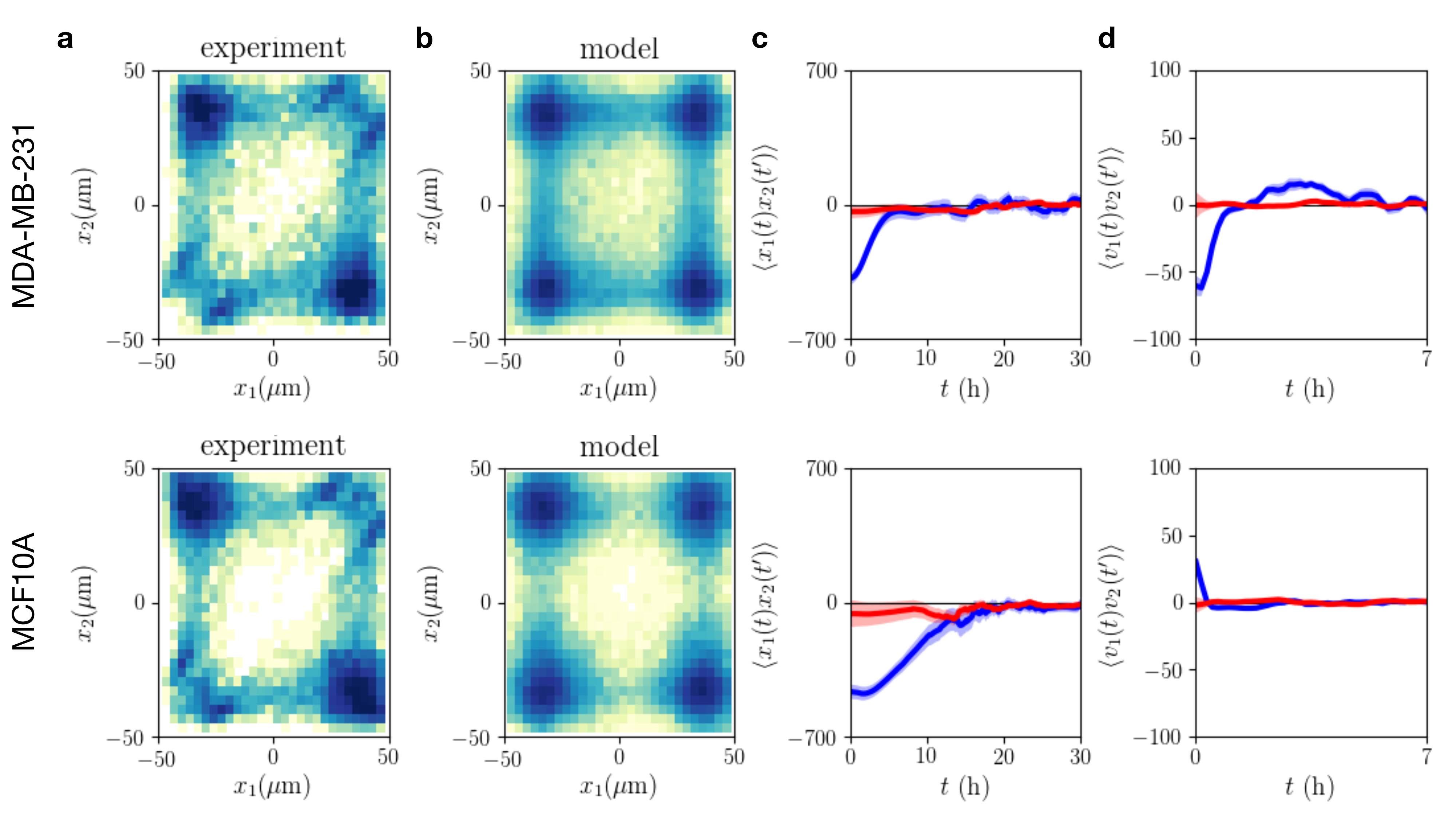}
	\centering 
		\caption{
		\textbf{Experimental and predicted dynamics for an inferred model without interactions.} 
		\textbf{a.} Experimental joint probability distribution $p(x_1,x_2)$. The colour bar corresponds to that shown in Fig.~\ref{fitting_fulldists}.
		\textbf{b.} Model prediction of the joint probability distribution $p(x_1,x_2)$. 
		\textbf{c.} Position cross-correlation functions for the experiment (blue) and model prediction (red).
		\textbf{d.} Velocity cross-correlation functions for same-side configurations.
		\textit{Top row:} MDA-MB-231 cells. \textit{Bottom row:}  MCF10A cells.
		 }
	\label{fitting_only_noint}
\end{figure}
%%%%%%%%%%%%%%%%%%%%%%%%%%%%%%%%%%%%

Next, we consider a model including only a cohesive term:
\begin{equation}
\dot{v}_i = F(x_i,v_i) + f(|\Delta x_{ij}|) \Delta x_{ij} + \sigma \eta_i(t)
\end{equation}
While this model can approximately capture the dynamics of MCF10A cells, except for the velocity cross-correlation function, it completely fails to describe the MDA-MB-231 statistics (Fig.~\ref{fitting_only_coh}). In fact it predicts that cells are more likely to occupy the same-side configuration, in qualitative disagreement with our experimental observations, likely due to the attractive nature of the cohesive interaction in MDA-MB-231 cells.

%%%%%%%%%%%%%%%%%%%%%%%%%%%%%%%%%%%%
%FIGURE 
\begin{figure}[h!]
	\includegraphics[width=0.9\textwidth]{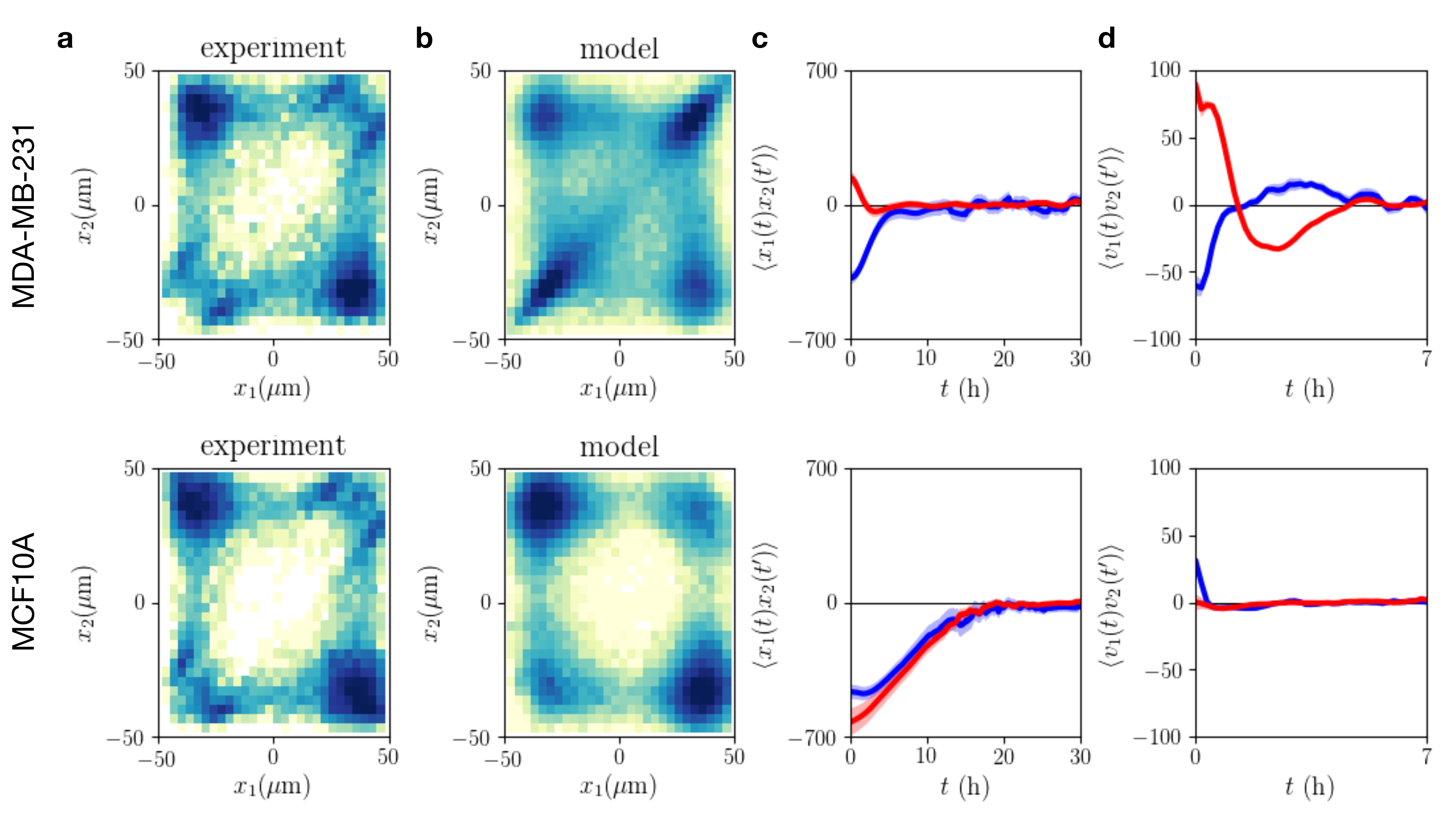}
	\centering
		\caption{
		\textbf{Experimental and predicted dynamics for an inferred model with only cohesive, but no friction interactions.} See Fig.~\ref{fitting_only_noint} for captions.
		 }
	\label{fitting_only_coh}
\end{figure}
%%%%%%%%%%%%%%%%%%%%%%%%%%%%%%%%%%%%

Finally, we consider a model including frictional interactions, but no cohesion:
\begin{equation}
\dot{v}_i = F(x_i,v_i) + \gamma(|\Delta x_{ij}|) \Delta v_{ij} + \sigma \eta_i(t)
\end{equation}
This model qualitatively fails to account for the MCF10A statistics (Fig.~\ref{fitting_only_al}): it predicts that cells are more likely to occupy the same-side configuration, likely due to the regular friction between MCF10A cells, which acts to slow cells down when they are close to each other.

In conclusion, we find that the simplest model within the class of models considered here, which can accurately capture the statistics of both MCF10A and MDA-MB-231 cell pairs, requires both cohesive and friction interactions.

%%%%%%%%%%%%%%%%%%%%%%%%%%%%%%%%%%%%
%FIGURE 
\begin{figure}[h!]
	\includegraphics[width=0.9\textwidth]{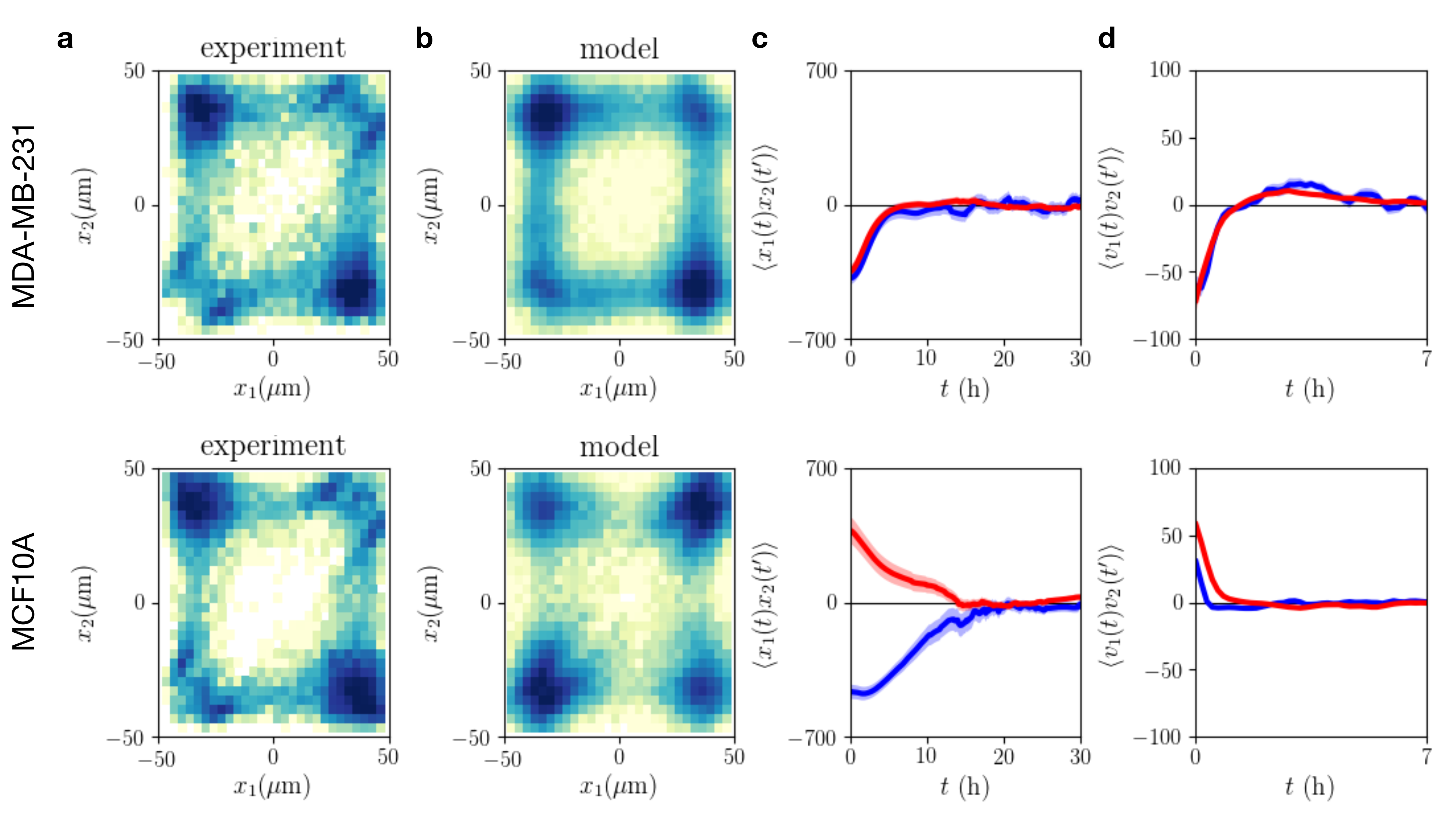}
	\centering
		\caption{
		\textbf{Experimental and predicted dynamics for an inferred model with only friction, but no cohesive interactions.} See Fig.~\ref{fitting_only_noint} for captions.
				 }
	\label{fitting_only_al}
\end{figure}
%%%%%%%%%%%%%%%%%%%%%%%%%%%%%%%%%%%%

%------------------------------------------------------------
\paragraph{Separation of single-cell and interaction terms}
\label{sec_single}

Here, we directly compare the single-cell term inferred from experiments with interacting cell pairs ($F(x_i,v_i)$ in Eq.~\eqref{eqn_eom}) to the deterministic term inferred from experiments in which only a single cell occupies the pattern~\cite{bruckner_stochastic_2019}, denoted $F_\mathrm{sc}(x,v)$. In Fig.~\ref{1cell}, the terms are compared side by side. Furthermore, we show the deterministic flow field $(\dot{x},\dot{v})=(v,F(x,v))$ superimposed for both experiments. These results indicate a remarkable similarity of the inferred terms, indicating that the contributions of single-cell dynamics (corresponding to the internal motility of the cell and its interaction with the local micro-environment placed by the micropattern) are not strongly affected by the presence of another cell.

%%%%%%%%%%%%%%%%%%%%%%%%%%%%%%%%%%%%
%FIGURE 
\begin{figure}[h!]
	\includegraphics[width=0.85\textwidth]{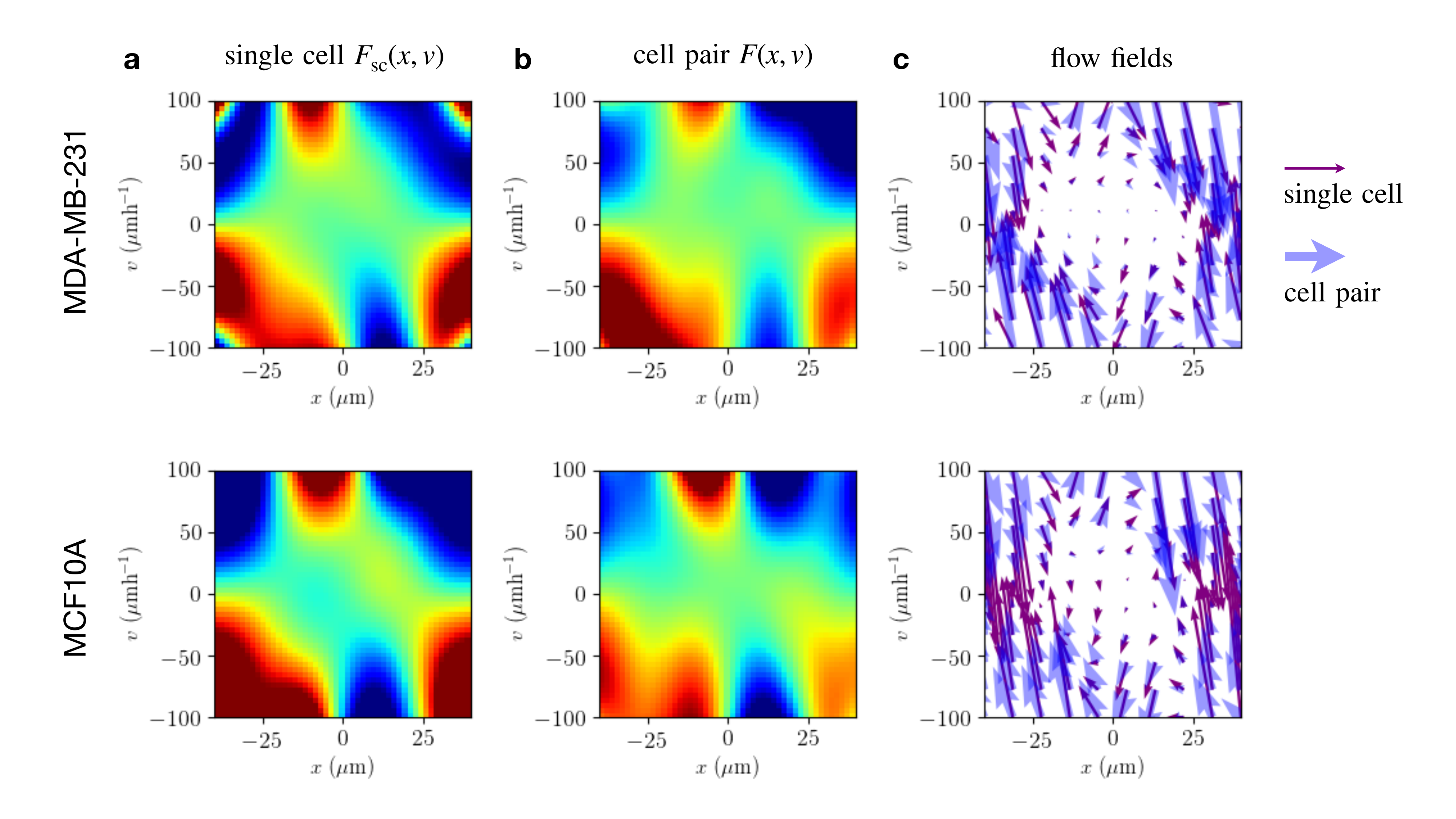}
	\centering
		\caption{
		\textbf{Disentangling single-cell and interaction contributions.} 
		\textbf{a}, Deterministic term $F_\mathrm{sc}(x,v)$ inferred from experiments with single cells confined to two-state micro-patterns~\cite{bruckner_stochastic_2019}, obtained by applying ULI with the same basis expansion as used for cell pair experiments (Eq.~\eqref{eqn_fourier}), without interaction terms. Plotted with the same colour scale as in Fig. 4 in the main text.
		\textbf{b}, Single-cell term $F(x,v)$ inferred from cell pair experiments (as shown in Fig. 4 in the main text).
		\textbf{c}, Direct comparison of the flow fields of both terms. Fat blue arrows: inferred from cell pair data, thin darkviolet arrows: inferred from single-cell experiments.
		 \textit{Top row:} MDA-MB-231 cells.
		\textit{Bottom row:} MCF10A cells.
				}
	\label{1cell}
\end{figure}
%%%%%%%%%%%%%%%%%%%%%%%%%%%%%%%%%%%%

\newpage

\section{Geometrical dynamics of nucleus translocation }
\label{sec:translocation}

Adapted from: \\
\textsc{Inferring geometrical dynamics of cell nucleus translocation}\\
Sirine Amiri$^*$, Yirui Zhang$^*$,  Andonis Gerardos, C\'ecile Sykes$^\dagger$ and Pierre Ronceray$^\dagger$ \\
arXiv:2312.12402 [cond-mat,physics:physics,q-bio].

\subsection{Introduction}

The cell nucleus, three-to-four times stiffer than the cytoskeleton and twice as viscous, has traditionally been regarded as a mechanically passive compartment housing genetic information \cite{guilak_viscoelastic_2000}. It is now established that in physiological conditions, the nucleus can experience large mechanical stresses that impact its shape and internal organisation, affecting, for example, gene transcription~\cite{uhler_regulation_2017}.
In particular, when cells migrate through complex environments, the nucleus happens to experience large deformations, for instance when passing through tight constrictions~\cite{davidson_nuclear_2014,wolf_physical_2013,patteson_vimentin_2019,estabrook_calculation_2021}. How these large deformations affect nucleus functioning and feed back into the behavior of the cell remain open questions. In fact, the overwhelming majority of cell migration studies focus on experiments on flat surfaces that were crucial to decipher the detailed mechanisms of cell motility~\cite{verkhovsky_self-polarization_1999,maiuri_actin_2015,keren_mechanism_2008}. However, the nucleus is only weakly altered in such experiments, which thus cannot be informative on the role of nuclear mechanics on cell motility, its passive mechanical resistance to deformation, and also the mechanosensory pathways through which these deformations feed back and actuate the cell behavior~\cite{jaalouk_mechanotransduction_2009,maurer_driving_2019,uhler_regulation_2017, venturini_nucleus_2020, lomakin_nucleus_2020}. 

Addressing this problem through \emph{in vivo} experimental observations of cell migration in tightly constraining environments such as the extracellular matrix and epithelial tissues  represents a tremendous challenge. Indeed, one would have to disentangle the complexity of the environment from that of the migrating cell. For this reason, here we study an \emph{in vitro} system of cells migrating in a microfabricated device that imposes three-dimensional mechanical constraints on spontaneously migrating eukaryotic cells~\cite{davidson_design_2015, gundersen_assembly_2018}. We therefore focus on the influence of the geometry on squeezed cell migration. Specifically, cells migrate in an array of pillars designed to impose constrictions of controlled size, which incur large deformations of the nucleus. Remarkably, we find that cells with a nucleus of diameter $\sim12$\unit{\micro\meter} in their rest state are able to spontaneously migrate through constrictions as tight as $2$\unit{\micro\meter}. We refer to this process as \emph{nuclear translocation}, in analogy with polymer translocation where a large macromolecule can pass through tight pores. Using bright-field and multichannel fluorescent imaging, we are able to track the trajectories of individual nuclei going through these constrictions. However, the analysis of the resulting trajectories poses multiple challenges due to their complexity, inherent stochasticity, and the limited amount of data: how does one extract quantitative models and mechanistic insights from such trajectories?

To tackle this challenge, we develop and apply here a data-driven approach to learn dynamical models directly from experimental nucleus trajectories. This contrasts with more traditional model-based approaches that postulate a model form and fit its parameters through the use of aggregate observables such as correlation functions: here we let the model emerge from the data, and the parameters are optimized directly on the entire data set. Such approaches have recently received a lot of attention, in particular due to the development of methods adapted to data-driven inference of deterministic dynamical models such as ordinary and partial differential equations~\cite{brunton_discovering_2016,champion_data-driven_2019}. These methods are well adapted for large-scale datasets such as tissue dynamics~\cite{romeo_learning_2021,schmitt_zyxin_2023}. Importantly however, single-cell dynamics are not deterministic: the inner complexity of these objects, coupled to the reliance to feedback pathways involving small numbers of signalling molecules, results in apparently erratic dynamics which is better captured by stochastic differential equations (SDEs)~\cite{kloeden_numerical_2010}. Data-driven approaches have been used to capture the dynamics of freely migrating cells~\cite{selmeczi_cell_2005,selmeczi_cell_2008,li_dicty_2011}, revealing a persistent random walk behavior. They have been used to quantify the dynamics of non-constraining confined cell migration~\cite{bruckner_learning_2021, bruckner_geometry_2022} and, recently, for constraining cell migration in an elastic environment~\cite{stoberl_nuclear_2023}. Newly introduced inference methods for SDEs~\cite{frishman_learning_2020,bruckner_inferring_2020} have made it possible to efficiently learn such dynamics and have resulted in insights in cell-cell interactions during confined migration that would not have been possible with pre-existing methods~\cite{bruckner_learning_2021}. However, to our knowledge, such methods have not been applied to cell migration with mechanical constraints that lead to large deformations of the nucleus. To this aim, we define and measure quantitative descriptors of the cell shape and state, then use \emph{Stochastic Force Inference} (SFI)~\cite{frishman_learning_2020} to construct a model that captures the dynamics of these shape descriptors. By including the constriction shape as an explicit input of the model, we are able to extrapolate the model to other constriction sizes. Our inference analysis explicitly takes into account the spatial constraints of nucleus translocation and is applicable to other experimental designs.

\subsection{Results}

\paragraph{Confined cell migration experiments}
We use a CRISPR-modified Mouse Embryonic Fibroblasts (MEFs) cell line that expresses nesprin-2 giant with a green fluorescent protein (GFP) sequence and lamin A/C with a red fluorescent protein (mCherry) sequence \cite{davidson_nesprin-2_2020}. The lamin biopolymer shell that lies right underneath the nuclear envelope is linked to the cytoskeleton through the LINC complex, which includes nesprins~\cite{vahabikashi_nuclear_2022, gruenbaum_lamins_2015,tapley_connecting_2013,crisp_coupling_2006,mellad_nesprins_2011}. Cells migrate through microfluidic devices that consist of a series of \SI{5}{\micro\meter} high pillar structures providing three sizes of constrictions (5, 3 and \SI{2}{\micro\meter}) and larger channels (\SI{15}{\micro\meter}) (Fig.\ref{fig:Figure1NT}a, b). Such migration devices are obtained by covalent assembly of a 3D-imprinted block of polydimethylsiloxane (PDMS) with a glass coverslip~\cite{gundersen_assembly_2018}. Cells are placed on one side of the device with culture medium, before the pillars. They exhibit global motion (on the $x$-axis) towards the other side, empty of cells but filled with culture medium (Fig.\ref{fig:Figure1NT}a). The apparent width of MEF cell nuclei (on the $y$-axis) is $12\pm$\SI{2}{\micro\meter} outside of constrictions. It is, therefore, larger than constriction sizes and smaller than the large channel of \SI{15}{\micro\meter}. Note that in all conditions, nuclei shapes are mostly cylindrical (on the $z$-axis), touching the bottom and the ceiling of the migration device (Fig.\ref{fig:Figure1NT}c). We confirm previous observations \cite{davidson_nesprin-2_2020} that during nucleus translocation through a constriction, nesprin signal intensity increases at the front of the nucleus while lamin signal does not.

    \begin{figure}[ht!] \centering
        \includegraphics[width=0.65\columnwidth]{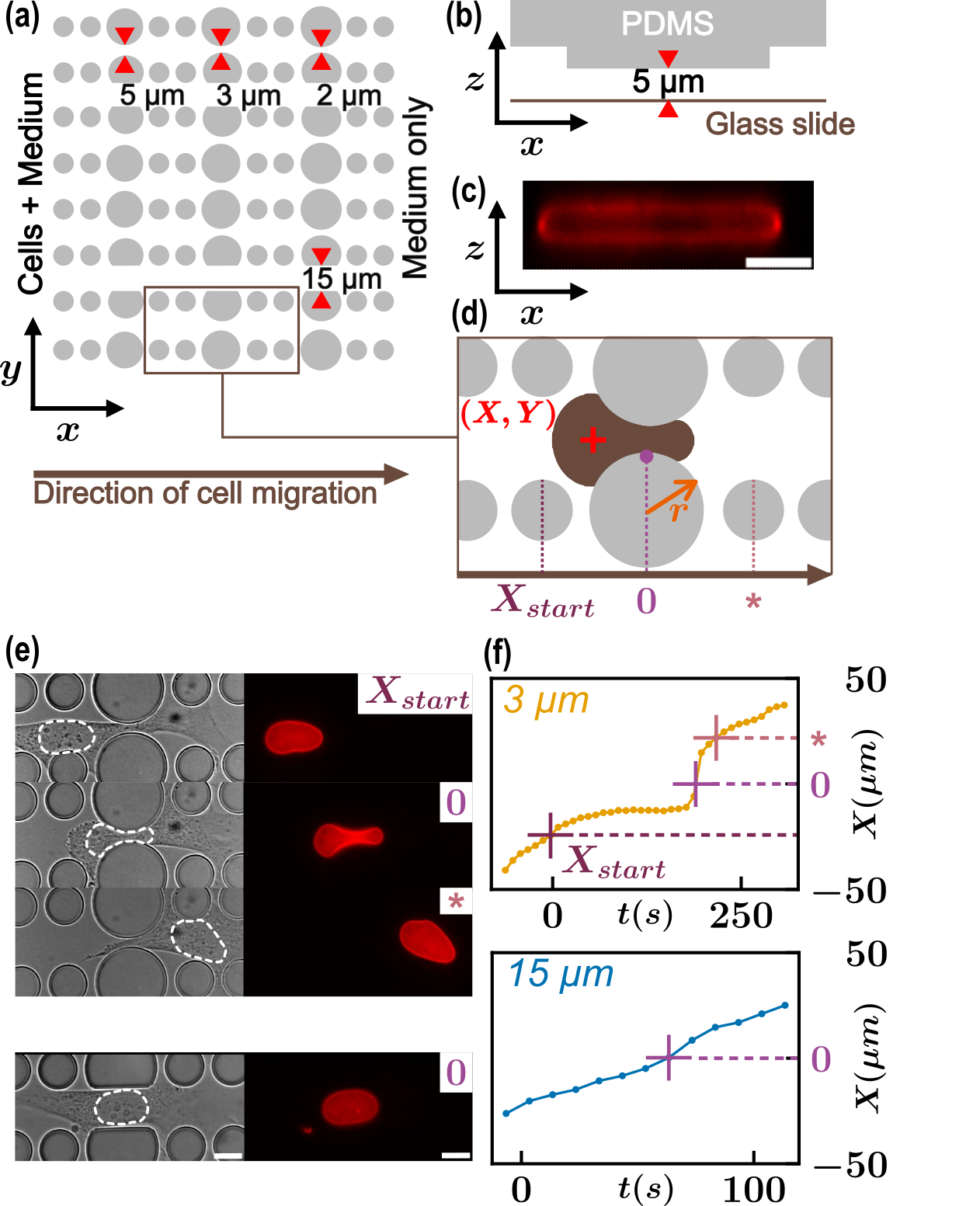}
        \caption{\textbf{\textit{CRISPR engineered MEFs are migrating in a microfluidic device made of constrictions.}} \textbf{(a)} Top view of the pattern used in the microfluidic device. It is composed of PDMS pillars of several widths in order to make three types of constrictions : 5, 3 and \SI{2}{\micro\meter} wide and a control channel of \SI{15}{\micro\meter}. \textbf{(b)} Side view of the microfluidic device. Height of the pillars is 5µm.\textbf{(c)} Side view ($x,z$) of lamin A/C signal in mCherry of an engineered MEF in the middle of a constriction. \textbf{(d)} Representation of the origin points used for a nucleus (\textit{in brown}) trajectory. ($X,Y$) is position of the nucleus (\textit{red star}). \textbf{(e)} Epifluorescence images of an engineered MEF crossing a \SI{3}{\micro\meter} constriction (\textit{top}) and a \SI{15}{\micro\meter} large channel (\textit{bottom}). Left is the transmission signal and right mCherry signal (for Lamin A/C). \textbf{(f)} Examples of trajectories of one cell through a \SI{3}{\micro\meter} constriction (\textit{top}) and one cell through \SI{15}{\micro\meter} large channel(\textit{bottom}). Annotated points corresponds to specific positions illustrated in \textbf{(d)}. Scale bars : \SI{10}{\micro\meter}.} 
        \label{fig:Figure1NT}
    \end{figure}

\paragraph{Extracting cell nucleus trajectories}
 We observe the movement of cell nuclei in the horizontal ($x,y$) plane when cells migrate between vertical $z$-oriented pillars (Fig.\ref{fig:Figure1NT}a,b). Nuclei are deformed when they translocate through narrow constrictions \cite{davidson_design_2015}. A constriction is defined by two facing big pillars of radius $r$ (Fig.\ref{fig:Figure1NT}d). For each constriction, we define the spatial origin ($x=0,y=0$) at the center of the constriction. The position of a nucleus is defined by its surface barycenter ($X,Y$) detected through the lamin signal (Fig.\ref{fig:Figure1NT}d, right). An image is taken every \SI{10}{\min}. This time interval was optimized to limit fluorescence bleaching and cell phototoxicity.

For each nucleus trajectory, we define $X_{\text{start}}$ as the middle of the small pillar that precedes the specific constriction (Fig.\ref{fig:Figure1NT}d). The same definition is adapted to the 15\unit{\micro\meter} channels with a truncated-disk pillar defining the "constriction". The start of a trajectory (at time $t=0$ min) is defined either by $X = X_{\text{start}}$ or by its interpolated value using a constant nuclear speed between the two available positions closest to $X_{\text{start}}$. The end of a trajectory is determined by the earliest of i) the end of the overall acquisition, ii) the start of a new trajectory in a new constriction, and iii) half an hour before the cell starts to divide or die. We exclude any trajectory corresponding to cells undergoing adherent cell-cell contact for more than an hour to exclusively address here the migration of individual cells. Examples of recorded images of a nucleus translocating through a \SI{3}{\micro\meter} constriction and a nucleus migrating through a large \SI{15}{\micro\meter} channel are given in Fig.\ref{fig:Figure1NT}e. The corresponding trajectories and origin points are displayed in Fig.\ref{fig:Figure1NT}f. We do not observe nuclear rupture during this deformation, contrarily to other mechanical studies of cell nuclei~\cite{denais_nuclear_2016, lomakin_nucleus_2020, pfeifer_gaussian_2022}.

A typical nucleus trajectory $X(t)$ through a \SI{3}{\micro\meter} constriction has a sigmoid-like shape, with a plateau soon after $X_{\text{start}}$ when the nucleus reaches the entrance of the constriction, and a sharp acceleration when it manages to pass through the center of the constriction at $X = 0$, followed by an unconstrained motion (Fig.\ref{fig:Figure1NT}f, top). A nucleus trajectory in a large \SI{15}{\micro\meter} channel displays a smooth movement (Fig.\ref{fig:Figure1NT}f, bottom) at almost constant velocity. However a fraction of cells do not translocate before the end of the trajectory recording. They are nevertheless included in our data set to avoid any statistical bias in the analysis. Overall, nucleus trajectories show some variability, both in the duration of the plateau and in the velocity of free migration, as can be seen in Fig.\ref{fig:Figure2NT}a. 

\paragraph{Data-driven modeling from geometric quantities}
We propose here an alternative approach to maintain the overdamped dynamics, which is more physical, and approximate $\Pi$ with available information. 
Indeed, we have access to more than just the nuclear center $X$: using the lamin signal, we can track the precise contour of the nucleus and extract a richer set of geometrical quantities. In particular, when the cell engages into the constriction, the nucleus starts elongating and protruding toward the narrow part of the constriction, as schematized in Fig.\ref{fig:Figure2NT}b. When exiting the constriction, the protrusion points backwards, and the nucleus progressively recovers is oval shape.

From these observations, we define two variables to characterize nucleus deformations.  First, to account for the geometrical shape change of the nucleus, we define its \emph{protrusion vector} $P=X_{c}-X$, with $X_{c}$ the barycenter of the contour of the nucleus. The quantity $P$ gives a measure of how much and in which direction the nucleus boundary protrudes relative to the center of mass. A positive (resp. negative) value of $P$ corresponds to a forward (resp. backward) extension of the nucleus relative to the center of mass (see Fig.\ref{fig:Figure2NT}b). 
Second, we characterize the relative ($x,y$) stretch by defining the aspect ratio of the cell nucleus $R$. A perfectly circular disk would correspond to $R=1$, whereas an ellipsoid oriented towards the $x$-axis (resp. $y$-axis) would correspond to $R>1$ (resp. $R<1$). Note that the reference shape of the nucleus is an oval, oriented towards the $x$ direction, and corresponds to a minimum value of $R$. When the nucleus starts to squeeze into the constriction, $R$ increases to a maximum value reached right in the middle of the constriction (see Fig.\ref{fig:Figure2NT}b).

The quantities $P$ and $R$ describe two different and complementary aspects of nucleus deformation. As illustrated in Fig.\ref{fig:Figure2NT}b, $P$ does not distinguish a dumbbell from an oval shape, whereas $R$ does; $R$ cannot distinguish a front protrusion from a back protrusion, whereas $P$ does. Note that our model, being overdamped, does not explicitly include the direction of movement. The breakdown of symmetry along the x axis could have been included in two ways: by including a shape descriptor internal to the cell that directly captures this orientation (e.g. by tracking the lamellipodium too), or by including terms in the dynamical model that break this symmetry externally (e.g. modelling a nutrient gradient in a chemotaxis model). We choose the latter here, motivated both by the fact that we aim to present a self-consistent dynamical model of the nucleus -- the shape of which does not reflect its direction of motion -- without referring to the extended cytoplasmic structures, and the fact that chemotaxis is indeed suggested by the fact that cells consistently direct themselves from the cell-rich to the medium-rich side of the microfluidic device. Examples of the time series $P$ and $R$ against $X$ when cells go through a $3$\unit{\micro\meter}-constriction (left) or a $15$\unit{\micro\meter}-channel (right) are displayed Fig.\ref{fig:Figure2NT}c. Whereas $P$ and $R$ are constant in a $15$\unit{\micro\meter} large channel, they are significantly affected by the $3$\unit{\micro\meter} constriction. These complementary geometrical data will allow us to infer a quantitative model for nuclear translocation dynamics.

\begin{figure}[tb]
  \begin{minipage}[c]{0.5\textwidth}
\includegraphics[width=\textwidth]{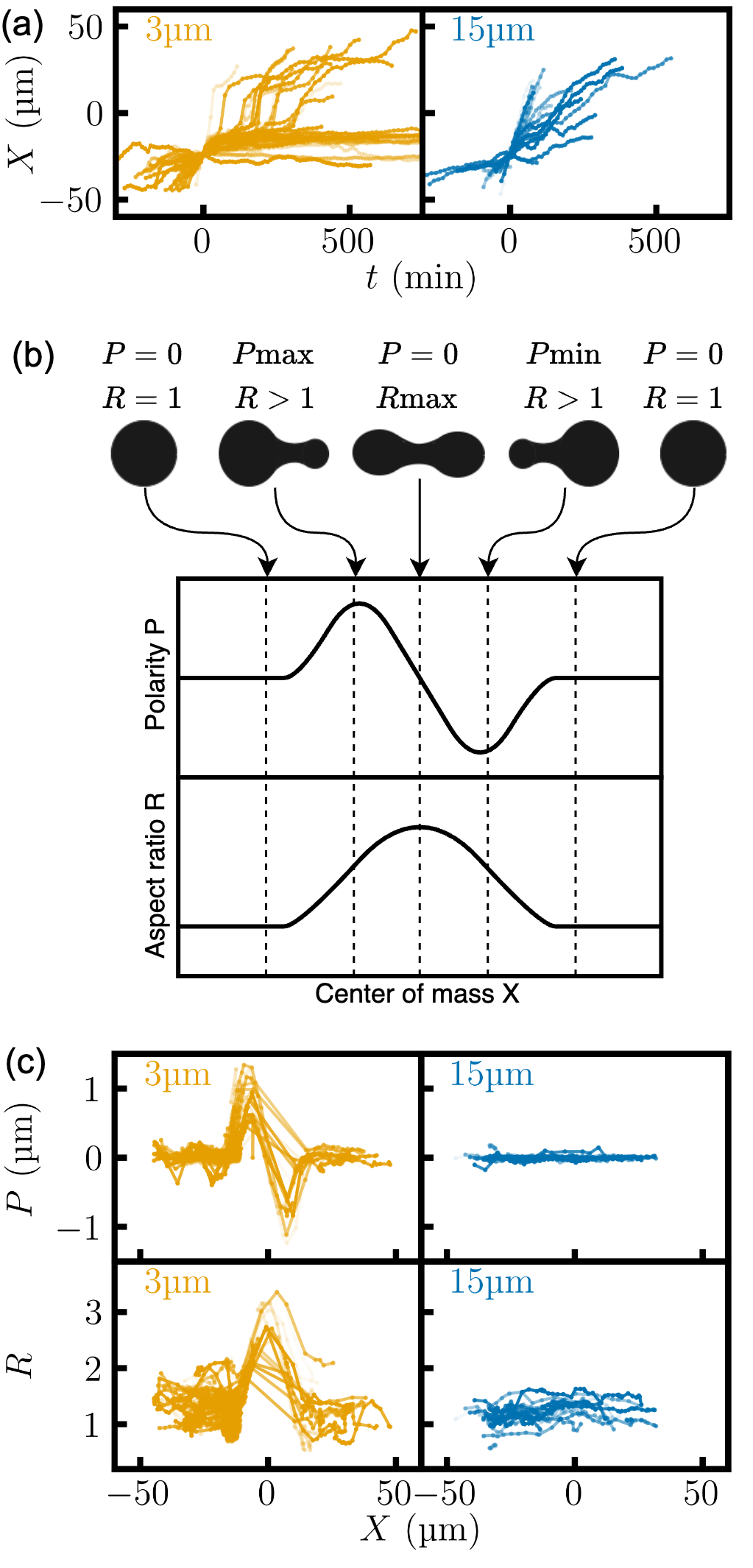}
  \end{minipage}\hfill
  \begin{minipage}[c]{0.4\textwidth}
\caption{\textbf{Experimental trajectories of cell nucleus translocation}. \textbf{(a)} Time series of nucleus position $X$ during translocation. \textbf{(b)} Schematic of characteristic nuclei shapes and their corresponding position during the translocation. \textbf{(c)} Protrusion vector $P$ and aspect ratio $R$ change along the translocation process. In \textbf{(a)} and \textbf{(c)}, results from $3$µm and $15$µm constraints are shown for comparison.}
\label{fig:Figure2NT}
  \end{minipage}\hfill
\end{figure}

%========================================
\paragraph{Inferring coupled dynamics of position and geometry}
Our aim is to obtain a data-driven, quantitative, autonomous description of nuclear translocation using the position $X$ and geometric descriptors $P$ and $R$.
More precisely, for each recorded nucleus trajectory in constraints $2, 3, 5$ and \SI{15}{\micro\meter}, we extract three time series $\{X_{t}, P_{t}, R_{t}\}$ at acquisition times $t = 0, \Delta t, 2\Delta t\dots$. These data serve as the input in our inference analysis, from which we aim to extract coupled SDEs capturing the continuous-time dynamics of $(X_t, P_t, R_t)$. We postulate that including the geometric quantities $P_t$ and $R_t$, on top of the nucleus position $X_t$, makes it possible to identify a set of such equations that is both autonomous (\emph{i.e.} that does not couple to the dynamics of other, unobserved quantities, in contrast to the approach of Ref.~\cite{bruckner_geometry_2022}) and physically first-order (\emph{i.e.} that does not introduce emergent inertia as a polarity model, in contrast to most pre-existing literature~\cite{li_dicty_2011,bruckner_stochastic_2019,bruckner_learning_2021,bruckner_learning_2023}).

To achieve this, we analyze the time series using a recently introduced framework, Stochastic Force Inference (SFI)~\cite{frishman_learning_2020}. SFI allows us to reconstruct first-order SDEs from such time series by employing a data-efficient quasi-maximum-likelihood linear regression algorithm. In practice, it consists of approximating the drift term with estimators formed by a linear combination of basis functions. Here, we start from a relatively large basis that we construct based on symmetries and our physical understanding of the quantities we model, and that include a systematic expansion of the geometrical features of the system -- \emph{i.e.} the $x$-dependent constriction width. We then iteratively reduce this basis to select an appropriately minimal model for the dynamical equations we aim to learn. More specifically, our starting model is:
\begin{equation}
\begin{split}
\dot{X}_t
    = &\overbrace{C_{X} + \alpha_{X} P_t + \beta_{X}(R_t -R_t^{-1})}^{\mathrm{internal\ driving}} + f_{X}(X_t, r) \\
    &+ \sqrt{2D_{X}}\cdot\eta_{X}(t),    
\end{split}
\label{eqn:generic-1}
\end{equation}
\begin{equation}
\begin{split}
    \dot{P}_t %&= f_{P}(X, P, R) + \sqrt{2D_{P}}\cdot\eta_{P}(t) \\
    &= \overbrace{C_P + \alpha_{P} P_t + \beta_{P}(R_t-R_t^{-1})}^{\mathrm{internal\ dynamics}} + f_{P}(X_t, r) \\ &+ \sqrt{2D_{P}}\cdot\eta_{P}(t),
\end{split}
\label{eqn:generic-2}
\end{equation}
\begin{equation}
\begin{aligned}
    \dot{R}_t %&= f_{R}(X, P, R) + \sqrt{2D_{R}}\cdot\eta_{R}(t) \\
    &= \overbrace{C_R + \alpha_{R} P_t + \beta_{R} (R_t - {R}_t^{-1})}^{\mathrm{internal\ dynamics}} +  f_{R}(X_t, r) \\&+ \sqrt{2D_{R}}\cdot\eta_{R}(t). 
\end{aligned}
\label{eqn:generic-3}
\end{equation}
Eq.\ref{eqn:generic-1} connects to the general form presented in Eq.\ref{eqn:CM-x} by approximating the polarity $\Pi$ with a linear combination of three terms: $C_{X}$, a constant drift representing the propensity of cells to migrate in the $x$ direction, physically motivated by the imbalance in cell populations between the two sides of the device; $\alpha_{X} P$ which is a vector-like term coupling the direction of motion and protrusion direction; and $\beta_{X}(R -R^{-1})$ by which the nucleus shape modulates the self-propulsion velocity around its rest shape $R=1$. The remainder, $f_{X}$, captures the effect of the environment, and thus depends on the position $X_t$ -- we omit, for simplicity, the possibility that it depends on the geometry. Similarly, the dynamics of $P$ is described by Eq.\ref{eqn:generic-2} (resp.~$R$ by Eq.\ref{eqn:generic-3}) with the same decomposition into internal dynamics and external influence, and we use the same basis functions. Note that we use the combination $(R-R^{-1})$ to reflect the fact that the aspect ratio $R$ is a ratio of lengths which should remain positive at all times, and has average value of $1$ in the absence of external constraints.

In a complex or unknown environment, the drifts $f_{X}$, $f_P$ and $f_R$ representing the influence of the environment on $X, P$ and $R$ would have to be expanded on a generic basis. Here, however, we take advantage of the fact that the geometry of the channel is known to simplify inference and allow for extrapolation of the model to other constriction sizes. Specifically, we include the radius $r$ of the pillars that form the constriction (see Fig.~\ref{fig:Figure1NT}d) as an explicit parameter of the inference, and construct our basis functions using the channel width $w(X, r)$ as well as the normal to the pillar $\hat{n}(X, r) = (n_{x}(X, r), n_{y}(X, r))$. 
Using $w(X, r), n_{x}(X, r)$ and $n_{y}(X, r)$ as ingredients, we approximate the environmental drift the nucleus experiences and reacts to, $f_{X}(X, r), f_{P}(X, r)$ and $f_{R}(X, r)$ in the constraint formed by pillars of radius $r$. 
Integrating the pillar radius as a control parameter into these functions allows us to infer a single model for the whole experimental data set of different constriction sizes, including the reference case where the channel does not have a constriction. It makes the model more straightforward and easier to interpret and allows us to use the data more efficiently.
As the influence of the pillars on the cell nucleus is expected to increase with decreasing channel width, we expand this geometrical influence in an inverse power series of the channel width in the basis, up to third-order, \emph{i.e.,} $1/w$,  $1/w^{2}$ and  $1/w^{3}$, which we multiply by geometrical quantities $1$, $n_x$ and $n_y$ that capture distinct features of the constriction. 

The SFI algorithm provides estimators for the coefficients of the drift field as a  linear combination of these basis functions. The initial model consists of the complete set of basis functions (in total $36$).
A challenge to the use of stochastic inference techniques on cell migration data is that the time interval between frames $\Delta t$ is typically of the same order as the typical translocation time, and cannot be easily decreased as more frequent imaging would incur phototoxicity. 
To overcome this problem, we introduce an improvement on the SFI algorithm to accommodate large time steps, which uses a trapezoidal integration scheme that results in lower discretization biases than previous methods.

%========================================
\paragraph{Model Selection algorithm}
%========================================
The learned model consisting of the full set of basis functions is constructed through physically motivated systematic expansion, and as such it is not minimal, which potentially leads to overfitting the data and precludes physical interpretation. To overcome this difficulty and obtain a more interpretable model, we improve this model through a sparsity-enforcing algorithm that consists in iteratively deleting the least statistically significant terms until a threshold significance is reached. This inference workflow, as schematized in Fig.\ref{fig:workflow}, differs from popular sparse learning algorithms that include a penalization based on the values of the coefficients~\cite{brunton_discovering_2016,boninsegna_sparse_2018,callaham_nonlinear_2021}, which would not be appropriate here due to the fact that coefficients have distinct physical dimensions. 

More specifically, this workflow consists of three iterative steps: \textit{infer}, \textit{bootstrap}, and \textit{update}. The first step \emph{infer} uses the SFI algorithm to learn coefficients using the current set of basis functions.
In the second step \emph{bootstrap}, we assess the statistical significance of each inferred coefficient using the bootstrap method, running the inference again on sets of trajectories sampled with replacement and using the standard deviation of the coefficients as a confidence interval~\cite{tibshirani_introduction_1994}. The significance of each basis function for our model is quantified by their signal-to-noise ratio -- \emph{i.e.} the ratio between the absolute value of the mean of the coefficient and its standard deviation.
If one of these ratios is below a chosen significance threshold of $3$ (corresponding to a $3\sigma$ rule), we move to the third step \emph{update}, where we simplify the model by removing the least significant function from the basis, and iterate the process. The outcome of this process is a final, minimal model where all terms are statistically significant.

\begin{figure}[tb] \centering
\includegraphics[width=0.55\textwidth]{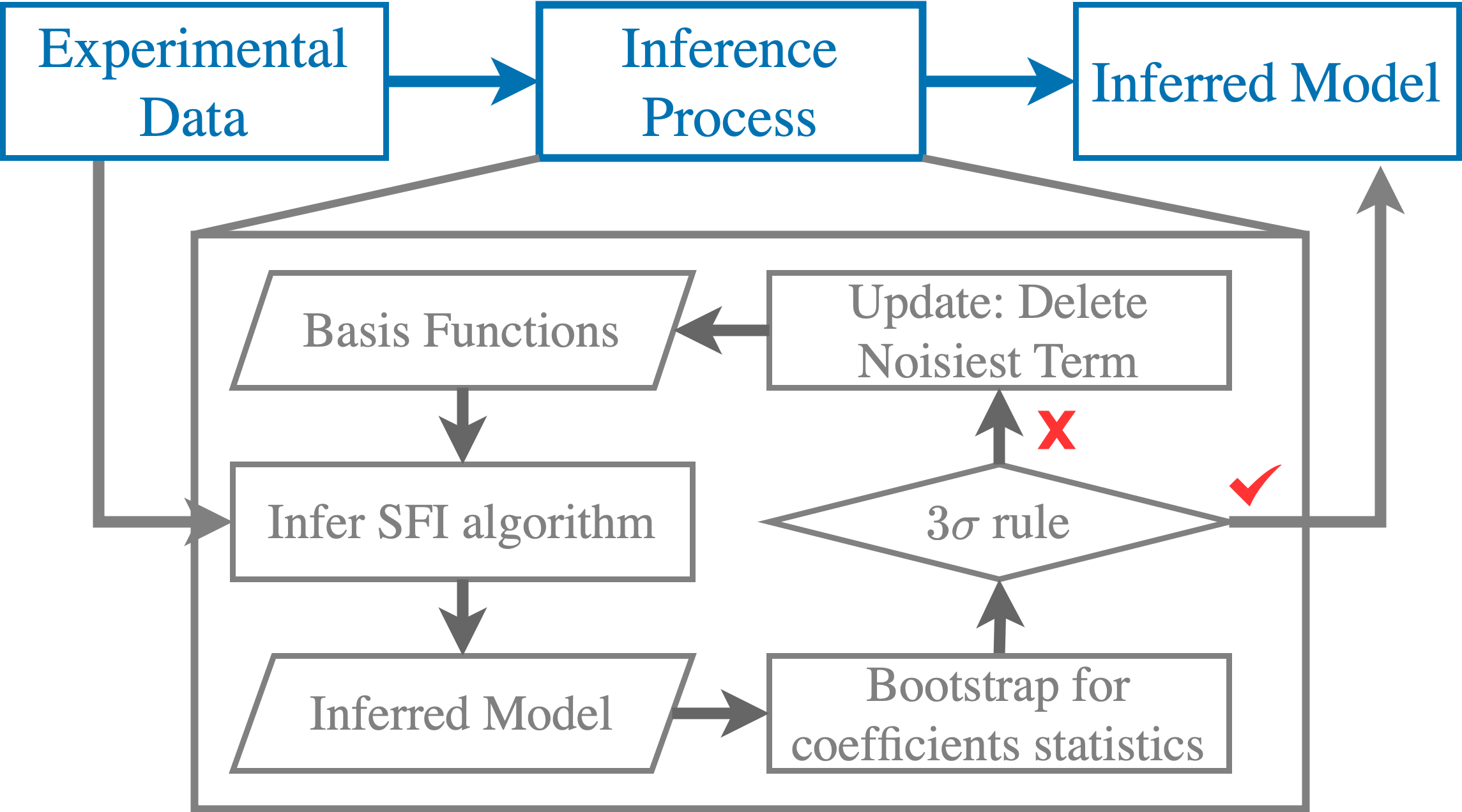}
\caption{\textbf{Schematic of the inference workflow}. To start, write down an initial model -- an overdamped Langevin equation of the problem at hand. Propose the basis functions that form the drift part of the equation and fix any known coefficients.  \textbf{Step 1. Infer}, input the experimental data and the candidate model into the SFI algorithm to obtain the most probable value of the unknown coefficients. \textbf{Step 2. Evaluate}, using bootstrap to obtain the mean and the standard deviation for each coefficient. We evaluate the significance of each coefficient against the $3\sigma$ rule. If one or more coefficients fails this test, then \textbf{Step 3. Update}, update the model by removing the noisiest term. Repeat this process until a final model is reached, where any further elimination would deteriorate the model.}
\label{fig:workflow}
\end{figure}

\paragraph{Resulting model for nuclear translocation dynamics}

Applying this inference workflow to the whole nuclear translocation data set, we obtain the following model: 
\begin{equation}
    \begin{aligned}
        \dot{X}_t = & C_X + \alpha_{X} P_t + \beta_{X}(R_t -R_t^{-1}) + \dfrac{a_{X}}{w^2(X_t, r)} +\\ & b_{X}\dfrac{n_{y}(X_t, r)}{w^{2}(X_t, r)} + \sqrt{2D_{X}}\cdot\eta_{X}(t),
    \end{aligned}
\label{eqn:final-1}
\end{equation}
\begin{equation}
    \begin{aligned}
        \dot{P}_t = & \alpha_{P} P_t + \beta_{P} (R_t-R^{-1}_t)  + a_{P} \dfrac{1}{w(X_t, r)} + b_{P} \dfrac{1}{w^2(X_t, r)} \\
        &+ c_{P} \dfrac{n_{X}(X_t, r)}{w^2(X_t, r)} + \sqrt{2D_{P}}\cdot\eta_{P}(t),
    \end{aligned}
\label{eqn:final-2}
\end{equation}
\begin{equation}
    \begin{aligned}
        \dot{R}_t = & C_R + \alpha_{R} P_t + \beta_{R} (R_t - R_t^{-1}) +  a_{R}\dfrac{1}{w(X_t, r)} \\
        &+ \sqrt{2D_{R}}\cdot\eta_{R}(t),
    \end{aligned}
\label{eqn:final-3}
\end{equation}
with a total of 14 drift terms. The values and standard deviations of the corresponding coefficients, as well as the inferred diffusion constants, are shown in Table \ref{table: coeffs}.

\begin{table}[b]
\centering
\begin{tabular}{|c||c|c|c|}\hline
 Corr. term&Coeffs.& Value& Unit\\\hline
 $\cdot$&$C_{X}$ & \num{2.6\pm0.5e-1} &\unit{\micro\meter\min^{-1}}
\\
 $P$&$\alpha_{X}$ & \num{-1.7\pm0.2e-1} &\unit{\min^{-1}}
\\
 $R-1/R$&$\beta_{X}$ &\num{6.7\pm0.7e-2} & \unit{\micro\meter\min^{-1}}
\\
 $1/w^2$&$a_{X}$ & \num{-4.7\pm1.1e1} &\unit{\micro\meter^{3}\min^{-1}}
\\
 $n_y/w^2$&$b_{X}$ & \num{4.8\pm1.1e1} & \unit{\micro\meter^{3}\min^{-1}}
\\
 $\eta_X$& $D_{X}$ & \num{5.8\pm0.8e-2} &\unit{\micro\meter^{2}\min{-1}} \\
\hline%------------------------------------------------------------
 $P$&$\alpha_{P}$ & \num{-3.2\pm0.6e-2} &\unit{\min^{-1}}
\\
 $R-1/R$&$\beta_{P}$ & \num{3.0\pm0.7e-3} &\unit{\micro\meter\min^{-1}}
\\
$1/w$ &$a_{P}$ & \num{7.8\pm1.3e-2} &\unit{\micro\meter^{2}\min^{-1}}
\\
$1/w^2$ &$b_{P}$ & \num{-1.0\pm0.2} &\unit{\micro\meter^{3}\min^{-1}}
\\
$n_x/w^2$&$c_{P}$ &\num{-3.1\pm0.5} &\unit{\micro\meter^{3}\min^{-1}}\\
$\eta_P$& $D_{P}$ & \num{8.0\pm1.4e-4}  &\unit{\micro\meter^{2}\min{-1}} \\
 \hline%------------------------------------------------------------
$\cdot$ & $C_{R}$ & \num{-5.7\pm1.2e-3} &\unit{\min^{-1}} 
\\
 $P$&$\alpha_{R}$ & \num{2.2\pm0.3e-2} &\unit{\micro\meter^{-1}\min^{-1}} 
\\
 $R - 1/R$&$\beta_{R}$ & \num{-8.7\pm1.4e-3} & \unit{\min^{-1}} 
\\
 $1/w$&$a_{R}$ & \num{1.1\pm0.2e-1} & \unit{\micro\meter\min^{-1}}
\\
 $\eta_R$&$D_{R}$ & \num{8.6\pm1.8e-4} & \unit{\min^{-1}} \\
 \hline%------------------------------------------------------------
\end{tabular}
\caption{Inferred coefficients for the minimal model, with corresponding terms in Eqs.~\ref{eqn:final-1}-\ref{eqn:final-3}. The confidence intervals correspond to the standard deviation obtained through bootstrapping.} 
\label{table: coeffs}
\end{table}

 A representative selection of trajectories $X(t)$ from the experiment and simulation is given in Fig.\ref{fig:Figure4}a. Averaged trajectories $(P, X)$ and $(R, X)$ are given for experiments and simulations in Fig.\ref{fig:Figure4}b ($N=\num{1000}$ simulated trajectories for the averaged quantities). Position-binned curves $P(X)$ and $R(X)$ at different constraints can be differentiated in both simulated and experimental data: the deformation and protrusion increase significantly as the constraint becomes smaller. The starting and ending points of the experimental and simulated curves $P(X)$ and $R(X)$ coincide. Additionally, a qualitative agreement can be seen in their dynamics.
 
% XXX  \textit{To complete / Improve the end}
 
Physically, the fact that $C_X>0$ indicates an average propensity of cells to migrate towards the nutrient-rich region. Interestingly, we find that $\alpha_X<0$ and $\beta_X>0$: when entering the constricted region, the cell slows down as the nucleus first protrudes, then accelerates as it elongates. The fact that $\alpha_P<0$ (resp. $\beta_R < 0$) confirms that in the absence of external forces these quantities relax back to the equilibrium shape $P=0$ (resp. $R=R{min}$). Regarding the $x$-dependent external forces, $f_X$ exhibits a repulsive term $a_X/w^2$ that slows the cell near the entrance of the constriction, and an attractive term $b_X n_y/w^2$ that accelerates it near $x=0$, \emph{i.e.} once it is engaged in the constriction. The protrusion force $f_P$ exhibits a term $c_P n_x/w^2$ that is odd under reflection symmetry and drives the rapid change of sign of the protrusion $P$ as the nucleus crosses the tightest point of the constriction. Finally, the dynamics of $R$ is captured by a single, elongation-driving term $a_R/w$ with $a_R > 0$; the relaxation back to the equilibrium value at the constriction exit is accelerated by the coupling $\alpha_R P$ with negative $P$ values. All in all, this model thus recapitulates with a few terms the directed migration of the cells through the channel, and the way the nuclei stall when reaching the constriction entrance, then protrude, elongate, and finally pop through rapidly. In the final stage, the protrusion reverts and points backward, leading to a rapid relaxation of the aspect ratio and the exit from the constriction.

\begin{figure}[tb]
  \begin{minipage}[c]{0.5\textwidth}
    \includegraphics[width=\textwidth]{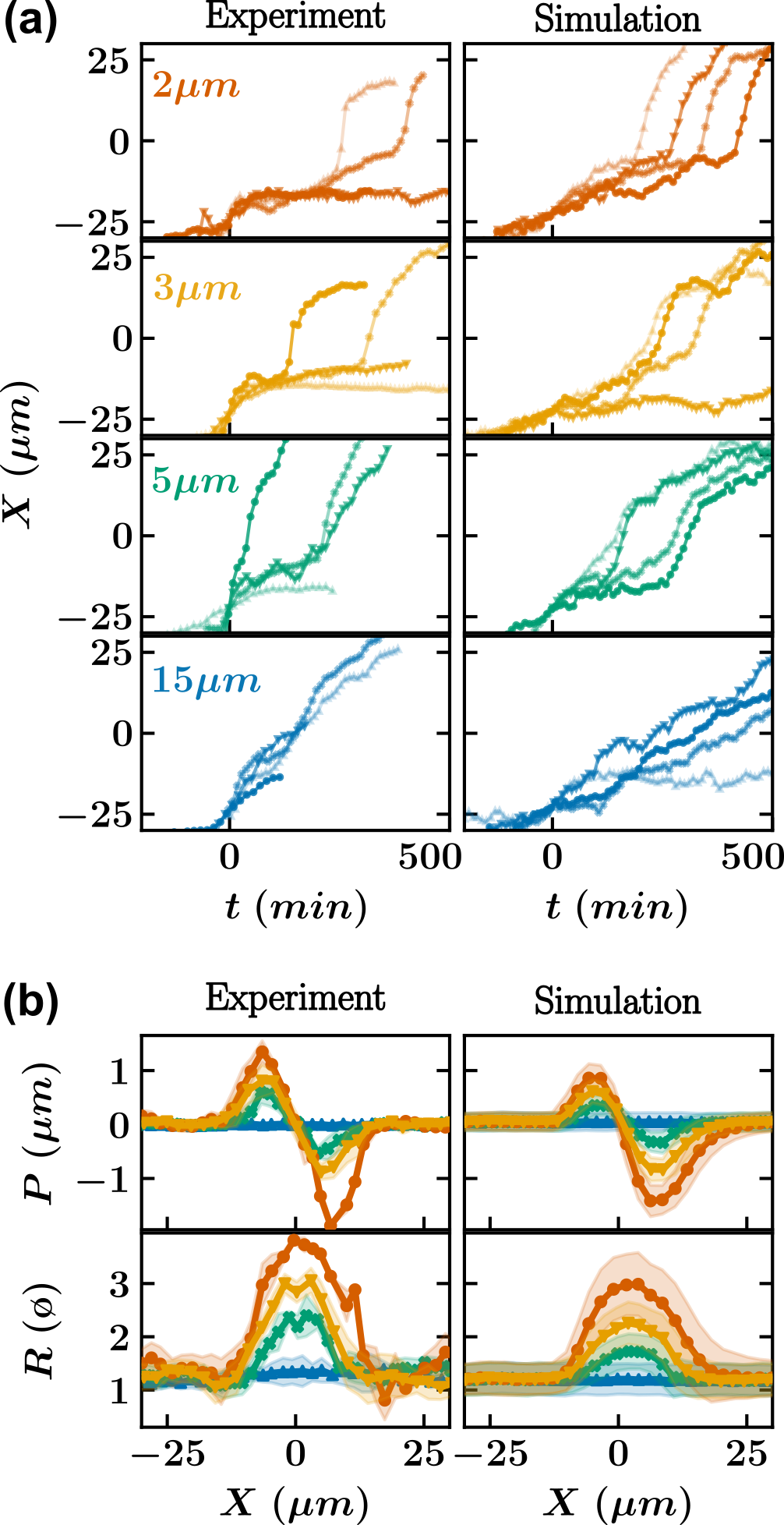}
  \end{minipage}\hfill
  \begin{minipage}[c]{0.4\textwidth}
\caption{\textbf{Simulation results of the reduced model.} \textbf{(a)} Comparison of four representative time series of cell nuclei position $X(t)$ from the experimental data and simulation. \textbf{(b)} Experiment and simulation comparison of the averaged trajectories of boundary polarity against nuclei position $P(X)$ and nuclei aspect ratio against nuclei position $R(X)$.\label{fig:Figure4}}
  \end{minipage}\hfill
\end{figure}

Our inference method also provides us with a physically interpretable estimate of the diffusion coefficients of the nucleus position $D_{X} \sim $\SI{5.8e-2}{\micro\meter^{2}\min^{-1}}. This value is several orders of magnitude above the equilibrium expectations from the Stokes-Einstein equation for a purely passive particle in the highly viscous cellular environment, $D_{\mathrm{Stokes-Einstein}} \sim $\SI{8e-7}{\micro\meter^{2}\min^{-1}}, reflecting the fact that cellular motion is activity-driven. Note that for simplicity of the analysis, we have assumed constant diffusion coefficients and Gaussian white noise. To investigate further these assumptions, larger amounts of data with a higher time resolution would be needed.

Finally, while the learned model provides good agreement in terms of capturing the dynamical geometric change of the nucleus during translocation, with a single parametric model encompassing the multiple constriction widths available, we note that it also presents some limitations. Indeed, this model is trained on a population of cells, and neglects any cell-to-cell variability due, \emph{e.g.}, to different sizes, genetic expression levels and age of the cells. This inherent variability manifests itself in a different way from the dynamical stochasticity captured here by the diffusion terms. Taking into account such cell-to-cell variability is a major challenge, as the amount of data available for each cell is small: data-efficient methods such as SFI~\cite{frishman_learning_2020} or Underdamped Langevin Inference~\cite{bruckner_inferring_2020} provide a promising avenue towards this, but single-event processes such as nuclear translocation studied here remain intractable with these approaches. A further difficulty comes from the limited frame rate, which leads to trajectories that appear to "tunnel through" right at the end of the passage through the constriction (as evidenced by long straight lines connecting data points in Fig~\ref{fig:Figure2NT}c) and lower the resolution of the translocation event. These challenges preclude the quantitative prediction of, \emph{e.g.}, mean translocation times, using the learned model.

\paragraph{Predictivity of the model}
As our learned model takes the constriction geometry as an explicit parameter, we can extrapolate it to predict nuclear translocation dynamics in other constriction sizes defined by the radius of the pillars ($r$). To assess the validity of this approach, we first test it on geometries for which experimental results are available: to this aim, we perform again the inference while masking one of our four constriction sizes (2, 3, 5 and \SI{15}{\micro\meter}). We then use the model inferred from the other three constriction widths to make predictions on the fourth geometry, which includes interpolations (when masking the 3 and 5\unit{\micro\meter} sets) and extrapolations (when masking 2 and 15\unit{\micro\meter} data). We then compare in Fig.\ref{fig:Figure5NT}a the prediction on the fourth, masked geometry with the actual experimental results. The good general agreement confirms the validity of this approach, and shows the usefulness of the learned model to predict behavior in geometries not used to train it. Note, however, that the simulation results exhibit smoother geometrical deformation than the experiments, in particular past the constriction. This discrepancy could be due to the small number of data points during the short time interval when the nuclei pass through the constraint.

Next, we extrapolate the model to other constriction sizes that were not studied experimentally: using the inferred model trained on the full data set, we simulate trajectories and compute the average geometric quantities $P(X)$ and $R(X)$ in constraints ranging from \SI{1.0}{\micro\meter} to \SI{15}{\micro\meter} (Fig.\ref{fig:Figure5NT}b). Each curve is obtained by averaging over \num{1000} simulated trajectories.  We observe a continuous increase in the maximum aspect ratio and geometric polarity as the constraint gets smaller. These predictions could be used for future experiment design, as a way to explore parameter space and focus experiments on the regions of interest.

\begin{figure*}[tb!]
\includegraphics[width=\textwidth]{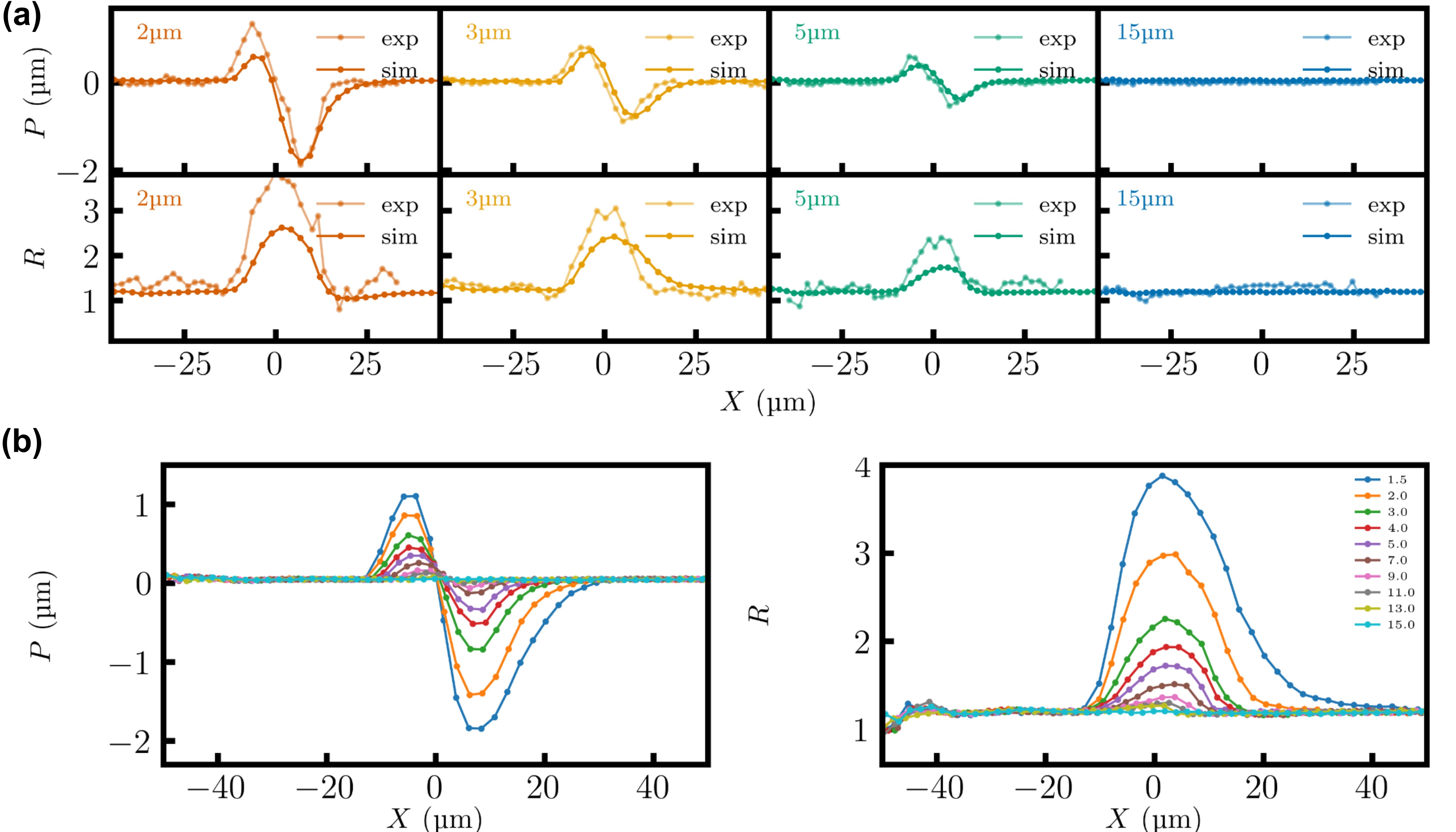}
\caption{\textbf{Predictivity of the inferred model to other values of pillar radius $r$.} \textbf{(a)} Extrapolation from partial data compared with the experimental data. Comparison of the boundary polarity $P$ against nuclei position $X$ (first row) and aspect ratio against $X$ (second row). \textbf{(e, f)} Extrapolation over a range of constraints with constriction size listed in the legend in \unit{\micro\meter}.}
\label{fig:Figure5NT}
\end{figure*}

\subsection{Discussion}

Here, we have studied the spontaneous migration of individual cells in a microfluidic device that exerts tight three-dimensional constraints mimicking physiological scenarii where cells are able to migrate in strongly confined environments.  Strikingly, cells can pass through constrictions much smaller than the rest diameter of their nucleus, leading to large  deformations of the nucleus during translocation ~\cite{davidson_nuclear_2014,davidson_design_2015}. This controlled experimental setup differs from previous studies of 2D confined cell migration without three-dimensional constraints~\cite{bruckner_stochastic_2019}, in which the nucleus is not significantly deformed. We segment and track cell nuclei to obtain trajectories that we use to quantify the dynamics of this nuclear translocation process. To this aim, we employ a data-driven approach that captures the stochastic nature of the motion and shape changes of the nucleus during cell motility in strongly constraining environments. In contrast with previous works where only the nucleus position was used~\cite{bruckner_learning_2021}, leading to effectively inertial dynamics, we include shape descriptors in our model that provide a proxy for the unobserved polarity of the cell. The outcome is an optimized set of overdamped equations that quantitatively captures the joint dynamics of nuclear position, protrusion and elongation as coupled stochastic differential equations. Importantly, the geometry of the environment is an explicit parameter of the resulting model, which allows for predictions and extrapolation to other constriction sizes. 

Our data-driven pipeline to infer these SDEs includes three main methodological developments compared to pre-existing methods.  First and most importantly, we symbolically include the geometry of the constriction in the inferred model, which allows us to train a single model on data for several constriction sizes. The resulting model can be successfully extrapolated to other constriction sizes and, potentially, other geometries. This contrasts with previous methods where the geometry is hard-coded into the inferred model, which both prevents extrapolation and precludes interpretability~\cite{bruckner_stochastic_2019,bruckner_learning_2021}. Second, we have introduced a sparsity-enforcing algorithm which simply consists in removing statistically insignificant terms from the set of basis functions, leveraging bootstrap estimates of significance to simplify the learned model. The desired level of significance (chosen here to be a standard $3\sigma$) is the only parameter of this technique, contrarily to popularly used sparse inference methods which include penalization terms with hyperparameters that require fine-tuning~\cite{brunton_discovering_2016,boninsegna_sparse_2018,callaham_nonlinear_2021}. Third, we propose a modification of the Stochastic Force Inference algorithm that consists in using trapezoidal integration for normalization matrices. While minor, this modification significantly improves performance of the method when the time step $\Delta t$ is large. The combination of these improvements allows us to robustly infer a minimal model for the geometrical kinematics of nuclear translocation.

Nuclear translocation involves a complex set of molecular mechanisms that enables cells to sense their mechanical environment and adapt their internal forces. Our study paves the way towards a data-driven understanding of this process,  where the nucleus is considered as an actor of the dynamical process, rather than a passive tracer lagging behind. In the future, this description could be enriched with other cell state descriptors, in particular with the spatial distribution of cytoskeletal and nuclear components, such as protein complexes involved in the mechanotransduction process. A challenge towards this, however, consists in selecting appropriate quantitative descriptors to include in the dynamical model: for instance, while nesprin -- the mechanical linking protein between cytoskeleton and nucleus -- is observed to accumulate at the front of the nucleus, this is not recapitulated by a polarity defined in terms of the first moment of the protein distribution.

\subsection{Appendices }\label{sec:NT-MM}

\paragraph{A1. Cell Culture}
Mouse Embryonic Fibroblasts (MEF) were CRISPR-modified to create a new cell line: MEFs SYNE2-GFP LMNA-mCh as described and validated in \cite{davidson_nesprin-2_2020}. Cells are cultured at 37°C in a humidified incubator with 5\% CO\textsubscript{2}, in DMEM (Dulbecco's Modified Eagle Medium - Gibco) supplemented with 10\% (v/v) Fetal Bovine Serum (FBS – Gibco). 

\paragraph{A2. Migration Devices}
The epoxy mold (R123/R614 - Soloplast) we used was replicated from a polydimethylsiloxane (PDMS) imprinted piece coming from the lab of Jan Lammerding (Cornell University, USA). A mix of PDMS (using a 10:1 ratio polymer:crosslinker) is vacuumed for 20 minutes to avoid bubbles, then poured into the epoxy mold and let to cure for 4 hours in a 60°C oven. Imprinted PDMS pieces are cut using a scalpel and biopsy punches (2mm and 5mm in diameter). Glass coverslips are soaked overnight in a 0.2M solution of HCl and rinsed with H\textsubscript{2}O and ethanol, dried with Kim wipes. To form a migration device, an imprinted PDMS piece and a treated glass coverlip are placed in a plasma cleaner for 1 minute and gently sticked together. This process creates covalent bonds between the PDMS and the glass \cite{borok_pdms_2021}. Devices are then directly put on a 100°C hot plate for 5 minutes to help the sticking process.

\paragraph{A3. Cell migration experiment}
Microfluidic devices are sterilized and rinsed under a microbiological safety post: first once with ethanol ($\sim$250\unit{\micro\liter}) then twice with Phosphate-buffered saline (PBS - Gibco) and twice with DMEM (Gibco) supplemented with 10\% (v/v) FBS. Cells are suspended at a concentration of 10 millions per mL in DMEM (Gibco) supplemented with 10\% (v/v) FBS. They are seeded in the device by adding 5\unit{\micro\liter} of the suspended solution in one of the two small ports of the device. After 6 hours, enough cells are in the constricted region of the device and acquisition can start. For that, cell medium is changed to DMEM without phenol red and with HEPES (15 mM) (Gibco), supplemented with 10\% FBS (Gibco), 100 units/mL penicillin, and 100 µg/mL streptomycin (Life Technologies).

\paragraph{A4. Image Acquisition}
Timelapse acquisitions are performed on an epifluorescence microscope (Nikon Ti-E) equipped with a sCMOS camera (2048 ORCA Flash 4.0 V2, Hamamatsu or Prime BSI, Teledyne), a perfect focus system, a 60x oil objective (Nikon), and a temperature and gas control chamber (set on 37°C, air at 5\% CO\textsubscript{2}). Images are taken every 10 minutes.

\paragraph{A5. Image Analysis}
Movies are analyzed using Image J/Fiji and Python. The projected nucleus surface is detected by using the "analyze particles" function on a threshold (median filter to 5.0 radius, normalized by 0.4\% and autolocal threshold "Bernsen" 5) applied on the mCherry image (corresponding to a lamin A/C signal). The nucleus contour is defined by a band of 1\unit{\micro\meter} width created from the detected nucleus projected surface ("reduce" and "make band" functions).

\paragraph{A6. Definition of the spatial origin}
The origin of the $x$ axis is set at the center of the constriction pillar (2, 3, \SI{5}{\micro\meter}) or half pillar (\SI{15}{\micro\meter}). The origin of the $y$ axis is set at the top center of the bottom constriction pillar (2, 3, \SI{5}{\micro\meter}) or the top center of the fitted circle to the bottom half pillar (\SI{15}{\micro\meter}).

\paragraph{A7. Definition of $X, X_c, P, R$}
The position of a nucleus is defined by its surface barycenter ($X, Y$),  specifically $X=\sum_{i=0}^{n}x_{i}/n$ and $Y=\sum_{i}y_i/n$ with $(x_{i},y_i)$ the coordinates of each pixel $i$ of the nucleus surface and $n$ the number of pixels in the nucleus surface. 
The $x$ coordinate of the center of the nucleus contour $X_c$, is defined as $X_c = \sum_{i=0}^{n_c}x_{c,i}/n_c$, with $x_{c,i}$ the $x$ coordinates of each pixel of the nucleus contour and $n_c$ the number of pixels in the nucleus contour.
The nucleus protrusion vector is defined as $P = X_{c} - X$.
The aspect ratio of the nucleus is defined by $R=R_{x}/R_{y}$ where $R_{x} = \sqrt{\sum_{i=0}^{n} (x_{i} - X)^{2}/n}$ and $R_{y} = \sqrt{\sum_{i=0}^{n} (y_{i} - Y)^{2}/n}$.

\paragraph{A8. Complementary expressions for basis functions} 
The direct and indirect effects of the environment, $f_{ext}(X, r), f_{P}(X, r)$ and $f_{R}(X, r)$ are approximated by combining the width function of the channel $w(X,r)$ with the normal vector calculated from the shape of the pillar which is a circle of radius r. The normal vector ${\hat{n}}(X,r) = (n_{x}(X,r), n_{y}(X,r))$ of a circle of radius r is given by
\begin{align}
n_{x}(x,r) &= \dfrac{x}{r}, \qquad -x^{*} < x < x^{*}, \\
n_{x}(x,r) &= 0, \qquad \mathrm{otherwise} \\
n_{y}(x,r) &= \dfrac{\sqrt{r^{2} - x^{2}}}{r}, \qquad -x^{*} < x < x^{*}, \\
n_{y}(x,r) &= \dfrac{\sqrt{r^{2} - x^{*2}}}{r} \qquad \mathrm{otherwise}.
\end{align}
where $x^*=\sqrt{r^2-r_s^2}$ with $r_s$ the small pillar radius. These quantities are schematized in Fig.\ref{fig:enter-label}. 
The channel width $w(x)$ is given by
\begin{align}
w(x,r) = H+2r_{\mathrm{s}} - 2 \sqrt{r^{2} - x^{2}} &\qquad -x^{*} < x < x^{*}, \\
w(x,r) = H &\qquad \mathrm{otherwise},  
\end{align}
where $H=15$ is the channel height (note that we neglect the texture of the small pillars here, as they do not constrict the nucleus). 
\begin{figure}[h!]
    \centering
    \includegraphics[width=0.45\textwidth]{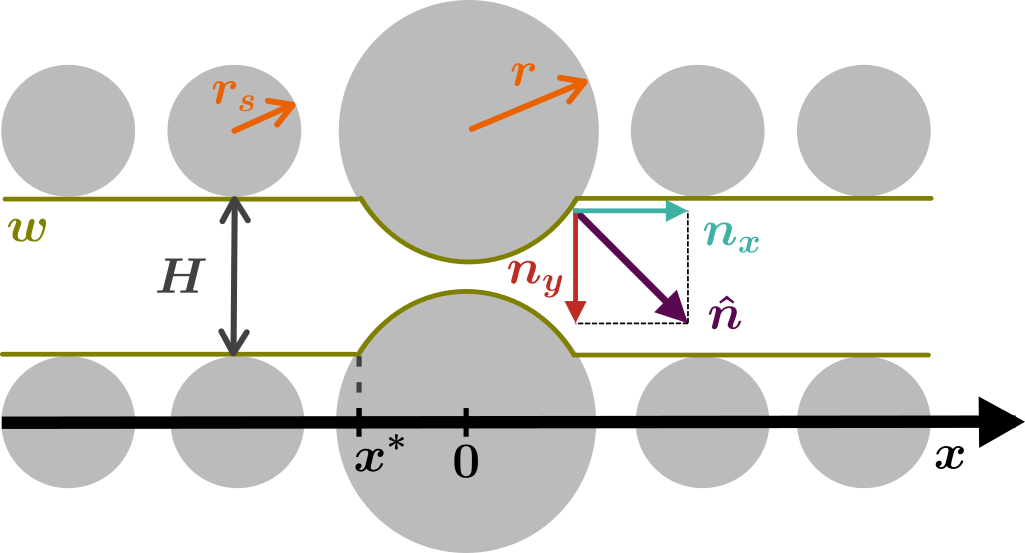}
    \caption{\textbf{Schematics of the geometric quantities describing the pillar shape used to construct force estimators.} Note that, for one series the centers of the pillars are aligned on $x$.}
    \label{fig:enter-label}
\end{figure}

\paragraph{A9. Full expression of the initial model}
The full model consisting of all basis functions, constructed by systematic expansion of the model over physically relevant variables, consists in linear combinations of the following basis functions:
\begin{eqnarray}
      &  \{ 1, P, (R-R^{-1}), \\
       & 1/w(X,r),n_x(X,r)/w(X,r),n_y(X,r)/w(X,r),\\ &1/w^2(X,r),n_x(X,r)/w^2(X,r),n_y(X,r)/w^2(X,r), \\&1/w^3(X,r),n_x(X,r)/w^3(X,r),n_y(X,r)/w^3(X,r)\}
\end{eqnarray}

and has 39 parameters ($3\times 12$ for the drift, $3$ for the diffusion). The model inference and reduction framework described in Fig~\ref{fig:workflow} retains only 14 significant terms.

\paragraph{A10. Improvement of the SFI algorithm for large time intervals $\Delta t$}
To learn the dynamics of our system, characterized in this paragraph by the vector $\mathbf{x}_t\equiv (X_t,P_t,R_t)$, it is essential to estimate its discrete time derivative using $\Delta \mathbf{x}_{t} = \mathbf{x}_{t + \Delta t}- \mathbf{x}_{t}$. A challenge on applying SFI on cellular dynamics data is that the time interval $\Delta t$ between frames is large (10 minutes) and of the same order of magnitude as the typical translocation time; it cannot be easily reduced due to phototoxicity: with the previously introduced algorithm~\cite{frishman_learning_2020}, this incurs $O(\Delta t)$ biases on the estimators. To adapt the method to this challenge, we propose a modification, which results in much smaller $O(\Delta t^2)$ biases.

Specifically, we focus in the derivation on approximate time differences $\Delta \mathbf{x}_t$ rather than the infinitesimal time difference $d \mathbf{x}_t$. Writing the dynamics in a generic form $\frac{\mathrm{d}\mathbf{x}}{\mathrm{d}t} = \mathbf{f}(\mathbf{x}_t) + \sqrt{2\mathbf{D}}\eta_t$, we have for discrete time increments:
\begin{align} %XXX CHECK
    \frac{\Delta \mathbf{x}_t}{\Delta t} &= \frac{1}{\Delta t} \int_t^{t+\Delta t} \frac{d\mathbf{x}_{t'}}{dt} dt' \\
    &=  \frac{1}{\Delta t} \int_t^{t+\Delta t}  \left[ \mathbf{f}(\mathbf{x}_{t'}) + \sqrt{2\mathbf{D}}\eta_{t'} \right] \mathrm{d}t'
\end{align}
SFI consists in approximating the unknown deterministic drift field $\mathbf{f}(\mathbf{x})$ by a linear combination of basis functions $\mathbf{f}(\mathbf{x}) = \sum_{\alpha} F_{\alpha} \mathbf{b}_{\alpha}(\mathbf{x})$ where $F_{\alpha}$ are the coefficients to infer and $\mathbf{b}_{\alpha}(\mathbf{x})$ are the basis functions. Thus, we can project the above equation on one of the basis functions $\mathbf{b}_{\gamma}(\mathbf{x}_t)$ and derive its average in the Itô convention:
\begin{equation}
    \left \langle \frac{\Delta  \mathbf{x}_t}{\Delta t} \mathbf{b}_{\gamma}(\mathbf{x}_t) \right \rangle =  \sum_{\alpha}  F_{\alpha} \left \langle \mathbf{b}_{\gamma}(\mathbf{x}_t) \frac{1}{\Delta t} \int_{t}^{t+\Delta t}\!\! \mathbf{b}_{\alpha}(\mathbf{x}_{t'})  dt' \right \rangle
    \label{eq:projected_dynamic}
\end{equation}
where $ \langle \cdot\rangle$ represents the expectation over many realisations of the noise $\eta_t$, conditioned on the initial value $\mathbf{x}_t$. Since we only measure $\mathbf{x}$ at discrete times $t, t+\Delta t, \dots$, we need to approximate $\int_{t}^{t+\Delta t} b_{\alpha}(\mathbf{x}_{t'})  dt' \approx \frac{\Delta t}{2}(b_{\alpha}(\mathbf{x}_{t}) + b_{\alpha}(\mathbf{x}_{t + \Delta t}))$. Importantly, this \emph{trapezoidal integration rule} is a more accurate approximation than the Riemann sum approximation $b_{\alpha}(\mathbf{x}_{t})\Delta t$ used in Ref.~\cite{frishman_learning_2020}. We note that the trapezoidal approximation concerns only the right-hand side of Eq.~\ref{eq:projected_dynamic}, which is a regular integral, and not the left-hand side which is a stochastic integral and remains in the It\^o convention.

By now averaging Eq.\ref{eq:projected_dynamic} over all data points $\{\mathbf{x}_{t_i}\}_{i=1,N}$, represented by $\langle \cdot\rangle_t$, we obtain the \emph{trapezoidal} approximation for the normalization matrix $\left\langle \left(\frac{1}{\Delta t} \int_{t}^{t+\Delta t} \mathbf{b}_{\alpha}(\mathbf{x}_{t'})  dt'\right)\mathbf{b}_{\gamma}(\mathbf{x}_{t})\right\rangle$:
\begin{equation}
    \mathbf{B}_{\alpha\gamma} = \left\langle\frac{1}{2} \left(\mathbf{b}_{\alpha}(\mathbf{x}_{t}) + \mathbf{b}_{\alpha}(\mathbf{x}_{t + \Delta t})\right)\mathbf{b}_{\gamma}(\mathbf{x}_{t}) \right\rangle_t
\end{equation}
which we use in the corrected It\^o estimator of $F_\alpha$ for large $\Delta t$:
\begin{equation}
    \hat{F}_{\alpha} = \sum_{\gamma} {\mathbf{B}^{-1}}_{\alpha\gamma}
    \left\langle \frac{\Delta  \mathbf{x}_{t}}{\Delta t} \mathbf{b}_{\gamma}(\mathbf{x}_{t})\right\rangle_t.
\end{equation}
Finally, similarly to Ref.~\cite{frishman_learning_2020}, we modify the integration convention of the stochastic integral towards the Stratonovich convention, in order to remove biases due to measurement noise, yielding the estimator used throughout this article:
\begin{equation}
    \hat{F}_{\alpha} =\sum_{\gamma} {\mathbf{B}^{-1}}_{\alpha\gamma}
    \left\langle \frac{\Delta  \mathbf{x}_{t_i}}{\Delta t} \left[ \frac{\mathbf{b}_{\gamma}(\mathbf{x}_{t_i})+\mathbf{b}_{\gamma}(\mathbf{x}_{t_{i+1}})}{2}\right] - \mathbf{d}_{t_i} \nabla \mathbf{b}_\gamma(\mathbf{x}_{t_i}) \right\rangle_t
\end{equation}
where we use the instantaneous noise-corrected diffusion estimator~\cite{vestergaard_optimal_2014} 
\begin{equation}
\mathbf{d}_{t} = \left[ (\Delta \mathbf{x}_t + \Delta \mathbf{x}_{t-\Delta t})^2 + 2 \Delta \mathbf{x}_t \Delta \mathbf{x}_{t-\Delta t}\right]/4\Delta t \end{equation}
The use of the trapezoidal method for discrete differences thus results in a lower-order discretization bias compared to the original SFI method, and enables accurate inference with the data set considered in this article.

\paragraph{A11. Bootstrap methods for coefficient error estimation}
We estimate the mean and standard deviation of each coefficient and the diffusion constant using the bootstrap method. More specifically, we sample with replacements the set of small 5-consecutive-points trajectories to generate an ensemble of trajectories~\cite{tibshirani_introduction_1994}. For each sample, the drift coefficients are estimated with the above-mentioned modified SFI algorithm, while the diffusion coefficients are estimated using the method introduced by Vestergaard \emph{et al.}~\cite{vestergaard_optimal_2014,frishman_learning_2020}. 
We compute the average and standard deviation of coefficients over 20 bootstrapped data sets obtained from the initial set of trajectories, and use the resulting standard deviations as indicators of the confidence interval for our assessment of the statistical significance of these coefficients.

\chapter{Perspectives}
\label{chap:perspectives}

\emph{In this last Chapter, I first describe the scientific process
  that led to the work presented here, then discuss future prospects.}

\paragraph{2017-2020: Entropy production inference.}
The results presented in this Thesis are the fruits of a research
effort that was sparked during an Aspen workshop in 2017, while I was
an independent postdoctoral fellow at the Princeton Center for
Theoretical Science (PCTS). At that time, discussions with Ben
Machta~\cite{machta_dissipation_2015} and, in particular, Chase
Broedersz~\cite{battle_broken_2016}, got me strongly interested in
out-of-equilibrium stochastic processes in the context of biophysical
systems, but also made me realize that there was a lack of appropriate
tools to extract the information from available experimental data in
this field. A few months later, over a coffee break, I discussed this
subject with Anna Frishman and got her interested in the problem. This
was the start of an intense collaboration which had us spend hundreds
of hours together in front of the large blackboards of Jadwin Hall
and, over the course of the following six months, resulted in a first
form of the methods and results presented in
chapters~\ref{chap:capacity} and \ref{chap:SFI}. Our initial question
was: based only on experimental trajectories of a stochastic system,
can we tell if, and how much, it is out-of-equilibrium?

This was thus thought, at first, as a way to measure entropy production
from trajectories -- the realization that you could infer drifts the
same way was originally a serendipitous epiphany after I took the
wrong stochastic convention in implementing Anna's formula. At that
time, many articles started appearing on arXiv about ways to measure
entropy production from
currents~\cite{ghanta_fluctuation_2017,seara_entropy_2018,gonzalez_experimental_2019},
uncertainty
relations~\cite{barato_thermodynamic_2015,li_quantifying_2019}, and
other methods... A stream of articles that was continued in the
following years. Much less was said at that time, however, about what could be
\emph{done} with such methods. With Anna Frishman and Chase Broedersz,
and thanks to the support of PCTS, we decided to organize a workshop
gathering the theorists who developed such methods, and the
experimentalists who produced data in systems we thought to be of
interest for nonequilibrium trajectory analysis. We provocatively
chose to title this
event\footnote{\url{https://pcts.princeton.edu/events/2018/why-measure-entropy-production}}
``Why measure entropy production?'', and asked the speakers to present
not only their results, but also their views on what could be actually
learned with entropy production inference, and for which
systems. There was also a long, final group discussion on the subject.

This event was, in my opinion, a very educational failure: the title
question was barely tackled by anyone among the speakers or in the
audience. The key problem that arose from these discussions is the
following: biological systems are strongly out-of-equilibrium, and so
in most cases the binary, qualitative question ``is the system at
equilibrium?'' is trivial and uninformative. A better question would
be the quantitative ``how strongly is it out-of-equilibrium?'' or,
alternatively, ``how much does it dissipate?''. However, tackling this
question from the entropy production inference angle is, in most
cases, fruitless: because these methods capture probability fluxes in
lower-dimensional representations of a high dimensional system, they
are by design unable to capture significantly more than $\sim 1k_B$
per data point. In available experimental data about, \emph{e.g.},
cytoskeletal fluctuations, this yields entropy production rates in the
range $0.1-100k_B/s$. This is an infinitesimal fraction of the total
dissipation in the cell, which is $>10^7 k_B/s$ for just a single
\emph{E. Coli} cell~\cite{belliveau_fundamental_2021} and mostly
occurs at the unobserved molecular scale, making it hopeless to
quantitatively capture entropy production from such trajectories. It
is not even clear that the measured entropy production correlates
positively with the total activity inside cells.  My feeling after
that workshop was that there was a disconnection between the intense
work of the statistical physics community on \emph{how} to measure
entropy production in biological systems, and the actual systems where
such methods could lead to interesting biophysical insights\footnote{I have since then become interested again in the field due to new developments in multi-point estimators for non-equilibrium trajectories~\cite{muenker_onsager_2022,knotz_mean_2023,ronceray_two_2023}, which have been promisingly applied to \emph{in vivo} trajectories.}. This led
me to re-center my goals towards what was originally a side result:
model inference. The entropy production estimators were included in
the paper with Anna Frishman (\Chap{chap:SFI}), and then used in
collaboration with Chase Broedersz in the track-free entropy
production project (\Chap{chap:BDB}), but I did not pursue further
this work. 

\paragraph{2019-2023: Stochastic Force Inference.}
In contrast, the problem of inferring stochastic differential
equations from data became more and more interesting to me as I worked
on it, and started to attract attention when we presented it. Indeed,
many experimentalists had -- and still have -- complex, stochastic
trajectories on their hands, and the prospect to have new methods to
extract useful mechanistic information from their data  appealed to them. In the
years following the arXiv release of SFI in 2018, then its subsequent
publication in 2020, I engaged in dozens of discussions and data
analysis projects. These included, to mention but a few, many flavors
of active and passive colloidal systems, actin filaments \emph{in
  vitro}, genomic loci in eukaryotic cells, motile bacteria, muscular
sarcomeres, migrating cells, fish schools, pedestrian traffic, shaken grains, and
more. While few of these discussions led to publication, each was
instructive and helped me understand better the actual needs of the
community.

At the same time, a small number of research groups started using SFI
on their own, and including them in their own pipelines. This included
work of colloid-surface interactions~\cite{lavaud_stochastic_2021},
dusty plasmas~\cite{yu_extracting_2022}, and directed network
reconstruction~\cite{cheng_efficient_2022}. Exchanging with these
users provided useful feedback that helped me target improvements for
the next versions of SFI and ULI.

In 2021, I started my group at Aix-Marseille Universit\'e as a Turing
Center for Living Systems (Centuri) group leader, first at the Center for
Theoretical Physics then, as I joined CNRS, at the Centre
Interdisciplinaire de Nanosciences de Marseille. At that time, it had
become clear to me that this research line had become both more
promising and more enjoyable than the other projects that I pursued
during my postdoc -- which included membraneless biomolecular
condensates and biological fiber networks mechanics -- and so I
decided to center my group around stochastic inference.

\paragraph{Since 2021: The inference group.}
Thanks to funding from my Centuri starting package, then more recently
from the European Research Council (ERC), I was able to form a
research group working on the development and applications of
stochastic inference methods. I now briefly present the projects of
each of the group members. The first four projects are funded by
Centuri, while the last is funded by my ERC starting grant.

\emph{Yirui Zhang (postdoc 2021-2023)} was in charge of the theory
part in the collaboration on nuclear translocation presented in
\Sec{sec:translocation}. She performed the data analysis, designed and
implemented the custom model selection method, and contributed to
writing the paper.

\emph{Andonis Gerardos (PhD student 2021-today)} started by working on
field-theoretical inference, extending the work on SDEs to stochastic
\emph{partial} differential equations. He became interested in model
selection, and has been proposing a new way principled way to enforce
sparsity constraints in stochastic inference. Finally, he proposed
multiple improvements to SFI to make it more robust, including the
trapezoidal method introduced in \Sec{sec:translocation}.

\emph{Arthur Co\"et (PhD student 2022-today)} is a marine biologist
that I co-supervise with Mar Benavides. He works on the dynamics of
bacteria colonizing ``marine snow'' particles -- aggregates of organic
matter that form near the ocean surface and slowly sediment. The part
of his project I am involved in consists in tracking such bacteria in
\emph{in vitro} reconstituted systems, then analyzing these tracks to
understand the range and biological mechanisms by which this
colonization occurs.

\emph{Jo\~ao Valeriano (PhD student 2023-today)} works on extending
stochastic inference techniques to partially observed
systems. Reconstructing the dynamics of the hidden variables, and
inferring their coupling with observed quantities, is a major
challenge that he is tackling using innovative filtering techniques.

\emph{Florian Goirand (postdoc 2024-today),} finally, works on the
extension of stochastic inference techniques towards a better
representation of the models, and will apply them to cell migration
problems.

\paragraph{Since 2023: SuperStoc.}
I now benefit from a European Research Council (ERC) Starting Grant,
which will fund and shape my group's research in the next five
years. This project, titled ``Superstoc -- Super-resolved stochastic
inference: learning the dynamics of soft biological matter'', aims at
expanding, improving and applying my methods. It is organized as the
development of a microscope -- a theoretical one, aiming at
\emph{seeing more} in experimental trajectories.  Real biological
trajectories are not ideal: they are \textbf{short} and sparse due to
experimental constraints such as phototoxicity, they are
\textbf{noisy} due to microscopy and tracking imperfections, and they
are \textbf{partial} as you never observe all degrees of freedom of
the system.  To address these challenges, this project is separated in
four parts:
\begin{itemize}
\item \textbf{\textsc{Resolve} -- } focuses on designing the
  high-robustness estimators that will serve as a \emph{lens} to
  resolve dynamical information in the \textbf{noisy} input
  trajectories. The goal is to capture precise models by filtering
  both measurement noise and stochastic diffusion to resolve subtle
  deterministic drifts.
\item \textbf{\textsc{Reveal} -- } aims at improving the
  \emph{penetration depth} of the method to reveal the hidden
  structures of the system from \textbf{partial} observations. These
  hidden variables couple with observed and thus encode memory effects
  in complex biological systems.
\item \textbf{\textsc{Represent} -- } targets the \emph{sensor} that
  integrates input signals and efficiently represents \textbf{short}
  trajectories into interpretable models. I will improve these learned
  functions to make the best use of the available information: simple
  physical models for "small data", complex neural networks for "big
  data".
\item \textbf{\textsc{Realize} -- } consists in implementing the
  method for practical use, then \emph{pointing this microscope} at
  biological data to discover new physics.
\end{itemize}
The outcome will be a robust and universal algorithmic toolbox to
achieve super-resolved stochastic inference from biological data. To
ensure that these methods are adapted to concrete needs, I will
integrate theoretical development and practical applications within my
group.

My aim it to make it easy and efficient to connect trajectories to
models, and thus to fill a gap in research on the dynamics of
biological matter. Such data-driven modeling will lead to quantitative
characterization and mechanistic insights on complex systems with
minimal amounts of data. While this research is primarily geared
towards biological matter, its potential impact extends to inert soft
matter, neuroscience, climate and population dynamics -- fields where
important insights lie hidden behind sparse and noisy data.  My
long-term goal is thus to provide a keystone method filling the
inference gap in the scientific pipeline that bridges between the
complex dynamics of experimental biology and the abstraction of
physical theories describing it.

\begin{figure*}[h]
  \centering
  \includegraphics[width=0.3\textwidth]{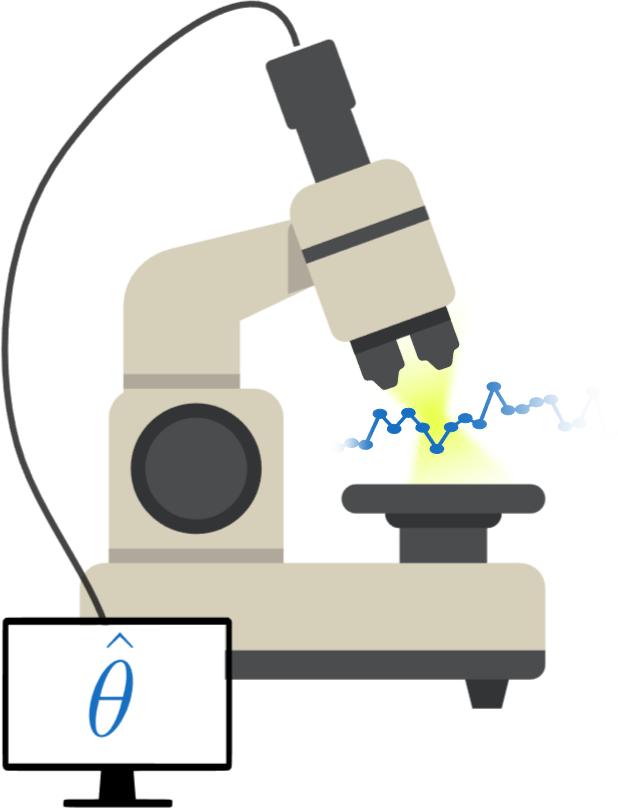}
\end{figure*}

%% ----------------------------------------------------------------
% Now begin the Appendices, including them as separate files

\addtocontents{toc}{\vspace{2em}} % Add a gap in the Contents, for aesthetics

\appendix % Cue to tell LaTeX that the following 'chapters' are Appendices

\addtocontents{toc}{\vspace{2em}}  % Add a gap in the Contents, for aesthetics
\backmatter

%\begin{multicols}{2}
%  \scriptsize
%  \setstretch{0.98}
\printbibliography
% \end{multicols}

 \end{document}